\documentclass[]{aa} 
\usepackage{graphicx}
\usepackage{soul}
\usepackage{txfonts,color}
\usepackage{amsmath}
\usepackage{nccmath}
\usepackage{array}
\usepackage{siunitx,booktabs}
\usepackage{lscape}
\usepackage{dcolumn}
\usepackage{mathtools}
\usepackage{multirow}
\usepackage{tabulary}

\newcommand{\jkfour}[1]{{\textcolor{black}{#1}}}
\newcommand{\mz}[1]{{\textcolor{black}{#1}}}
\newcommand{\comobj}{{\textcolor{black}{GMFs$+$nearby clouds}}}

\newcommand{\cellh}{1.8cm}

\begin{document} 

\title{Star-forming content of the giant molecular filaments in the Milky Way}

\author{M. Zhang
          \inst{1,2}
          \and
	J. Kainulainen
	\inst{3,1}
	\and
    M. Mattern
    \inst{4}
    \and
    M. Fang
    \inst{5}
    \and
    Th. Henning
    \inst{1}
    }

\institute{Max-Planck-Institut f\"{u}r Astronomie,
              K\"{o}nigstuhl 17, D-69117 Heidelberg, Germany\\
              \email{zhang@mpia.de}
         \and
             Purple Mountain Observatory, and Key Laboratory for Radio Astronomy, Chinese Academy of Sciences, Nanjing 210008, China
         \and
			Dept. of Space, Earth and Environment, Chalmers University of Technology, Onsala Space Observatory, 439 92 Onsala, Sweden
		\and
             Max-Planck-Institut f\"{u}r Radioastronomie (MPIfR), Auf dem H\"{u}gel 69, D-53121 Bonn, Germany
             \and
             Department of Astronomy, University of Arizona, 933 North Cherry Avenue, Tucson, AZ 85721, USA
             }

   \date{}

\abstract
   {Observations have discovered numerous giant molecular filaments (GMFs) in the Milky Way. Their role in the Galactic star formation and Galaxy-scale evolution of dense gas is unknown.} 
   {We investigate systematically the star-forming content of all currently known GMFs. This allows us to estimate the star formation rates (SFRs) of the GMFs and to establish relationships between the SFRs and the GMF properties.}
   {We identify and classify the young stellar object (YSO) population of each GMF using multi-wavelength photometry from near{-} to far{-}infrared. We estimate the total SFRs assuming a universal and fully sampled initial mass function and luminosity function.}
   {We uniformly estimate the physical properties of {57} GMFs. The GMFs show correlations between the $^{13}$CO line width, mass, and size, similar to Larson's relations. We identify 36\,{394} infrared excess sources {in 57 GMFs} and obtain SFRs for 46 GMFs. The median SFR surface density ($\Sigma_{\textrm{SFR}}$) and star formation efficiency (SFE) of GMFs are 0.62\,M$_{\sun}$\,Myr$^{-1}$\,pc$^{-2}$ and 1\%, similar to the nearby star-forming clouds. {The star formation rate per free-fall time of GMFs is {between} 0.002$-$0.05 with {the} median value of 0.02. }
We also find a strong {correlation} between SFR and dense gas mass {that is defined as gas mass above a visual extinction of 7 mag}, {which suggests that the SFRs of the GMFs {scale} similarly with dense gas as those of nearby molecular clouds.}  
{We also find a strong correlation between the mean SFR per unit length and dense gas mass per unit length. The origin of this scaling remains unknown, calling for further studies \jkfour{that can link the structure of GMFs to their SF activity and explore the differences between GMFs and other molecular clouds}.}} 
   {}

   \maketitle

\section{Introduction}

Galactic molecular clouds {are known to commonly} exhibit filamentary structures \citep[e.g.,][]{barnard27,ldn62,se79,bally87,abe94,perault96,cam99,hatchell05,mye09}. 
Recently, the \textit{Herschel} results {have highlighted the ubiquity of filaments in the interstellar medium (ISM) and their importance} in the star formation process \citep[for a recent review, see][]{and14-pp6}. {Indeed, studying the impact of filamentary morphology on star formation has become a central topic in current star formation/ISM studies, especially in nearby molecular clouds in which the link between the cloud structure and star formation can be well resolved} \citep[e.g.,][]
{and10,men10,schmalzl10,arz11,hill11,juvela12b,hacar13,pal13,alv14,classy-serpenss,classy-serpens,classy-overview,kony15,roc15,hacar16,kai16,lev16,planck3216,planck3316,stutz2016-slingshot,kai17}. 

{Recently, observational studies have discovered and identified a growing number of large-scale filaments with lengths up to $\sim$100 pc. These objects may be linked to the Galaxy-scale distribution of dense gas and trace features such as the central potential well of spiral arms and “spurs” sheared off from spiral arms} \citep[e.g.,][]{nessie10,kainulainen-snake,goodman14,sample-gmf,sample-wang15,sample-bones,sample-agmf,sample-li16,sample-wang16}. 
{This raises immediately several questions. How relevant are the giant filamentary structures for Galactic star formation? Do they trace Galaxy-scale star formation patterns or trends? Does star formation in them proceed in a similar manner as in other molecular clouds, or does their filamentary nature, or location in the Galaxy, cause a different star formation activity?}

{Giant filaments have only been identified in the Milky Way since the past few years, and therefore, we still lack a systematic picture of their basic properties and star formation. \jkfour{In general, there is no clear definition for what constitutes `a giant filament'; different studies in literature adopt different definitions. Broadly speaking, a common nominator for these definitions is that the cloud must show clearly elongated shape \emph{over some column density regime}. For example, }the giant filaments can be identified based on the mid-infrared dark features against the Galactic infrared background emission \citep{sample-gmf,sample-bones,sample-agmf} or on the far-infrared or millimeter continuum emission features \citep{sample-wang15,sample-li16,sample-wang16}. \jkfour{In the former case the selection leads to clouds whose high column density regions show strong elongation; in the latter case the elongation can be more prominent at lower column densities. The filament candidates identified from column density data are usually considered as continuous clouds only if they show coherence in position-position-velocity space. The} velocity coherence can be \jkfour{established} with various tracers such as CO \citep{sample-gmf,sample-wang15,sample-bones,sample-agmf}, CS and/or NH$_3$ \citep{sample-li16}, and HCO$^{+}$ and/or N$_2$H$^+$ \citep{sample-wang16}. \jkfour{As a result of this wide variety of definition techniques, the known giant filaments do not form a uniformly defined sample. Regardless of this ambiguity, it is interesting to consider the basic properties of the objects that fall into this group and compare them with other molecular clouds.}}

\jkfour{Based on studies exploiting} the above methods, the giant filaments have lengths from $\sim$10 to $\sim$500 pc \citep{li13} and masses from $\sim$10$^3$ to $\sim$10$^5$\,$M_{\sun}$. The location of the filaments with respect to the Galactic spiral arms is unclear; efforts have been made to differentiate arm- and inter-arm filaments \citep{sample-wang15,sample-bones,sample-li16,sample-wang16}, but the results are sensitive to the uncertainties of the spiral arm models \citep{vallee08,reid14}.
Some studies have tried to 
link the giant filaments to star formation \citep{henning10,kim15,gong16,xiong17}.
\mz{For example, \citet{kim15} identified $\sim$300 YSOs in G53.2 that is a filamentary molecular cloud with length of $\gtrsim$45 pc. Comparison of the YSO population with other nearby star-forming clusters and infrared dark clouds (IRDCs) indicated that G53.2 is similar to the nearby star-forming clusters in age, but at a later evolutionary stage than IRDCs. In contrast, \citet{samal15} investigated the star formation activity of a filamentary dark cloud that is part of a $\sim$20 pc filamentary cloud and found a high protostar fraction ($\sim$70\%), which suggests that the filament could still be at a very early evolutionary stage. These studies towards individual long filamentary clouds suggest that giant filaments could exhibit different star formation activities. To estimate the range of SFRs in the giant filaments and to understand the origin of the differing star formation activities, investigation of the star-forming content in a large number of giant filaments is necessary.}

In this paper, we present the first systematic investigation of the star-forming content of the Galactic giant molecular filaments (GMFs). 
{We gather together all currently known giant filaments from different studies for a homogeneous analysis. {\jkfour{As mentioned before,} `giant filament' is a general and not \jkfour{a unique} designation, \jkfour{even if the individual studies use well-defined identification criteria}. \jkfour{Several different names are} used for this type of large-scale filamentary structures, based on the different identification methods, e.g., the GMFs \citep{sample-gmf,sample-agmf}, the cold giant filaments \citep{sample-wang15}, and the `bones' or `skeleton' of the Galaxy \citep{goodman14,sample-bones}. {\jkfour{Together, these samples form the current census of large filamentary structures in the Milky Way, even if the census remains non-uniformly defined. For simplicity,} we refer to our objects as GMFs{, simply reflecting the fact that} we estimate their properties mainly based on $^{13}$CO molecular line data.}} We identify and classify the YSO population of} 
each GMF using archival multi-wavelength photometric catalogs from near{-} 
to far{-}infrared. {We further estimate the total star-forming content of the GMFs, 
based on the assumption of a universal and fully-sampled initial mass function (IMF). {This allows us to estimate the star formation rates (SFRs) of the GMFs and correlate the SFRs with the properties of the gas distribution in the GMFs. The resulting relationships between gas and star formation are then compared with literature values and star formation relations.} In Section~\ref{sect:data}, we summarize the data used in this paper. In Section~\ref{sect:method}, we describe the methods that are used to obtain the physical properties of the GMFs 
{and} to identify YSOs, and we estimate the SFR of each GMF. We investigate the spatial distribution of Class I sources and compare the SFRs of the GMFs with those of nearby star-forming clouds in Section~\ref{sect:results}. In Section~\ref{sect:discussion}, we 
{compare star formation within GMFs 
with the canonical} 
star formation relations. 
We summarize our main conclusions in Section~\ref{sect:conclusions}.

\section{Data and point source catalogs}\label{sect:data}


We use the archival multi-band photometric catalogs from near-infrared (NIR), mid-infrared (MIR), and far-infrared (FIR) surveys to identify YSO candidates in each Galactic GMF. We also obtain the physical parameters such as mass and column density of each GMF using the $^{13}$CO molecular line data. Table~\ref{tablesummary} summarizes these surveys and their purpose and the following sections introduce these data more closely.

\begin{table*}
\scriptsize
\caption{Summary of data used in this paper}
\label{tablesummary}
\centering
\setlength{\tabcolsep}{0.5em}
\begin{tabular}{l| l| l | l| p{5.3cm}| l}
\hline\hline
Band & Surveys & Spatial resolution & Sensitivity & Purpose & Ref.\\
\hline 
NIR & UKIDSS-GPS & $\sim$1\arcsec & $K$~$=$~18.05\,mag (5$\sigma$) & identifying YSOs & \citet{gps-ukidss}\\
NIR & VVV & $\sim$1\arcsec&  $K_s $~$\sim$~18\,mag & identifying YSOs & \citet{vvv}\\
NIR & 2MASS & $\sim$4\arcsec& $K_s$~$=$~14.3\,mag (10$\sigma$) & identifying YSOs & \citet{2mass}\\
MIR & GLIMPSE&$\sim$2\arcsec& [8.0]~$=$~13.0\,mag (3$\sigma$)& identifying YSOs & \citet{glimpse}\\
MIR & MIPSGAL&$\sim$6\arcsec~at 24\,$\mu$m& $\sim$2\,mJy at 24\,$\mu$m & identifying YSOs & \citet{mipsgal}\\
MIR & AllWISE& $\sim$6$-$12\arcsec& variable\tablefootmark{a}& identifying YSOs that are saturated on \textit{Spitzer} images & \citet{wise}\\
MIR & RMS  &$\sim$18.3\arcsec& $\sim$3 Jy at 21\,$\mu$m (50\% completeness)\tablefootmark{b}
& including massive YSOs & \citet{rms-survey}\\
FIR & Hi-Gal& $\sim$6$-$35\arcsec& $\sim$12.4\,mJy at 70\,$\mu$m (1$\sigma$)& identifying protostellar objects & \citet{higal-survey}\\
mm  & GRS &46\arcsec&  0.27\,K per 0.2\,km\,s$^{-1}$ ($^{13}$CO) & obtaining gas properties of GMFs & \citet{grs}\\
mm  & ThrUMMS& 72\arcsec& 0.7\,K per 0.34\,km\,s$^{-1}$ ($^{13}$CO) & obtaining gas properties of GMFs & \citet{thrumms}\\
\hline                                   
\end{tabular}
\tablefoot{
\tablefoottext{a}{The sensitivity variation can be found in the Explanatory Supplement to the AllWISE Data Release Products.}
\tablefoottext{b}{The actual sensitivity is a function of Galactic Longitude and Latitude.}
}
\end{table*}

\subsection{Near infrared data}

We use the archival catalog from the UKIRT infrared deep sky survey \citep[UKIDSS;][]{ukidss} {to obtain near-infrared data for the objects} in the northern Galactic plane. For the {objects} in the southern Galactic plane, there is the archival data from the VISTA variables in the V\'{i}a L\'{a}ctea survey \citep[VVV;][]{vvv}. However, the released catalog of VVV survey is based on aperture photometry, which is clearly unsatisfactory for the crowded fields such as the innermost Galactic center. Thus we downloaded and stacked the archival multi-epoch images of VVV survey and then performed PSF photometry. Finally we obtain our own near-infrared photometric catalog for the GMFs in the southern Galactic plane. Due to the saturation problem of UKIDSS and VVV surveys, we also use the 2MASS point source catalog \citep{2mass} for the bright sources.

\subsubsection{The UKIDSS/GPS catalog} 

The UKIDSS Galactic Plane Survey \citep[GPS;][]{gps-ukidss} 
covers the northern Galactic plane at Galactic latitudes -5\degr $< b <$ 5\degr~in the $J, H, K$ filters with the UKIRT Wide Field Camera \citep[WFCAM;][]{wfcam}. 
Compared with 2MASS survey, UKIDSS/GPS is deeper and has better spatial resolution. The median 5$\sigma$ depths are $J =$ 19.77, $H = $19.00, and $K = $18.05\,mag \citep{warren07}. Details about the photometric system, calibration and data processing can be found in \citet{hewett06}, \citet{hodgkin09}, 
\citet{irwin08}, and \citet{hambly08}.
We use the point source catalog from Date Release 10 Plus (DR10PLUS\footnote{\url{http://wsa.roe.ac.uk/dr10plus_release.html}}). Because the sources with $J < $13.25, $H < $12.75, or $K < 12.0$\,mag are affected by saturation \citep{gps-ukidss}, we replace these sources with the photometric results from 2MASS point source catalog \citep{2mass}. 


\subsubsection{PSF photometry for VVV survey}


The VVV survey is an ESO public survey, which uses VIRCAM \citep[VISTA InfraRed CAMera;][]{vircam1,vircam2} equipped on the VISTA telecope to observe $\sim$562 square degrees in the Galactic bulge (-10 $\leq$ $l$ $\leq$ 10, -10 $\leq$ $b$ $\leq$ 5) and part of the adjacent Galactic plane (-65 $\leq$ $l$ $\leq$ -10, -2 $\leq$ $b$ $\leq$ 2) in five bands, $Z, Y, J, H, K_s$, as well as time coverage spanning over five years \citep{vvv, vvvdr1}. 
The VVV data were processed using VISTA data flow system (VDFS) pipeline at the Cambridge Astronomical Survey Unit (CASU\footnote{\url{http://casu.ast.cam.ac.uk/vistasp/}}) \citep{lewis10,vvvdr1}. The process of VVV data reduction, calibration, and photometry can be found in \citet{vvvdr1}. Here we should note that the archival photometric catalog from VVV data release is based on aperture photometry 
which is apparently unsatisfactory for the crowded fields. 

We improved an automatic pipeline to perform the PSF photometry on the VVV survey imaging data. This pipeline is based on DAOPHOT algorithm \citep{daophot} and is written in IDL and Python. We also make this pipeline be able to run in multi-core mode, which can significantly decrease the calculating time of PSF fitting. Figure~\ref{flowchart-pipeline} shows the flow chart of the pipeline. 
In this paper, we use the data from VSA\footnote{\url{http://horus.roe.ac.uk/vsa/}} VVV data release 4 (VVVDR4). In order to detect more faint sources, we stacked all multi-epoch images for 
each filter of $J$, $H$, and $K_s$. 
We mosaiced multi-epoch images 
chip by chip after excluding the chip images obtained in bad weather condition and then split each stacked chip into the pieces of $\sim$1000 $\times$ 1000 square pixels. The PSF photometry is performed on each piece with PyRAF\footnote{\url{http://www.stsci.edu/institute/software_hardware/pyraf}} which is a command Python scripting language for running IRAF\footnote{\url{http://iraf.noao.edu/}} tasks. Source detection is conducted using DAOFIND task in IRAF with the threshold of $S/N > $~3. Then we use DAOPHOT PSF task to model the PSF function and ALLSTAR task to obtain the photometric results. The absolute photometric calibration was obtained by a comparison of the instrumental magnitudes of relatively isolated, unsaturated bright sources with the counterparts in the VVV archival CASU catalog. The 2MASS filters are also slightly different from the VISTA filters. To obtain the photometric results in 2MASS photometric system, we calculated the transformations from VISTA to 2MASS system tile by tile using the method suggested by \citet{soto13}. The saturated sources in our final catalog are also replaced by 2MASS magnitudes.
\begin{center}
\begin{figure}
\includegraphics[width=1.0\linewidth]{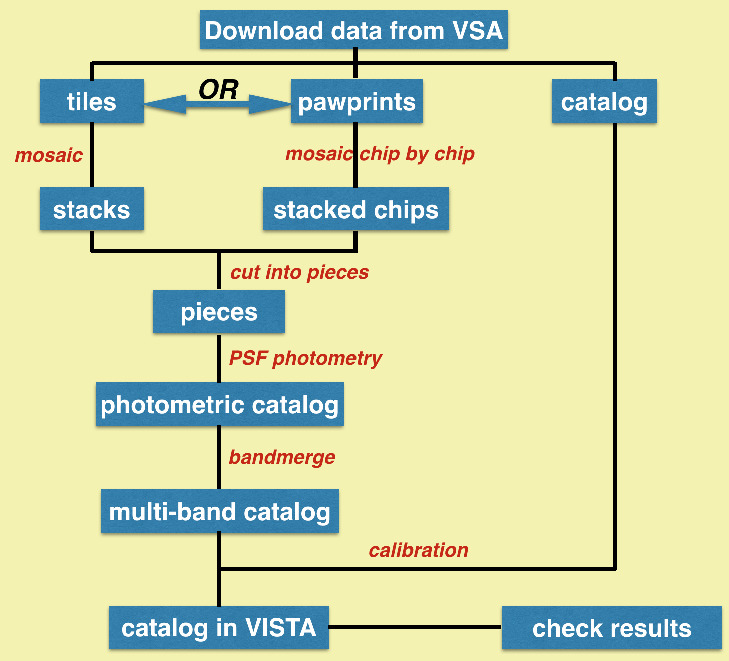}
\caption{The flow chart of our automatic PSF-fitting pipeline that is used to do the PSF photometry on the VVV survey imaging data.}
\label{flowchart-pipeline}
\end{figure}
\end{center}

Figure \ref{counts} shows the $K_s$ magnitude distribution of the sources detected by PSF photometry (black line) and sources in VVV archival CASU catalog (red line) in two tile regions, of which `b332' is a tile in Galactic bulge while `d095' is a tile in Galactic disk. Obviously, our PSF photometry can reach $\sim$1-2 mag deeper than aperture photometry in $K_s$ band.

\begin{figure*}
\centering
\includegraphics[width=1.0\linewidth]{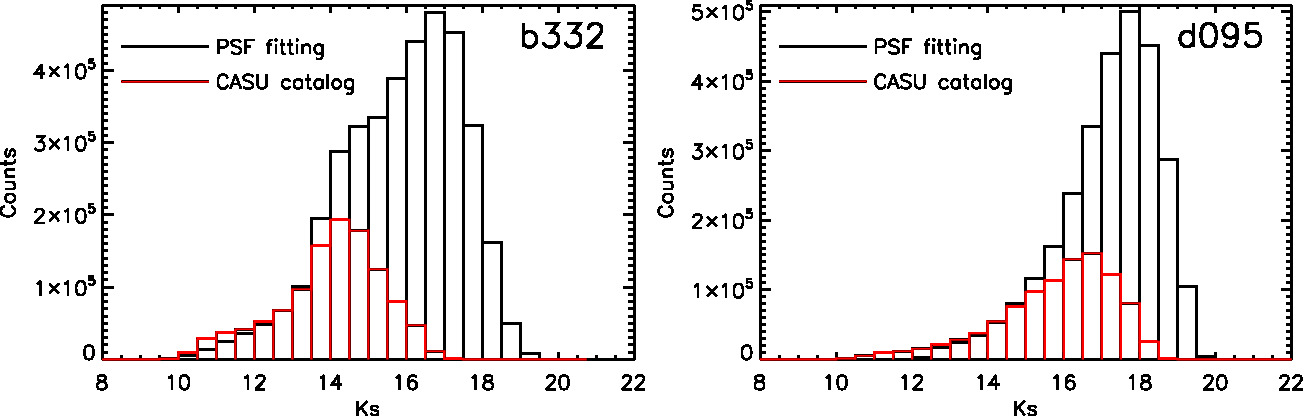}
\caption{The $K_s$ magnitude distributions of the sources detected by PSF photometry (black line) and sources in VVV CASU catalog (red line) for tile `b332' (\textit{left panel}) and tile `d095' (\textit{right panel}).}
\label{counts}
\end{figure*}

\subsection{Mid-infrared data}

We use the archival catalogs from the \textit{Spitzer} galactic plane surveys, GLIMPSE \citep{glimpse,glimpse09} and MIPSGAL \citep{mipsgal}, to trace the mid-infrared emission of the point sources. In this paper, we used the merged GLIMPSE catalog released by \citet{glimpse-catalog} and the high quality 24\,$\mu$m point source catalog released by \citet{mipsgal-catalog}. We also use the bright sources from the AllWISE \citep{wise} catalog as a supplement due to the saturation problem of \textit{Spitzer} surveys. However, some bright source such as the massive young stellar objects (MYSOs) are still saturated on the WISE images. To include the MYSOs in our final catalog, we use the massive protostar catalog from the Red MSX Source (RMS) survey \citep{rms-survey}.
\subsection{Far-infrared Herschel Hi-GAL catalog}
The \textit{Herschel} Infrared Galactic plane survey \citep[Hi-GAL;][]{higal-survey} is a \textit{Herschel} key project that initially aimed at observing the inner Galactic plane of $|l|$ $<$ 60\degr, $|b|$ $<$ 1\degr~and was subsequently extended to the entire Galactic plane \citep{higal-osurvey}. Hi-Gal uses the Photodetector Array Camera and Spectrometer \citep[PACS;][]{pacs} and the Spectral and Photometric Imaging REceiver \citep[SPIRE;][]{spire} in parallel mode (pMode) to sample the spectral energy distribution (SED) at 70, 160, 250, 350, and 500\,$\mu$m simultaneously with the spatial resolution of $\sim$6\arcsec, 12\arcsec, 18\arcsec, 24\arcsec, and 35\arcsec,~
respectively. 
The details about the Hi-GAL observing strategy, data reduction, calibration, source detection and photometry can be found in \citet{higal-survey}, \citet{higal-pipeline}, and \citet{higal-catalog}. In this paper, we use the compact source catalog from Hi-GAL DR1 which is the first public data release and limited to the inner Galactic plane \citep[-70\degr $\leq l \leq$ 68\degr, $|b|$ $\leq$ 1\degr;][]{higal-catalog}. 

\subsection{Multiband photometric catalog}
We use STILTS \citep{stilts} to merge the catalogs in different wavelength bands. \citet{higal-catalog} presented five single band catalogs at 70, 160, 250, 350, and 500\,$\mu$m. We firstly crossmatched these catalogs to obtain a merged Hi-GAL catalog. \citet{dunham08} suggested 70\,$\mu$m as a crucial wavelength for the embedded protostars and found a tight correlation between the protostellar luminosity and the flux at 70\,$\mu$m. Thus the detection at 70\,$\mu$m of the compact sources has been used to distinguish protostars from starless cores \citep{bontemps10,konyves10,kony15,gian12,higal-osurvey,ryql13,gacz13}. The objective of this paper is to identify young stellar objects (YSOs). Thus here we only care about the compact sources with 70\,$\mu$m detections in the Hi-GAL photometric catalogs. Therefore, we used the 70\,$\mu$m compact source catalog as the reference catalog. Any other Hi-GAL catalog is matched with the reference catalog with the tolerances of 12\arcsec, 18\arcsec, 24\arcsec, and 35\arcsec, corresponding to the spatial resolutions at 160, 250, 350, and 500\,$\mu$m, individually. The merged Hi-GAL catalog only includes the compact sources with 70\,$\mu$m detections. Secondly, we matched the MIPSGAL catalog with the GLIMPSE catalog using a tolerance of 2\arcsec~to obtain an merged \textit{Spitzer} catalog and then matched the merged Hi-GAL catalog with this merged \textit{Spitzer} catalog using a tolerance of 6\arcsec. To add the near-infrared catalog, we checked the counts of matched sources between the near-infrared catalog and the \textit{Spitzer}+\textit{Herschel} catalog using different tolerances and finally selected a tolerance of 0.6\arcsec~to match the near-infrared  catalog with the merged \textit{Spitzer}+\textit{Herschel} catalog. This multiband photometric catalog will be used for the subsequent YSO identification.

\subsection{Molecular line data}
We use 
$^{13}$CO (J = 1$-$0) data cubes from the GRS \citep{grs} and ThrUMMS \citep{thrumms} surveys to estimate the physical parameters of the GMFs such as the column density, mass, and velocity dispersion.

\subsubsection{GRS data}
The Boston University-FCRAO Galactic Ring Survey \citep[GRS;][]{grs} use the Five College Radio Astronomy
Observatory (FCRAO) 14 m telescope to survey the partial Galactic plane of 18\degr~$< l <$ 55\fdg7, $|b|$ $\lesssim$ 1\degr~ and 14\degr~$< l <$ 18\degr, $|b| \lesssim $0.5\degr~at the $^{13}$CO (J = 1$-$0) frequency ($\nu_0$ = 110.2 GHz) with a beam size of 46\arcsec. The velocity resolution of GRS is about 0.2\,km\,s$^{-1}$ and the RMS noise is about 0.27K per 0.2\,km\,s$^{-1}$. The main beam efficiency of the antenna is 0.48. The GRS data are available online\footnote{\url{http://www.bu.edu/galacticring/}}. We use the FITS data cubes of GRS data release.

\subsubsection{ThrUMMS data}
The Three-mm Ultimate Mopra Milky Way Survey \citep[ThrUMMS;][]{thrumms} use 22 m Mopra telescope to survey the fourth Galactic quadrant of 300\degr~$< l <$ 360\degr, $|b| \lesssim $1\degr~in the $J = 1-0$ lines of $^{12}$CO, $^{13}$CO, C$^{18}$O, and CN with a spatial resolution of 72\arcsec~and a velocity resolution of $\sim$0.3\,km\,s$^{-1}$. The sensitivities for 
$^{13}$CO are about 
0.7\,K channel$^{-1}$. 
The ThrUMMS data is also available online\footnote{\url{http://www.astro.ufl.edu/~peterb/research/thrumms/rbank/}}. We use the data release 4 (DR4) FITS cubes for 
$^{13}$CO. 

\section{Methods}\label{sect:method}
In the section we describe the sample selection of the GMFs and {the determination of their physical} parameters (Section~\ref{sect:sample}), YSO identification and classification (Section~\ref{sect:youngstar}) and estimation of star formation rate in each GMF (Section~\ref{sect:sfr}).

\subsection{\textcolor{black}{Sample of GMFs}}
\label{sect:sample}

{ We construct our sample by combining the samples of giant filaments identified by} \citet{sample-gmf}, \citet{sample-wang15}, \citet{sample-bones}, \citet{sample-agmf}, \citet{sample-li16}, and \citet{sample-wang16}, including 121 Galactic giant filaments with the length of $>$ 10\,pc. \jkfour{Table~\ref{tablesample} summarize the sample information.} Note that \citet{sample-li16} presented a large filament sample based on the data from APEX Telescope Large Area Survey of the Galaxy \citep[ATLASGAL;][]{atlasgal}. We only include the velocity coherent filaments with lengths of $>$ 10 pc of their sample. 

\begin{table*}
\scriptsize
\caption{Summary of giant filaments identified with different dataset and criteria.}
\label{tablesample}
\centering
\setlength{\tabcolsep}{0.2em}
\setlength\extrarowheight{2pt}
\begin{tabular}{|c |c|c|c| c | c| c| c| c|}
\hline
Filament&\multicolumn{4}{c|}{Identification}&\multicolumn{3}{c|}{Velocity coherence confirmation}&Reference\\[2pt]
\cline{2-5}\cline{6-8}
samples&Dataset&resolution&Criteria&$N_{\textrm{H}_2}$ regime&Dataset&Spectral lines&Criteria&\\[2pt]
\hline                       

\parbox[t][\cellh][c]{2cm}{\centering Giant Molecular\\ Filament (GMF)}&\parbox[t][\cellh][c]{2cm}{\centering GLIMPSE,\\UKIDSS-GPS,\\2MASS} &\parbox[t][\cellh][c]{1cm}{\centering $\sim$1-2$^{\prime\prime}$}& \parbox[t][\cellh][c]{3.5cm}{\centering visual inspection for long absorption or extinction feature,\\ projected length $>$ 1\degr,\\ allowing for gaps}&\parbox[t][\cellh][c]{1cm}{\centering $A_V$$\gtrsim$10}&\parbox[t][\cellh][c]{1.5cm}{\centering GRS,\\ ThrUMMS}&\parbox[t][\cellh][c]{1cm}{\centering $^{13}$CO}&\parbox[t][\cellh][c]{3.5cm}{\centering  continuous velocity variations,\\ without steep velocity jumps}&\parbox[t][\cellh][c]{1cm}{\centering 1, 4}\\ \hline

\parbox[t][\cellh][c]{2cm}{\centering could giant filament (CGF)}&\parbox[t][\cellh][c]{2cm}{\centering Herschel Hi-Gal}&\parbox[t][\cellh][c]{1cm}{\centering $\sim$25-35$^{\prime\prime}$}& \parbox[t][\cellh][c]{3cm}{\centering visual inspection for `skinny long' emission feature,\\ aspect ratio $\gg$ 10,\\ lower temperature than surroundings}&\parbox[t][\cellh][c]{2cm}{\centering $A_V\sim$ 5} &\parbox[t][\cellh][c]{1.5cm}{\centering GRS}&\parbox[t][\cellh][c]{1cm}{\centering $^{13}$CO}&\parbox[t][\cellh][c]{3cm}{\centering  continuous velocity,\\without velocity broken}&\parbox[t][\cellh][c]{1cm}{\centering 2}\\\hline

\parbox[t][\cellh][c]{2cm}{\centering Milky Way bone (bone)}&\parbox[t][\cellh][c]{2cm}{\centering GLIMPSE,\\MIPSGAL}&\parbox[t][\cellh][c]{1cm}{\centering $\sim$2-6$^{\prime\prime}$} & \parbox[t][\cellh][c]{3cm}{\centering visual inspection for `largely continuous filamentary extinction features',\\ parallel to Galactic plane,\\within 20 pc of mid-plane,\\aspect ratio$\geqslant$50}&\parbox[t][\cellh][c]{1cm}{\centering $A_V\sim$ 20}&\parbox[t][\cellh][c]{1.5cm}{\centering HOPS,\\MALT90,\\BGPS,\\GRS,\\ThrUMMS}&\parbox[t][\cellh][c]{1cm}{\centering NH$_3$,\\N$_2$H$^{+}$,\\HCO$^{+}$,\\$^{13}$CO}&\parbox[t][\cellh][c]{3cm}{\centering $\Delta v$ $<$ 3\,km\,s$^{-1}$ per 10\,pc}&\parbox[t][\cellh][c]{1cm}{\centering 3}\\\hline

\parbox[t][\cellh][c]{2cm}{\centering ATLASGAL filament (AF)}&\parbox[t][\cellh][c]{2cm}{\centering ATLASGAL}& \parbox[t][\cellh][c]{1cm}{\centering $\sim$20$^{\prime\prime}$} &\parbox[t][\cellh][c]{3cm}{\centering extracting skeletons with DisPerSE\tablefootmark{a},\\ aspect ratio $>$ 3}&\parbox[t][\cellh][c]{1.5cm}{\centering $A_V\gtrsim$10}&\parbox[t][\cellh][c]{1.2cm}{\centering ATLASGAL follow-up molecular line observations\tablefootmark{b}}&\parbox[t][\cellh][c]{1cm}{\centering NH$_3$,\\N$_2$H$^{+}$,\\CS,\\$^{13}$CO}&\parbox[t][\cellh][c]{3cm}{\centering $\sigma_{v} <$ 10\,km\,s$^{-1}$ between connected compact sources}&\parbox[t][\cellh][c]{1cm}{\centering 5}\\\hline

\parbox[t][\cellh][c]{2cm}{\centering MST filament (MF)}&\parbox[t][\cellh][c]{2cm}{\centering BGPS}&\parbox[t][\cellh][c]{1cm}{\centering $\sim$33$^{\prime\prime}$} & \parbox[t][\cellh][c]{3cm}{\centering MST algorithm,\\projected length $\geqslant$ 10\,pc,\\linearity $>$ 1.5}&\parbox[t][\cellh][c]{1.5cm}{\centering $A_V\gtrsim$ 10}&\parbox[t][\cellh][c]{1.5cm}{\centering BGPS}&\parbox[t][\cellh][c]{1cm}{\centering N$_2$H$^{+}$,\\HCO$^{+}$}&\parbox[t][\cellh][c]{3cm}{\centering $\Delta v$ $<$ 2\,km\,s$^{-1}$ between connected clumps}&\parbox[t][\cellh][c]{1cm}{\centering 6}\\\hline

\end{tabular}
\tablebib{
(1)~\citet{sample-gmf}; (2) \citet{sample-wang15}; (3) \citet{sample-bones}; (4) \citet{sample-agmf}; (5) \citet{sample-li16}; (6) \citet{sample-wang16}.
}
\tablefoot{
\tablefoottext{a}{\citet{disperse}.}
\tablefoottext{b}{Please see the Table 3 of \citet{sample-li16} for the complete velocity references.}
}
\end{table*}

{Our agglomerate sample of 121 giant filaments has been identified using different methods and different datasets. Thus their physical parameters such as mass are also calculated based on  different gas tracers and methods. To obtain a systematic investigation of the filament properties, we re-estimated their parameters using a uniform method.}


Of 121 giant filaments, 82 are covered by the GRS or ThrUMMS survey data. {We also exclude} three filaments that show self-absorption, four filaments that are not fully covered by $^{13}$CO data, and {18} redundant filaments that are identified several times by different papers. {We {integrate the $^{13}$CO data cubes over the velocity range of each filament to obtain maps of integrated intensity}. Assuming a uniform excitation temperature of 10\,K and $^{13}$CO abundance of [$^{13}$CO/H$_2$]$\sim$1.2$\times$10$^{-6}$ \citep{frerking82,wilson94}, }we obtain {$^{13}$CO-based} column density maps for {56} giant filaments. {The uncertainties of these column density maps are mainly from the variation of excitation temperature and $^{13}$CO abundance.} Note that we use the extinction map constructed with the technique {of} \citet{kt13} for the {"Nessie" filament} {\citep{mattern18}}. {The details about how to obtain the column density maps and associated uncertainties can be found in Appendix~\ref{ap1} and \ref{ap4}}. 


We use the extinction contour level of $A_V =$ 3\,mag to define the boundaries of the GMFs. {We note as a caveat for the later comparisons that this value is} slightly higher than the values used by \citet{lada10} ($A_V \sim$ 1\,mag) and \citet{heiderman10,evans14} ($A_V =$~2\,mag). We also checked each {GMF} and made sure that {its boundary can morphologically separate the} filament from the surrounding diffuse gas. {As examples, Figs.~\ref{fig:CFG47}-\ref{fig:GMF341} (left panels) show the column density maps of seven GMFs; the column density maps of the other GMFs can be found in Appendix~\ref{ap5}.} 


For the Nessie, the high resolution extinction map has projection effects due to the lack of information along the line-of-sight although the Nessie has been confirmed as a velocity coherent filament \citep{nessie10}. To crop the extinction map to the Nessie, we introduce a polygon around the cloud. The area selection is mainly based on by eye inspection of the derived column density map with orientation on the $A_V = $3 mag contour and the observations published by \citet{nessie10} {\citep{mattern18}}.
%

{{We obtain the dense gas mass by} integrating the $^{13}$CO-based extinction map above $A_V=$~7\,mag in each GMF.}
{{Note that the definitions of dense gas vary in literature.} \citet{evans14} adopted the extinction contour of $A_V =$ 8\,mag to calculate the $M_{\textrm{dense}}$. \citet{sample-gmf} and \citet{sample-agmf} use the continuum millimeter observations such as ATLASGAL to trace the dense gas. The threshold that they adopted based on the sensitivity of ATLASGAL corresponds to $A_V \sim$~10\,mag. {Our choice is motivated by the choice 
of \citet{lada10}}. They used the extinction contour of $A_K =$ 0.8\,mag to measure the $M_{\textrm{dense}}$, corresponding to $A_V \approx$ 7\,mag if using the extinction law suggested by \citet{xue16}.}

Based on the column density map and the defined boundary of each object, we {estimated the basic physical parameters of the {57} GMFs (see Appendix~\ref{ap2} for details).} Table~\ref{table1} lists {these parameters}. 
{Note that the uncertainties listed in Table~\ref{table1} do not include the distance uncertainty.}
%

\begin{figure*}
\includegraphics[width=1.0\linewidth]{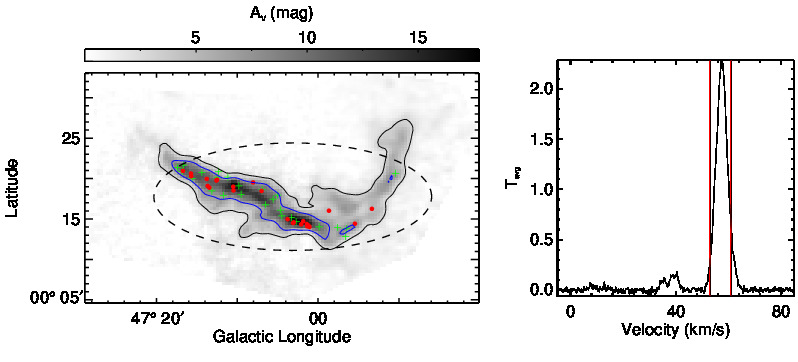}
\caption{Overview of GMF 39 (CFG047.06+0.26). \textit{Left:} the background is CO-based extinction map. The black and blue contours represent the visual extinction of $A_V=$3 and 7\,mag, individually. The dashed ellipses are obtained through fitting the pixels inside the regions with $A_V>$3\,mag. The identified YSOs are labeled with red filled circles (Class I) and green pluses (Class II); \textit{right:} the $^{13}$CO average spectrum for the region with $A_V>$3\,mag. The red vertical lines mark the GMF velocity range.}
\label{fig:CFG47}
\end{figure*}

\begin{figure*}
\includegraphics[width=1.0\linewidth]{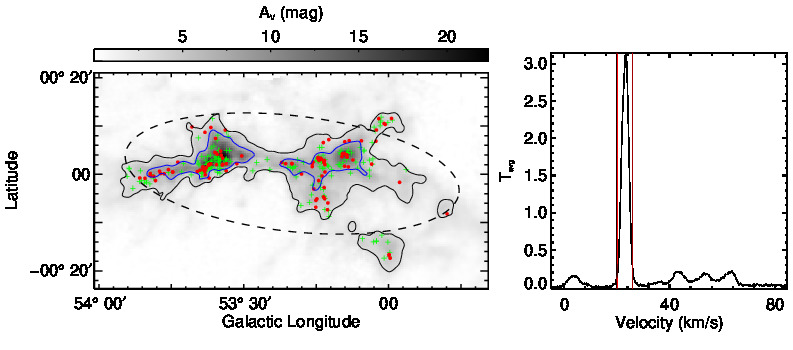}
\caption{Overview of GMF 43 (GMF54.0-52.0). Others are same as Fig.~\ref{fig:CFG47}.}
\label{fig:GMF54}
\end{figure*}

\begin{figure*}
\includegraphics[width=1.0\linewidth]{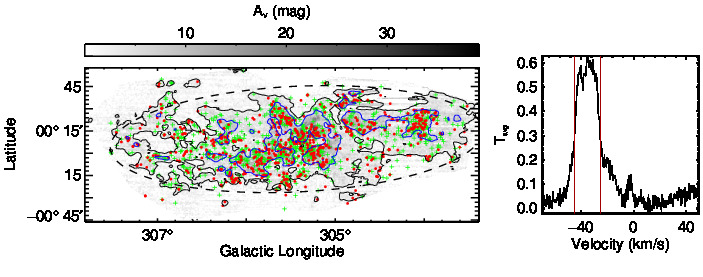}
\caption{Overview of GMF 44 (GMF307.2-305.4). Others are same as Fig.~\ref{fig:CFG47}.}
\label{fig:GMF307}
\end{figure*}

\begin{figure*}
\includegraphics[width=1.0\linewidth]{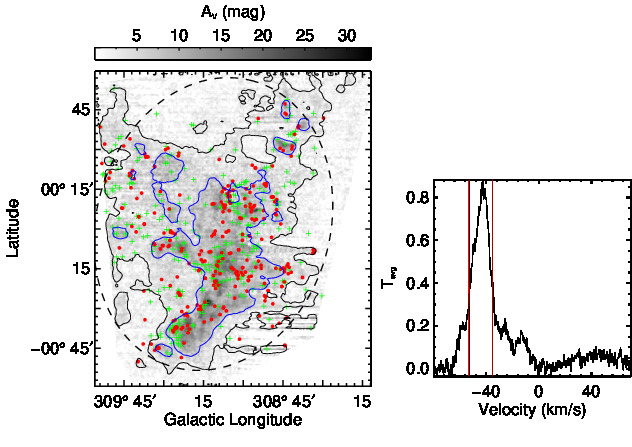}
\caption{Overview of GMF 45 (GMF309.5-308.7). Others are same as Fig.~\ref{fig:CFG47}.}
\label{fig:GMF309}
\end{figure*}

\begin{figure*}
\includegraphics[width=1.0\linewidth]{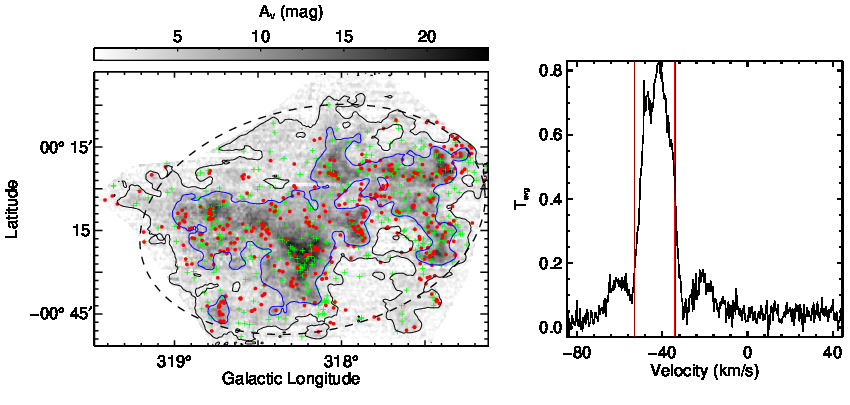}
\caption{Overview of GMF 46 (GMF319.0-318.7). Others are same as Fig.~\ref{fig:CFG47}.}
\label{fig:GMF319}
\end{figure*}

\begin{figure*}
\includegraphics[width=1.0\linewidth]{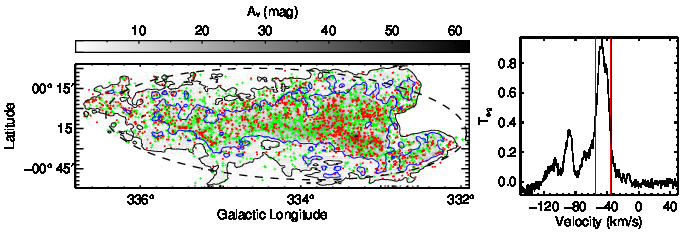}
\caption{Overview of GMF 52 (GMF335.6-333.6a). Others are same as Fig.~\ref{fig:CFG47}.}
\label{fig:GMF335a}
\end{figure*}

\begin{figure*}
\includegraphics[width=1.0\linewidth]{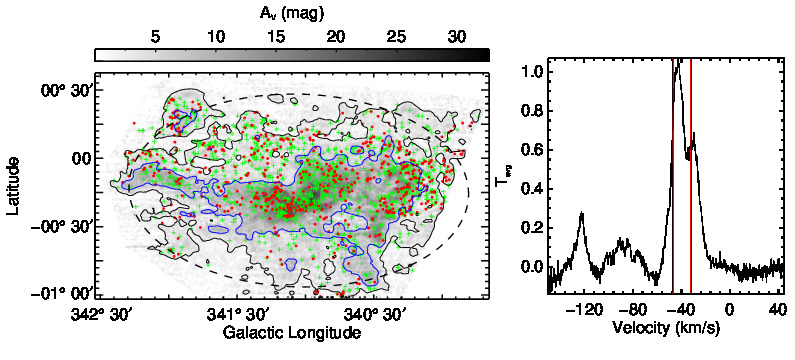}
\caption{Overview of GMF 54 (GMF341.9-337.1). Others are same as Fig.~\ref{fig:CFG47}.}
\label{fig:GMF341}
\end{figure*}

\subsection{YSO identification and classification}\label{sect:youngstar}


\subsubsection{YSO identification}\label{sect:yso}
YSOs usually show excessive infrared emission that can be used to distinguish them from field stars. Based on the slope of SEDs, \citet{lada84} and \citet{lada87} developed a `standard' empirical classification scheme to indicate the different evolutionary stages of YSOs. Some authors use the infrared spectral index to identify YSO candidates in the star-forming regions \citep{mallick13,kim15}. YSOs can be also identified by the multi-color criteria (i.e., color-color and color-magnitude diagrams). In the present work, we combine several methods that are based on the infrared multi-color criteria to identify YSO candidates \citep{gutermuth08,robitaille08,gutermuth09,ven13,koenig14,saral15}. This method uses the SEDs of sources from 1 to 500\,$\mu$m and can efficiently mitigate the effects of contamination. 
The details about this method can be found in the Appendix~\ref{ap3}. In general, we searched through several tens of millions sources and selected {36394} 
sources with the infrared excess in {57} GMFs, of which {4821} 
($\sim$13\%) are AGB candidates. The {31573} 
YSO candidates include {5611} 
Class I candidates, {22609} 
Class II candidates, {384} 
transition disk candidates, {2874} 
prostellar objects, and 95 massive YSO candidates (see Appendix~\ref{ap3} for the criteria of different classifications). 

Note that the identification of the bona-fide YSOs needs the spectroscopic observations, which are not available here. However, in the subsequent context we will simply use ``YSOs'' and ``Class I/II sources'' to refer ``YSO candidates'' and ``Class I/II candidates'' for convenience although they are just candidates.




\subsubsection{Correcting for extinction}\label{sect:av}
Considering that our objects are located at large distances and associated with dense gas, it is \textcolor{black}{necessary} to correct the flux densities of the YSOs for extinction. Note that we only wish to correct for the {diffuse} foreground extinction rather than the local extinction from dense cores surrounding protostars \citep{evans09,dunham15}. We use the method suggested by \citet{fang13} and \citet{mypub15} to de-redden the photometry of YSOs.


\begin{itemize}
\item[1.]{For the sources with $J$, $H$, $K_s$ detections, the extinction is obtained by employing the $JHK_s$ color-color diagram. A detailed description can be found in \citet{fang13} and \citet{mypub15}. Here we only summarize a few aspects of this scheme. The location of each YSO in the $JHK_s$ color–color diagram depends on both its intrinsic colors and its extinction. Figure~\ref{ccd} shows the $J-H$ versus $H-K_s$ color-color diagram of the YSOs in the GMF {39} (CFG047.06$+$0.26). Given the different origins of intrinsic colors of YSOs, the color–color diagram is divided into three subregions. In region 1, the intrinsic color of $[J-H]_0$ is simply assumed to be 0.6; in region 2, the intrinsic color of a YSO is obtained from the intersection between the reddening vector and the locus of main sequence stars \citep{bb}; in region 3, the intrinsic color is derived from where the reddening vector and the classical T Tauri star (CTTS) locus \citep{tt} intersects. Then the extinction values of YSOs are estimated from observed and intrinsic colors with the extinction law of \citet{xue16}.}
\item[2.]{For other sources (outside these three regions or without detections in $JHK_s$ bands), their extinction is estimated with the median extinction values of surrounding Class II sources and transition disks that have extinction measurements in step 1.}
\end{itemize}
\begin{figure}
\includegraphics[width=1.0\linewidth]{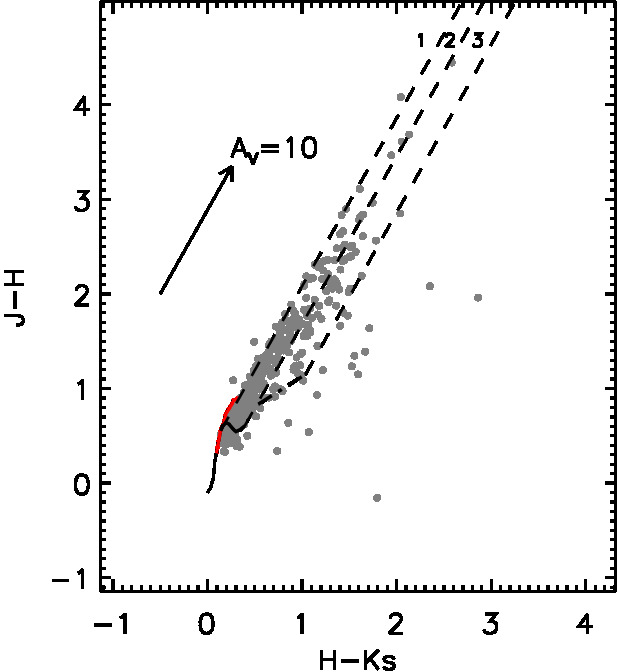}
\caption{the $H-K_{s}$ versus $J-H$ color–color diagram for the YSOs in the GMF {39} (CFG047.06$+$0.26). The solid curves show the intrinsic colors for the main-sequence stars (black) and giants \citep[red;][]{bb}, and the dash–dotted line is the locus of T Tauri stars from \citet{tt}. The dashed lines show the reddening direction, and the arrow shows the reddening vector. The extinction law we adopted is from \citet{xue16}. Note that the dashed lines separate the diagram into three regions marked with numbers 1, 2, and 3 in the figure. We use different methods to estimate the extinction of YSOs in different regions (see the text for details).}
\label{ccd}
\end{figure}
With extinctions obtained using above method, we de-redden the SED of each YSO using the extinction law of \citet{xue16}.

\subsubsection{YSO classification}


{We re-classify the identified YSOs after the extinction correction using the de-reddened SEDs of the YSOs.} There are two main YSO classification schemes, one is based on the spectral index \citep[$\alpha$,][]{greene94} and the other employs the bolometric temperature \citep[$T_{\textrm{bol}}$,][]{chen95}. There are both advantages and disadvantages for these two classification schemes.

The source's spectral index, {i.e., the slope of the source's SED defined as}
\begin{equation*}
\alpha = \frac{d\,\textrm{log}(\lambda S_{\lambda})}{d\,\textrm{log}(\lambda)},
\end{equation*}
where $S_{\lambda}$ is the flux density at wavelength $\lambda$, can be used to {classify} the evolutionary state of the source. By fitting the de-reddened SEDs from 2 to 24\,$\mu$m, the YSOs can be classified as Class I, `Flat spectrum', Class II, and Class III sources based on the scheme suggested by \citet{greene94}.

The bolometric temperature of a source that is defined as \citep{boltemp93}
\begin{equation*}
T_{\textrm{bol}} = 1.25 \times 10^{-11} \frac{\int_{0}^{\infty} \nu S_{\nu}d\nu}{\int_{0}^{\infty} S_{\nu}d\nu}~~~~\textrm{K},
\end{equation*}
where $S_{\nu}$ is the source's flux density at frequency $\nu$, can be also used to determine the evolutionary stage of a source. In practice, we calculate these integrals using the trapezoid rule to integrate over the finitely sampled SEDs following the suggestion of \citet{dunham08,dunham15}. Based on their bolometric temperature, YSOs can be classified as Class 0, Class I, and Class II sources according to the classification scheme of \citet{chen95}.


Figure~\ref{ysoclass} shows the de-reddened $\alpha_{2-24\mu m}$ that is calculated based on the SED from 2 to 24\,$\mu$m and $T_{\textrm{bol}}$ of YSOs in the sample of GMF {39} (CFG047.06$+$0.26). The dashed lines mark the classification criteria. 
\begin{figure}
\includegraphics[width=0.9\linewidth]{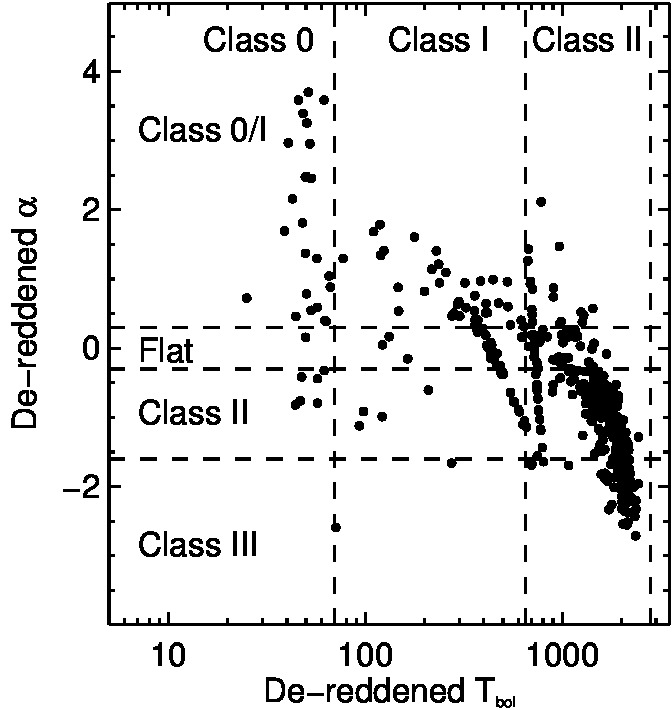}
\caption{The de-reddened $\alpha$ and $T_{\textrm{bol}}$ of YSOs in the GMF {39} (CFG047.06$+$0.26). The dashed lines show the boundaries of YSO classification schemes \citep{greene94,chen95}.}
\label{ysoclass}
\end{figure}

The spectral index of $\alpha_{2-24\mu m}$ has been widely used for YSO classification. However, $\alpha_{2-24\mu m}$ is not sensitive for the protostars that are identified based on the emission at 70\,$\mu$m: some of these protostars only have weak emission or are even invisible at 2$-$24\,$\mu$m.

The bolometric temperature $T_{\textrm{bol}}$ is more appropriate for the embedded objects and can be used to isolate Class 0 sources \citep{evans09}. However, \citet{dunham15} found that the measured $T_{\textrm{bol}}$ of YSOs with disks is mostly determined by the spectral type of the central source and the value of $T_{\textrm{bol}}$ is correlated with the range of SED that is used to calculate $T_{\textrm{bol}}$ (see Appendix C of \citealt{dunham15} for details), which means that the $T_{\textrm{bol}}$ measured with different ranges of SEDs of YSOs with the similar spectral types would be clustered into different narrow $T_{\textrm{bol}}$ ranges. This effect can be also used to explain the patterns in Fig.~\ref{ysoclass}. As \citet{dunham15} pointed, $T_{\textrm{bol}}$ is not a good discriminator between Class II and Class III.

{Taking into account the above issues, we} decided to combine $\alpha_{2-24\mu m}$ and $T_{\textrm{bol}}$ to classify the YSOs: 
\begin{itemize}
\item[1]{The protostellar objects that are identified based on the emission at 70\,$\mu$m are classified as Class I sources.}
\item[2]{The YSOs that have $\alpha_{2-24\mu m}$ measurements are classified as Class I ($\alpha_{2-24\mu m}$ $\geq$ 0.3) and Flat ($-$0.3$\leq$ $\alpha_{2-24\mu m} <$ 0.3) sources. The YSOs without $\alpha_{2-24\mu m}$ measurements but with $T_{\textrm{bol}}$ $\leq$ 650 K are also classified as Class I sources.}
\item[3]{Excluding the Class I and Flat sources classified in step 1 and 2, the remaining YSOs are classified as Class II sources if they a) have $\alpha_{2-24\mu m}$ measurements and $-$1.6 $\leq$ $\alpha_{2-24\mu m}$ $<$ $-$0.3 or b) have not $\alpha_{2-24\mu m}$ measurements but 650 $<$ $T_{\textrm{bol}}$ $\leq$ 2800 K.}
\item[4] {The remaining YSOs are classified as Class III sources.}
\end{itemize}
Finally, each of YSOs can be classified as Class I, Flat, Class II, or Class III source. 

The status of Flat sources is not very clear and {they are generally considered} to be a transitional class. \citet{heiderman15} presented an HCO$^{+}$ survey of Class 0$+$I and Flat YSOs in the Gould Belt clouds and they defined a Class 0$+$I$+$Flat source that is associated with HCO$^{+}$ emission as a stage 0$+$I source which consists of a star and disk embedded in a dense, infalling envelope (i.e., a bona fide protostar). \citet{heiderman15} found that $\sim$50\% of Flat sources are associated with HCO$^{+}$ emission. They also found that the fraction of sources that are associated with HCO$^{+}$ emission is $\gtrsim$70\% for the sources with bolometric luminosity ($L_{\textrm{bol}}$) of $\gtrsim$ 1 $L_{\sun}$, which means that the bright Flat sources have high probability to be bona fide protostars. 

Considering the distances of our GMFs ($>$ 2 kpc), we can only detect the bright Flat sources that have high probability to be the bona fide protostars. Therefore, in this paper we simply {group together the Flat sources as Class I sources and refer to them together as `Class I sources'}.

\subsubsection{Estimating and excluding contamination}\label{sect:contamination}
Considering the distances of the GMFs ($>$ 2 kpc), our YSOs are contaminated by foreground sources such as the foreground AGBs and the foreground YSOs which are associated with the molecular clouds that are located between us and the GMFs. 
In Section~\ref{sect:yso}, we have isolated a bunch of AGBs using the multi-color criteria (see Appendix~\ref{ap3}). However, because AGB stars can mimic the colors of YSOs, it is difficult to exclude AGBs only based on the color criteria. Simply assuming an universal spatial distribution for the Galactic AGB stars \citep{jackson02,robitaille08}, the contamination fraction of foreground AGBs mainly depends on the distances of the GMFs. On the other hand, the contamination fraction of foreground YSOs mainly depends on the number of molecular clouds that are located before the GMFs, which is related with the distances and the Galactic longitudes and latitudes of the GMFs.




We use the $A_V$ values obtained in Section~\ref{sect:av} and the 3D extinction map to isolate the foreground contamination. By comparing the 2MASS photometry to the Stellar Population Synthesis Model of the Galaxy \citep{besanmodel}, \citet{marshall06} established a 3D extinction map of the inner Galaxy ($|l| <$ 100\degr, $|b| <$ 10\degr). Using their extinction map, we can estimate the foreground extinction in different lines of sight towards a GMF based on its distance. If the extinction value of a YSO measured in Section~\ref{sect:av} is lower than the corresponding foreground extinction, this YSO would have high probability to be a foreground contamination. We checked the YSOs in each GMF and marked the possible foreground contamination using this method. The fraction of foreground contamination in different GMFs is $\sim$5\%$-${96}\%~in Class I sources with a median value of $\sim${31}\%~and $\sim$8\%$-${100}\%~in Class II sources with a median value of $\sim${28}\%.

Our YSOs are also contaminated by the background sources, including the extragalactic objects, background AGBs, and background YSOs which are associated with the molecular clouds that are located behind the GMFs. 

For the extragalactic contamination, on the one hand, the YSO identification method suggested by \citet{gutermuth09} and \citet{koenig14} can efficiently mitigate the contamination of star-forming galaxies and AGNs. On the other hand, the contamination due to galaxies should be negligible since we are observing through the Galactic plane \citep[contamination fraction $\lesssim$ 2\%;][]{robitaille08,jose16}. Thus we think that the extragalactic contamination is not important in our YSOs. 

The residual contamination of background AGBs is estimated with the control fields.
For each GMF, we select five nearby fields with weak $^{13}$CO emission as the control fields and apply the method described in Section~\ref{sect:yso} to all the control fields to select YSOs. We assume that there is no YSOs in each control field. Thus all selected `YSOs' in the control fields are actually contamination of AGBs (if neglecting the extragalactic contamination). Using the distance of corresponding GMF, we can separate the AGBs in the control fields into `foreground' and `background'. The foreground AGBs can be excluded with the method mentioned above. 
With an assumption of a uniform distribution for AGB stars, we can estimate the number of residual background AGBs in each GMF using the mean value of the surface density of background AGBs in five control fields. Combining the {numbers of} background AGBs identified by color criteria and estimated using control fields, we found that the fraction of background contamination is $\sim${0.5}\%$-$37\%~for Class I sources with a median value of {12}\% and $\sim${1}\%$-$25\%~for Class II sources with a median value of 10\%.

The background YSOs are difficult to remove without the information of radial velocities of YSOs. If there are one or more molecular clouds located behind the GMF, we obviously overestimate the number of YSOs for this GMF.

{After excluding the possible contamination, we finally obtain 7\,028 Class I sources and 11\,526 Class II sources in 57 GMFs.}
{Figures.~\ref{fig:CFG47}-\ref{fig:GMF341} (left panels) show the YSO spatial distributions in seven GMFs and the distributions of YSOs in other GMFs can be found in Appendix~\ref{ap5}.}
\subsection{Estimating star formation rates and efficiencies}\label{sect:sfr}

In order to calculate the star formation rate (SFR) and efficiency (SFE) of each GMF, we must estimate the total mass of YSOs. However, considering the distances of the GMFs, we obviously miss many low-mass young stars. We can {obtain an} estimate {of} the mass of undetected low-mass YSO population by assuming an universal initial mass function (IMF) or luminosity function (LF). In the present work, we use different methods to estimate the total mass of Class I and Class II populations, based on which the SFR and SFE of each GMF can be obtained.
\subsubsection{Mass of Class II populations}\label{sect:massc2}
We use the de-reddened photometry of Class II sources (excluding foreground and background contamination) in each GMF to estimate the flux completeness. Figure~\ref{com-c2} shows the $K_s$ absolute magnitude histogram of Class II sources in GMF {46} (GMF319.0$-$318.7). We simply adopt the peak position of histogram as the completeness of $K_s$ band ($\sim$1 mag for GMF {46}). If there are not enough Class II sources 
to construct a histogram, we use the median value of $K_s$ magnitudes as the completeness. The $K_s$ completeness of Class II sources in different GMFs are $\sim$$-$2.6$-$2 mag, mainly depending on the cloud distances. To transfer the $K_s$ completeness to mass completeness, we need to establish a relation between SED fluxes and masses for Class II sources. 

\begin{figure}
\includegraphics[width=1.0\linewidth]{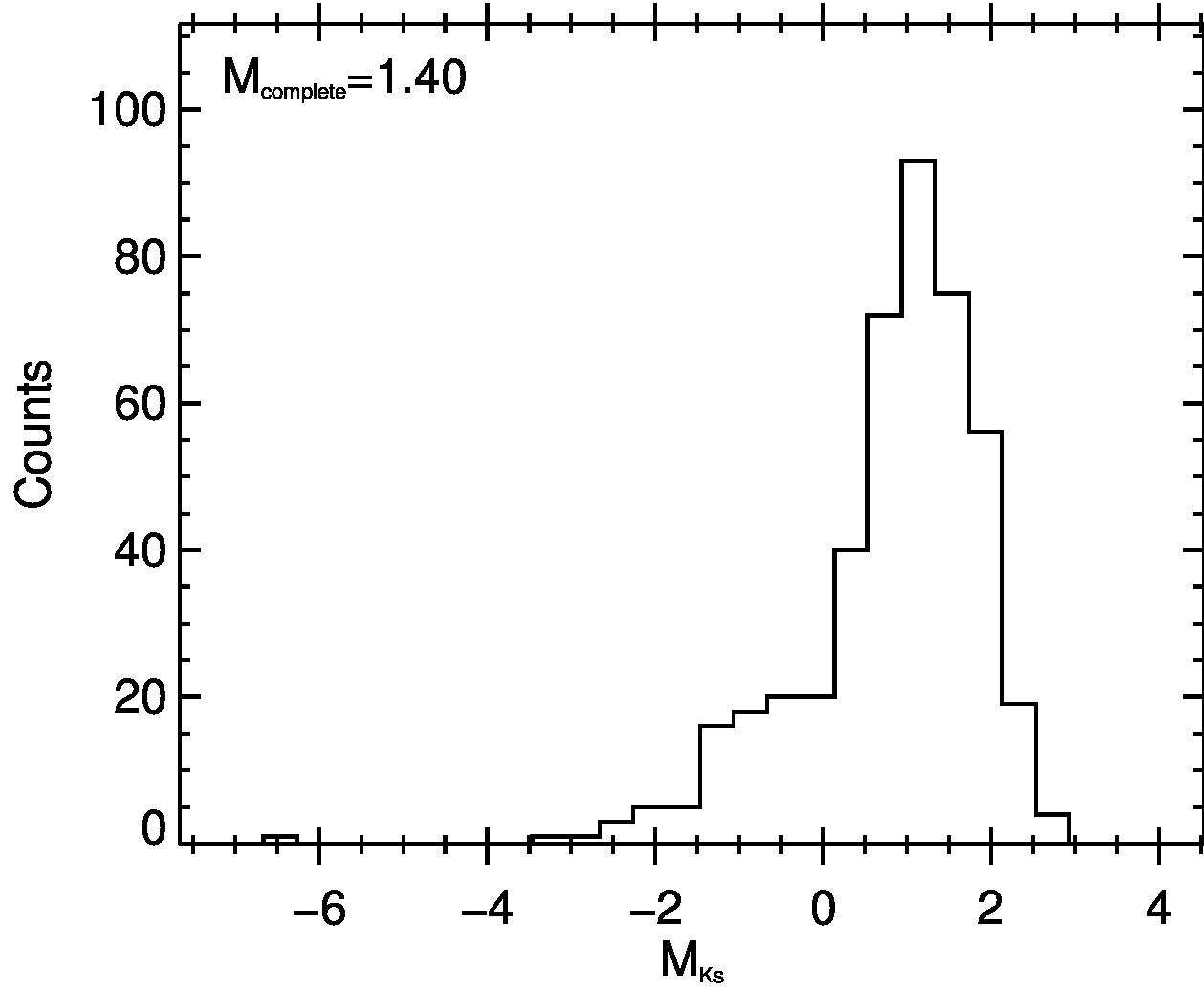}
\caption{$K_s$ absolute magnitude ($M_{K_s}$) histogram of Class II sources in GMF {46} (GMF319.0$-$318.7).}
\label{com-c2}
\end{figure}

\citet{tobitaille06} presented a grid of radiation transfer models of YSOs ($\sim$20, 000 YSO models), covering a wide range of stellar masses, disk masses, envelope masses, and accretion rates. For each YSO model, SEDs are calculated for 10 inclination angles. The model SEDs are also convolved to commonly used filters to generate broadband fluxes within 50 different apertures (100$-$100 000 AU). In our analysis, we use the aperture of $\sim$45 000 AU, corresponding to $\sim$3\arcsec~at a distance of 15 kpc. Using other large apertures do not change our results \citep{heyer16}. \citet{tobitaille06} also defined three evolutionary stages of YSOs based on the accretion rates of envelopes and disks relative to the stellar masses: Stage I has $\dot{M}_{\textrm{env}}/M_{*} > 10^{-6}$ yr$^{-1}$; Stage II has $\dot{M}_{\textrm{env}}/M_{*} < 10^{-6}$ yr$^{-1}$ and $M_{\textrm{disk}}/M_{*} > 10^{-6}$; and Stage III has $\dot{M}_{\textrm{env}}/M_{*} < 10^{-6}$ yr$^{-1}$ and $M_{\textrm{disk}}/M_{*} < 10^{-6}$.

\citet{and13} investigated the disk and stellar masses of Class II sources in Taurus and they found a roughly linear scaling between disk and host star masses, with a typical disk-to-star mass ratio of $\sim$0.2\%$-$0.6\%. We extract the Stage II models with $0.001<M_{\textrm{disk}}/M_{*}<0.01$, $0.08<M_{*}<7\,\textrm{M}_{\sun}$, and 30\degr$<$ inclination angle $<$60\degr. Figure~\ref{kmass} shows the distribution of these models in the $M_{K_s}-M_{*}$ plane. Obviously, there is a correlation between stellar mass and $K_s$ band absolute magnitude for Class II sources. We use a 4 orders polynomial function to fit the points and then establish a relation between $M_{K_s}$ and $M_{*}$. The gray region in Fig.~\ref{kmass} shows the fitting uncertainties (1$\sigma$), which is also adopted as the uncertainty of the relation. \citet{fang13} investigated the YSOs in L1641 using the optical spectroscopy and estimate the stellar masses of YSOs based on effective temperatures and bolometric luminosities with several different pre-main sequence (PMS) evolutionary models. We plot the $K_s$ absolute magnitudes and stellar masses obtained with \citet{s00} PMS models of the Classic T Tauri stars (CTTS) from \citet{fang13} in Fig.~\ref{kmass} with green dots. Most of CTTS are located in the gray region of Fig.~\ref{kmass}, which confirms that $M_{K_s}-M_{*}$ relation for Class II sources established by us is also consistent with the observational results.

\begin{figure}
\includegraphics[width=1.0\linewidth]{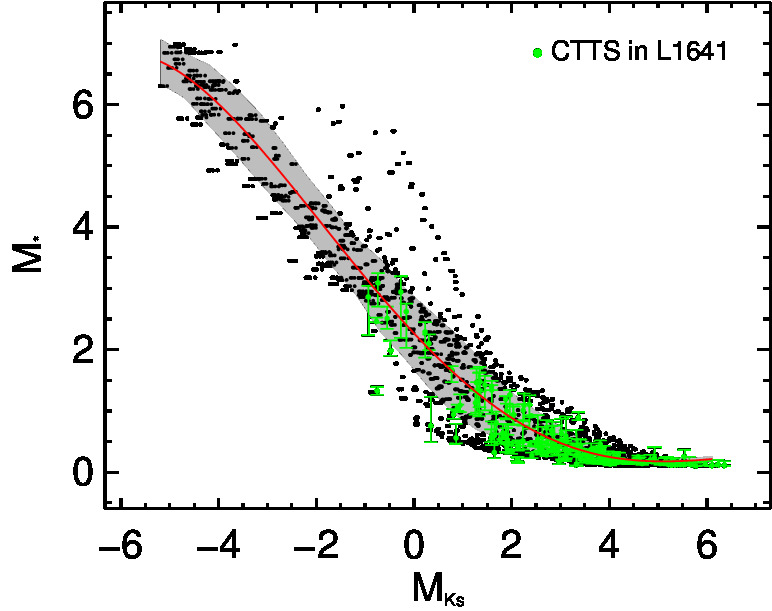}
\caption{The relation between stellar mass and $K_s$ absolute magnitude of Class II source. The black dots represent the \citet{tobitaille06} Stage 2 models with 0.001$<M_{\textrm{disk}}/M_{*}<$0.01, 0.08$<M_{*}<$7\,M$_{\sun}$, and 30\degr$<$ inclination angle $<$60\degr. The red curve shows the robust polynomial fitting while the gray region shows the 1$\sigma$ uncertainty of the fitting. The CTTS in L1641 from \citet{fang13} are marked with green filled circles.}
\label{kmass}
\end{figure}

Using above $M_{K_s}-M_{*}$ relation, the flux completeness of Class II sources in each GMF can be transfered to mass completeness. The obtained mass completeness ($M_{\textrm{comp}}$) of Class II sources in different GMFs are from 0.9 to 4.7 $M_{\sun}$, with the average uncertainty of $\pm${0.6} $M_{\sun}$.

The number ($N_{\textrm{ClassII}}$) of Class II sources brighter than the completeness limit 
is used to estimate the total number and mass of Class II sources by extrapolating from $M_{\textrm{comp}}$ down to the hydrogen burning limit (0.08 $M_{\sun}$) with the Kroupa IMF \citep{imf}. 

\subsubsection{Mass of Class I populations}\label{sect:massc1}
Protostars are in the main accretion phase and the luminosities of protostars are also correlated with the accretion process. However, the accretion rate onto a protostar is still under debate \citep{dunhampp6}. On the other hand, the intrinsic stellar parameters of protostars are also poorly constrained. The pre-main sequence evolution, especially during the first few Myr, is much less well constrained \citep[][and references therein]{dunhampp6}. Therefore, it is difficult to construct a relation between luminosity and mass for Class I sources.

We also extracted some \citet{tobitaille06} Stage I models with 0.001~$<M_{\textrm{disk}}/M_{*}<$~0.1, 0.08~$<M_{*}<$~7\,M$_{\sun}$, 30\degr~$<$ inclination angle $<$~60\degr, and 10$^{-8}~<{\dot{M}_{\textrm{disk}}}<$~10$^{-5}$ \citep{cg12,ant14,heyer16}. However, we did not find any correlation between $M_{*}$ and $M_{K_s}$ for Class I sources. Therefore, in the present work we try to use LF to estimate the total number of Class I sources in each GMF.

\citet{kry12} identified protostars in nine star-forming molecular clouds within 1 kpc and constructed the combined LFs for low-mass star-forming clouds and high-mass star-forming clouds (LLF and HLF hereafter) as shown in Fig.~\ref{lftemplate}. The red vertical lines in Fig.~\ref{lftemplate} show the luminosity completeness of LLF and HLF ($L_{cut,LLF}$ and $L_{cut,HLF}$). \citet{kry12} found significant difference between LLF and HLF: HLF peaked near 1\,$L_{\sun}$ with a tail extending towards luminosities above 100\,$L_{\sun}$ while LLF peaked below 1\,$L_{\sun}$ without $>$100\,$L_{\sun}$ tail. Assuming the universal LFs for low-mass star-forming regions and high-mass star-forming regions, LLF and HLF constructed by \citet{kry12} as the templates can be used to estimate the total number of Class I sources in our GMFs.

\begin{figure}
\includegraphics[width=1.0\linewidth]{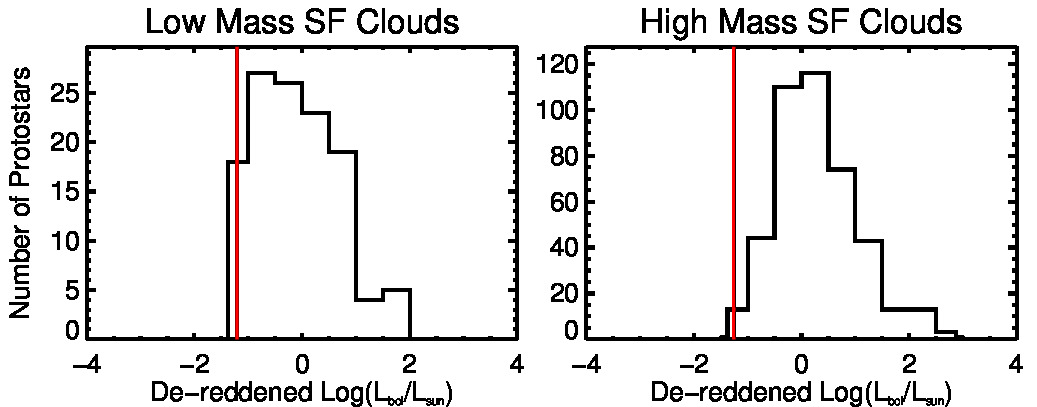}
\caption{De-reddened luminosity functions constructed by \citet{kry12} for the combined low-mass star-forming clouds (\textit{left}) and combined high-mass star-forming clouds (\textit{right}). The red vertical lines show the de-reddened $L_{cut}$.}
\label{lftemplate}
\end{figure}

In each GMF, we calculate the bolometric luminosities of Class I sources using the trapezoid rule to integrate over the finitely sampled de-reddened SEDs \citep{dunham08,dunham15}. We found that all our GMFs harbored Class I sources with luminosities of $>$100\,$L_{\sun}$, indicating that all our samples are high-mass star-forming molecular clouds. Thus we assume an universal LF that is the same as HLF for all the GMFs.



Because most of Class I sources are identified based on \textit{Spitzer} photometric catalog ($<$20\%~Class I sources are identified based on emission at 70 $\mu$m), the luminosity completeness of Class I sources ($L_{cut}$) is estimated based on different \textit{Spitzer} bands. We plot the histograms of photometry at 8\,$\mu$m and 24\,$\mu$m for Class I sources in each GMF, and then adopt the peaks as the magnitude completeness ($m_8$ and $m_{24}$). Using the method presented by \citet{kry12}, $m_8$ and $m_{24}$ can be converted to the luminosity completeness and finally the higher value is adopted as $L_{cut}$. Figure~\ref{lfc1} shows the de-reddened LF of Class I sources in GMF {46} (GMF319.0$-$318.7) and the corresponding $L_{cut}$ is marked with the red line.

\begin{figure}
\includegraphics[width=1.0\linewidth]{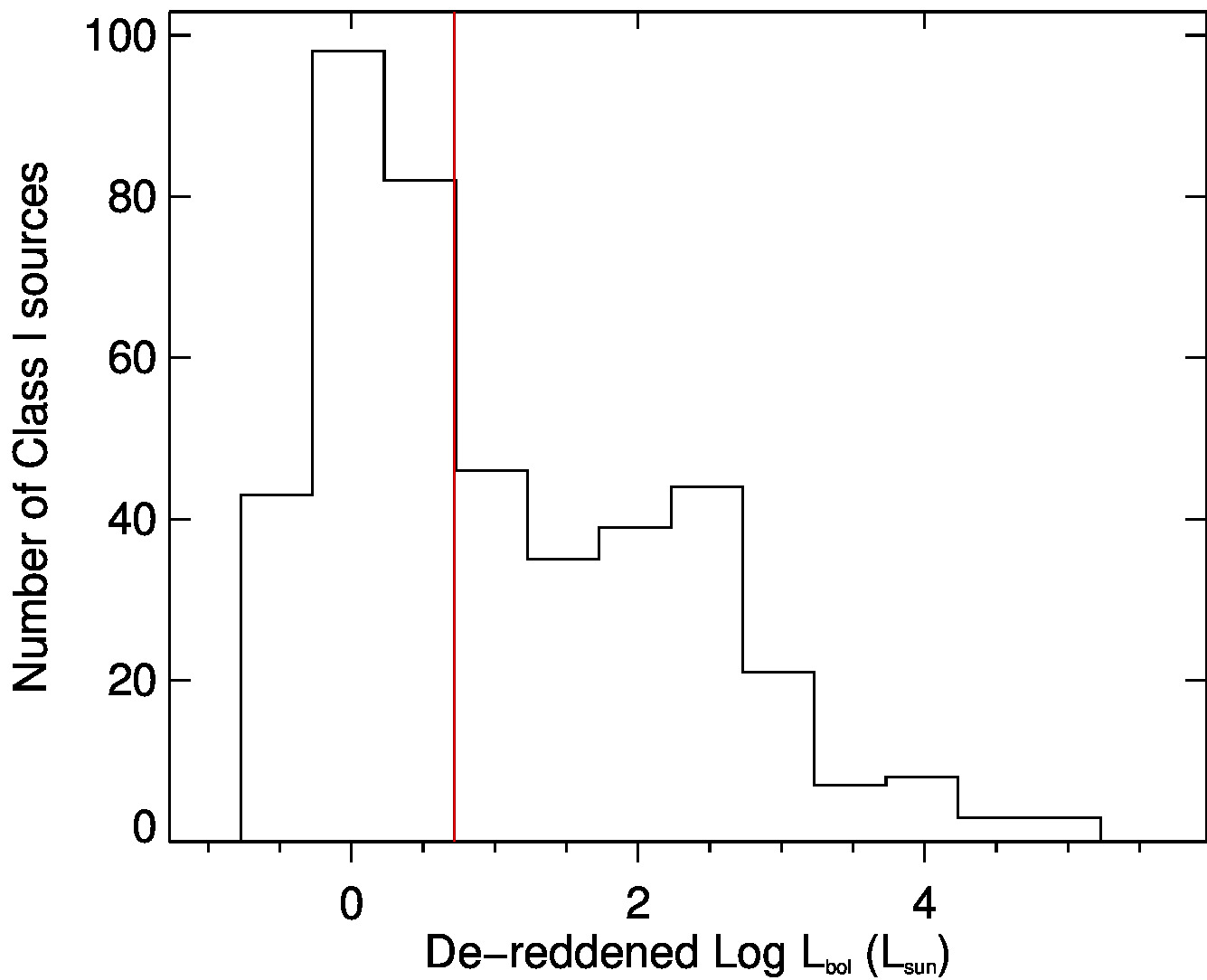}
\caption{De-reddened luminosity function of Class I sources in GMF {46} (GMF319.0$-$318.7). The red vertical line shows the de-reddened $L_{cut}$.}
\label{lfc1}
\end{figure}

The number ($N_{\textrm{ClassI}}$) of Class I sources brighter than $L_{cut}$ 
is used to estimate the total number of Class I sources by extrapolating from $L_{cut}$ down to $L_{cut,HLF}$ with HLF template. The $L_{cut,HLF}=$0.055$L_{\sun}$ is similar as $L_{cut,Orion}=$0.047$L_{\sun}$ \citep{kry12}. The mass completeness of protostars in Orion is about 0.2\,$M_{\sun}$ \citep{willis13}. Thus the total number of Class I sources estimated above is also complete down to $\sim$0.2\,$M_{\sun}$. Using the total number of Class II sources that is obtained by extrapolating down to 0.2\,$M_{\sun}$ with Kroupa IMF, we can calculate the number ratio of Class I to Class II sources, based on which the total number of Class I sources down to 0.08\,$M_{\sun}$ ($N_{\textrm{tot}}^{\textrm{c1}}$) can be estimated. The uncertainty of total number of Class II sources also contribute to the final uncertainty of $N_{\textrm{tot}}^{\textrm{c1}}$. The average uncertainty of log$(N_{\textrm{tot}}^{\textrm{c1}})$ in all GMFs is about $\pm$0.13 dex.

During above calculating process, we assume an universal LF for all high-mass star-forming molecular clouds. \citet{kry12} examined the luminosity functions as a function of the local surface density of YSOs in the Orion molecular cloud and they found a significant difference between the luminosity functions of protostars in regions of high and low stellar density, the former of which is biased towards more luminous sources. This result has also been confirmed in the Cygnus-X star-formating complex by \citet{kry14}, indicating the variations of protostellar LF due to the different surrounding environment. 
\citet{kry12} constructed two LFs for protostars in the low and high stellar surface density regions of Orion. We found that using different LFs to estimate total number of protostars could introduce an additional uncertainty of about $\pm$0.2 dex to the log$(N_{\textrm{tot}}^{\textrm{c1}})$, which results in an additional uncertainty of about 0.04 dex to the log(SFR). 

Assuming a mean mass of 0.5 $M_{\sun}$, the total mass of Class I populations can be obtained based on $N_{\textrm{tot}}^{\textrm{c1}}$.

\subsubsection{SFRs and SFEs}\label{sect:sfrsfe}

{We calculate two estimates for the SFRs of the GMFs based on their YSO content. The first reflects the average SFR during the Class I + Class II lifetimes, and we call this measure simply SFR. The second reflects the average SFR during the Class I lifetime only, and we call this measure the \emph{current} SFR, or cSFR. Both quantities are calculated by dividing the total mass of YSOs in respective classes by the relevant time-scale. We adopt 2 Myr \citep{evans09} and 0.54 Myr \citep{heiderman15} as the Class I$+$II and Class I lifetimes, respectively}.
{We also calculate SFE analogously with the SFR with} 
\begin{equation*}
\textrm{SFE} = \frac{M_{\textrm{YSO}}}{M_{\textrm{YSO}}+M_{\textrm{cloud}}},\end{equation*}
where $M_{\textrm{YSO}}$ is the total mass of {Class I + Class II} sources in each GMF.
{Similarly, we calculate the SFE of the dense gas with}
\begin{equation*}
\textrm{SFE}_{\textrm{dense}} = \frac{M_{\textrm{YSO}}}{M_{\textrm{YSO}}+M_{\textrm{dense}}}.
\end{equation*}

\mz{There are {11} GMFs that have not enough Class I or Class II sources to estimate the total mass of YSOs, of which {10} GMFs are located at the distances of $>8$\,kpc and one GMF (GMF\,10, i.e., F24 in Table~\ref{table1}) has only four Class I sources with the luminosities of $>L_{cut}$ although it is located at a distance of 4.7\,kpc}. We also note that the dense gas mass fraction ($M_{\textrm{dense}}/M_{\textrm{cloud}}$) of GMF\,10 is only $\sim$0.4\%. Therefore, we finally obtain the SFRs and SFEs for only 46 GMFs. The uncertainties of SFRs and SFEs are from the uncertainties of total mass of YSOs. Note that we obtain the probability distributions of SFRs and SFEs with {a} Monte Carlo method {and use these distributions to define the uncertainties} (see Appendix~\ref{ap4}). The uncertainty introduced due to the variations of protostellar LF during estimating the total mass of Class I sources is not included. 
Table~\ref{table2} shows the SFR and SFE obtained {for} each GMF. {Note that the uncertainties listed in Table~\ref{table2} do not include the distance uncertainty (see Appendix~\ref{ap4} for details).}

{Even though} we have excluded some {contamination from the identified YSO population} 
{(see Section~\ref{sect:contamination}), YSOs on the background of the cloud, possibly associated with other molecular clouds along the line of sight, may still contaminate the total YSO estimate}. 
{Specifically, }the SFRs and SFEs could be overestimated for the GMFs that 
{exhibit} other {CO} velocity components
{, and thus possibly other star-froming regions, along the} line of sight. We checked each GMF and 
{flagged the ones} 
that are overlapped with other GMFs (Table~\ref{table2}). We also examined the $^{13}$CO average spectrum of each GMF and calculated the flux ratio of GMF to other velocity components. High flux ratio implies that the GMF is the dominating component along the line of sight. We use this ratio as a parameter ($R$) 
{as one measure of} the reliability of obtained SFR and SFE. Note that we did not obtain the $R$ value for Nessie due to the lack of sensitive $^{13}$CO data.
{Figures~\ref{fig:CFG47}-\ref{fig:GMF341} (right panels) show the $^{13}$CO average spectra for seven GMFs with $R>$~1.}

All above calculations of SFRs and SFEs are based on the assumption that the identified YSOs are individual objects. Obviously, the presence of unresolved clustering 
{are not accounted for in} our estimations. {However, }\citet{morales17} analyzed near-infrared UKIDSS observations of a sample of GLIMPSE-selected YSO candidates in the Galactic plane \citep{robitaille08} and found that $\sim$87\%~of YSO candidates have only one dominant UKIDSS counterpart. {Based on this,} \citet{morales17} suggested that the YSO clustering within the GLIMPSE resolution is not important for the GLIMPSE-selected YSOs with intermediate to high masses{, and hence,} no significant corrections are needed for estimates of the SFR based on the assumption that the GLIMPSE YSOs are individual objects. In our YSO sample, $\sim$80\%~are identified based on the \textit{Spitzer} data while $\sim$20\% are selected using AllWISE and \textit{Herschel} Hi-GAL data with the spatial resolution of $\sim$6\arcsec. To estimate the {possible} influence of YSO clustering on 
{our SFRs}, we 
{inspect a worst-case scenario in which} 
 all YSOs identified using AllWISE and \textit{Herschel} data are in fact YSO clusters. \citet{gutermuth09} investigated 36 nearby YSO clusters and found that a typical young cluster has 26 members with a surface density of 60 pc$^{-2}$. Thus each AllWISE or Hi-GAL selected YSO can be decomposed into several members within the resolution of 6\arcsec~based on its distance, and then `intrinsic' number of YSOs can be obtained 
 {for each} GMF. We found that the YSO clustering has small influence on the SFR estimates for the GMFs located at the distances of $<$ 4.5 kpc. However, for the GMFs with the distances of $>$ 4.5 kpc, we could significantly underestimate the SFRs by a factor of 1.5-7. For the GMFs with the distance of $<$5.5 kpc, we could underestimate the SFRs within a factor of up to 2.

\section{Results}\label{sect:results}

\subsection{Gas properties of the GMFs}\label{sect:gasper}

Figure~\ref{sample-distribution} shows the spatial distribution of {57} GMFs. We marked their positions on the $^{13}$CO integrated intensity images of the GRS and ThrUMMS survey data with the red ellipses.

\begin{figure*}
\includegraphics[width=1.0\linewidth]{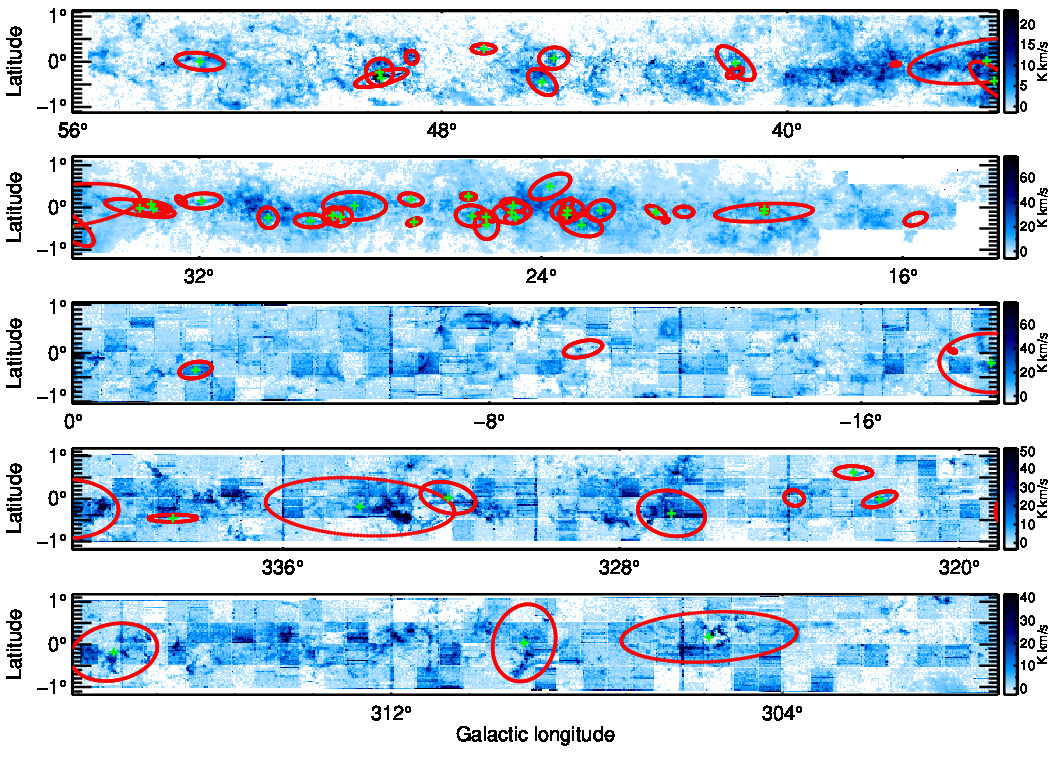}
\caption{The spatial distribution of {57} Galactic GMFs. The backgrounds are $^{13}$CO intensity maps of the GRS and ThrUMMS data, integrated over the entire observed velocity range. The red ellipses mark the positions of {57} GMFs. We obtained the star formation rates of 46 GMFs (see Sect.~\ref{sect:sfr} for details) that are labeled with green pluses.}
\label{sample-distribution}
\end{figure*}

Figure~\ref{sample-mcpara} shows the distributions of physical parameters of {57} GMFs. The median length of {57} GMFs is {67} pc, with a minimum of {17} pc and a maximum of {268} pc, {where the length of GMFs is defined as the major axis obtained through fitting the column density map of a GMF with an ellipse (see Section~\ref{sect:sample})}. 
For the samples originally identified by \citet{sample-gmf}, \citet{sample-wang15}, and \citet{sample-agmf} based on the same GRS and ThrUMMS data, the lengths estimated using our method described in Section~\ref{sect:sample} are similar as that 
{given} by the corresponding 
papers. However, for the samples originally identified by \citet{sample-li16} and \citet{sample-wang16} based on the dense gas tracers, the lengths estimated using $^{13}$CO data with our method are significantly longer
. \citet{sample-bones} mainly focus on the dense part of the giant filament and the lengths 
{given} by them are usually shorter than the lengths obtained by us. 
The heliocentric distances of {57} GMFs are in the range of [2,13] kpc, with a median value of {5.1} kpc. 35 ($\sim${61}\%) of {57} 
{GMFs} are located at the distances of $<$ 5.5 kpc and 22 ($\sim${39}\%) are located at the distances of $<$ 4.5 kpc.

{To provide a rough description of the GMF shapes, we compute the ratio of the major to minor axis of the ellipses that were fit to the column density maps of the GMFs and obtain the values are between 1-6.7 with the median value of $\sim$2. We call this value an `aspect ratio', however, we immediately emphasize that an ellipsoid is not a good approximation of the morphology of the GMFs; a cylindric model that identifies and traces the crest of the filament would be more accurate. Such definition is beyond this paper; our goal is only to provide a simple descriptive statistic. As a result of this definition, our aspect ratios are smaller than given for filaments in the literature by works that adopt more refined techniques to describe the filamentary morphology \citep[e.g.,][]{sample-wang15,sample-wang16}. Some works in literature studying molecular clouds in general have adopted a description of molecular cloud shapes that is similar to ours. For example, \citet{gmc} uses a similar technique to describe shapes of giant molecular clouds (GMCs). They find an average aspect ratio of $\sim$1.5 for all GMCs, a value which is slightly smaller than the median aspect ratio of our GMFs. An interpretation of this is that our aspect ratio reflects mostly the elongation of the relatively low column density GMC inside which the denser gas shows a high-aspect ratio, filamentary morphology.}

\begin{figure*}
\includegraphics[width=1.0\linewidth]{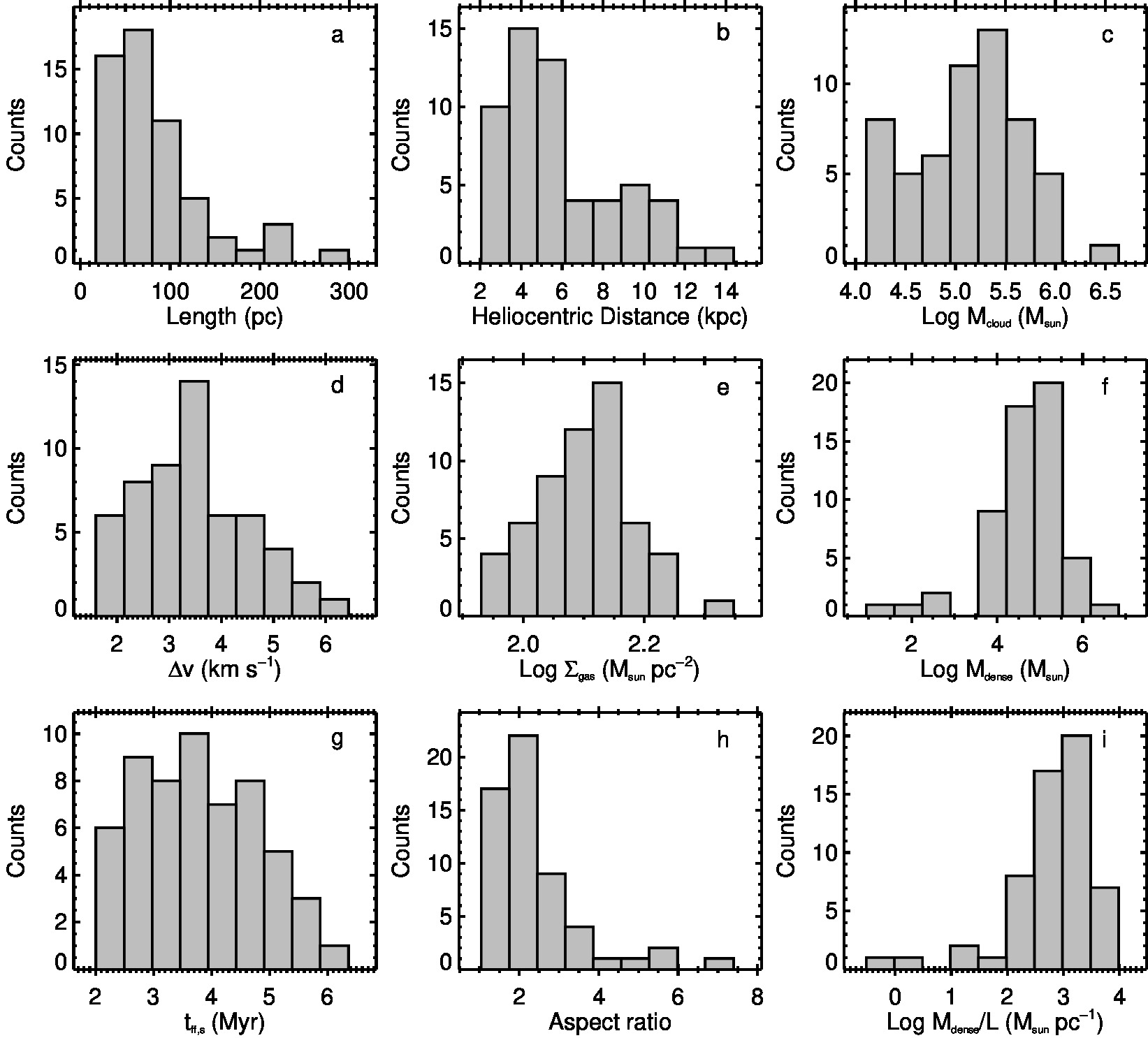}
\caption{Distributions of (a) filament length; (b) heliocentric distance; (c) logarithm of cloud mass $M_{\textrm{cloud}}$; (d) line width $\Delta v$; (e) logarithm of gas surface density $\Sigma_{\textrm{gas}}$; (f) logarithm of dense gas mass $M_{\textrm{dense}}$; {(g) free-fall time obtained based on spherical morphology; (h) {aspect ratio}; and (i) logarithm of dense gas mass per unit length}.}
\label{sample-mcpara}
\end{figure*}

We obtained the average $^{13}$CO line width ($\Delta$v) of each GMF based on its moment 2 map (see Appendix~\ref{ap2}). Table~\ref{table1} lists the values of $\Delta v$ for {56} GMFs (except Nessie). These values are in the range of [1.6, 5.9] km s$^{-1}$ with a median value of {3.4} km s$^{-1}$. \citet{evans14} compiled a list of linewidths for $\sim$29 nearby molecular clouds and they obtained the median linewidth of 1.5 km s$^{-1}$, with a minimum of 0.8 km s$^{-1}$ and a maximum of 3 km s$^{-1}$. The value of {3.4} km s$^{-1}$ is almost twice of the median linewidth of nearby low-mass star-forming regions, however, similar to the median value of $\sim$4 km s$^{-1}$ estimated in the IRDCs \citep{simon06b,du08}. This is not surprising because most of GMFs are also identified based on the IRDCs.

The median values of $M_{\textrm{cloud}}$ and $\Sigma_{\textrm{gas}}$ are {1.5}$\times$10$^5$ M$_{\sun}$ and {128} M$_{\sun}$ pc$^{-2}$, respectively, which are slightly lower than the typical values of the local giant molecular clouds \citep[$\sim$2$\times$10$^5$ M$_{\sun}$ and $\sim$170 M$_{\sun}$ pc$^{-2}$;][]{solomon87}. Note that we use the extinction contour of $A_V =$ 3 mag to estimate the cloud mass, which is inclined to underestimate the total mass of the GMFs.

In Fig.~\ref{larson}, we {show} the line widths ($\Delta v$) versus the sizes ($R$) and masses ($M_{\textrm{cloud}}$). Here the size is the equivalent radius obtained using the area of the GMFs. {Note that the error bars in this and the subsequent figures include distance errors (see Appendix~\ref{ap4} for details).} {We first tested the relationships for correlations without considering the uncertainties and find significant correlations between $\Delta v$ and $R$, $M_{\textrm{cloud}}$ with Pearson $r=$ 0.67 for $\Delta v$ versus R and 0.71 for $\Delta v$ versus $M_{\textrm{cloud}}$. As a comparison, a significant (3$\sigma$) correlation requires Pearson $|r|>$ 3/$\sqrt{N_s-1}$ where $N_s$ is the sample size (56 here) \citep{vut16}. The direct least-square linear fitting without considering uncertainties gives the slopes of 0.35$\pm$0.05 for $\Delta v$ versus R and 0.17$\pm$0.02 for $\Delta v$ versus $M_{\textrm{cloud}}$. 
We then calculate the linear regression coefficients using the Bayesian method developed by \citet{kelly07} that allows the inclusion of uncertainties along both axes during the fitting process. Note that the fitting routine requires symmetric uncertainties; for asymmetric uncertainties we simply use the maximum of the two asymmetric errors. Based on the posterior distributions as shown in the insert plots of Fig.~\ref{larson}, the 95\%~confidence intervals of correlation coefficients are about [0.65, 1.0] for $\Delta v$ versus $R$ and [0.61, 1.0] for $\Delta v$ versus $M_{\textrm{cloud}}$, respectively. These indicate significant correlations between $\Delta v$ and $R$, $M_{\textrm{cloud}}$. The posterior median estimates of the slopes are 0.42$\pm$0.10 for $\Delta v$ versus $R$ and 0.22$\pm$0.06 for $\Delta v$ versus $M_{\textrm{cloud}}$, individually, which are slightly larger than the slopes obtained without considering uncertainties. These values are also consistent with the canonical relationships obtained by \citet{larson81} and \citet{solomon87}.}

\begin{figure*}
\includegraphics[width=1.0\linewidth]{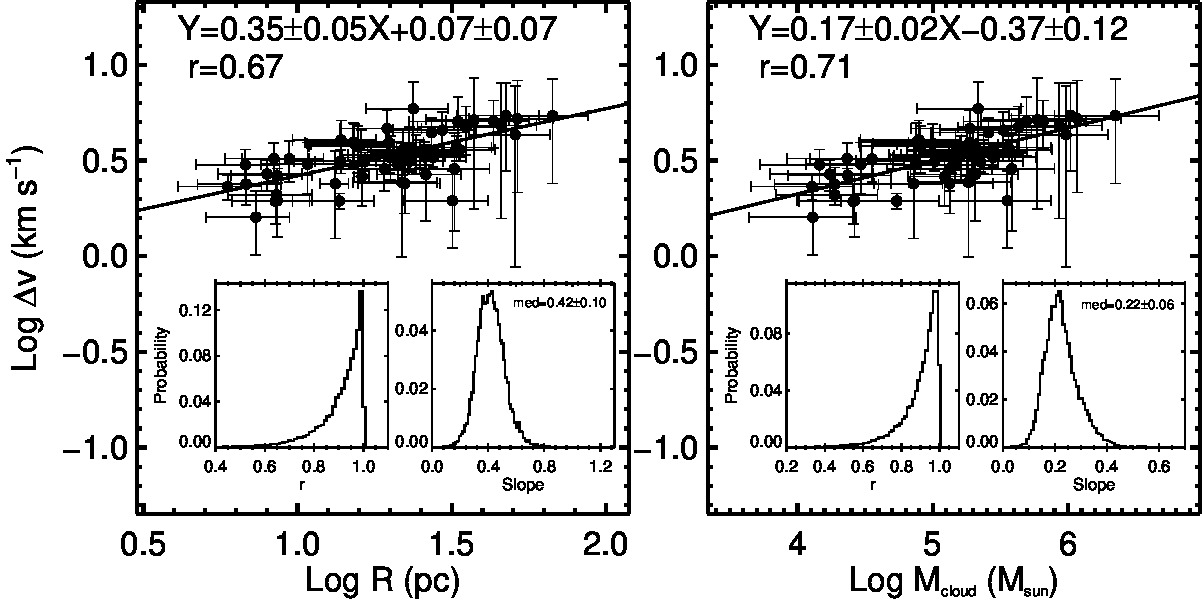}
\caption{$^{13}$CO line widths vs. sizes (\textit{left}) and masses (\textit{right}) of the GMFs in log-log plane. The solid lines show the best linear fittings to the points {without considering the uncertainties}. The correlation coefficients and the fitting results {that are obtained without considering the uncertainties} are also marked in {the top region of} panels. {The insert small plots in each panel show the probability distributions of correlation coefficients and fitting slopes that are obtained after including uncertainties on both axes using the Bayesian linear regression method developed by \citet{kelly07}.}}
\label{larson}
\end{figure*}
\subsection{\textcolor{black}{The observed YSO content of GMFs}}

Table~\ref{table2} lists the number of Class I and Class II sources detected in each GMF. In total, we identified {18\,395} 
YSOs (ClassI$+$II) in 46 GMFs 
and {$\sim$38\% of the YSOs} are Class I sources.

The spatial distribution of YSOs is a powerful diagnostic
of their formation and early evolution \citep{kraus08}. {We next consider shortly the spatial distribution of the YSOs. We only discuss the Class I sources, because they are still possibly close to their formation sites due to their short lifetime} \citep{lada13}. The investigation of the clustering of stars at the time of their formation is helpful to establish the connections between the star formation process and the molecular cloud structures. 

The degree of clustering of protostars can be quantified with the two-point correlation function (TPCF). In this part, 
we calculate the TPCF using the estimator presented by \citet{landy93}, following the method suggested by \citet[][see their Section 3.3 for details]{kai17}. The TPCF, $\xi (r)$, describes the excess probability distribution of the sources with different separations ($r$) comparing to a random distribution. The $\xi(r) =$ 0 indicates that the distribution can not distinguish from a random distribution while $\xi(r) >$ 0 and $\xi(r) <$ 0 indicate excess and deficit of separations, individually. We only use the Class I sources with the luminosities higher than the luminosity completeness ($L_{cut}$) that is obtained in Section~\ref{sect:massc1} to calculate the TPCF in each GMF. We finally obtained the TPCFs of Class I sources in {six} `reliable' GMFs that have $R >$ 1, $N_{\textrm{obs}}^{\textrm{c1}} (> L_{cut}) >$ 50, distance $<$ 4.5 kpc, and are not overlapped with other GMFs based on in total 1683 Class I sources. {Figures~\ref{fig:GMF54}$-$
\ref{fig:GMF341} 
show the spatial distributions of Class I sources in these six GMFs and Fig.~\ref{tpcf} shows the TPCFs.} 
\begin{figure*}
\includegraphics[width=1.0\linewidth]{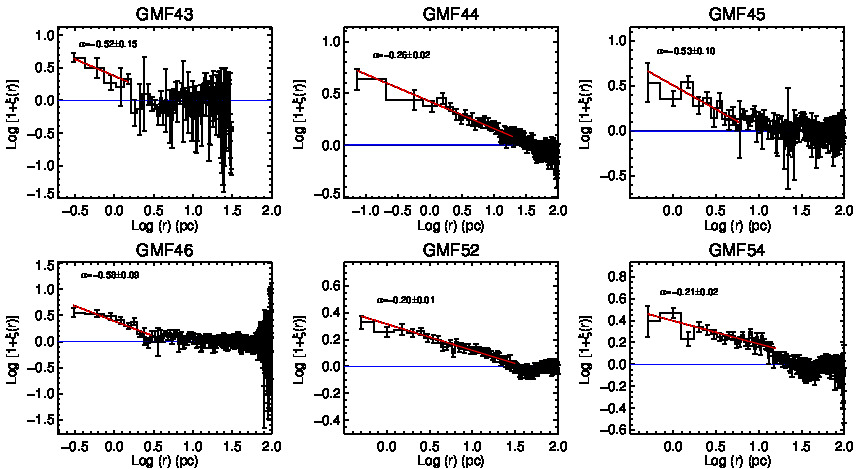}
\caption{Two-point correlation functions for the Class I sources in 6 GMFs. The title on each panel corresponds to the ID number of Table~\ref{table2}. The red solid lines show the linear fits to the roughly straight line part of the data and the fitting slopes are labeled in each panel.}
\label{tpcf}
\end{figure*}

The TPCFs of Class I sources in these six GMFs show 1) obvious clustering of protostars at the small stellar separations; 2) a monotonous decrease with stellar separations in the clustering regime. We try to fit the {linear} part of TPCFs with a single power law. The obtained power law indices ($\alpha$) are from -0.58 to {-0.20}.

The power law TPCFs of YSOs have been observed in some Galactic and extragalactic star-forming regions and are usually interpreted as hierarchical or fractal distributions that could be related to the hierarchical structures of the gas \citep{gomez93,larson95,simon97,kraus08,gou14,koenig14}. The power law index is also related to the 2D fractal dimension $D_2$ as $D_2 = \alpha + $2 \citep{larson95}.
Therefore, the single power law TPCFs in our samples can be also interpreted as fractal clustering of YSOs with the fractal dimensions of $D_2$$\sim$1.42$-${1.80}. We note that this range of $D_2$ is consistent with the range of $D_2\sim$1.2$-$1.9 that is obtained in the nearby star-forming regions \citep{nakajima98,alf11}. 
\subsection{\textcolor{black}{Star formation rates and efficiencies}}\label{sect:result-sfrsfe}

Table~\ref{table2} lists the SFRs and SFEs obtained {for} 46 GMFs. Due to the large uncertainties of SFR estimations for the distant GMFs (see Section~\ref{sect:sfrsfe}), in this part we only apply the statistical analysis to the 34 GMFs with the distances of $<$5.5 kpc. 

{Figure \ref{hist_sf} shows the combined probability distributions of SFR, SFE, and $\Sigma_{\textrm{SFR}}$ (see Appendix~\ref{ap4}) for these 34 GMFs.}
The median values of SFR, $\Sigma_{\textrm{SFR}}$, and SFE for these 34 GMFs are {630} M$_{\sun}$ Myr$^{-1}$, 0.62 M$_{\sun}$ Myr$^{-1}$ pc$^{-2}$, and 1\%, individually.

\begin{figure*}
\includegraphics[width=1.0\linewidth]{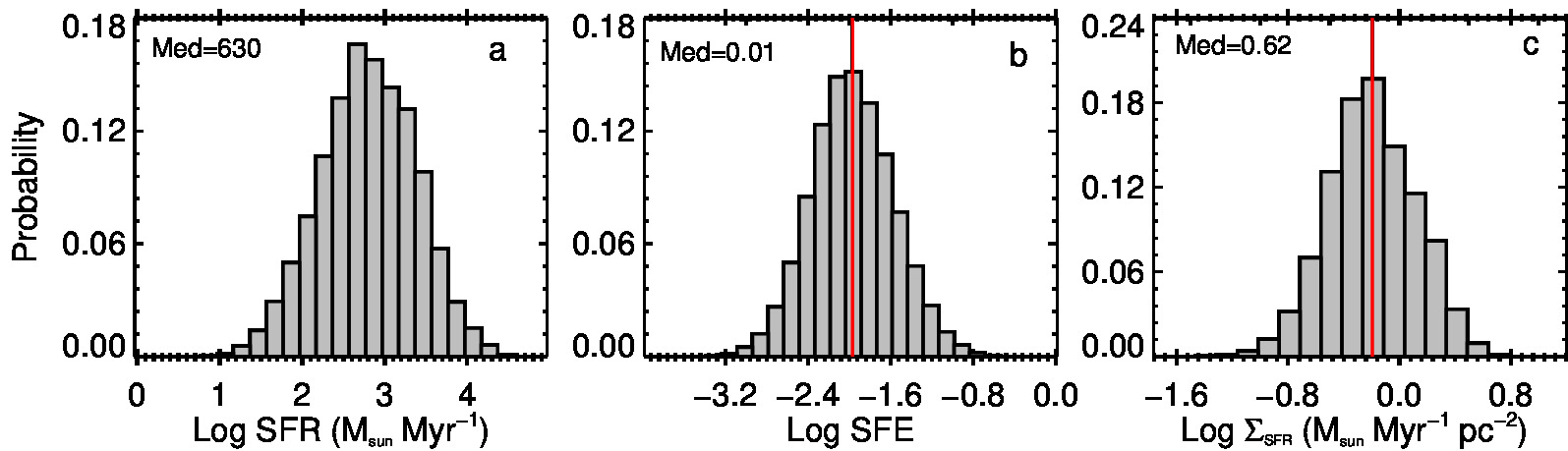}
\caption{{Probability distributions }of (a) star formation rate; (b) star formation efficiency; (c) surface density of star formation rate for 34 GMFs with the distances of $<$5.5 kpc. {The red vertical lines show the median values of SFE and $\Sigma_{\textrm{SFR}}$} that are re-estimated with Class I+Flat+II sources in the nearby star-forming regions (see the text for details).}
\label{hist_sf}
\end{figure*}

\citet{evans09} obtained the $\Sigma_{\textrm{SFR}}$ in five nearby star-forming regions and found that the values of $\Sigma_{\textrm{SFR}}$ are in the range of 0.65$-$3.2 M$_{\sun}$ Myr$^{-1}$ pc$^{-2}$, with a median value of 1.3 M$_{\sun}$ Myr$^{-1}$ pc$^{-2}$. \citet{evans14} re-investigated the star formation rates in 29 nearby star-forming clouds and obtained $\Sigma_{\textrm{SFR}}$ in the range of 0.06$-$3.88 M$_{\sun}$ Myr$^{-1}$ pc$^{-2}$ with a median value\footnote{We have excluded the nearby star-forming clouds with $\Sigma_{\textrm{SFR}} =$ 0 during the statistics. Thus this value of 0.9 is significantly higher than the value of 0.48 offered in Table 1 of \citet{evans14}.} of 0.9 M$_{\sun}$ Myr$^{-1}$ pc$^{-2}$. 
 The SFEs obtained in five nearby star-forming clouds by \citet{evans09} are $\sim$3$-$6\%, with a median value of $\sim$5\%. Based on a larger sample, \citet{evans14} found that SFEs in nearby star-forming regions are in the range of [0.2\%, 7.8\%], with a median value of 1.8\%. 
At a first glance, the $\Sigma_{\textrm{SFR}}$ and SFEs of the GMFs are on average significantly lower than that of the molecular clouds in the Gould Belt. However, we must remind that there is a systematic difference between the SFRs obtained in the GMFs and nearby star-forming regions due to the different methods that are used to estimate the SFRs.

\citet{evans09} and \citet{evans14} use all YSOs detected in the nearby star-forming regions 
to estimate the SFRs, including Class I, Flat, Class II, and Class III sources. However, due to the lack of strong infrared excess, the Class III sources could be incomplete in some clouds. Moreover, the optical spectroscopical observations in Serpens and Lupus found that $\sim$40-100\% of Class III sources are actually background AGB stars \citep{oliveria09,remero12}. Considering the high contamination fraction in the Galactic plane, we only use Class I+Flat and Class II sources detected in the GMFs to estimate the total number of YSOs. 

To compare the GMFs with nearby star-forming regions, we re-calculate the SFRs of the nearby clouds based on the latest version of YSO samples in the \textit{c2d} and Gould Belt surveys presented by \citet{dunham15}. We obtain the extinction map for each nearby molecular cloud with 2MASS \citep{2mass} point source catalog using PNICER \citep{pnicer} that is 
{an improved version of} the well-known NICER and NICEST algorithms \citep{nice,nicer,nicest}{, providing a} suite for measuring extinction 
{towards} individual sources and producing extinction maps with an unsupervised machine learning algorithm. {The uncertainties of extinction maps are mainly from the 2MASS photometric uncertainties and the scatters of intrinsic colors.} We use the extinction contour of $A_V=$ 3 mag to define the boundaries of nearby molecular clouds {and the gas mass and dense gas mass are obtained through integrating down to $A_V=$~3, and 7\,mag in 2MASS extinction maps. The uncertainties of mass estimates are mainly from the uncertainties of extinction maps and distances.} SFRs {of nearby clouds are estimated} with Class I$+$Flat$+$II sources within the cloud boundaries based on the relation of 
$\textrm{SFR}=0.25N(\textrm{YSOs}) ~\textrm{M}_{\sun}~\textrm{Myr}^{-1}.$ {The uncertainties of SFRs are based on counting statistics. 
We use the method suggested by \citet{evans14} to estimate the uncertainties of the parameters of nearby clouds. The details about this method can be found in the Appendix B of \citet{evans14}.} 

Table~\ref{table3} lists the SFRs obtained in 24 nearby molecular clouds. 
The median values of SFEs and $\Sigma_{\textrm{SFR}}$ 
in the nearby star-forming regions are 1.1\%~and 0.64 M$_{\sun}$ Myr$^{-1}$ pc$^{-2}$, respectively, {which are also marked with red vertical lines in Fig.~\ref{hist_sf}.} \mz{Therefore, on average, the SFE and $\Sigma_{\textrm{SFR}}$ in the GMFs are similar to that in nearby star-forming regions. However, there could still be systematic differences for SFE because we use $^{13}$CO-based column density maps and extinction maps to estimate $M_{\textrm{cloud}}$ 
for the GMFs and nearby star-forming regions, respectively. \citet{ripple13} compared the column density distribution derived using dust extinction with that derived using $^{13}$CO emission in the Orion molecular cloud. They found that $^{13}$CO provided a reliable tracer of H$_2$ mass within the area with strong self-shielding ($A_V>$~3\,mag).} 
{However, \citet{ripple13} used $^{12}$CO data to estimate the excitation temperature and then combined the excitation temperature measurements with the $^{13}$CO emission to derive the column densities{, which means that they used a slightly more refined method compared to ours (described in Section~\ref{sect:sample}). Therefore, it is not clear if their conclusion applies as such to the masses we derive. We cannot totally exclude the caveat of systematic differences in the M$_\mathrm{cloud}$ and SFE values for GMFs and nearby clouds.}}

Most of the GMFs are associated with the spiral arms of the Milky Way \citep{sample-wang15,sample-bones,sample-agmf,sample-li16,sample-wang16}. The nearby star-forming clouds appear to be located in an inter-arm branch between two major spiral arms of the Galaxy \citep{xu13,molinari14}. Whether the spiral arms trigger the star formation is still under debate. Based on the extragalactic observations, some studies suggested that there are significant enhancements of star formation in the spiral arms \citep{seigar02,silva12}, but some other studies found no significant difference for star formation efficiencies in the spiral arm and inter-arm regions \citep{foyle10,wall16}. 
{In the Milky Way, several studies also }found no significant difference of star formation properties in Galactic spiral arm and inter-arm regions \citep{eden13,ragan16}. {Our results favor the interpretation that the star formation activities in the spiral arm and inter-arm regions are similar.}

Recently, \citet{simugf} investigated the evolution of GMFs using a high-resolution section of a spiral galaxy simulation. They found that the GMFs are only detected in the inter-arm regions and are pressure confined rather than gravitationally bounded as a whole. In their simulation, when the GMFs enter into the spiral arm regions, they begin to break into short filament sections due to the stellar feedback from star formation or differential force from the gravitational potential. \citet{simugf} did not find a significant increase of star formation events once the GMFs enter the spiral arm, at least in their selected GMF sample. Based on this scenario, we {would} expect the similar star formation activity between the GMFs in spiral arm regions and the molecular clouds in inter-arm regions.
\section{\textcolor{black}{Discussion}}
\label{sect:discussion}

\subsection{Star formation relations}\label{sflaw}

{
We next consider the relationships between the star formation and gas properties in the GMFs and compare them with those from the literature.} We only include in the analysis the 34 GMFs with distances less than 5.5 kpc. 
{\mz{{In some analyses below, we consider two separate samples: GMFs and \comobj. However, our conclusions are mainly based on GMFs alone, because of the possibility that there may be systematic differences between the physical parameters, such as gas mass and free-fall time, of the GMFs and nearby clouds. Note that to minimize this possibility, we have re-derived the parameters that are listed in Table~\ref{table3} for the nearby clouds, following a procedure similar to the GMFs (see Section~\ref{sect:result-sfrsfe} for details).}} {To search for correlations between the parameters, we always use two methods: 1) the Pearson correlation coefficient  (does not considering uncertainties); 2) Bayesian linear regression method that makes use of uncertainties along both axes \citep{kelly07}.}}



Figure~\ref{sflaw1} {(panel a)} shows the {relationship between} $\Sigma_{\textrm{SFR}}$ and $\Sigma_{\textrm{gas}}$ for the GMFs and nearby star-forming clouds. 
The {figure also shows the} relations {derived} by \citet{kennicutt98} and \citet{bigiel08}. 
{The GMFs fall slightly above these relations \citep[as do the nearby star-forming regions, cf.,][]{heiderman10}}. 
%
{Without considering uncertainties}, we find Pearson $r$ values of {0.08} and {0.40} for GMFs and \comobj, respectively. {Note that a significant correlation needs $r>$~0.52 for GMFs ($N_s=$~34) and $r>$~0.40 for \comobj~($N_s=$~58). If considering uncertainties, the Bayesian linear regression method \citep{kelly07} can give the posterior distributions of correlation coefficients as shown in the insert plots of Fig.~\ref{sflaw1}. The 95\%~confidential intervals of correlation coefficients are [-0.9, 0.9] for GMFs and [0.11, 0.77] with a median value of 0.51 for \comobj. 
Thus, there are no convincing correlations between $\Sigma_{\textrm{SFR}}$ and $\Sigma_{\textrm{gas}}$ for GMFs, but there could be a weak correlation for \comobj.}

\begin{figure*}
\includegraphics[width=1.0\linewidth]{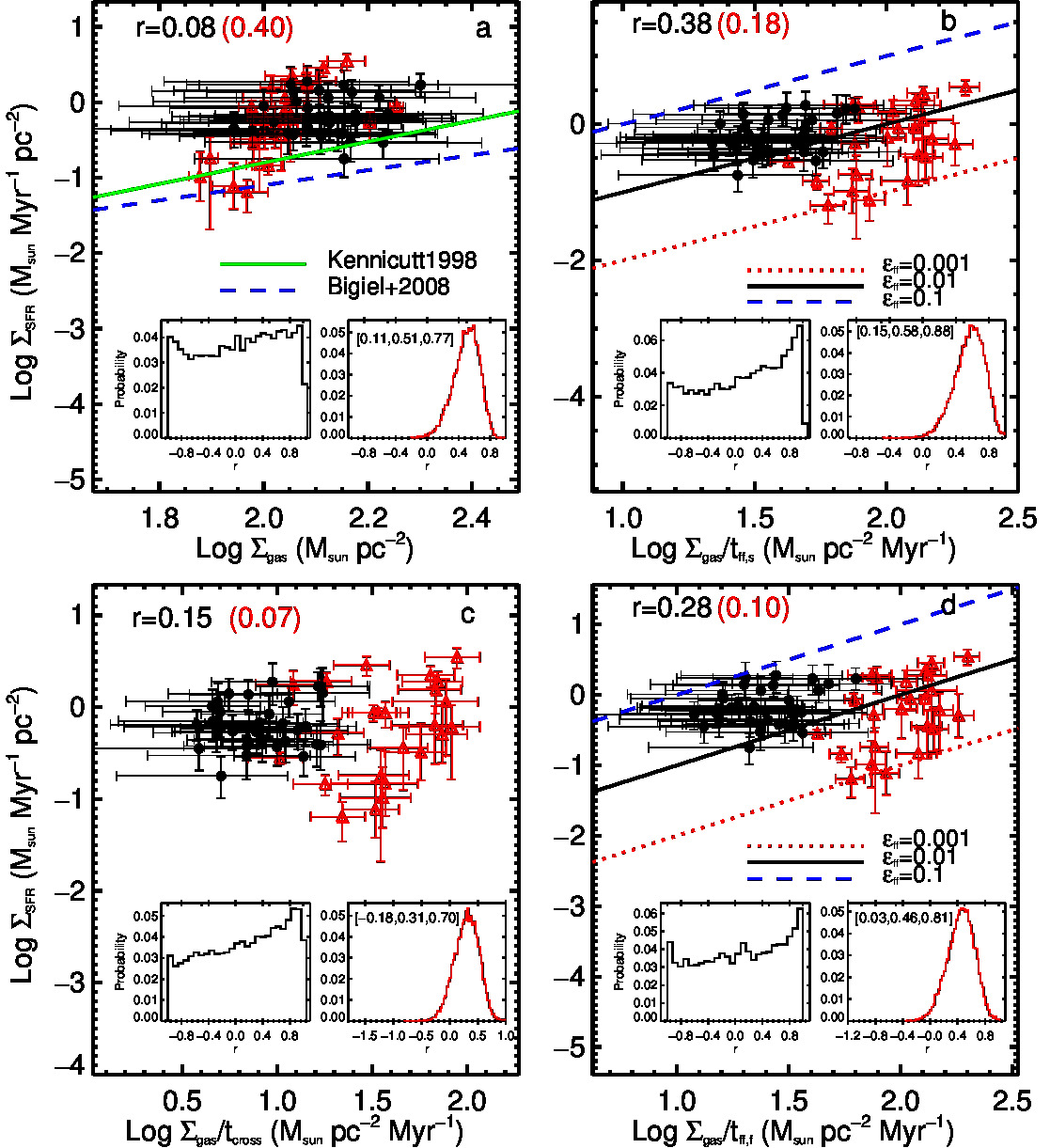}
\caption{Surface density of SFR, $\Sigma_{\textrm{SFR}}$, as a function of (a) $\Sigma_{\textrm{gas}}$, (b) $\Sigma_{\textrm{gas}}$/$t_{\textrm{ff,s}}$, (c) $\Sigma_{\textrm{gas}}$/$t_{\textrm{cross}}$, and (d) $\Sigma_{\textrm{gas}}$/$t_{\textrm{ff,f}}$ for the Galactic GMFs (black filled circles) and the nearby star-forming regions (red triangles). The lines in (a) correspond to the star formation relations suggested by \citet[][green solid]{kennicutt98} and \citet[][blue dashed]{bigiel08}. The lines in (b) and (d) correspond to star formation law suggested by \citet{krumholz12} with $\epsilon_{\textrm{ff}} =$ 0.001, 0.01, and 0.1, respectively. The {Pearson} correlation coefficients between the different sets of parameters {obtained without considering uncertainties} are marked on top region of each panel and the black fonts are for GMFs while the red fonts are for \comobj. {The insert plots in each panel show probability distributions of correlation coefficients ($r$) obtained with Bayesian linear regression method by \citet{kelly07} for GMFs (black line) and \comobj~(red line).}}
\label{sflaw1}
\end{figure*}

{We next discuss the relation between star formation and gas surface density calculated per various time-scales.} \citet{krumholz12} {considered} a volumetric star formation law, $\rho(\textrm{SFR}) \propto \rho_{\textrm{gas}}^x$, in which the SFR is proportional to the gas mass per free-fall time (i.e., $x=$1.5). Because the volume density is difficult to measure, they only consider the projected quantities:
\begin{equation*}
\Sigma_{\textrm{SFR}} = \epsilon_{\textrm{ff}}\frac{\Sigma_{\textrm{gas}}}{t_{\textrm{ff}}}
\end{equation*}
where $\epsilon_{\textrm{ff}}$ is a dimensionless measure of SFR (i.e., SFR$_{\textrm{ff}}$). \citet{krumholz12} found that a set of SFR measurements, from nearby star-forming regions to distant sub-millimeter galaxies, can be well fitted by this relation when $\epsilon_{\textrm{ff}}\sim$1\%. We plot $\Sigma_{\textrm{SFR}}$ versus $\Sigma_{\textrm{gas}}/t_{\textrm{ff,s}}$ and $\Sigma_{\textrm{gas}}/t_{\textrm{ff,f}}$ in Fig.~\ref{sflaw1} {(panels b and d)} for GMFs and \comobj, where $t_{\textrm{ff,s}}$ and $t_{\textrm{ff,f}}$ are free-fall time calculated based on spherical and filamentary morphology assumptions (see Appendix~\ref{ap2}), respectively.
Most of GMFs and nearby star-forming regions are located between $\epsilon_{\textrm{ff}} =$ 0.001 and 0.1. The mean values of SFR$_{\textrm{ff,s}}$ and SFR$_{\textrm{ff,f}}$ are {0.02} and {0.03} for the GMFs, respectively. 
These values are higher than the 
value of 0.01 
obtained by \citet{krumholz12}. 
We find no {convincing} correlation between $\Sigma_{\textrm{SFR}}$ and $\Sigma_{\textrm{gas}}$/$t_{\textrm{ff}}$ for GMFs. {However, there could be also a weak correlation between $\Sigma_{\textrm{SFR}}$ and $\Sigma_{\textrm{gas}}$/$t_{\textrm{ff}}$ for \comobj.} 

The crossing time of molecular clouds can {be hypothesized to be a relevant} timescale, replacing the $t_{\textrm{ff}}$ in the volumetric star formation law. \citet{evans14} defined $t_{\textrm{cross}}$ $=$ $\frac{size}{\langle \Delta v \rangle}$, where $size$ is the equivalent diameter and $\langle \Delta v \rangle$ is the mean linewidth of the molecular clouds. They found no {significant} correlation between $\Sigma_{\textrm{SFR}}$ and $\Sigma_{\textrm{gas}}$/$t_{\textrm{cross}}$ for the nearby star-forming regions. Figure~\ref{sflaw1} {(panel c)} shows the relation between $\Sigma_{\textrm{SFR}}$ and $\Sigma_{\textrm{gas}}$/$t_{\textrm{cross}}$ for the GMFs and nearby star-forming regions. We do not find a significant correlation between $\Sigma_{\textrm{SFR}}$ and $\Sigma_{\textrm{gas}}$/$t_{\textrm{cross}}$ for GMFs or \comobj. 

{Here we must note as a caveat that although we do not detect convincing correlations between $\Sigma_{\textrm{SFR}}$ and $\Sigma_{\textrm{gas}}$, $\Sigma_{\textrm{gas}}$/$t_{\textrm{ff}}$ or $\Sigma_{\textrm{gas}}$/$t_{\textrm{cross}}$ for GMFs, it does not mean that there is no possibility of strong correlations between these quantities. If we see the posterior distributions of correlation coefficients obtained with Bayesian linear regression method \citep{kelly07} for GMFs in Fig.~\ref{sflaw1} (insert plots), they are close to a uniform distribution rather than a distribution peaked at zero. In other words, the uncertainties of $\Sigma_{\textrm{SFR}}$, $\Sigma_{\textrm{gas}}$, $\Sigma_{\textrm{gas}}$/$t_{\textrm{ff}}$, and $\Sigma_{\textrm{gas}}$/$t_{\textrm{cross}}$ in GMFs are too large to decide whether there are correlations between them.} 

Another star formation relation {that highlights the role of \emph{dense gas}} has been developed{. This relation originates from the works of} \citet{solomon88} and \citet{gao04}, based on the investigation of galaxies, which suggests that the SFR traced by infrared luminosity is tightly correlated with dense gas traced by HCN luminosity. \citet{wu05} found that this relation can be extended to the Galactic dense cores, indicating a global star-forming law from local star-forming regions to the distant galaxies. \citet{kai09} showed that the Solar neighborhood clouds with a higher number of YSOs systematically have a higher relative amount of gas at high column densities. \citet{lada10} investigated the star formation activity in {several} nearby clouds and found that {the} SFR estimated with the YSO counting method {correlates best with specifically} the mass {at high column densities} ($A_{K} >$ 0.8 mag), which is agreement with the star formation relation suggested by \citet{solomon88}, \citet{gao04}, and \citet{wu05}. {\citet{shimajiri17} found that this linear SFR$-M_{\textrm{dense}}$ relation obtained in the nearby clouds can be also extended to the external galaxies after improving the dense gas mass estimates of external galaxies with a new HCN conversion factor, which indicates a quasi-universal star formation efficiency in the dense molecular gas of galaxies on a
wide range of scales.} \citet{lada12} argued that their star formation law is only compatible with the above volumetric law \citep{krumholz12} if $x=$1 and $\rho_{\textrm{gas}} > \rho_{\textrm{th}}$, where $\rho_{\textrm{th}}$ is a volume density threshold that roughly corresponds to the surface density threshold of $A_{K} =$ 0.8 mag. \citet{kai14} showed that also the volume density of gas follows these relations, i.e., the relative amount of gas at high volume densities correlates with the SFR of the clouds; this directly evidences a volumetric star formation law and the impact of dense gas for star formation.  

Our results enable extending the above analyses from the Solar neighborhood to a wider, Galactic context. {{As a caveat, we again note that we derive the dense gas mass by integrating the $^{13}$CO-based extinction maps down to $A_V=$~7\,mag. However, it is possible that $^{13}$CO is not a good tracer of dense gas} due to the optical depth effects and depletion \citep{pineda08,goodman09}. Compared with the mass estimates in nearby clouds based on the near-infrared extinction mapping, we could significantly underestimate the dense gas mass for the GMFs. To {avoid problems due to this}, we only use GMFs to investigate the star formation relations rather than combine the GMFs and nearby clouds together in the subsequent part.}

Figure~\ref{sflaw2} shows the relation between SFR and $M_{\textrm{dense}}$ \mz{(panel a) and the relation between SFR and $M_{\textrm{cloud}}$ (panel b)} for the GMFs. 
There are strong correlations for SFR versus $M_{\textrm{dense}}$ and SFR versus $M_{\textrm{cloud}}$ for the GMFs. {With Bayesian linear regression method considering uncertainties \citep{kelly07}, we also obtain the probability distributions of intrinsic scatters ($\sigma$) for SFR$-M_{\textrm{dense}}$ and SFR$-M_{\textrm{cloud}}$ relations. The posterior median estimates of $\sigma$ are 0.12$\pm$0.07 for both SFR$-M_{\textrm{dense}}$ and SFR$-M_{\textrm{cloud}}$ relations. Thus, we do not find a better correlation of SFR with the dense gas mass than with the total cloud mass, as found by \citet{lada10}. This difference quite possibly originates from different definitions of the total gas mass; we use the level of $A_V=$~3\,mag to define boundaries of molecular clouds, while \citet{lada12} uses $\sim$1 mag. {Indeed, we re-analysed the nearby clouds using a higher threshold for total gas (see measurements in Table~\ref{table3}) and found significant correlations between SFR and $M_{\textrm{dense}, A_V>7}$, $M_{\textrm{cloud}, A_V>3}$. The intrinsic scatters are 0.37$\pm$0.08 for SFR$-M_{\textrm{dense}}$ relation and 0.46$\pm$0.10 for SFR$-M_{\textrm{cloud}}$. In conclusion, the SFR does not correlate better with dense gas mass than the total gas mass in nearby clouds either, if we use $A_V=$~3\,mag to define the boundary of the clouds.}} 

{It is possible that including information on the dense gas would improve the correlation between the $\Sigma_{\textrm{SFR}}$ and surface density of gas discussed earlier (Fig. \ref{sflaw1}, panel a). The SFR$-$$M_{\textrm{dense}}$ relation can be transformed into a relation in the $\Sigma_{\textrm{SFR}}-\Sigma_{\textrm{gas}}$ plane by introducing a parameter $f_{\textrm{DG}} = M_{\textrm{dense}}/M_{\textrm{cloud}}$ \citep{lada12}. Figure~\ref{sflaw2} (panel c) shows the relation between $\Sigma_{\textrm{SFR}}$ and $f_{\textrm{DG}}\Sigma_{\textrm{gas}}$ for the GMFs. 
{{The Pearson $r$ (does not consider uncertainties) is 0.18. Using the Bayesian linear regression method by \citet{kelly07}}, the 95\%~confidential interval of the correlation coefficient is [-0.4, 0.9] with a median value of 0.33. Thus we find no significant}
correlation between $\Sigma_{\textrm{SFR}}$ and $f_{\textrm{DG}}\Sigma_{\textrm{gas}}$ for {GMFs}.} {However, due to the large uncertainties we can not exclude the possibility that there is a correlation between $\Sigma_{\textrm{SFR}}$ and $f_{\textrm{DG}}\Sigma_{\textrm{gas}}$ for {GMFs}.}



\begin{figure*}
\includegraphics[width=1.0\linewidth]{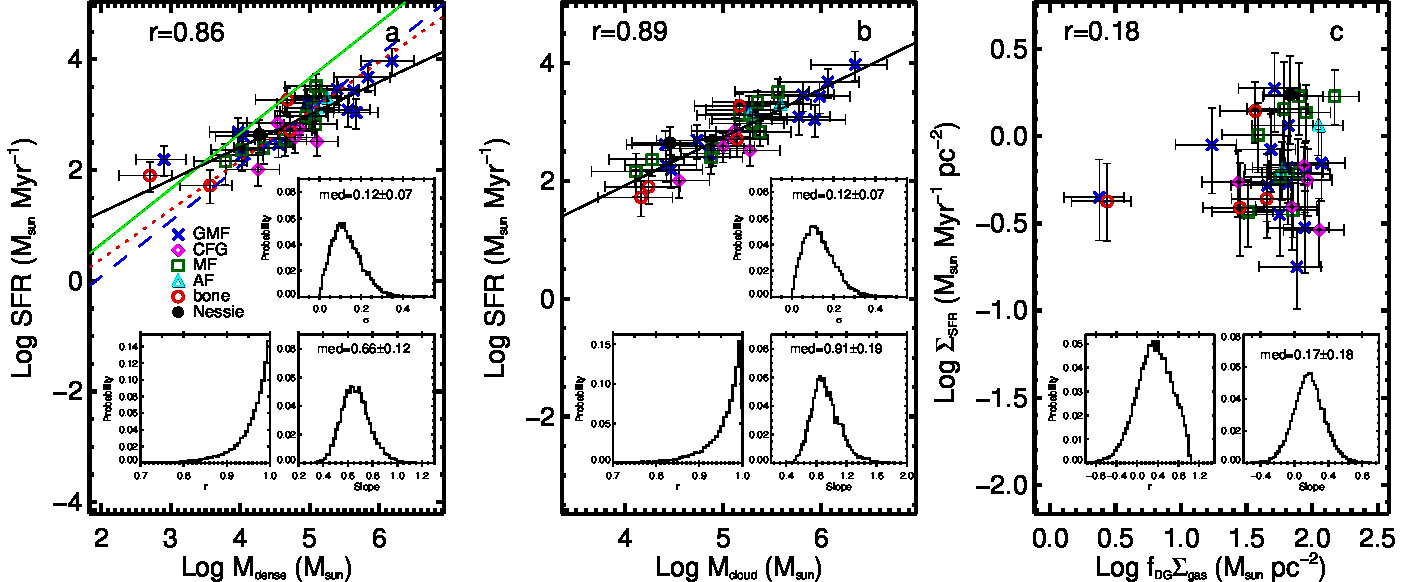}
\caption{{{SFR and $\Sigma_{\textrm{SFR}}$ as a function of }(a) {dense gas mass}; (b) cloud mass; and (c) $f_{\textrm{DG}}\Sigma_{\textrm{gas}}$ for GMFs. \jkfour{Note that the GMFs from different original identification publications are marked with different symbols. The abbreviations correspond to the column one of Table~\ref{tablesample}.} The straight lines  represent Lada's equation \citep[green solid line,][]{lada10}, the relation from \citet{wu05} (blue dashed line), fitting result from \citet{evans14} (red dotted line), and our linear fitting results without considering uncertainties (black solid line). The Pearson correlation coefficients between the different sets of parameters obtained without considering uncertainties are marked on top region of each panel. The insert plots in each panel show the probability distributions of correlation coefficients ($r$), fitting slopes, and intrinsic scatters ($\sigma$) obtained with Bayesian linear regression method by \citet{kelly07}.}}
\label{sflaw2}
\end{figure*}

{Overall, the star formation rate per free-fall time of GMFs is in the range of 0.002$-$0.05 with a median value of 0.02, which is roughly consistent with the results of \citet{krumholz12}. However, the large uncertainties of $\Sigma_{\textrm{SFR}}$, $\Sigma_{\textrm{gas}}$, $f_{\textrm{DG}}$$\Sigma_{\textrm{gas}}$, and $\Sigma_{\textrm{gas}}/t_{\textrm{ff}}$ do not allow us to obtain any robust conclusion about whether there are correlations between these parameters; for this purpose, more accurate measurements of SFR and gas mass are needed. Covering wider dynamic ranges of the parameters would also be helpful in studying the relationships. 
However, we detect a significant correlation between SFR and $M_{\textrm{dense}}$ for GMFs, which indicates that the gas mass above a visual extinction of 7 mag can be used as a predictor of SFR.}

\subsection{What controls SFRs of the molecular clouds?}\label{sect:sublinear}

We found in the previous section a strong correlation between SFR and gas mass for GMFs. {The Bayesian linear regression gives the relationship} 
\begin{equation} 
\textrm{SFR}~(\textrm{M}_{\sun}\,\textrm{Myr}^{-1}) \propto M_{\textrm{dense}}^{(0.66 \pm 0.12)} ~(\textrm{M}_{\sun}).
\label{eq:sfr_mdense_us}
\end{equation}
{This is shallower than the approximately linear relationship found in the nearby clouds \citep{lada10}}
\begin{equation}
{\textrm{{SFR}}_{\textrm{{Lada}}}}~(\textrm{M}_{\sun}\,\textrm{Myr}^{-1})=0.046M_{\textrm{dense}}~(\textrm{M}_{\sun}),
\label{ladaequation}
\end{equation}
or in the Galactic dense cores \citep{wu05}
\begin{equation}
\textrm{{SFR}}_{\textrm{{Wu}}}~(\textrm{M}_{\sun}\,\textrm{Myr}^{-1})=0.012 M_{\textrm{dense}}~(\textrm{M}_{\sun}).
\label{wuequation}
\end{equation}
The slope we derive is {closer} to that derived by \citet{evans14}, who re-investigated the SFRs in the nearby star-forming regions with a uniform YSO sample. They obtained a relation of
\begin{equation}
\textrm{{SFR}}_{\textrm{{Evans}}}~(\textrm{M}_{\sun}\,\textrm{Myr}^{-1})=0.041M_{\textrm{dense}}^{0.89} (\textrm{M}_{\sun}). 
\label{evansequation}
\end{equation}


{However, the role of uncertainties in the parameters of Eq.~\ref{ladaequation}, \ref{wuequation}, and~\ref{evansequation} is unclear and thus it is not possible to address the significance of the differences between Eq.~\ref{eq:sfr_mdense_us} and Eq.~\ref{ladaequation}, \ref{wuequation}, \ref{evansequation}.} 

\citet{evans14} do not find a significant correlation in SFR/$M_{\textrm{dense}}$ versus $M_{\textrm{dense}}$ data (see their Fig.~6), which suggests a linear relationship, in agreement with \citet{lada10}. {We also investigated the relationship between SFR/$M_{\textrm{dense}}$ and $M_{\textrm{dense}}$ in the GMFs (Fig.~\ref{sflaw2p2})}.
{There is a significant negative correlation between SFR/$M_{\textrm{dense}}$ and $M_{\textrm{dense}}$ for GMFs. The 95\%~confidential interval of correlation coefficient obtained with Bayesian linear regression method by \citet{kelly07} is [-1, -0.67].}
{Therefore, the SFR$-M_{\textrm{dense}}$ relation in GMFs could be a sub-linear rather than a linear relation.}
{{We also investigated this relation using different sub-samples such as the GMFs with distances of $<$3.5, 4, and 5 kpc or GMFs with $R>$1 and found that the slopes of SFR$-M_{\textrm{dense}}$ relations are $<$1 (considering 1$\sigma$ uncertainties) in all cases (see Appendix~\ref{apsubsamples} for details), which confirms that the detection of this approximate sub-linear SFR$-M_{\textrm{dense}}$} relation in GMFs is not due to sample selection effects.} \jkfour{However, we emphasize that the SFR$-M_{\textrm{dense}}$ relation detected in GMFs is still roughly consistent with the relations suggested by \citet{wu04,lada10,evans14} if considering 3$\sigma$ uncertainties of the slope.}
\begin{figure}
\includegraphics[width=1.0\linewidth]{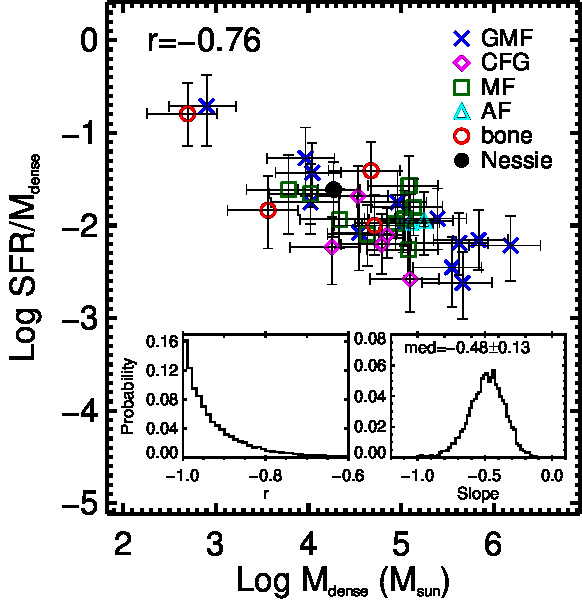}
\caption{{SFR/$M_{\textrm{dense}}$ as a function of dense gas mass for the GMFs. \jkfour{Note that the GMFs from different original identification publications are marked with different symbols. The abbreviations correspond to the column one of Table~\ref{tablesample}.} The Pearson $r$ obtained without considering uncertainties is marked on the top of the figure. The insert plots show the probability distributions of correlation coefficient and fitting slope obtained with Bayesian linear regression method by \citet{kelly07}.}}
\label{sflaw2p2}
\end{figure}


{We shortly speculate about the role of filamentary morphology in the star formation of GMFs. Note that in this work, we have not defined the length of the GMFs accurately; we have only measured the length of the GMFs through the fit of an ellipsoid to the column density map of the GMF. }
\mz{For cylindrical structures, the \emph{mass per unit length}, or \emph{line mass}, is an important parameter that is linked to the physical state of the filament \citep[e.g.,][]{stod63,ostriker64,fp00,fis12,recchi14}. With this in mind, we investigated} 
{the \emph{line density} relation between star formation and dense gas, i.e., SFR/$L$ vs. $M_\mathrm{dense} / L$, as shown in Fig. \ref{sflaw3} (panel a). $L$ represents the length of clouds that is the major axis obtained through fitting the extinction map of a cloud with an ellipse. We find a {significant} correlation between the line densities of star formation rate and dense gas mass. The correlation coefficients are 0.72 {without considering uncertainties. If considering uncertainties, the 95\%~confidential interval of correlation coefficient obtained with the Bayesian linear regression method is [0.58, 0.99].} 
For all data points, we obtain the relation
\begin{equation}
\frac{\textrm{SFR}}{L} \propto \left( \frac{M_{\textrm{dense}}}{L} \right) ^{(0.55 \pm 0.12)}.
\label{eq:0}
\end{equation}
{In the context of GMCs in general, it has been found that the internal structure, or morphology, plays an important role in predicting the SFRs of GMCs.}
For instance, \citet{lada13,lada17} investigated the star formation activity in several well-studied local molecular clouds and found that the number of protostars is correlated with the area of the cloud above a given extinction $A_{\textrm{K}}$, denoted as $S'(A_{\textrm{K}})$. Obviously, $S'(A_{\textrm{K}})$ reflects the cloud structure. 
\citet{kai14} used the probability distributions of gas density as the measure of the internal cloud structure, and showed that the SFR is correlated with the mass above an H$_2$ number density threshold of $\sim$5000 cm$^{-3}$. 
\citet{hacar13} showed that filamentary clouds contain intricate sub-structures, called fibers, some of which are "fertile" and others "non-fertile" from the perspective of star formation. 
{Applying such approaches to GMFs would be a good starting point to further study the relationship indicated by Eq.~\ref{eq:0}}

\jkfour{In Figs~\ref{sflaw2}-\ref{sflaw3}, we plot GMFs that are originally identified with different dataset and criteria (see Table~\ref{tablesample}) with different symbols. Nessie is also highlighted in the figures. To show different symbols more clearly, we re-plot these figures without error bars in Fig.~\ref{fig:sflaw_sample}. It seems that different samples from different original publications follow the similar SFR$-M$ or SFR/$L-M_{\textrm{dense}}/L$ relations. We do not find obvious difference or trends between different samples (i.e., symbols). Due to the large uncertainties of measurements and limited number of samples, we did not try to do any further statistics.}

\begin{figure}
\includegraphics[width=1.0\linewidth]{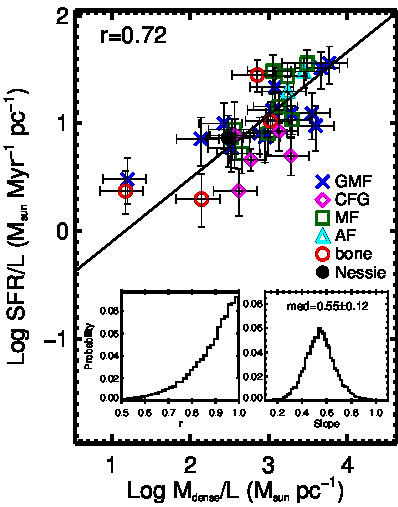}
\caption{{{SFR per unit length vs. dense gas mass per unit length for GMFs.} \jkfour{Note that the GMFs from different original identification publications are marked with different symbols. The abbreviations correspond to the column one of Table~\ref{tablesample}.} {The black line shows the linear fittings obtained without considering uncertainties. The Pearson $r$ obtained without considering uncertainties is also marked on the top of the figure. The insert plots show the probability distributions of correlation coefficient and fitting slope obtained with Bayesian linear regression method by \citet{kelly07}.}}}
\label{sflaw3}
\end{figure}

Another important property that could correlate with star formation is the age of molecular cloud. A reasonable assumption is that young cloud has low SFR and then star formation accelerates. Based on a large sample of giant molecular clouds (GMCs), \citet{lee16} investigated the SFRs of $\sim$200 star-forming complexes (SFCs) with \textit{WMAP} free-free emission fluxes and found that there is a wide spread in the SFE and SFR$_{\textrm{ff}}$ of SFCs. These large scatters can be explained with a time-variable SFR$_{\textrm{ff}}$ model and \citet{lee16} suggested that star formation is a dynamic process on GMC scales in the Milky Way. However, cloud age is difficult to determine. With the accurate censuses of YSOs in nearby molecular clouds, relative numbers of YSO classes have been used as age estimators \citep[e.g.,][]{sadavoy2013phd,evans14,stutz2015}. Following this approach, we adopt the number ratio of Class II to Class I sources, $N_{\textrm{ClassII}}/N_{\textrm{ClassI}}$, as an age indicator. \mz{Figure~\ref{ysoratio} shows the relation between SFR/$L$ and $N_{\textrm{ClassII}}/N_{\textrm{ClassI}}$ for the GMFs (\textit{left panel}) and nearby star-forming clouds (\textit{right panel}).}
~{Based on the Bayesian linear regression method \citep{kelly07}, the 95\%~confidence intervals of the correlation coefficients are [-0.59, 0.99] with a median value of 0.57 for GMFs and [-0.01, 0.96] with a median value of 0.82 for nearby clouds. Note that a significant correlation needs $r>$~0.52 ($N_s=$~34) for GMFs and $r>$~0.75 ($N_s=$~17) for nearby clouds. Therefore, }we did not find significant correlation between SFR/$L$ and $N_{\textrm{ClassII}}/N_{\textrm{ClassI}}$ for the GMFs. 
However, there could be a marginal correlation between SFR/$L$ and $N_{\textrm{ClassII}}/N_{\textrm{ClassI}}$ for the nearby star-forming regions, which indicates that 
the SFR of nearby star-forming regions could be correlated with their age: evolved clouds are inclined to have higher SFR per unit length. However, the uncertainty due to the small number of Class I sources in most of nearby clouds is still large. 

\begin{figure*}
\includegraphics[width=1.0\linewidth]{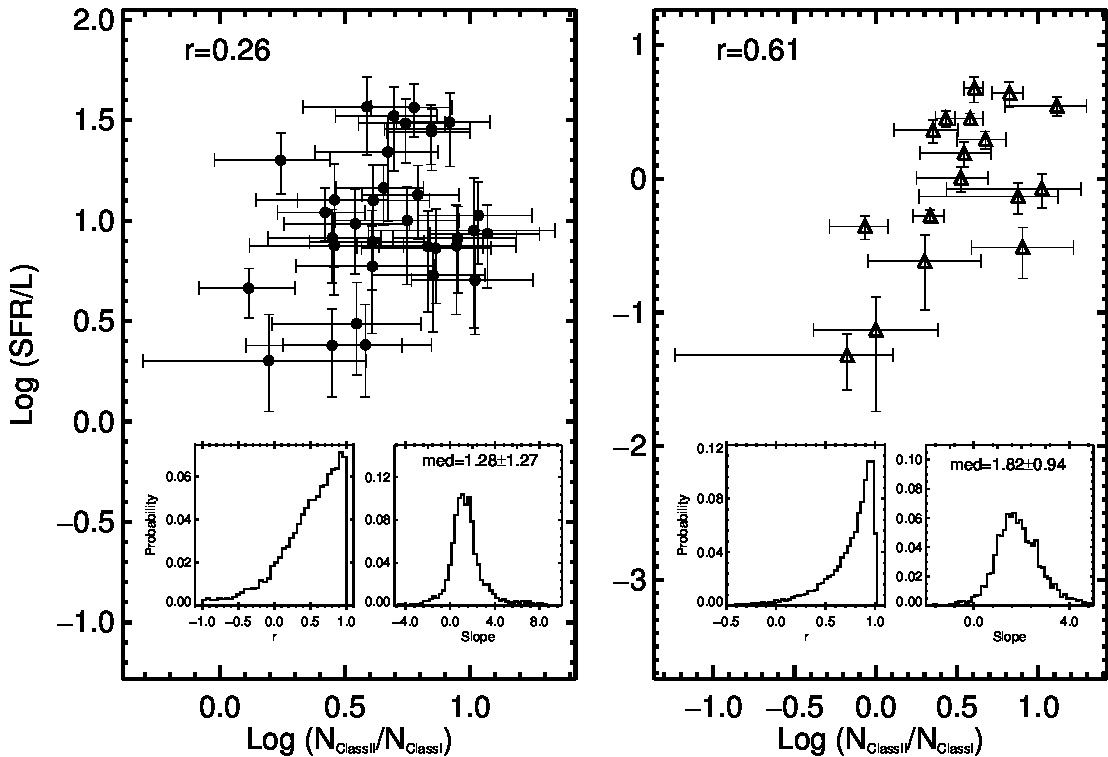}
\caption{{SFR per unit length} vs. {Class II to Class I YSO fraction} for the GMFs (\textit{left panel}) and the nearby star-forming clouds (\textit{right panel}). 
{The Pearson $r$ values obtained without considering uncertainty are marked on the top of panels. The insert plots in each panel show the probability distributions of correlations coefficient and fitting slope obtained with Bayesian linear regression method by \citet{kelly07}.}
}
\label{ysoratio}
\end{figure*}

Overall, {the significant correlation between SFR and $M_{\textrm{dense}}$ in GMFs suggests that $M_{\textrm{dense}}$ can be a parameter that control the star formation of molecular clouds. However, we find that the SFR$-M_{\textrm{dense}}$ relation in GMFs may be slightly shallower than that in nearby clouds suggested by \citet{lada10}. We also detect a significant correlation between SFR per unit length and dense gas mass per unit length for GMFs. {The origin of this correlation is still unknown, calling for further studies linking the structure of GMFs to their SF activity. }} 

\section{\textcolor{black}{Conclusions}}\label{sect:conclusions}

In this paper, we investigate the properties {of {57} previously identified Galactic giant molecular filaments (GMFs) using a uniform method. We also }estimate the SFRs of 46 GMFs {by identifying and classifying their} YSO {populations and using simple assumptions to correct for the completeness}. The main results are summarized as follows.



\begin{itemize}

\item[1]{The {57} GMFs {we analyze} are located at the distances of 2$-$13 kpc{, and their median length, gas mass ($M_{\textrm{cloud}}$), and gas surface density ($\Sigma_{\textrm{gas}}$) are {67} pc,} {1.5}$\times$10$^5$\,M$_{\sun}$ and {128}\,M$_{\sun}$\,pc$^{-2}$, respectively, {where the gas mass is obtained by integrating the column density maps down to $A_V=$~3\,mag}. The median {mean 
$^{13}$CO line width ($\Delta v$) is {3.4}\,km\,s$^{-1}$, similar {to infrared dark clouds and higher than the value in nearby star-forming regions}. {We find significant correlations between the line widths and sizes, and the line widths and masses, of the GMFs, similar to Larson's relations.}}} 

\item[2]{We identify {36\,394} sources with infrared excess in {57} GMFs. 
We found that the fractions of foreground contamination in the GMFs are 5$-${96}\%~for Class I and 8$-${100}\%~for Class II sources. The fractions of background contamination are {0.5}$-$37\%~and {1}$-$25\%~for Class I and II sources, respectively. \mz{After removing the possible contamination, we obtained {7\,028} Class I sources and {11\,526} Class II sources in {57} GMFs.}} 

\item[3]{The two-point correlation functions (TPCFs) of Class I sources in six GMFs show strong clustering at small scales. Below those scales, the TPCFs can be described by a single power law, which can be interpreted as a hierarchical or fractal distribution of protostars. The obtained power-law indices are between $[-0.58, {-0.20}]$, corresponding to the {2D} fractal dimensions, {$D_2$}, between $[1.4,1.8]$. {These are similar to the values obtained for nearby clouds, $D_2 \sim$~1.2$-$1.9 \citep{nakajima98,alf11}.}}


\item[4]{Using the number of identified YSOs, we estimate the SFRs for 46 GMFs. For the 34 GMFs closer than 5.5 kpc with reliable SFR determinations, the median values of the SFR surface densities and SFEs are 
0.62\,M$_{\sun}$\,Myr$^{-1}$\,pc$^{-2}$, and 1\%, similar to the values of 
0.64\,M$_{\sun}$\,Myr$^{-1}$\,pc$^{-2}$ and 1.1\%~obtained for the nearby star-forming clouds. This suggests that the giant filaments are similar in their star-forming properties to the nearby, Solar neighbourhood star-forming regions. This, in turn, implies similar star formation activity per gas mass in spiral arm and inter-arm environments, if most GMFs are associated with the spiral arms.}

\item[5]{
{The star formation rate per free-fall time of GMFs is in the range of $\sim$0.002-0.05 with a median value of $\sim$0.02, which is roughly consistent with the results from \citet{krumholz12}. We also test the correlations between $\Sigma_{\textrm{SFR}}$ vs. $\Sigma_{\textrm{gas}}$ and $\Sigma_{\textrm{SFR}}$ vs. $\Sigma_{\textrm{gas}}$ per free-fall time. However, the large uncertainties of these parameters hamper any definite conclusion.} 
{However}, the correlations between the SFR and gas mass above the extinction contours of $A_\mathrm{V}=\{3, 7\}$ mag indicate that the mass of dense gas is an important parameter {in predicting SFRs of molecular clouds}.} 

\item[6]{We find a significant correlation between SFR per unit length and {dense gas mass per unit length }
{for the GMFs, {i.e., $\frac{\textrm{SFR}}{L}\propto \frac{M_{\textrm{dense}}}{L}^{(0.55\pm 0.12)}$}, where $M_{\textrm{dense}}$ is the mass above the extinction contour of $A_V=$~7\,mag. } 
{While the origin of this correlation is elusive, it suggests that there is a link between the (mean) line mass and the specific star formation rate of GMFs. What exactly sets the strength of the relationship is an interesting topic for future study.}}
\end{itemize}

In summary, we perform the first systematic investigation of star-forming content in the Galactic GMFs with a uniform method. Because the SFR surface density and SFE of GMFs are, on average, similar to that of nearby star-forming clouds, our results {favor} the {view} that the GMFs are not special objects in the Galaxy. Our method can be also applied to a more general sample of Galactic GMCs to estimate their SFRs. In the near future, a comprehensive investigation of star-forming content towards the Galactic molecular clouds, including GMFs and other GMCs, will be  helpful to further assess whether star formation in the GMFs proceeds in a similar manner as in other molecular clouds. {Further studies should also be directed to establish if the giant filaments form a class of objects that in some fundamental way differs from other molecular clouds. Doing this would require developing a uniform way to define giant filaments; such definition is currently missing and hampering the studies of filamentary structures in general.} 

\begin{acknowledgements}
This project has received funding from the European Union's Horizon 2020 research and innovation programme under grant agreement No~639459 (PROMISE). This work was supported by the National Natural
Science Foundation of China (grants No. 11503086). This work is based in part on observations made with the Spitzer Space Telescope, which is operated by the Jet Propulsion Laboratory, California Institute of Technology under a contract with NASA. This publication makes use of data products from the Wide-field Infrared Survey Explorer, which is a joint project of the University of California, Los Angeles, and the Jet Propulsion Laboratory/California Institute of Technology, funded by the National Aeronautics and Space Administration. This paper made use of information from the Red MSX Source survey database at \url{http://rms.leeds.ac.uk/cgi-bin/public/RMS_DATABASE.cgi} which was constructed with support from the Science and Technology Facilities Council of the UK. This publication makes use of molecular line data from the Boston University-FCRAO Galactic Ring Survey (GRS). The GRS is a joint project of Boston University and Five College Radio Astronomy Observatory, funded by the National Science Foundation under grants AST-9800334, AST-0098562, \& AST-0100793.    
\end{acknowledgements}

\bibliographystyle{aa} 
\bibliography{ref}

\clearpage

\clearpage

\longtab{
\begin{landscape}
\scriptsize
\centering
\setlength\tabcolsep{4pt}
\setlength{\extrarowheight}{2.5pt}
\begin{longtable}{S[table-format=2.0]lS[table-format=3.3]S[table-format=2.3]S[table-format=4.2]S[table-format=3.1]S[table-format=2.1]S[table-format=2.1]cccccccccS[table-format=3.0]c}
\caption{\label{table1} Physical parameters of the Galactic giant molecular filaments}\\
\hline\hline
\multicolumn{1}{c}{(1)} & \multicolumn{1}{c}{(2)} & \multicolumn{1}{c}{(3)} & \multicolumn{1}{c}{(4)} & \multicolumn{1}{c}{(5)} & \multicolumn{1}{c}{(6)} & \multicolumn{1}{c}{(7)} & \multicolumn{1}{c}{(8)} & \multicolumn{1}{c}{(9)} & \multicolumn{1}{c}{(10)} & \multicolumn{1}{c}{(11)} & \multicolumn{1}{c}{(12)} & \multicolumn{1}{c}{(13)} & \multicolumn{1}{c}{(14)} & \multicolumn{1}{c}{(15)} & \multicolumn{1}{c}{(16)} & \multicolumn{1}{c}{(17)} & \multicolumn{1}{c}{(18)}& \multicolumn{1}{c}{(19)} \\
\multicolumn{1}{c}{ID} & \multicolumn{1}{c}{Name} & \multicolumn{1}{c}{$l$} & \multicolumn{1}{c}{$b$} & \multicolumn{1}{c}{Area} & \multicolumn{1}{c}{Length} & \multicolumn{1}{c}{Aspect} & \multicolumn{1}{c}{Distance} & \multicolumn{1}{c}{$v_{\textrm{min}}$, $v_{\textrm{\textrm{LSR}}}$, $v_{\textrm{max}}$} & \multicolumn{1}{c}{$\Delta$$v$} & \multicolumn{1}{c}{Log($M_{\textrm{cloud}}$)} & \multicolumn{1}{c}{Log($M_{\textrm{dense}}$)} & \multicolumn{1}{c}{$t_{\textrm{ff,s}}$} & \multicolumn{1}{c}{$t_{\textrm{ff,f}}$}   & \multicolumn{1}{c}{$\Sigma_{\textrm{gas}}$}  & \multicolumn{1}{c}{$\frac{M_{\textrm{cloud}}}{Length}$} & \multicolumn{1}{c}{$R_{\textrm{gal}}$} & \multicolumn{1}{c}{$z$} & \multicolumn{1}{c}{Reference}\\
\multicolumn{1}{c}{GMF} & &\multicolumn{1}{c}{(\degr)} & \multicolumn{1}{c}{(\degr)} & \multicolumn{1}{c}{(pc$^2$)} & \multicolumn{1}{c}{(pc)} &\multicolumn{1}{c}{ratio}& \multicolumn{1}{c}{(kpc)} & \multicolumn{1}{c}{(km\,s$^{-1}$)} & \multicolumn{1}{c}{(km\,s$^{-1}$)} & \multicolumn{1}{c}{(M$_{\sun}$)} & \multicolumn{1}{c}{(M$_{\sun}$)} & \multicolumn{1}{c}{(Myr)}  & \multicolumn{1}{c}{(Myr)} & \multicolumn{1}{c}{(10$^{2}$ M$_{\sun}$ pc$^{-2}$)}  & \multicolumn{1}{c}{(10$^3$ M$_{\sun}$ pc$^{-1}$)} &\multicolumn{1}{c}{(kpc)} & \multicolumn{1}{c}{(pc)} & \\
\hline
\endfirsthead
\caption{continued.}\\
\hline\hline
\multicolumn{1}{c}{(1)} & \multicolumn{1}{c}{(2)} & \multicolumn{1}{c}{(3)} & \multicolumn{1}{c}{(4)} & \multicolumn{1}{c}{(5)} & \multicolumn{1}{c}{(6)} & \multicolumn{1}{c}{(7)} & \multicolumn{1}{c}{(8)} & \multicolumn{1}{c}{(9)} & \multicolumn{1}{c}{(10)} & \multicolumn{1}{c}{(11)} & \multicolumn{1}{c}{(12)} & \multicolumn{1}{c}{(13)} & \multicolumn{1}{c}{(14)} & \multicolumn{1}{c}{(15)} & \multicolumn{1}{c}{(16)} & \multicolumn{1}{c}{(17)} & \multicolumn{1}{c}{(18)}& \multicolumn{1}{c}{(19)} \\
\multicolumn{1}{c}{ID} & \multicolumn{1}{c}{Name} & \multicolumn{1}{c}{$l$} & \multicolumn{1}{c}{$b$} & \multicolumn{1}{c}{Area} & \multicolumn{1}{c}{Length}& \multicolumn{1}{c}{Aspect} & \multicolumn{1}{c}{Distance} & \multicolumn{1}{c}{$v_{\textrm{min}}$, $v_{\textrm{LSR}}$, $v_{\textrm{max}}$} & \multicolumn{1}{c}{$\Delta$$v$} & \multicolumn{1}{c}{Log($M_{\textrm{cloud}}$)} & \multicolumn{1}{c}{Log($M_{\textrm{dense}}$)} & \multicolumn{1}{c}{$t_{\textrm{ff,s}}$} & \multicolumn{1}{c}{$t_{\textrm{ff,f}}$}   & \multicolumn{1}{c}{$\Sigma_{\textrm{gas}}$}  & \multicolumn{1}{c}{$\frac{M_{\textrm{cloud}}}{Length}$} & \multicolumn{1}{c}{$R_{\textrm{gal}}$} & \multicolumn{1}{c}{$z$} & \multicolumn{1}{c}{Reference}\\
\multicolumn{1}{c}{GMF} & &\multicolumn{1}{c}{(\degr)} & \multicolumn{1}{c}{(\degr)} & \multicolumn{1}{c}{(pc$^2$)} & \multicolumn{1}{c}{(pc)} &\multicolumn{1}{c}{ratio}& \multicolumn{1}{c}{(kpc)} & \multicolumn{1}{c}{(km\,s$^{-1}$)} & \multicolumn{1}{c}{(km\,s$^{-1}$)} & \multicolumn{1}{c}{(M$_{\sun}$)} & \multicolumn{1}{c}{(M$_{\sun}$)} & \multicolumn{1}{c}{(Myr)} & \multicolumn{1}{c}{(Myr)} & \multicolumn{1}{c}{(10$^2$ M$_{\sun}$ pc$^{-2}$)}  & \multicolumn{1}{c}{(10$^3$ M$_{\sun}$ pc$^{-1}$)} &\multicolumn{1}{c}{(kpc)} & \multicolumn{1}{c}{(pc)} & \\
\hline
\endhead
\hline
\endfoot
1    &  G015.653$-$0.224         &  15.726         &    -0.261         &  1554.18        &  109.6          &  2.0            &  11.5           &  [54.9, 58.2, 61.4]              &2.40$\pm$0.73                      &  5.16$_{-0.27}^{+0.18}$             &  3.99$_{-0.27}^{+0.18}$             &  4.6$_{-0.9}^{+1.6}$                &  8.2$_{-1.6}^{+2.9}$                &  0.92$_{-0.43}^{+0.47}$             &  1.31$_{-0.61}^{+0.67}$             &  4.1              &-61         &5     \\
2    &  F19                      &  18.988         &    -0.039         &  108.85         &  17.2           &  1.6            &  4.4            &  [59.4, 61.4, 63.3]              &2.31$\pm$0.35                      &  4.14$_{-0.28}^{+0.18}$             &  3.82$_{-0.28}^{+0.18}$             &  2.0$_{-0.4}^{+0.7}$                &  2.8$_{-0.6}^{+1.0}$                &  1.28$_{-0.61}^{+0.68}$             &  0.81$_{-0.38}^{+0.43}$             &  4.4              &10          &6     \\
3    &  GMF20.0$-$17.9           &  19.000         &    -0.118         &  1771.27        &  142.2          &  5.6            &  3.5            &  [37.0, 43.5, 50.0]              &5.90$\pm$2.27                      &  5.37$_{-0.27}^{+0.18}$             &  5.06$_{-0.27}^{+0.18}$             &  4.0$_{-0.8}^{+1.4}$                &  7.8$_{-1.5}^{+2.7}$                &  1.33$_{-0.62}^{+0.69}$             &  1.66$_{-0.77}^{+0.85}$             &  5.2              &9           &1     \\
4    &  F20                      &  20.772         &    -0.102         &  2687.48        &  92.0           &  1.8            &  11.7           &  [53.3, 57.7, 62.2]              &4.56$\pm$1.09                      &  5.55$_{-0.27}^{+0.18}$             &  5.22$_{-0.27}^{+0.18}$             &  4.4$_{-0.9}^{+1.6}$                &  6.2$_{-1.3}^{+2.2}$                &  1.33$_{-0.61}^{+0.70}$             &  3.89$_{-1.80}^{+2.05}$             &  4.9              &-29         &6     \\
5    &  G021.173$-$0.312         &  21.170         &    -0.307         &  231.71         &  33.2           &  2.4            &  13.0           &  [32.5, 35.1, 37.6]              &2.64$\pm$0.23                      &  4.41$_{-0.27}^{+0.19}$             &  3.94$_{-0.27}^{+0.19}$             &  2.6$_{-0.5}^{+0.9}$                &  4.1$_{-0.9}^{+1.5}$                &  1.11$_{-0.51}^{+0.60}$             &  0.77$_{-0.36}^{+0.42}$             &  6.0              &-81         &5     \\
6    &  BC\_021.25$-$0.15        &  21.403         &    -0.109         &  199.42         &  34.7           &  2.4            &  3.9            &  [63.5, 66.0, 68.5]              &2.69$\pm$0.50                      &  4.28$_{-0.27}^{+0.18}$             &  2.74$_{-0.27}^{+0.18}$             &  2.7$_{-0.5}^{+0.9}$                &  4.5$_{-0.9}^{+1.6}$                &  0.95$_{-0.44}^{+0.50}$             &  0.55$_{-0.25}^{+0.29}$             &  4.9              &7           &3     \\
7    &  G022.519$-$0.025         &  22.650         &    -0.072         &  2319.65        &  74.9           &  1.4            &  6.6            &  [104.3, 108.3, 112.3]           &4.43$\pm$0.85                      &  5.45$_{-0.27}^{+0.18}$             &  5.04$_{-0.27}^{+0.18}$             &  4.5$_{-0.9}^{+1.6}$                &  5.9$_{-1.2}^{+2.1}$                &  1.21$_{-0.57}^{+0.64}$             &  3.76$_{-1.76}^{+1.99}$             &  3.4              &0           &5     \\
8    &  F21                      &  23.082         &    -0.409         &  1645.35        &  80.2           &  2.0            &  4.6            &  [74.4, 76.9, 79.4]              &3.14$\pm$0.37                      &  5.39$_{-0.28}^{+0.19}$             &  5.17$_{-0.28}^{+0.19}$             &  3.7$_{-0.8}^{+1.3}$                &  5.5$_{-1.1}^{+2.0}$                &  1.48$_{-0.70}^{+0.80}$             &  3.03$_{-1.43}^{+1.64}$             &  4.5              &-19         &6     \\
9    &  F23                      &  23.406         &    -0.073         &  2341.38        &  78.5           &  1.5            &  5.9            &  [95.0, 97.5, 100.0]             &3.29$\pm$0.36                      &  5.48$_{-0.28}^{+0.18}$             &  5.12$_{-0.28}^{+0.18}$             &  4.3$_{-0.9}^{+1.6}$                &  5.9$_{-1.2}^{+2.1}$                &  1.29$_{-0.61}^{+0.68}$             &  3.86$_{-1.81}^{+2.04}$             &  3.7              &1           &6     \\
10   &  F24                      &  23.409         &    0.051          &  230.33         &  32.2           &  2.1            &  4.7            &  [82.1, 83.6, 85.1]              &2.09$\pm$0.07                      &  4.31$_{-0.27}^{+0.18}$             &  1.90$_{-0.27}^{+0.18}$             &  2.9$_{-0.6}^{+1.0}$                &  4.5$_{-0.9}^{+1.6}$                &  0.89$_{-0.42}^{+0.47}$             &  0.64$_{-0.30}^{+0.34}$             &  4.4              &17          &6     \\
11   &  F22                      &  23.422         &    -0.175         &  798.92         &  36.4           &  1.1            &  5.4            &  [98.3, 101.8, 105.3]            &3.85$\pm$0.73                      &  5.20$_{-0.28}^{+0.19}$             &  5.07$_{-0.28}^{+0.19}$             &  2.7$_{-0.5}^{+1.0}$                &  3.2$_{-0.6}^{+1.2}$                &  2.00$_{-0.96}^{+1.08}$             &  4.40$_{-2.11}^{+2.38}$             &  4.0              &-5          &6     \\
12   &  CFG024.00+0.48           &  23.802         &    0.484          &  1377.31        &  94.4           &  2.1            &  5.2            &  [93.0, 96.0, 99.0]              &3.39$\pm$0.57                      &  5.16$_{-0.28}^{+0.18}$             &  4.58$_{-0.28}^{+0.18}$             &  4.2$_{-0.8}^{+1.5}$                &  7.1$_{-1.4}^{+2.5}$                &  1.05$_{-0.49}^{+0.55}$             &  1.53$_{-0.72}^{+0.81}$             &  4.2              &55          &2     \\
13   &  F26                      &  24.597         &    -0.232         &  603.84         &  45.2           &  1.9            &  5.1            &  [92.6, 98.0, 103.4]             &4.05$\pm$1.07                      &  4.94$_{-0.28}^{+0.19}$             &  4.68$_{-0.28}^{+0.19}$             &  2.9$_{-0.6}^{+1.0}$                &  4.2$_{-0.9}^{+1.5}$                &  1.43$_{-0.67}^{+0.77}$             &  1.91$_{-0.90}^{+1.03}$             &  4.3              &-8          &6     \\
14   &  F27                      &  24.648         &    0.048          &  859.56         &  56.7           &  1.9            &  5.6            &  [106.2, 108.7, 111.2]           &3.13$\pm$0.36                      &  5.05$_{-0.27}^{+0.19}$             &  4.72$_{-0.27}^{+0.19}$             &  3.3$_{-0.7}^{+1.2}$                &  4.9$_{-1.0}^{+1.7}$                &  1.31$_{-0.61}^{+0.72}$             &  1.99$_{-0.93}^{+1.09}$             &  4.0              &15          &6     \\
15   &  BC\_24.95$-$0.17         &  24.681         &    -0.119         &  1406.85        &  68.2           &  1.9            &  4.3            &  [44.5, 47.0, 49.5]              &3.01$\pm$0.53                      &  5.21$_{-0.27}^{+0.19}$             &  4.71$_{-0.27}^{+0.19}$             &  4.1$_{-0.8}^{+1.4}$                &  5.8$_{-1.2}^{+2.0}$                &  1.14$_{-0.53}^{+0.62}$             &  2.35$_{-1.09}^{+1.27}$             &  4.8              &5           &3     \\
16   &  BC\_025.24$-$0.45        &  25.265         &    -0.411         &  1232.38        &  49.9           &  1.2            &  4.3            &  [54.7, 57.4, 60.2]              &3.28$\pm$0.54                      &  5.17$_{-0.27}^{+0.19}$             &  4.74$_{-0.27}^{+0.19}$             &  3.8$_{-0.8}^{+1.3}$                &  4.8$_{-1.0}^{+1.7}$                &  1.21$_{-0.56}^{+0.65}$             &  2.98$_{-1.38}^{+1.59}$             &  4.8              &-16         &3     \\
17   &  F28                      &  25.271         &    -0.225         &  145.62         &  32.3           &  3.0            &  3.9            &  [61.3, 63.8, 66.2]              &2.38$\pm$0.53                      &  4.32$_{-0.28}^{+0.18}$             &  4.06$_{-0.28}^{+0.18}$             &  2.1$_{-0.4}^{+0.7}$                &  3.6$_{-0.7}^{+1.3}$                &  1.43$_{-0.67}^{+0.74}$             &  0.64$_{-0.30}^{+0.33}$             &  5.1              &0           &6     \\
18   &  F29                      &  25.599         &    -0.199         &  1668.62        &  66.5           &  1.6            &  4.9            &  [89.3, 93.3, 97.3]              &3.71$\pm$1.04                      &  5.34$_{-0.28}^{+0.19}$             &  4.98$_{-0.28}^{+0.19}$             &  4.0$_{-0.8}^{+1.4}$                &  5.4$_{-1.1}^{+2.0}$                &  1.31$_{-0.62}^{+0.71}$             &  3.28$_{-1.55}^{+1.79}$             &  4.5              &-4          &6     \\
19   &  G025.762+0.241           &  25.681         &    0.253          &  597.03         &  48.5           &  2.2            &  7.7            &  [107.8, 110.3, 112.8]           &3.14$\pm$0.44                      &  4.91$_{-0.27}^{+0.19}$             &  4.64$_{-0.27}^{+0.19}$             &  3.0$_{-0.6}^{+1.1}$                &  4.5$_{-0.9}^{+1.6}$                &  1.36$_{-0.63}^{+0.73}$             &  1.67$_{-0.78}^{+0.89}$             &  3.6              &38          &5     \\
20   &  BC\_26.94$-$0.30         &  26.945         &    -0.341         &  142.13         &  27.0           &  2.4            &  4.6            &  [64.0, 67.9, 71.9]              &3.00$\pm$0.61                      &  4.20$_{-0.28}^{+0.18}$             &  3.60$_{-0.28}^{+0.18}$             &  2.3$_{-0.5}^{+0.8}$                &  3.7$_{-0.7}^{+1.3}$                &  1.11$_{-0.53}^{+0.58}$             &  0.59$_{-0.28}^{+0.31}$             &  4.7              &-13         &3     \\
21   &  F30                      &  27.017         &    0.187          &  725.56         &  49.0           &  2.0            &  5.0            &  [88.9, 93.0, 97.1]              &3.94$\pm$1.02                      &  4.92$_{-0.28}^{+0.19}$             &  4.37$_{-0.28}^{+0.19}$             &  3.4$_{-0.7}^{+1.3}$                &  4.9$_{-1.0}^{+1.8}$                &  1.14$_{-0.54}^{+0.62}$             &  1.69$_{-0.79}^{+0.91}$             &  4.5              &28          &6     \\
22   &  F31                      &  28.335         &    0.037          &  3376.17        &  110.5          &  2.3            &  4.2            &  [76.2, 79.2, 82.2]              &3.70$\pm$0.55                      &  5.59$_{-0.27}^{+0.19}$             &  5.12$_{-0.27}^{+0.19}$             &  5.0$_{-1.0}^{+1.8}$                &  7.3$_{-1.5}^{+2.6}$                &  1.16$_{-0.54}^{+0.63}$             &  3.53$_{-1.65}^{+1.92}$             &  5.1              &17          &6     \\
23   &  CFG028.68$-$0.28         &  28.663         &    -0.225         &  730.35         &  46.8           &  1.3            &  4.9            &  [84.0, 88.0, 92.0]              &3.87$\pm$1.04                      &  5.04$_{-0.28}^{+0.19}$             &  4.83$_{-0.28}^{+0.19}$             &  3.0$_{-0.6}^{+1.1}$                &  4.2$_{-0.8}^{+1.6}$                &  1.50$_{-0.72}^{+0.81}$             &  2.34$_{-1.12}^{+1.26}$             &  4.7              &-6          &2     \\
24   &  G028.854$-$0.238         &  28.850         &    -0.191         &  1965.92        &  60.5           &  1.3            &  7.4            &  [92.0, 95.2, 98.3]              &3.47$\pm$0.58                      &  5.42$_{-0.28}^{+0.19}$             &  5.07$_{-0.28}^{+0.19}$             &  4.1$_{-0.8}^{+1.4}$                &  5.1$_{-1.0}^{+1.8}$                &  1.33$_{-0.62}^{+0.71}$             &  4.31$_{-2.02}^{+2.31}$             &  4.0              &-18         &5     \\
25   &  CFG029.18$-$0.34         &  29.373         &    -0.317         &  1194.19        &  67.3           &  2.4            &  5.2            &  [89.0, 93.5, 98.0]              &4.65$\pm$1.12                      &  5.31$_{-0.27}^{+0.19}$             &  5.13$_{-0.27}^{+0.19}$             &  3.2$_{-0.7}^{+1.1}$                &  4.7$_{-1.0}^{+1.7}$                &  1.69$_{-0.79}^{+0.93}$             &  3.00$_{-1.40}^{+1.65}$             &  4.6              &-15         &2     \\
26   &  G030.315$-$0.154         &  30.372         &    -0.244         &  1237.58        &  46.1           &  1.1            &  5.3            &  [100.2, 103.7, 107.1]           &3.93$\pm$0.65                      &  5.31$_{-0.27}^{+0.19}$             &  5.14$_{-0.27}^{+0.19}$             &  3.2$_{-0.7}^{+1.1}$                &  4.0$_{-0.8}^{+1.4}$                &  1.67$_{-0.78}^{+0.89}$             &  4.48$_{-2.09}^{+2.38}$             &  4.6              &-10         &5     \\
27   &  G32.02+0.06              &  31.948         &    0.152          &  1265.53        &  92.4           &  2.7            &  5.6            &  [91.5, 95.5, 99.6]              &3.38$\pm$0.86                      &  5.24$_{-0.26}^{+0.18}$             &  4.97$_{-0.26}^{+0.18}$             &  3.6$_{-0.7}^{+1.3}$                &  6.1$_{-1.3}^{+2.1}$                &  1.38$_{-0.63}^{+0.73}$             &  1.89$_{-0.86}^{+1.00}$             &  4.7              &26          &1     \\
28   &  G032.401+0.082           &  32.433         &    0.160          &  779.72         &  54.0           &  1.9            &  11.3           &  [39.0, 42.9, 46.7]              &3.75$\pm$1.02                      &  4.95$_{-0.27}^{+0.19}$             &  4.48$_{-0.27}^{+0.19}$             &  3.5$_{-0.7}^{+1.2}$                &  5.2$_{-1.0}^{+1.8}$                &  1.14$_{-0.53}^{+0.62}$             &  1.65$_{-0.76}^{+0.89}$             &  6.2              &27          &5     \\
29   &  F34                      &  33.086         &    -0.030         &  2082.55        &  114.3          &  3.6            &  7.8            &  [96.1, 98.9, 101.8]             &3.38$\pm$0.58                      &  5.32$_{-0.27}^{+0.18}$             &  4.44$_{-0.27}^{+0.18}$             &  4.8$_{-1.0}^{+1.7}$                &  8.0$_{-1.6}^{+2.9}$                &  1.00$_{-0.47}^{+0.53}$             &  1.83$_{-0.86}^{+0.96}$             &  4.6              &1           &6     \\
30   &  G033.104+0.068           &  33.138         &    0.076          &  1697.01        &  136.7          &  5.7            &  9.3            &  [80.5, 83.5, 86.5]              &3.34$\pm$0.51                      &  5.22$_{-0.28}^{+0.19}$             &  4.47$_{-0.28}^{+0.19}$             &  4.6$_{-0.9}^{+1.6}$                &  8.9$_{-1.8}^{+3.2}$                &  0.98$_{-0.46}^{+0.52}$             &  1.21$_{-0.57}^{+0.65}$             &  5.1              &14          &5     \\
31   &  G033.685$-$0.020         &  33.410         &    -0.018         &  3920.35        &  206.8          &  4.9            &  7.1            &  [98.3, 104.7, 111.1]            &4.72$\pm$1.34                      &  5.68$_{-0.29}^{+0.18}$             &  5.29$_{-0.29}^{+0.18}$             &  5.1$_{-1.0}^{+1.8}$                &  9.8$_{-2.0}^{+3.5}$                &  1.22$_{-0.59}^{+0.64}$             &  2.31$_{-1.11}^{+1.21}$             &  4.6              &5           &5     \\
32   &  GMF38.1$-$32.4b          &  35.086         &    -0.456         &  588.44         &  70.2           &  2.7            &  2.9            &  [43.0, 44.5, 46.0]              &1.94$\pm$0.18                      &  4.77$_{-0.28}^{+0.19}$             &  4.01$_{-0.28}^{+0.19}$             &  3.5$_{-0.7}^{+1.2}$                &  6.3$_{-1.3}^{+2.2}$                &  1.00$_{-0.47}^{+0.54}$             &  0.84$_{-0.40}^{+0.45}$             &  6.2              &-3          &1     \\
33   &  GMF38.1$-$32.4a          &  35.265         &    0.021          &  5925.63        &  228.3          &  3.8            &  3.5            &  [50.0, 55.0, 60.0]              &5.09$\pm$1.41                      &  5.86$_{-0.28}^{+0.18}$             &  5.43$_{-0.28}^{+0.18}$             &  5.6$_{-1.1}^{+2.0}$                &  10.3$_{-2.0}^{+3.7}$               &  1.21$_{-0.57}^{+0.63}$             &  3.14$_{-1.48}^{+1.64}$             &  5.8              &19          &1     \\
34   &  G037.410$-$0.070         &  37.434         &    -0.058         &  222.36         &  36.2           &  2.5            &  10.0           &  [53.0, 56.5, 60.0]              &3.24$\pm$0.66                      &  4.41$_{-0.28}^{+0.18}$             &  3.82$_{-0.28}^{+0.18}$             &  2.5$_{-0.5}^{+0.9}$                &  4.2$_{-0.8}^{+1.5}$                &  1.15$_{-0.54}^{+0.59}$             &  0.71$_{-0.33}^{+0.36}$             &  6.1              &-8          &5     \\
35   &  GMF41.0$-$41.3           &  41.214         &    -0.034         &  363.26         &  52.1           &  2.2            &  2.7            &  [34.0, 38.0, 42.0]              &3.02$\pm$0.90                      &  4.50$_{-0.27}^{+0.19}$             &  2.94$_{-0.27}^{+0.19}$             &  3.3$_{-0.7}^{+1.2}$                &  5.8$_{-1.2}^{+2.0}$                &  0.88$_{-0.41}^{+0.47}$             &  0.61$_{-0.28}^{+0.33}$             &  6.6              &18          &1     \\
36   &  F38                      &  41.230         &    -0.261         &  1143.55        &  62.5           &  2.1            &  8.6            &  [56.7, 59.9, 63.0]              &2.86$\pm$0.67                      &  5.16$_{-0.27}^{+0.18}$             &  4.81$_{-0.27}^{+0.18}$             &  3.7$_{-0.7}^{+1.3}$                &  5.3$_{-1.1}^{+1.8}$                &  1.26$_{-0.58}^{+0.67}$             &  2.31$_{-1.06}^{+1.22}$             &  6.0              &-33         &6     \\
37   &  F39                      &  45.449         &    0.083          &  3467.22        &  102.7          &  1.4            &  8.4            &  [52.7, 59.2, 65.7]              &5.07$\pm$1.72                      &  5.73$_{-0.27}^{+0.19}$             &  5.52$_{-0.27}^{+0.19}$             &  4.4$_{-0.9}^{+1.5}$                &  6.1$_{-1.3}^{+2.1}$                &  1.53$_{-0.72}^{+0.83}$             &  5.18$_{-2.42}^{+2.82}$             &  6.5              &20          &6     \\
38   &  F40                      &  45.710         &    -0.470         &  3552.13        &  97.2           &  1.6            &  7.1            &  [57.1, 60.1, 63.1]              &3.60$\pm$0.46                      &  5.61$_{-0.28}^{+0.19}$             &  5.05$_{-0.28}^{+0.19}$             &  5.1$_{-1.1}^{+1.8}$                &  6.9$_{-1.4}^{+2.5}$                &  1.14$_{-0.54}^{+0.61}$             &  4.16$_{-1.99}^{+2.23}$             &  6.1              &-47         &6     \\
39   &  CFG047.06+0.26           &  47.049         &    0.299          &  276.87         &  44.4           &  2.6            &  4.4            &  [53.0, 57.0, 61.0]              &3.23$\pm$0.79                      &  4.58$_{-0.28}^{+0.19}$             &  4.29$_{-0.28}^{+0.19}$             &  2.4$_{-0.5}^{+0.9}$                &  4.2$_{-0.9}^{+1.5}$                &  1.39$_{-0.66}^{+0.75}$             &  0.87$_{-0.41}^{+0.47}$             &  6.2              &40          &2     \\
40   &  G048.629+0.096           &  48.678         &    0.100          &  1521.24        &  60.8           &  1.1            &  10.8           &  [11.9, 16.5, 21.1]              &3.59$\pm$0.94                      &  5.30$_{-0.28}^{+0.19}$             &  4.90$_{-0.28}^{+0.19}$             &  3.9$_{-0.8}^{+1.4}$                &  5.1$_{-1.0}^{+1.9}$                &  1.31$_{-0.63}^{+0.70}$             &  3.27$_{-1.57}^{+1.74}$             &  8.2              &22          &5     \\
41   &  CFG049.21$-$0.34         &  49.357         &    -0.375         &  823.80         &  121.5          &  4.1            &  5.4            &  [66.0, 69.0, 72.0]              &2.61$\pm$0.77                      &  5.09$_{-0.27}^{+0.18}$             &  4.86$_{-0.27}^{+0.18}$             &  3.1$_{-0.6}^{+1.1}$                &  6.8$_{-1.4}^{+2.4}$                &  1.50$_{-0.69}^{+0.79}$             &  1.02$_{-0.47}^{+0.54}$             &  6.3              &-19         &2     \\
42   &  F42                      &  49.396         &    -0.241         &  1847.46        &  62.6           &  1.0            &  5.4            &  [57.0, 60.0, 63.0]              &3.49$\pm$0.79                      &  5.42$_{-0.28}^{+0.18}$             &  5.12$_{-0.28}^{+0.18}$             &  3.9$_{-0.8}^{+1.4}$                &  5.0$_{-1.0}^{+1.9}$                &  1.41$_{-0.68}^{+0.74}$             &  4.17$_{-2.00}^{+2.19}$             &  6.3              &-7          &6     \\
43   &  GMF54.0$-$52.0           &  53.330         &    0.006          &  220.30         &  41.5           &  2.9            &  2.0            &  [20.0, 23.0, 26.0]              &1.94$\pm$0.47                      &  4.43$_{-0.27}^{+0.20}$             &  4.06$_{-0.27}^{+0.20}$             &  2.5$_{-0.5}^{+0.9}$                &  4.4$_{-0.9}^{+1.6}$                &  1.21$_{-0.57}^{+0.69}$             &  0.64$_{-0.30}^{+0.37}$             &  7.3              &23          &1     \\
44   &  GMF307.2$-$305.4         &  305.433        &    0.162          &  8179.46        &  226.7          &  3.5            &  3.1            &  [-45.0, -35.0, -25.0]           &4.31$\pm$3.44                      &  6.02$_{-0.27}^{+0.18}$             &  5.67$_{-0.27}^{+0.18}$             &  5.9$_{-1.2}^{+2.1}$                &  10.0$_{-2.0}^{+3.5}$               &  1.29$_{-0.60}^{+0.67}$             &  4.65$_{-2.16}^{+2.43}$             &  7.0              &29          &4     \\
45   &  GMF309.5$-$308.7         &  309.184        &    0.035          &  6640.08        &  120.5          &  1.3            &  3.7            &  [-53.0, -44.0, -35.0]           &4.82$\pm$3.25                      &  5.98$_{-0.29}^{+0.18}$             &  5.71$_{-0.29}^{+0.18}$             &  5.3$_{-1.1}^{+2.0}$                &  6.9$_{-1.4}^{+2.5}$                &  1.43$_{-0.69}^{+0.74}$             &  7.86$_{-3.80}^{+4.07}$             &  6.7              &21          &4     \\
46   &  GMF319.0$-$318.7         &  318.170        &    -0.180         &  4361.06        &  102.4          &  1.6            &  2.8            &  [-53.0, -43.5, -34.0]           &5.16$\pm$3.39                      &  5.81$_{-0.28}^{+0.19}$             &  5.58$_{-0.28}^{+0.19}$             &  4.7$_{-1.0}^{+1.7}$                &  6.3$_{-1.3}^{+2.3}$                &  1.47$_{-0.70}^{+0.78}$             &  6.25$_{-2.97}^{+3.33}$             &  6.5              &11          &4     \\
47   &  GMF324.5$-$321.4b        &  321.850        &    -0.014         &  227.25         &  31.9           &  2.6            &  2.2            &  [-35.0, -31.5, -28.0]           &1.95$\pm$0.69                      &  4.44$_{-0.27}^{+0.19}$             &  4.04$_{-0.27}^{+0.19}$             &  2.5$_{-0.5}^{+0.9}$                &  3.8$_{-0.8}^{+1.3}$                &  1.22$_{-0.56}^{+0.66}$             &  0.87$_{-0.40}^{+0.47}$             &  6.7              &20          &4     \\
48   &  G322.363+0.542           &  322.463        &    0.622          &  3200.51        &  155.7          &  3.0            &  9.9            &  [-58.6, -54.7, -50.7]           &1.95$\pm$0.84                      &  5.60$_{-0.28}^{+0.18}$             &  5.19$_{-0.28}^{+0.18}$             &  4.8$_{-0.9}^{+1.8}$                &  8.4$_{-1.7}^{+3.2}$                &  1.23$_{-0.59}^{+0.64}$             &  2.53$_{-1.21}^{+1.32}$             &  6.1              &109         &5     \\
49   &  G323.929+0.036           &  323.856        &    0.018          &  2152.63        &  83.2           &  1.3            &  10.0           &  [-63.9, -58.6, -53.3]           &2.67$\pm$1.14                      &  5.35$_{-0.28}^{+0.19}$             &  4.74$_{-0.28}^{+0.19}$             &  4.7$_{-1.0}^{+1.7}$                &  6.7$_{-1.4}^{+2.4}$                &  1.04$_{-0.49}^{+0.56}$             &  2.70$_{-1.28}^{+1.46}$             &  5.9              &4           &5     \\
50   &  G327.157$-$0.256         &  326.762        &    -0.349         &  3284.82        &  104.7          &  1.5            &  3.7            &  [-67.3, -61.4, -55.4]           &2.86$\pm$1.51                      &  5.62$_{-0.28}^{+0.19}$             &  5.29$_{-0.28}^{+0.19}$             &  4.7$_{-1.0}^{+1.7}$                &  6.8$_{-1.4}^{+2.4}$                &  1.28$_{-0.61}^{+0.70}$             &  4.01$_{-1.92}^{+2.19}$             &  5.6              &-5          &5     \\
51   &  GMF335.6$-$333.6b        &  332.086        &    0.024          &  1488.79        &  77.1           &  1.9            &  3.3            &  [-55.0, -50.0, -45.0]           &2.43$\pm$1.44                      &  5.30$_{-0.28}^{+0.19}$             &  4.99$_{-0.28}^{+0.19}$             &  3.8$_{-0.8}^{+1.4}$                &  5.7$_{-1.1}^{+2.0}$                &  1.33$_{-0.63}^{+0.71}$             &  2.57$_{-1.21}^{+1.37}$             &  5.6              &18          &4     \\
52   &  GMF335.6$-$333.6a        &  334.178        &    -0.191         &  14184.86       &  267.5          &  3.3            &  3.4            &  [-55.0, -45.0, -35.0]           &5.42$\pm$3.02                      &  6.39$_{-0.27}^{+0.18}$             &  6.22$_{-0.27}^{+0.18}$             &  5.9$_{-1.2}^{+2.1}$                &  9.3$_{-1.9}^{+3.4}$                &  1.73$_{-0.81}^{+0.90}$             &  9.20$_{-4.30}^{+4.78}$             &  5.5              &5           &4     \\
53   &  Nessie                   &  338.579        &  -0.459         &  275.07         &  62.7           &  6.7            &  3.1            &  \ldots    &\ldots      &4.49$_{-0.20}^{+0.13}$             &  4.30$_{-0.19}^{+0.13}$             &  2.7$_{-0.4}^{+0.7}$                &  5.6$_{-0.9}^{+1.4}$                &  1.13$_{-0.42}^{+0.40}$             &  0.49$_{-0.18}^{+0.18}$             &  5.6              &0           &1     \\
54   &  GMF341.9$-$337.1         &  341.045        &    -0.232         &  8404.29        &  151.8          &  1.8            &  3.5            &  [-47.0, -39.5, -32.0]           &5.22$\pm$3.07                      &  6.10$_{-0.27}^{+0.19}$             &  5.87$_{-0.27}^{+0.19}$             &  5.5$_{-1.1}^{+2.0}$                &  7.5$_{-1.5}^{+2.7}$                &  1.51$_{-0.71}^{+0.83}$             &  8.37$_{-3.92}^{+4.57}$             &  5.2              &2           &4     \\
55   &  G341.938+0.054           &  341.935        &    0.057          &  166.17         &  29.9           &  2.4            &  9.5            &  [-125.0, -120.9, -116.9]        &1.60$\pm$0.59                      &  4.15$_{-0.28}^{+0.19}$             &  0.98$_{-0.28}^{+0.19}$             &  2.8$_{-0.6}^{+1.0}$                &  4.4$_{-0.9}^{+1.6}$                &  0.85$_{-0.40}^{+0.47}$             &  0.47$_{-0.22}^{+0.26}$             &  3.0              &7           &5     \\
56   &  G349.876+0.099           &  350.059        &    0.093          &  6966.31        &  183.9          &  2.5            &  11.2           &  [-72.9, -62.2, -51.5]           &5.43$\pm$2.64                      &  6.05$_{-0.27}^{+0.19}$             &  5.85$_{-0.27}^{+0.19}$             &  5.1$_{-1.0}^{+1.8}$                &  8.0$_{-1.6}^{+2.9}$                &  1.62$_{-0.76}^{+0.87}$             &  6.15$_{-2.87}^{+3.28}$             &  3.3              &9           &5     \\
57   &  GMF358.9$-$357.4         &  357.761        &    -0.357         &  554.15         &  39.8           &  2.0            &  2.9            &  [0.0, 5.0, 10.0]                &2.39$\pm$1.15                      &  4.87$_{-0.28}^{+0.18}$             &  4.56$_{-0.28}^{+0.18}$             &  3.0$_{-0.6}^{+1.1}$                &  4.1$_{-0.8}^{+1.5}$                &  1.34$_{-0.64}^{+0.71}$             &  1.87$_{-0.89}^{+0.99}$             &  5.4              &0           &4     \\
\end{longtable}
\tablefoot{Columns are (1) GMF ID; (2) GMF name; (3) Galactic longitude of center of the ellipse that is obtained through fitting the region with $A_V>$~3\,mag in the $^{13}$CO-based extinction map; (4) Galactic latitude of center of the ellipse; (5) area encompassed by contour level of $A_V=$~3\,mag; (6) length of major axis of the ellipse; (7) aspect ratio of the ellipse, defining as the ratio of major axis to minor axis of the ellipse (8) distance adopted from literature; (9) velocity range ($v_{\textrm{min}}$ and $v_{\textrm{max}}$) and central velocity ($v_{\textrm{LSR}}$) adopted from literature or calculated based on $^{13}$CO (J=1-0) (see Appendix~\ref{ap1} for details); (10) mean line width of $^{13}$CO (J=1-0); (11) total gas mass obtained through integrating in the $^{13}$CO-based extinction map down to $A_V=$~3\,mag. The uncertainties are from the variations of $^{13}$CO abundance and extinction temperatures (see Appendix~\ref{ap4} for details).; (12) dense gas mass obtained through integrating down to $A_V=$~7\,mag. The uncertainties are from the variations of $^{13}$CO abundance and extinction temperatures. (13) free-fall time calculated based on spherical morphology assumption. The associated uncertainties are from mass uncertainties; (14) free-fall time calculated based on filamentary morphology assumption. The associated uncertainties are from mass uncertainties; (15) gas surface density ($M_{\textrm{cloud}}$/Area) whose uncertainties are from mass uncertainties; (16) mass per unit length, whose uncertainties are from mass uncertainties; (17) Galactocentric radius; (18) height above the Galactic midplane; (19) reference as shown below. Note that the associated uncertainties of the parameters listed in this table do not include the distance errors (also see Appendix~\ref{ap4} for more information).}
\tablebib{(1)~\citet{sample-gmf}; (2) \citet{sample-wang15}; (3) \citet{sample-bones}; (4) \citet{sample-agmf}; (5) \citet{sample-li16}; (6) \citet{sample-wang16}.}
\end{landscape}
}

\clearpage

\longtab{
\begin{landscape}
\setlength\tabcolsep{5pt}
\setlength{\extrarowheight}{2.5pt}
\scriptsize
\begin{longtable}{S[table-format=2.0]lS[table-format=3.0]S[table-format=3.0]S[table-format=3.0]cccccccS[table-number-alignment=left,table-space-text-pre=~~~]ccccc}
\caption{\label{table2} YSO content of the Galactic giant molecular filaments}\\
\hline\hline
\multicolumn{1}{c}{(1)} & \multicolumn{1}{c}{(2)} & \multicolumn{1}{c}{(3)} & \multicolumn{1}{c}{(4)} & \multicolumn{1}{c}{(5)} & \multicolumn{1}{c}{(6)} & \multicolumn{1}{c}{(7)} & \multicolumn{1}{c}{(8)} & \multicolumn{1}{c}{(9)} & \multicolumn{1}{c}{(10)} & \multicolumn{1}{c}{(11)} & \multicolumn{1}{c}{(12)} & \multicolumn{1}{c}{(13)} & \multicolumn{1}{c}{(14)} & \multicolumn{1}{c}{(15)} & \multicolumn{1}{c}{(16)} & \multicolumn{1}{c}{(17)} & \multicolumn{1}{c}{(18)} \\
\multicolumn{1}{c}{ID} & \multicolumn{1}{c}{Name} & \multicolumn{1}{c}{$N_{\textrm{obs}}^{\textrm{c1}}$} & \multicolumn{1}{c}{$N_{\textrm{obs}}^{\textrm{c2}}$} & \multicolumn{1}{c}{$N_{\textrm{$K_s$}}^{\textrm{c2}}$} & \multicolumn{1}{c}{$M_{\textrm{comp}}^{\textrm{c2}}$} & \multicolumn{1}{c}{Log($N_{\textrm{tot}}^{\textrm{c2}}$)} & \multicolumn{1}{c}{Log($N_{\textrm{tot}}^{\textrm{c1}}$)} & \multicolumn{1}{c}{Log($M_{\textrm{tot}}^{\textrm{c1}}$)} & \multicolumn{1}{c}{Log($M_{\textrm{tot}}^{\textrm{c2}}$)} & \multicolumn{1}{c}{Log(SFR)} & \multicolumn{1}{c}{Log(cSFR)} & \multicolumn{1}{c}{Log($\Sigma_{\textrm{SFR}}$)} & \multicolumn{1}{c}{Log($\frac{\textrm{SFR}}{Length}$)} & \multicolumn{1}{c}{Log(SFE)} & \multicolumn{1}{c}{Log(SFE$_{\textrm{dense}}$)} & \multicolumn{1}{c}{$R$} & \multicolumn{1}{c}{Overlap?} \\
\multicolumn{1}{c}{GMF} &  &  &  &  & \multicolumn{1}{c}{($M_{\sun}$)} &  &  & \multicolumn{1}{c}{($M_{\sun}$)} & \multicolumn{1}{c}{($M_{\sun}$)} & \multicolumn{1}{c}{($M_{\sun}$\,Myr$^{-1}$)} & \multicolumn{1}{c}{($M_{\sun}$\,Myr$^{-1}$)} & \multicolumn{1}{c}{($M_{\sun}$\,Myr$^{-1}$\,pc$^{-2}$)} & \multicolumn{1}{c}{($M_{\sun}$\,Myr$^{-1}$\,pc$^{-1}$)} &  &  &  &  \\
\hline
\endfirsthead
\caption{continued.}\\
\hline\hline
\multicolumn{1}{c}{(1)} & \multicolumn{1}{c}{(2)} & \multicolumn{1}{c}{(3)} & \multicolumn{1}{c}{(4)} & \multicolumn{1}{c}{(5)} & \multicolumn{1}{c}{(6)} & \multicolumn{1}{c}{(7)} & \multicolumn{1}{c}{(8)} & \multicolumn{1}{c}{(9)} & \multicolumn{1}{c}{(10)} & \multicolumn{1}{c}{(11)} & \multicolumn{1}{c}{(12)} & \multicolumn{1}{c}{(13)} & \multicolumn{1}{c}{(14)} & \multicolumn{1}{c}{(15)} & \multicolumn{1}{c}{(16)} & \multicolumn{1}{c}{(17)} & \multicolumn{1}{c}{(18)} \\
\multicolumn{1}{c}{ID} & \multicolumn{1}{c}{Name} & \multicolumn{1}{c}{$N_{\textrm{obs}}^{\textrm{c1}}$} & \multicolumn{1}{c}{$N_{\textrm{obs}}^{\textrm{c2}}$} & \multicolumn{1}{c}{$N_{\textrm{$K_s$}}^{\textrm{c2}}$} & \multicolumn{1}{c}{$M_{\textrm{comp}}^{\textrm{c2}}$} & \multicolumn{1}{c}{Log($N_{\textrm{tot}}^{\textrm{c2}}$)} & \multicolumn{1}{c}{Log($N_{\textrm{tot}}^{\textrm{c1}}$)} & \multicolumn{1}{c}{Log($M_{\textrm{tot}}^{\textrm{c1}}$)} & \multicolumn{1}{c}{Log($M_{\textrm{tot}}^{\textrm{c2}}$)} & \multicolumn{1}{c}{Log(SFR)} & \multicolumn{1}{c}{Log(cSFR)} & \multicolumn{1}{c}{Log($\Sigma_{\textrm{SFR}}$)} & \multicolumn{1}{c}{Log($\frac{\textrm{SFR}}{Length}$)} & \multicolumn{1}{c}{Log(SFE)} & \multicolumn{1}{c}{Log(SFE$_{\textrm{dense}}$)} & \multicolumn{1}{c}{$R$} & \multicolumn{1}{c}{Overlap?} \\
\multicolumn{1}{c}{GMF} &  &  &  &  & \multicolumn{1}{c}{($M_{\sun}$)} &  &  & \multicolumn{1}{c}{($M_{\sun}$)} & \multicolumn{1}{c}{($M_{\sun}$)} & \multicolumn{1}{c}{($M_{\sun}$\,Myr$^{-1}$)} & \multicolumn{1}{c}{($M_{\sun}$\,Myr$^{-1}$)} & \multicolumn{1}{c}{($M_{\sun}$\,Myr$^{-1}$\,pc$^{-2}$)} & \multicolumn{1}{c}{($M_{\sun}$\,Myr$^{-1}$\,pc$^{-1}$)} &  &  &  &  \\
\hline
\endhead
\hline
\endfoot
2 & F19 & 10 & 20 & 14 & 2.23$\pm$1.29 & 2.69$_{-0.69}^{+0.26}$ & 1.73$_{-0.32}^{+0.13}$ & 1.43$_{-0.32}^{+0.13}$ & 2.44$_{-0.48}^{+0.26}$ & 2.19$_{-0.49}^{+0.26}$ & 1.70$_{-0.32}^{+0.13}$ & 0.15$_{-0.49}^{+0.26}$ & 0.95$_{-0.49}^{+0.26}$ & $-$1.66$_{-0.44}^{+0.34}$ & $-$1.34$_{-0.39}^{+0.33}$ & 0.2 & Y\\
3 & GMF20.0$-$17.9 & 161 & 253 & 201 & 1.79$\pm$0.66 & 3.49$_{-0.30}^{+0.17}$ & 3.08$_{-0.08}^{+0.06}$ & 2.78$_{-0.08}^{+0.06}$ & 3.24$_{-0.32}^{+0.18}$ & 3.07$_{-0.22}^{+0.14}$ & 3.05$_{-0.08}^{+0.06}$ & -0.18$_{-0.22}^{+0.14}$ & 0.91$_{-0.22}^{+0.14}$ & $-$2.00$_{-0.30}^{+0.29}$ & $-$1.70$_{-0.30}^{+0.28}$ & 0.6 & Y\\
6 & BC\_021.25$-$0.15 & 15 & 21 & 18 & 2.20$\pm$0.61 & 2.35$_{-0.33}^{+0.23}$ & 1.94$_{-0.37}^{+0.16}$ & 1.64$_{-0.37}^{+0.16}$ & 2.08$_{-0.38}^{+0.24}$ & 1.92$_{-0.26}^{+0.18}$ & 1.91$_{-0.37}^{+0.16}$ & -0.38$_{-0.26}^{+0.18}$ & 0.38$_{-0.26}^{+0.18}$ & $-$2.05$_{-0.34}^{+0.30}$ & $-$0.62$_{-0.24}^{+0.21}$ & 0.3 & N\\
7 & G022.519$-$0.025 & 56 & 216 & 192 & 2.92$\pm$0.58 & 3.72$_{-0.20}^{+0.13}$ & 2.96$_{-0.21}^{+0.09}$ & 2.66$_{-0.21}^{+0.09}$ & 3.48$_{-0.21}^{+0.14}$ & 3.24$_{-0.19}^{+0.12}$ & 2.93$_{-0.21}^{+0.09}$ & -0.13$_{-0.19}^{+0.12}$ & 1.36$_{-0.19}^{+0.12}$ & $-$1.91$_{-0.24}^{+0.27}$ & $-$1.51$_{-0.24}^{+0.26}$ & 0.2 & N\\
8 & F21 & 127 & 300 & 248 & 2.52$\pm$0.59 & 3.84$_{-0.21}^{+0.13}$ & 3.05$_{-0.11}^{+0.06}$ & 2.75$_{-0.11}^{+0.06}$ & 3.59$_{-0.21}^{+0.13}$ & 3.35$_{-0.19}^{+0.11}$ & 3.01$_{-0.11}^{+0.06}$ & 0.13$_{-0.19}^{+0.11}$ & 1.44$_{-0.19}^{+0.11}$ & $-$1.74$_{-0.27}^{+0.27}$ & $-$1.53$_{-0.25}^{+0.28}$ & 0.3 & Y\\
9 & F23 & 144 & 337 & 265 & 3.03$\pm$0.56 & 3.97$_{-0.18}^{+0.11}$ & 3.27$_{-0.11}^{+0.05}$ & 2.97$_{-0.11}^{+0.05}$ & 3.72$_{-0.18}^{+0.11}$ & 3.49$_{-0.16}^{+0.09}$ & 3.24$_{-0.11}^{+0.05}$ & 0.12$_{-0.16}^{+0.09}$ & 1.60$_{-0.16}^{+0.09}$ & $-$1.69$_{-0.26}^{+0.28}$ & $-$1.35$_{-0.23}^{+0.27}$ & 0.2 & Y\\
11 & F22 & 81 & 149 & 105 & 2.86$\pm$0.60 & 3.60$_{-0.20}^{+0.13}$ & 2.89$_{-0.13}^{+0.07}$ & 2.58$_{-0.13}^{+0.07}$ & 3.36$_{-0.20}^{+0.14}$ & 3.12$_{-0.15}^{+0.12}$ & 2.85$_{-0.13}^{+0.07}$ & 0.22$_{-0.15}^{+0.12}$ & 1.56$_{-0.15}^{+0.12}$ & $-$1.78$_{-0.26}^{+0.29}$ & $-$1.65$_{-0.25}^{+0.28}$ & 0.3 & Y\\
12 & CFG024.00$+$0.48 & 50 & 113 & 96 & 2.35$\pm$0.63 & 3.32$_{-0.27}^{+0.19}$ & 2.77$_{-0.18}^{+0.07}$ & 2.46$_{-0.18}^{+0.07}$ & 3.07$_{-0.28}^{+0.19}$ & 2.87$_{-0.24}^{+0.16}$ & 2.73$_{-0.18}^{+0.07}$ & -0.27$_{-0.24}^{+0.16}$ & 0.89$_{-0.24}^{+0.16}$ & $-$1.99$_{-0.31}^{+0.29}$ & $-$1.42$_{-0.27}^{+0.29}$ & 0.4 & N\\
13 & F26 & 24 & 60 & 46 & 2.28$\pm$0.61 & 3.07$_{-0.27}^{+0.15}$ & 2.17$_{-0.23}^{+0.13}$ & 1.87$_{-0.23}^{+0.13}$ & 2.82$_{-0.28}^{+0.16}$ & 2.57$_{-0.27}^{+0.16}$ & 2.14$_{-0.23}^{+0.13}$ & -0.21$_{-0.27}^{+0.16}$ & 0.91$_{-0.27}^{+0.16}$ & $-$2.07$_{-0.29}^{+0.30}$ & $-$1.81$_{-0.29}^{+0.29}$ & 0.3 & Y\\
14 & F27 & 47 & 103 & 84 & 3.04$\pm$0.59 & 3.38$_{-0.22}^{+0.13}$ & 2.72$_{-0.13}^{+0.09}$ & 2.42$_{-0.13}^{+0.09}$ & 3.13$_{-0.23}^{+0.14}$ & 2.91$_{-0.21}^{+0.11}$ & 2.69$_{-0.13}^{+0.09}$ & -0.02$_{-0.21}^{+0.11}$ & 1.16$_{-0.21}^{+0.11}$ & $-$1.84$_{-0.26}^{+0.28}$ & $-$1.52$_{-0.25}^{+0.27}$ & 0.3 & Y\\
15 & BC\_24.95$-$0.17 & 134 & 334 & 282 & 2.35$\pm$0.60 & 3.78$_{-0.24}^{+0.14}$ & 2.99$_{-0.08}^{+0.06}$ & 2.69$_{-0.08}^{+0.06}$ & 3.53$_{-0.25}^{+0.14}$ & 3.29$_{-0.18}^{+0.12}$ & 2.95$_{-0.08}^{+0.06}$ & 0.14$_{-0.18}^{+0.12}$ & 1.46$_{-0.18}^{+0.12}$ & $-$1.62$_{-0.28}^{+0.28}$ & $-$1.15$_{-0.25}^{+0.26}$ & 0.3 & Y\\
16 & BC\_025.24$-$0.45 & 44 & 121 & 109 & 2.33$\pm$0.61 & 3.24$_{-0.24}^{+0.18}$ & 2.25$_{-0.18}^{+0.10}$ & 1.95$_{-0.18}^{+0.10}$ & 2.98$_{-0.25}^{+0.19}$ & 2.72$_{-0.24}^{+0.17}$ & 2.22$_{-0.18}^{+0.10}$ & -0.37$_{-0.24}^{+0.17}$ & 1.03$_{-0.24}^{+0.17}$ & $-$2.15$_{-0.30}^{+0.32}$ & $-$1.72$_{-0.32}^{+0.31}$ & 0.6 & Y\\
17 & F28 & 19 & 39 & 35 & 1.93$\pm$0.68 & 2.80$_{-0.38}^{+0.21}$ & 2.40$_{-0.37}^{+0.14}$ & 2.10$_{-0.37}^{+0.14}$ & 2.55$_{-0.41}^{+0.21}$ & 2.38$_{-0.25}^{+0.16}$ & 2.37$_{-0.37}^{+0.14}$ & 0.22$_{-0.25}^{+0.16}$ & 0.88$_{-0.25}^{+0.16}$ & $-$1.64$_{-0.30}^{+0.29}$ & $-$1.39$_{-0.30}^{+0.29}$ & 0.4 & Y\\
18 & F29 & 77 & 130 & 105 & 2.34$\pm$0.60 & 3.45$_{-0.26}^{+0.14}$ & 2.84$_{-0.11}^{+0.08}$ & 2.54$_{-0.11}^{+0.08}$ & 3.20$_{-0.26}^{+0.14}$ & 2.99$_{-0.18}^{+0.11}$ & 2.81$_{-0.11}^{+0.08}$ & -0.24$_{-0.18}^{+0.11}$ & 1.16$_{-0.18}^{+0.11}$ & $-$2.05$_{-0.26}^{+0.28}$ & $-$1.70$_{-0.27}^{+0.28}$ & 0.5 & Y\\
19 & G025.762$+$0.241 & 13 & 19 & 13 & 4.71$\pm$1.07 & 2.98$_{-0.39}^{+0.20}$ & 2.67$_{-0.33}^{+0.15}$ & 2.37$_{-0.33}^{+0.15}$ & 2.73$_{-0.40}^{+0.20}$ & 2.59$_{-0.21}^{+0.14}$ & 2.64$_{-0.33}^{+0.15}$ & -0.18$_{-0.21}^{+0.14}$ & 0.91$_{-0.21}^{+0.14}$ & $-$2.01$_{-0.27}^{+0.30}$ & $-$1.75$_{-0.27}^{+0.29}$ & 0.6 & N\\
20 & BC\_26.94$-$0.30 & 7 & 5 & 4 & 2.40$\pm$1.80 & 2.05$_{-0.48}^{+0.34}$ & 1.92$_{-0.26}^{+0.14}$ & 1.62$_{-0.26}^{+0.14}$ & 1.81$_{-0.50}^{+0.34}$ & 1.73$_{-0.25}^{+0.23}$ & 1.89$_{-0.26}^{+0.14}$ & -0.42$_{-0.25}^{+0.23}$ & 0.30$_{-0.25}^{+0.23}$ & $-$2.15$_{-0.35}^{+0.34}$ & $-$1.56$_{-0.36}^{+0.33}$ & 3.1 & N\\
21 & F30 & 25 & 43 & 36 & 2.27$\pm$0.61 & 2.91$_{-0.30}^{+0.17}$ & 2.11$_{-0.19}^{+0.09}$ & 1.81$_{-0.19}^{+0.09}$ & 2.66$_{-0.32}^{+0.17}$ & 2.42$_{-0.28}^{+0.15}$ & 2.08$_{-0.19}^{+0.09}$ & -0.44$_{-0.28}^{+0.15}$ & 0.73$_{-0.28}^{+0.15}$ & $-$2.19$_{-0.30}^{+0.31}$ & $-$1.65$_{-0.29}^{+0.29}$ & 0.8 & N\\
22 & F31 & 283 & 715 & 620 & 2.01$\pm$0.63 & 4.04$_{-0.29}^{+0.16}$ & 3.16$_{-0.05}^{+0.04}$ & 2.86$_{-0.05}^{+0.04}$ & 3.78$_{-0.31}^{+0.16}$ & 3.53$_{-0.22}^{+0.15}$ & 3.13$_{-0.05}^{+0.04}$ & 0.01$_{-0.22}^{+0.15}$ & 1.49$_{-0.22}^{+0.15}$ & $-$1.77$_{-0.30}^{+0.29}$ & $-$1.31$_{-0.29}^{+0.28}$ & 0.4 & Y\\
23 & CFG028.68$-$0.28 & 29 & 72 & 61 & 2.54$\pm$0.60 & 3.12$_{-0.27}^{+0.16}$ & 2.11$_{-0.21}^{+0.12}$ & 1.81$_{-0.21}^{+0.12}$ & 2.87$_{-0.28}^{+0.16}$ & 2.61$_{-0.27}^{+0.14}$ & 2.07$_{-0.21}^{+0.12}$ & -0.26$_{-0.27}^{+0.14}$ & 0.93$_{-0.27}^{+0.14}$ & $-$2.13$_{-0.30}^{+0.30}$ & $-$1.92$_{-0.28}^{+0.30}$ & 0.4 & Y\\
24 & G028.854$-$0.238 & 20 & 39 & 33 & 3.48$\pm$0.53 & 3.07$_{-0.22}^{+0.15}$ & 2.17$_{-0.23}^{+0.11}$ & 1.87$_{-0.23}^{+0.11}$ & 2.82$_{-0.23}^{+0.15}$ & 2.57$_{-0.19}^{+0.15}$ & 2.14$_{-0.23}^{+0.11}$ & -0.72$_{-0.19}^{+0.15}$ & 0.79$_{-0.19}^{+0.15}$ & $-$2.54$_{-0.30}^{+0.30}$ & $-$2.19$_{-0.27}^{+0.29}$ & 0.5 & Y\\
25 & CFG029.18$-$0.34 & 18 & 64 & 56 & 2.94$\pm$0.58 & 3.05$_{-0.29}^{+0.17}$ & 2.07$_{-0.37}^{+0.16}$ & 1.77$_{-0.37}^{+0.16}$ & 2.79$_{-0.30}^{+0.18}$ & 2.53$_{-0.27}^{+0.16}$ & 2.04$_{-0.37}^{+0.16}$ & -0.55$_{-0.27}^{+0.16}$ & 0.70$_{-0.27}^{+0.16}$ & $-$2.47$_{-0.33}^{+0.30}$ & $-$2.29$_{-0.29}^{+0.30}$ & 0.7 & N\\
26 & G030.315$-$0.154 & 101 & 200 & 178 & 2.45$\pm$0.60 & 3.62$_{-0.22}^{+0.14}$ & 2.94$_{-0.11}^{+0.07}$ & 2.63$_{-0.11}^{+0.07}$ & 3.38$_{-0.23}^{+0.14}$ & 3.15$_{-0.20}^{+0.12}$ & 2.90$_{-0.11}^{+0.07}$ & 0.06$_{-0.20}^{+0.12}$ & 1.48$_{-0.20}^{+0.12}$ & $-$1.87$_{-0.26}^{+0.29}$ & $-$1.70$_{-0.26}^{+0.28}$ & 0.4 & Y\\
27 & G32.02$+$0.06 & 39 & 80 & 65 & 2.41$\pm$0.64 & 3.28$_{-0.24}^{+0.16}$ & 2.46$_{-0.14}^{+0.07}$ & 2.16$_{-0.14}^{+0.07}$ & 3.04$_{-0.25}^{+0.16}$ & 2.79$_{-0.19}^{+0.15}$ & 2.43$_{-0.14}^{+0.07}$ & -0.31$_{-0.19}^{+0.15}$ & 0.82$_{-0.19}^{+0.15}$ & $-$2.16$_{-0.31}^{+0.30}$ & $-$1.88$_{-0.29}^{+0.29}$ & 1.3 & N\\
29 & F34 & 63 & 69 & 41 & 3.12$\pm$0.55 & 3.35$_{-0.18}^{+0.12}$ & 3.25$_{-0.13}^{+0.07}$ & 2.95$_{-0.13}^{+0.07}$ & 3.11$_{-0.18}^{+0.12}$ & 3.04$_{-0.10}^{+0.08}$ & 3.21$_{-0.13}^{+0.07}$ & -0.28$_{-0.10}^{+0.08}$ & 0.98$_{-0.10}^{+0.08}$ & $-$1.98$_{-0.22}^{+0.27}$ & $-$1.13$_{-0.21}^{+0.25}$ & 0.3 & Y\\
30 & G033.104$+$0.068 & 22 & 13 & 6 & 4.16$\pm$1.75 & 2.69$_{-0.59}^{+0.27}$ & 2.48$_{-0.20}^{+0.10}$ & 2.18$_{-0.20}^{+0.10}$ & 2.45$_{-0.62}^{+0.32}$ & 2.34$_{-0.42}^{+0.22}$ & 2.44$_{-0.20}^{+0.10}$ & -0.89$_{-0.42}^{+0.22}$ & 0.20$_{-0.42}^{+0.22}$ & $-$2.57$_{-0.37}^{+0.33}$ & $-$1.83$_{-0.32}^{+0.32}$ & 0.4 & Y\\
31 & G033.685$-$0.020 & 123 & 130 & 93 & 3.22$\pm$0.52 & 3.52$_{-0.18}^{+0.11}$ & 3.33$_{-0.08}^{+0.06}$ & 3.03$_{-0.08}^{+0.06}$ & 3.27$_{-0.18}^{+0.11}$ & 3.17$_{-0.11}^{+0.08}$ & 3.30$_{-0.08}^{+0.06}$ & -0.43$_{-0.11}^{+0.08}$ & 0.85$_{-0.11}^{+0.08}$ & $-$2.21$_{-0.23}^{+0.28}$ & $-$1.82$_{-0.23}^{+0.27}$ & 0.4 & Y\\
32 & GMF38.1$-$32.4b & 84 & 144 & 120 & 1.64$\pm$0.66 & 3.21$_{-0.34}^{+0.20}$ & 2.39$_{-0.10}^{+0.08}$ & 2.09$_{-0.10}^{+0.08}$ & 2.95$_{-0.38}^{+0.22}$ & 2.71$_{-0.28}^{+0.19}$ & 2.36$_{-0.10}^{+0.08}$ & -0.06$_{-0.28}^{+0.19}$ & 0.86$_{-0.28}^{+0.19}$ & $-$1.76$_{-0.36}^{+0.31}$ & $-$1.03$_{-0.31}^{+0.27}$ & 0.4 & Y\\
33 & GMF38.1$-$32.4a & 430 & 677 & 529 & 1.96$\pm$0.64 & 3.98$_{-0.31}^{+0.17}$ & 3.23$_{-0.04}^{+0.03}$ & 2.93$_{-0.04}^{+0.03}$ & 3.72$_{-0.33}^{+0.17}$ & 3.49$_{-0.22}^{+0.15}$ & 3.20$_{-0.04}^{+0.03}$ & -0.29$_{-0.22}^{+0.15}$ & 1.13$_{-0.22}^{+0.15}$ & $-$2.06$_{-0.33}^{+0.30}$ & $-$1.65$_{-0.31}^{+0.29}$ & 0.8 & Y\\
35 & GMF41.0$-$41.3 & 34 & 51 & 39 & 1.33$\pm$0.61 & 2.66$_{-0.37}^{+0.22}$ & 2.14$_{-0.15}^{+0.11}$ & 1.84$_{-0.15}^{+0.11}$ & 2.39$_{-0.44}^{+0.23}$ & 2.20$_{-0.26}^{+0.20}$ & 2.11$_{-0.15}^{+0.11}$ & -0.36$_{-0.26}^{+0.20}$ & 0.49$_{-0.26}^{+0.20}$ & $-$1.99$_{-0.37}^{+0.31}$ & $-$0.56$_{-0.23}^{+0.20}$ & 0.6 & N\\
37 & F39 & 29 & 13 & 6 & 4.01$\pm$0.88 & 2.65$_{-0.32}^{+0.22}$ & 2.63$_{-0.22}^{+0.11}$ & 2.33$_{-0.22}^{+0.11}$ & 2.41$_{-0.33}^{+0.22}$ & 2.38$_{-0.17}^{+0.13}$ & 2.60$_{-0.22}^{+0.11}$ & -1.16$_{-0.17}^{+0.13}$ & 0.37$_{-0.17}^{+0.13}$ & $-$3.04$_{-0.28}^{+0.30}$ & $-$2.84$_{-0.28}^{+0.29}$ & 3.4 & N\\
38 & F40 & 40 & 32 & 16 & 3.40$\pm$1.10 & 2.96$_{-0.42}^{+0.22}$ & 2.35$_{-0.17}^{+0.09}$ & 2.04$_{-0.17}^{+0.09}$ & 2.71$_{-0.44}^{+0.22}$ & 2.50$_{-0.31}^{+0.20}$ & 2.31$_{-0.17}^{+0.09}$ & -1.05$_{-0.31}^{+0.20}$ & 0.51$_{-0.31}^{+0.20}$ & $-$2.79$_{-0.38}^{+0.32}$ & $-$2.24$_{-0.34}^{+0.32}$ & 1.3 & N\\
39 & CFG047.06$+$0.26 & 24 & 25 & 21 & 2.40$\pm$0.62 & 2.49$_{-0.34}^{+0.23}$ & 1.94$_{-0.24}^{+0.13}$ & 1.64$_{-0.24}^{+0.13}$ & 2.23$_{-0.39}^{+0.24}$ & 2.03$_{-0.26}^{+0.20}$ & 1.90$_{-0.24}^{+0.13}$ & -0.41$_{-0.26}^{+0.20}$ & 0.38$_{-0.26}^{+0.20}$ & $-$2.25$_{-0.34}^{+0.31}$ & $-$1.95$_{-0.32}^{+0.31}$ & 8.2 & N\\
41 & CFG049.21$-$0.34 & 65 & 47 & 42 & 2.79$\pm$0.60 & 3.05$_{-0.23}^{+0.16}$ & 2.99$_{-0.17}^{+0.09}$ & 2.69$_{-0.17}^{+0.09}$ & 2.80$_{-0.24}^{+0.16}$ & 2.75$_{-0.15}^{+0.10}$ & 2.95$_{-0.17}^{+0.09}$ & -0.17$_{-0.15}^{+0.10}$ & 0.66$_{-0.15}^{+0.10}$ & $-$2.04$_{-0.24}^{+0.28}$ & $-$1.81$_{-0.23}^{+0.28}$ & 0.5 & Y\\
42 & F42 & 118 & 87 & 72 & 2.50$\pm$0.60 & 3.25$_{-0.20}^{+0.15}$ & 2.88$_{-0.09}^{+0.06}$ & 2.58$_{-0.09}^{+0.06}$ & 3.00$_{-0.20}^{+0.16}$ & 2.84$_{-0.14}^{+0.12}$ & 2.84$_{-0.09}^{+0.06}$ & -0.43$_{-0.14}^{+0.12}$ & 1.04$_{-0.14}^{+0.12}$ & $-$2.27$_{-0.25}^{+0.28}$ & $-$1.98$_{-0.28}^{+0.29}$ & 0.6 & Y\\
43 & GMF54.0$-$52.0 & 105 & 177 & 146 & 0.92$\pm$0.43 & 3.11$_{-0.33}^{+0.20}$ & 2.40$_{-0.08}^{+0.06}$ & 2.10$_{-0.08}^{+0.06}$ & 2.85$_{-0.38}^{+0.20}$ & 2.62$_{-0.32}^{+0.17}$ & 2.36$_{-0.08}^{+0.06}$ & 0.28$_{-0.32}^{+0.17}$ & 1.00$_{-0.32}^{+0.17}$ & $-$1.52$_{-0.32}^{+0.30}$ & $-$1.16$_{-0.29}^{+0.28}$ & 2.7 & N\\
44 & GMF307.2$-$305.4 & 732 & 990 & 922 & 1.25$\pm$0.60 & 3.89$_{-0.36}^{+0.21}$ & 3.47$_{-0.06}^{+0.04}$ & 3.17$_{-0.06}^{+0.04}$ & 3.63$_{-0.42}^{+0.21}$ & 3.46$_{-0.33}^{+0.18}$ & 3.44$_{-0.06}^{+0.04}$ & -0.45$_{-0.33}^{+0.18}$ & 1.10$_{-0.33}^{+0.18}$ & $-$2.26$_{-0.31}^{+0.31}$ & $-$1.90$_{-0.30}^{+0.30}$ & 3.3 & N\\
45 & GMF309.5$-$308.7 & 296 & 362 & 326 & 1.80$\pm$0.65 & 3.52$_{-0.40}^{+0.20}$ & 3.01$_{-0.07}^{+0.05}$ & 2.71$_{-0.07}^{+0.05}$ & 3.25$_{-0.36}^{+0.21}$ & 3.06$_{-0.25}^{+0.18}$ & 2.98$_{-0.07}^{+0.05}$ & -0.76$_{-0.25}^{+0.18}$ & 0.98$_{-0.25}^{+0.18}$ & $-$2.61$_{-0.33}^{+0.33}$ & $-$2.34$_{-0.35}^{+0.32}$ & 1.9 & N\\
46 & GMF319.0$-$318.7 & 429 & 498 & 449 & 1.40$\pm$0.63 & 3.58$_{-0.33}^{+0.20}$ & 2.99$_{-0.12}^{+0.08}$ & 2.69$_{-0.12}^{+0.08}$ & 3.32$_{-0.38}^{+0.23}$ & 3.11$_{-0.34}^{+0.18}$ & 2.96$_{-0.12}^{+0.08}$ & -0.53$_{-0.34}^{+0.18}$ & 1.10$_{-0.34}^{+0.18}$ & $-$2.39$_{-0.36}^{+0.32}$ & $-$2.17$_{-0.37}^{+0.31}$ & 3.5 & N\\
47 & GMF324.5$-$321.4b & 51 & 78 & 75 & 0.99$\pm$0.47 & 2.74$_{-0.47}^{+0.21}$ & 2.17$_{-0.15}^{+0.12}$ & 1.86$_{-0.15}^{+0.12}$ & 2.48$_{-0.38}^{+0.21}$ & 2.28$_{-0.34}^{+0.20}$ & 2.13$_{-0.15}^{+0.12}$ & -0.08$_{-0.34}^{+0.20}$ & 0.77$_{-0.34}^{+0.20}$ & $-$1.86$_{-0.34}^{+0.30}$ & $-$1.47$_{-0.30}^{+0.30}$ & 17.7 & N\\
48 & G322.363$+$0.542 & 6 & 11 & 8 & 3.92$\pm$1.19 & 2.72$_{-0.33}^{+0.21}$ & 1.80$_{-0.30}^{+0.19}$ & 1.50$_{-0.30}^{+0.19}$ & 2.48$_{-0.34}^{+0.22}$ & 2.22$_{-0.32}^{+0.19}$ & 1.77$_{-0.30}^{+0.19}$ & -1.28$_{-0.32}^{+0.19}$ & 0.03$_{-0.32}^{+0.19}$ & $-$3.06$_{-0.33}^{+0.33}$ & $-$2.66$_{-0.34}^{+0.32}$ & 10.0 & N\\
50 & G327.157$-$0.256 & 240 & 340 & 284 & 1.74$\pm$0.66 & 3.67$_{-0.30}^{+0.18}$ & 3.48$_{-0.08}^{+0.06}$ & 3.18$_{-0.08}^{+0.06}$ & 3.42$_{-0.31}^{+0.19}$ & 3.32$_{-0.17}^{+0.13}$ & 3.45$_{-0.08}^{+0.06}$ & -0.20$_{-0.17}^{+0.13}$ & 1.30$_{-0.17}^{+0.13}$ & $-$2.00$_{-0.26}^{+0.28}$ & $-$1.67$_{-0.27}^{+0.28}$ & 0.7 & N\\
51 & GMF335.6$-$333.6b & 301 & 477 & 374 & 1.62$\pm$0.66 & 3.70$_{-0.35}^{+0.18}$ & 3.07$_{-0.07}^{+0.04}$ & 2.77$_{-0.07}^{+0.04}$ & 3.45$_{-0.37}^{+0.18}$ & 3.23$_{-0.34}^{+0.18}$ & 3.04$_{-0.07}^{+0.04}$ & 0.06$_{-0.34}^{+0.18}$ & 1.34$_{-0.34}^{+0.18}$ & $-$1.77$_{-0.32}^{+0.31}$ & $-$1.48$_{-0.31}^{+0.30}$ & 0.5 & N\\
52 & GMF335.6$-$333.6a & 1524 & 2397 & 2012 & 1.80$\pm$0.65 & 4.45$_{-0.30}^{+0.17}$ & 3.91$_{-0.09}^{+0.07}$ & 3.61$_{-0.09}^{+0.07}$ & 4.19$_{-0.32}^{+0.17}$ & 3.99$_{-0.24}^{+0.15}$ & 3.87$_{-0.09}^{+0.07}$ & -0.16$_{-0.24}^{+0.15}$ & 1.57$_{-0.24}^{+0.15}$ & $-$2.10$_{-0.32}^{+0.30}$ & $-$1.93$_{-0.29}^{+0.30}$ & 1.6 & N\\
53 & Nessie & 51 & 137 & 98 & 1.47$\pm$0.65 & 3.18$_{-0.44}^{+0.22}$ & 2.28$_{-0.14}^{+0.09}$ & 1.97$_{-0.14}^{+0.09}$ & 2.92$_{-0.35}^{+0.23}$ & 2.67$_{-0.34}^{+0.21}$ & 2.24$_{-0.14}^{+0.09}$ & 0.23$_{-0.34}^{+0.21}$ & 0.87$_{-0.34}^{+0.21}$ & $-$1.52$_{-0.34}^{+0.26}$ & $-$1.34$_{-0.31}^{+0.26}$ & \ldots & N\\
54 & GMF341.9$-$337.1 & 605 & 1158 & 947 & 1.90$\pm$0.65 & 4.18$_{-0.29}^{+0.17}$ & 3.53$_{-0.05}^{+0.03}$ & 3.22$_{-0.05}^{+0.03}$ & 3.92$_{-0.31}^{+0.18}$ & 3.70$_{-0.28}^{+0.14}$ & 3.49$_{-0.05}^{+0.03}$ & -0.22$_{-0.28}^{+0.14}$ & 1.52$_{-0.28}^{+0.14}$ & $-$2.11$_{-0.32}^{+0.31}$ & $-$1.88$_{-0.30}^{+0.29}$ & 1.2 & N\\
57 & GMF358.9$-$357.4 & 56 & 97 & 70 & 1.85$\pm$0.66 & 2.97$_{-0.31}^{+0.20}$ & 2.18$_{-0.16}^{+0.10}$ & 1.88$_{-0.16}^{+0.10}$ & 2.71$_{-0.34}^{+0.21}$ & 2.47$_{-0.33}^{+0.18}$ & 2.15$_{-0.16}^{+0.10}$ & -0.27$_{-0.33}^{+0.18}$ & 0.87$_{-0.33}^{+0.18}$ & $-$2.09$_{-0.38}^{+0.31}$ & $-$1.79$_{-0.36}^{+0.32}$ & 0.9 & N\\
\end{longtable}
\tablefoot{Columns are (1) GMF ID; (2) GMF name; (3) number of detected Class I sources; (4) number of detected Class II sources; (5) number of Class II sources with the $K_s$ band detection; (6) mass completeness of Class II sources; (7) total number of Class II sources estimated assuming a universal IMF; (8) total number of Class I sources estimated assuming a universal luminosity function; (9) total mass of Class I sources; (10) total mass of Class II sources; (11) star formation rate; (12) current star formation rate obtained with only Class I sources; (13) surface density of star formation rate; (14) star formation rate per unit length; (15) star formation efficiency; (16) dense gas star formation efficiency; (17) $^{13}$CO flux ratio of GMF to other velocity components. High value implies that the GMF is the dominating component along the line of sight; (18) overlap with other GMFs? Note that the associated uncertainties of the parameters listed in this table do not include the distance errors (also see Appendix~\ref{ap4} for more information).}
\end{landscape}
}

\clearpage

\begin{table*}
\scriptsize
\caption{Properties of nearby molecular clouds}
\label{table3}
\centering
\begin{tabular}{l c r r c c c c c c c}
\hline\hline
\multicolumn{1}{l}{Cloud} & \multicolumn{1}{c}{Dist.\tablefootmark{a}} & \multicolumn{1}{c}{Length\tablefootmark{b}} & \multicolumn{1}{c}{SFR} & \multicolumn{1}{c}{$M_{\textrm{cloud}}$\tablefootmark{c}} & \multicolumn{1}{c}{$M_{\textrm{dense}}$\tablefootmark{c}} & \multicolumn{1}{c}{$\Sigma_{\textrm{SFR}}$} & \multicolumn{1}{c}{$\Sigma_{\textrm{gas}}$} & \multicolumn{1}{c}{$t_{\textrm{ff}}$\tablefootmark{d}} & \multicolumn{1}{c}{$\Delta v$\tablefootmark{a}} & \multicolumn{1}{c}{$t_{\textrm{cross}}$\tablefootmark{e}} \\ 
\multicolumn{1}{c}{} & \multicolumn{1}{c}{(pc)} & \multicolumn{1}{c}{(pc)} & \multicolumn{1}{c}{($M_{\sun}$\,Myr$^{-1}$)} & \multicolumn{1}{c}{($M_{\sun}$)} & \multicolumn{1}{c}{($M_{\sun}$)} & \multicolumn{1}{c}{($M_{\sun}$\,Myr$^{-1}$\,pc$^{-2}$)} & \multicolumn{1}{c}{($M_{\sun}$\,pc$^{-2}$)} & \multicolumn{1}{c}{(Myr)} & \multicolumn{1}{c}{(km\,s$^{-1}$)} & \multicolumn{1}{c}{(Myr)} \\ 
\hline
Aquila &260$\pm$55 &22.05$\pm$4.66 &105.00$\pm$5.12 &32160$\pm$14411 &19507.4$\pm$8734.9 &0.52$\pm$0.22 &160.03$\pm$4.00 &2.10$\pm$0.22 &3.00 &7.70$\pm$2.82\\
AurigaN &450$\pm$23 &26.11$\pm$1.33 &1.25$\pm$0.56 &1813$\pm$360 &200.2$\pm$41.0 &0.06$\pm$0.03 &92.91$\pm$8.97 &1.54$\pm$0.08 &1.70 &4.23$\pm$1.29\\
Auriga &450$\pm$23 &49.67$\pm$2.54 &26.25$\pm$2.56 &8965$\pm$1778 &1215.7$\pm$239.6 &0.28$\pm$0.04 &95.61$\pm$9.19 &2.25$\pm$0.12 &1.70 &9.28$\pm$2.82\\
Cepheus1 &300$\pm$34 &7.65$\pm$0.87 &7.75$\pm$1.39 &495$\pm$180 &211.0$\pm$76.6 &1.90$\pm$0.55 &121.13$\pm$16.62 &0.91$\pm$0.08 &1.90 &1.73$\pm$0.56\\
Cepheus3 &288$\pm$25 &4.35$\pm$0.38 &6.75$\pm$1.30 &342$\pm$98 &94.0$\pm$26.3 &2.24$\pm$0.58 &113.40$\pm$12.80 &0.88$\pm$0.06 &1.60 &1.77$\pm$0.55\\
Cepheus5 &200$\pm$66 &3.78$\pm$1.25 &2.25$\pm$0.75 &214$\pm$177 &51.6$\pm$44.3 &1.15$\pm$0.85 &109.83$\pm$18.58 &0.80$\pm$0.15 &1.60 &1.42$\pm$0.63\\
ChaIII &150$\pm$15 &6.89$\pm$0.69 &0.25$\pm$0.25 &284$\pm$82 &16.2$\pm$4.5 &0.08$\pm$0.08 &87.58$\pm$7.93 &1.01$\pm$0.07 &1.10 &2.67$\pm$0.84\\
ChaII &178$\pm$18 &5.77$\pm$0.58 &4.25$\pm$1.03 &537$\pm$151 &160.3$\pm$45.4 &0.87$\pm$0.28 &110.70$\pm$8.77 &1.00$\pm$0.06 &1.20 &2.99$\pm$0.95\\
ChaI &150$\pm$15 &4.33$\pm$0.43 &15.25$\pm$1.95 &694$\pm$196 &323.9$\pm$91.5 &2.86$\pm$0.68 &130.25$\pm$10.82 &0.94$\pm$0.06 &0.85 &4.42$\pm$1.40\\
CoronaAus. &130$\pm$11 &3.46$\pm$0.29 &8.00$\pm$1.41 &329$\pm$82 &190.9$\pm$49.6 &3.51$\pm$0.86 &144.57$\pm$11.88 &0.72$\pm$0.04 &1.50 &1.64$\pm$0.51\\
IC5146E &950$\pm$80 &8.76$\pm$0.74 &17.25$\pm$2.08 &1869$\pm$417 &124.0$\pm$27.3 &0.87$\pm$0.18 &94.81$\pm$5.22 &1.53$\pm$0.08 &2.45 &2.95$\pm$0.92\\
IC5146NW &950$\pm$80 &16.99$\pm$1.43 &7.50$\pm$1.37 &5249$\pm$1251 &1330.9$\pm$329.9 &0.15$\pm$0.04 &102.09$\pm$7.14 &1.87$\pm$0.10 &2.04 &5.72$\pm$1.78\\
LupusIII &200$\pm$40 &7.11$\pm$1.42 &6.00$\pm$1.22 &400$\pm$178 &111.3$\pm$49.8 &1.55$\pm$0.70 &103.37$\pm$4.84 &0.98$\pm$0.10 &2.11 &1.52$\pm$0.55\\
LupusIV &150$\pm$40 &4.13$\pm$1.10 &1.00$\pm$0.50 &184$\pm$107 &70.7$\pm$41.4 &0.61$\pm$0.45 &112.70$\pm$5.50 &0.75$\pm$0.10 &1.53 &1.36$\pm$0.55\\
LupusI &150$\pm$40 &8.16$\pm$2.18 &2.50$\pm$0.79 &393$\pm$238 &93.1$\pm$56.0 &0.64$\pm$0.40 &100.37$\pm$7.33 &0.99$\pm$0.14 &2.17 &1.49$\pm$0.60\\
LupusVI &150$\pm$40 &6.73$\pm$1.80 &0.50$\pm$0.35 &216$\pm$123 &14.3$\pm$8.1 &0.18$\pm$0.16 &78.95$\pm$2.85 &1.02$\pm$0.14 &1.20 &2.25$\pm$0.90\\
LupusV &150$\pm$40 &5.02$\pm$1.34 &0.25$\pm$0.25 &181$\pm$105 &\ldots &0.10$\pm$0.12 &75.45$\pm$3.59 &1.01$\pm$0.14 &1.20 &2.11$\pm$0.85\\
Musca &160$\pm$60 &7.46$\pm$2.80 &0.25$\pm$0.25 &166$\pm$134 &17.6$\pm$14.0 &0.15$\pm$0.18 &98.17$\pm$5.56 &0.81$\pm$0.15 &0.80 &2.65$\pm$1.27\\
OphNorth1 &130$\pm$20 &1.92$\pm$0.30 &0.25$\pm$0.25 &51$\pm$21 &13.6$\pm$5.7 &0.51$\pm$0.53 &105.18$\pm$10.87 &0.58$\pm$0.05 &0.80 &1.43$\pm$0.48\\
OphNorth3 &130$\pm$20 &8.15$\pm$1.25 &0.75$\pm$0.43 &225$\pm$89 &65.9$\pm$26.2 &0.36$\pm$0.24 &108.09$\pm$9.75 &0.82$\pm$0.07 &1.00 &2.35$\pm$0.79\\
OphNorth6 &130$\pm$20 &2.25$\pm$0.35 &0.25$\pm$0.25 &73$\pm$29 &10.9$\pm$4.3 &0.33$\pm$0.34 &95.50$\pm$9.17 &0.68$\pm$0.06 &0.85 &1.68$\pm$0.57\\
Ophiuchus &125$\pm$25 &13.44$\pm$2.69 &59.00$\pm$3.84 &4020$\pm$1862 &1587.7$\pm$741.6 &1.78$\pm$0.72 &120.99$\pm$7.66 &1.54$\pm$0.16 &0.94 &9.99$\pm$3.60\\
Perseus &250$\pm$30 &26.70$\pm$3.20 &75.50$\pm$4.34 &4764$\pm$1820 &1896.2$\pm$724.9 &1.91$\pm$0.47 &120.56$\pm$17.14 &1.61$\pm$0.15 &1.54 &6.65$\pm$2.15\\
Serpens &429$\pm$22 &15.19$\pm$0.78 &43.00$\pm$3.28 &8846$\pm$1173 &6262.6$\pm$839.9 &0.87$\pm$0.11 &179.83$\pm$5.42 &1.40$\pm$0.04 &2.16 &5.29$\pm$1.61\\
\hline
\end{tabular}
\tablefoot{
\tablefoottext{a}{The distance and mean linewidth of each cloud are adopted directly from \citet{evans14}.}
\tablefoottext{b}{We fit the region with $A_V>$ 3\,mag in the extinction map with an ellipse for each cloud and use the major axis as the length of cloud.}
\tablefoottext{c}{The gas mass and dense gas mass of each cloud are obtained through integrating the extinction map down to $A_V=$ 3\,mag ($A_V=$ 6\,mag for Serpens) and down to $A_V=$ 7\,mag, respectively.}
\tablefoottext{d}{The free-fall time is calculated based on the spherical assumption.}
\tablefoottext{e}{The crossing time of each cloud is obtained as $t_{\textrm{cross}}=size/\Delta v$, where $size$ is twice of equivalent radius $r=$sqrt(Area/$\pi$).}
}
\tablebib{
The distance uncertainties are adopted from following papers: Aquila \citep{straizys03}; AurigaN and Auriga \citep{lada09}; Cepheus 1, 3, and 5 \citep{kirk09}; Cham I and III \citep{belloche11}; Cham II \citep{whittet97}; CoronaAus. \citep{corabook08}; IC5146E and NW \citep{harvey08}; Lupus I, III-VI \citep{lupusbook}; Musca \citep{knude98,franco91,cox16}; Ophiuchus \citep{degeus89}; OphNorth1, 3, and 6 \citep{wilking08,hatchell12}; Perseus \citep{disper}; Serpens \citep{dzib10,dzib11}.
}
\end{table*}

\clearpage

\longtab{
\begin{landscape}
\setlength\tabcolsep{5pt}
\setlength{\extrarowheight}{2.5pt}
\scriptsize
\begin{longtable}{S[table-format=2.0]lcccS[table-number-alignment=center,table-space-text-post=~~~~~~~~]ccccccc}
\caption{\label{table4} Parameters and associated uncertainties of the Galactic giant molecular filaments}\\
\hline\hline
\multicolumn{1}{c}{(1)} & \multicolumn{1}{c}{(2)} & \multicolumn{1}{c}{(3)} & \multicolumn{1}{c}{(4)} & \multicolumn{1}{c}{(5)} & \multicolumn{1}{c}{(6)} & \multicolumn{1}{c}{(7)} & \multicolumn{1}{c}{(8)} & \multicolumn{1}{c}{(9)} & \multicolumn{1}{c}{(10)} & \multicolumn{1}{c}{(11)} & \multicolumn{1}{c}{(12)} & \multicolumn{1}{c}{(13)} \\
\multicolumn{1}{c}{ID} & \multicolumn{1}{c}{Name} & \multicolumn{1}{c}{Log(SFR)} & \multicolumn{1}{c}{Log(SFR$_{\textrm{ff,s}}$)}&\multicolumn{1}{c}{Log(SFR$_{\textrm{ff,f}}$)}&\multicolumn{1}{c}{Log($\Sigma_{\textrm{SFR}}$)} & \multicolumn{1}{c}{Log($\frac{\textrm{SFR}}{Length}$)} & \multicolumn{1}{c}{Log($M_{\textrm{cloud}}$)}&\multicolumn{1}{c}{Log($M_{\textrm{dense}}$)}&\multicolumn{1}{c}{Log($\Sigma_{\textrm{gas}}$/t$_{\textrm{ff,s}}$)}&\multicolumn{1}{c}{Log($\Sigma_{\textrm{gas}}$/t$_{\textrm{ff,f}}$)}&\multicolumn{1}{c}{Log($\frac{M_{\textrm{cloud}}}{Length}$)}&\multicolumn{1}{c}{Log($\frac{M_{\textrm{dense}}}{Length}$)}\\
\multicolumn{1}{c}{GMF} &  & \multicolumn{1}{c}{($M_{\sun}$\,Myr$^{-1}$)} &  & &\multicolumn{1}{c}{($M_{\sun}$\,Myr$^{-1}$\,pc$^{-2}$)} & \multicolumn{1}{c}{($M_{\sun}$\,Myr$^{-1}$\,pc$^{-1}$)} & \multicolumn{1}{c}{($M_{\sun}$)} & \multicolumn{1}{c}{($M_{\sun}$)} & \multicolumn{1}{c}{($M_{\sun}$\,pc$^{-2}$\,Myr$^{-1}$)}&\multicolumn{1}{c}{($M_{\sun}$\,pc$^{-2}$\,Myr$^{-1}$)}&\multicolumn{1}{c}{($M_{\sun}$\,pc$^{-1}$)} & \multicolumn{1}{c}{($M_{\sun}$\,pc$^{-1}$)} \\
\hline
\endfirsthead
\caption{continued.}\\
\hline\hline
\multicolumn{1}{c}{(1)} & \multicolumn{1}{c}{(2)} & \multicolumn{1}{c}{(3)} & \multicolumn{1}{c}{(4)} & \multicolumn{1}{c}{(5)} & \multicolumn{1}{c}{(6)} & \multicolumn{1}{c}{(7)} & \multicolumn{1}{c}{(8)} & \multicolumn{1}{c}{(9)} & \multicolumn{1}{c}{(10)} & \multicolumn{1}{c}{(11)} & \multicolumn{1}{c}{(12)} & \multicolumn{1}{c}{(13)} \\
\multicolumn{1}{c}{ID} & \multicolumn{1}{c}{Name} & \multicolumn{1}{c}{Log(SFR)} & \multicolumn{1}{c}{Log(SFR$_{\textrm{ff,s}}$)}&\multicolumn{1}{c}{Log(SFR$_{\textrm{ff,f}}$)}&\multicolumn{1}{c}{Log($\Sigma_{\textrm{SFR}}$)} & \multicolumn{1}{c}{Log($\frac{\textrm{SFR}}{Length}$)} & \multicolumn{1}{c}{Log($M_{\textrm{cloud}}$)}&\multicolumn{1}{c}{Log($M_{\textrm{dense}}$)}&\multicolumn{1}{c}{Log($\Sigma_{\textrm{gas}}$/t$_{\textrm{ff,s}}$)}&\multicolumn{1}{c}{Log($\Sigma_{\textrm{gas}}$/t$_{\textrm{ff,f}}$)}&\multicolumn{1}{c}{Log($\frac{M_{\textrm{cloud}}}{Length}$)}&\multicolumn{1}{c}{Log($\frac{M_{\textrm{dense}}}{Length}$)}\\
\multicolumn{1}{c}{GMF} &  & \multicolumn{1}{c}{($M_{\sun}$\,Myr$^{-1}$)} &  & &\multicolumn{1}{c}{($M_{\sun}$\,Myr$^{-1}$\,pc$^{-2}$)} & \multicolumn{1}{c}{($M_{\sun}$\,Myr$^{-1}$\,pc$^{-1}$)} & \multicolumn{1}{c}{($M_{\sun}$)} & \multicolumn{1}{c}{($M_{\sun}$)} & \multicolumn{1}{c}{($M_{\sun}$\,pc$^{-2}$\,Myr$^{-1}$)}&\multicolumn{1}{c}{($M_{\sun}$\,pc$^{-2}$\,Myr$^{-1}$)}&\multicolumn{1}{c}{($M_{\sun}$\,pc$^{-1}$)} & \multicolumn{1}{c}{($M_{\sun}$\,pc$^{-1}$)} \\
\hline
\endhead
\hline
\endfoot
2 & F19 & 2.16$_{-0.42}^{+0.31}$ & $-$1.64$_{-1.45}^{+0.48}$ & $-$1.51$_{-1.45}^{+0.48}$ & 0.16$_{-0.36}^{+0.27}$ & 0.94$_{-0.36}^{+0.25}$ & 4.11$_{-0.45}^{+0.31}$ & 3.79$_{-0.45}^{+0.31}$ & 1.81$_{-0.43}^{+0.28}$ & 1.68$_{-0.43}^{+0.28}$ & 2.89$_{-0.33}^{+0.23}$ & 2.56$_{-0.33}^{+0.23}$\\
3 & GMF20.0$-$17.9 & 3.05$_{-0.27}^{+0.21}$ & $-$1.70$_{-0.54}^{+0.40}$ & $-$1.41$_{-0.54}^{+0.40}$ & -0.18$_{-0.20}^{+0.18}$ & 0.91$_{-0.19}^{+0.15}$ & 5.34$_{-0.45}^{+0.32}$ & 5.02$_{-0.45}^{+0.32}$ & 1.54$_{-0.42}^{+0.28}$ & 1.25$_{-0.42}^{+0.28}$ & 3.19$_{-0.31}^{+0.23}$ & 2.88$_{-0.31}^{+0.23}$\\
6 & BC\_021.25$-$0.15 & 1.90$_{-0.29}^{+0.24}$ & $-$1.91$_{-0.55}^{+0.42}$ & $-$1.69$_{-0.55}^{+0.42}$ & -0.37$_{-0.23}^{+0.22}$ & 0.37$_{-0.21}^{+0.19}$ & 4.24$_{-0.44}^{+0.32}$ & 2.70$_{-0.44}^{+0.32}$ & 1.55$_{-0.41}^{+0.28}$ & 1.33$_{-0.41}^{+0.28}$ & 2.72$_{-0.32}^{+0.23}$ & 1.17$_{-0.32}^{+0.23}$\\
8 & F21 & 3.34$_{-0.26}^{+0.19}$ & $-$1.46$_{-0.57}^{+0.40}$ & $-$1.29$_{-0.57}^{+0.40}$ & 0.14$_{-0.16}^{+0.16}$ & 1.44$_{-0.15}^{+0.13}$ & 5.35$_{-0.46}^{+0.31}$ & 5.13$_{-0.46}^{+0.31}$ & 1.61$_{-0.45}^{+0.30}$ & 1.43$_{-0.45}^{+0.30}$ & 3.46$_{-0.32}^{+0.23}$ & 3.24$_{-0.32}^{+0.23}$\\
11 & F22 & 3.11$_{-0.24}^{+0.19}$ & $-$1.65$_{-0.42}^{+0.43}$ & $-$1.57$_{-0.42}^{+0.43}$ & 0.23$_{-0.16}^{+0.15}$ & 1.56$_{-0.14}^{+0.12}$ & 5.17$_{-0.43}^{+0.32}$ & 5.04$_{-0.43}^{+0.32}$ & 1.88$_{-0.44}^{+0.30}$ & 1.80$_{-0.44}^{+0.30}$ & 3.62$_{-0.32}^{+0.24}$ & 3.49$_{-0.32}^{+0.24}$\\
12 & CFG024.00$+$0.48 & 2.85$_{-0.27}^{+0.21}$ & $-$1.66$_{-0.55}^{+0.42}$ & $-$1.43$_{-0.55}^{+0.42}$ & -0.26$_{-0.20}^{+0.18}$ & 0.89$_{-0.18}^{+0.16}$ & 5.12$_{-0.44}^{+0.32}$ & 4.54$_{-0.44}^{+0.32}$ & 1.41$_{-0.43}^{+0.28}$ & 1.18$_{-0.43}^{+0.28}$ & 3.16$_{-0.32}^{+0.24}$ & 2.57$_{-0.32}^{+0.24}$\\
13 & F26 & 2.55$_{-0.30}^{+0.23}$ & $-$1.90$_{-0.43}^{+0.43}$ & $-$1.74$_{-0.43}^{+0.43}$ & -0.21$_{-0.21}^{+0.19}$ & 0.91$_{-0.20}^{+0.16}$ & 4.90$_{-0.43}^{+0.31}$ & 4.64$_{-0.43}^{+0.31}$ & 1.70$_{-0.43}^{+0.29}$ & 1.54$_{-0.43}^{+0.29}$ & 3.25$_{-0.31}^{+0.24}$ & 2.99$_{-0.31}^{+0.24}$\\
15 & BC\_24.95$-$0.17 & 3.27$_{-0.25}^{+0.21}$ & $-$1.30$_{-0.57}^{+0.41}$ & $-$1.15$_{-0.57}^{+0.41}$ & 0.15$_{-0.18}^{+0.17}$ & 1.45$_{-0.16}^{+0.14}$ & 5.17$_{-0.46}^{+0.32}$ & 4.68$_{-0.46}^{+0.32}$ & 1.46$_{-0.42}^{+0.29}$ & 1.30$_{-0.42}^{+0.29}$ & 3.35$_{-0.33}^{+0.24}$ & 2.85$_{-0.33}^{+0.24}$\\
16 & BC\_025.24$-$0.45 & 2.71$_{-0.30}^{+0.24}$ & $-$1.86$_{-0.67}^{+0.44}$ & $-$1.76$_{-0.67}^{+0.44}$ & -0.36$_{-0.22}^{+0.19}$ & 1.02$_{-0.20}^{+0.18}$ & 5.14$_{-0.44}^{+0.32}$ & 4.71$_{-0.44}^{+0.32}$ & 1.51$_{-0.43}^{+0.28}$ & 1.41$_{-0.43}^{+0.28}$ & 3.45$_{-0.32}^{+0.23}$ & 3.03$_{-0.32}^{+0.23}$\\
17 & F28 & 2.37$_{-0.28}^{+0.23}$ & $-$1.61$_{-0.61}^{+0.44}$ & $-$1.37$_{-0.61}^{+0.44}$ & 0.23$_{-0.22}^{+0.19}$ & 0.87$_{-0.20}^{+0.17}$ & 4.28$_{-0.44}^{+0.32}$ & 4.02$_{-0.44}^{+0.32}$ & 1.85$_{-0.45}^{+0.28}$ & 1.61$_{-0.45}^{+0.28}$ & 2.78$_{-0.33}^{+0.23}$ & 2.53$_{-0.33}^{+0.23}$\\
18 & F29 & 2.97$_{-0.26}^{+0.20}$ & $-$1.75$_{-0.44}^{+0.40}$ & $-$1.62$_{-0.44}^{+0.40}$ & -0.23$_{-0.17}^{+0.16}$ & 1.16$_{-0.15}^{+0.13}$ & 5.30$_{-0.44}^{+0.32}$ & 4.94$_{-0.44}^{+0.32}$ & 1.53$_{-0.44}^{+0.29}$ & 1.39$_{-0.44}^{+0.29}$ & 3.49$_{-0.32}^{+0.24}$ & 3.13$_{-0.32}^{+0.24}$\\
20 & BC\_26.94$-$0.30 & 1.72$_{-0.33}^{+0.28}$ & $-$2.08$_{-0.52}^{+0.46}$ & $-$1.88$_{-0.52}^{+0.46}$ & -0.41$_{-0.28}^{+0.26}$ & 0.30$_{-0.26}^{+0.23}$ & 4.16$_{-0.44}^{+0.32}$ & 3.57$_{-0.44}^{+0.32}$ & 1.69$_{-0.44}^{+0.28}$ & 1.48$_{-0.44}^{+0.28}$ & 2.74$_{-0.32}^{+0.23}$ & 2.14$_{-0.32}^{+0.23}$\\
21 & F30 & 2.40$_{-0.29}^{+0.22}$ & $-$1.95$_{-0.42}^{+0.42}$ & $-$1.80$_{-0.42}^{+0.42}$ & -0.43$_{-0.20}^{+0.18}$ & 0.72$_{-0.19}^{+0.16}$ & 4.88$_{-0.42}^{+0.32}$ & 4.33$_{-0.42}^{+0.32}$ & 1.53$_{-0.45}^{+0.29}$ & 1.37$_{-0.45}^{+0.29}$ & 3.20$_{-0.32}^{+0.24}$ & 2.65$_{-0.32}^{+0.24}$\\
22 & F31 & 3.51$_{-0.28}^{+0.22}$ & $-$1.36$_{-0.59}^{+0.43}$ & $-$1.19$_{-0.59}^{+0.43}$ & 0.01$_{-0.22}^{+0.18}$ & 1.49$_{-0.20}^{+0.15}$ & 5.56$_{-0.44}^{+0.32}$ & 5.09$_{-0.44}^{+0.32}$ & 1.37$_{-0.42}^{+0.29}$ & 1.20$_{-0.42}^{+0.29}$ & 3.53$_{-0.32}^{+0.23}$ & 3.05$_{-0.32}^{+0.23}$\\
23 & CFG028.68$-$0.28 & 2.59$_{-0.28}^{+0.22}$ & $-$1.95$_{-0.64}^{+0.41}$ & $-$1.81$_{-0.64}^{+0.41}$ & -0.25$_{-0.21}^{+0.18}$ & 0.93$_{-0.18}^{+0.15}$ & 5.00$_{-0.44}^{+0.32}$ & 4.79$_{-0.44}^{+0.32}$ & 1.70$_{-0.44}^{+0.29}$ & 1.56$_{-0.44}^{+0.29}$ & 3.34$_{-0.32}^{+0.24}$ & 3.13$_{-0.32}^{+0.24}$\\
25 & CFG029.18$-$0.34 & 2.52$_{-0.27}^{+0.22}$ & $-$2.26$_{-0.58}^{+0.42}$ & $-$2.09$_{-0.58}^{+0.42}$ & -0.54$_{-0.21}^{+0.18}$ & 0.70$_{-0.18}^{+0.16}$ & 5.27$_{-0.43}^{+0.31}$ & 5.10$_{-0.43}^{+0.31}$ & 1.73$_{-0.46}^{+0.29}$ & 1.56$_{-0.46}^{+0.29}$ & 3.46$_{-0.32}^{+0.23}$ & 3.28$_{-0.32}^{+0.23}$\\
26 & G030.315$-$0.154 & 3.13$_{-0.25}^{+0.20}$ & $-$1.65$_{-0.60}^{+0.41}$ & $-$1.56$_{-0.60}^{+0.41}$ & 0.06$_{-0.17}^{+0.17}$ & 1.48$_{-0.15}^{+0.13}$ & 5.27$_{-0.45}^{+0.32}$ & 5.10$_{-0.45}^{+0.32}$ & 1.72$_{-0.42}^{+0.28}$ & 1.63$_{-0.42}^{+0.28}$ & 3.62$_{-0.32}^{+0.23}$ & 3.45$_{-0.32}^{+0.23}$\\
32 & GMF38.1$-$32.4b & 2.69$_{-0.32}^{+0.26}$ & $-$1.51$_{-0.59}^{+0.44}$ & $-$1.25$_{-0.59}^{+0.44}$ & -0.05$_{-0.28}^{+0.21}$ & 0.86$_{-0.25}^{+0.20}$ & 4.73$_{-0.41}^{+0.31}$ & 3.97$_{-0.41}^{+0.31}$ & 1.46$_{-0.41}^{+0.29}$ & 1.21$_{-0.41}^{+0.29}$ & 2.90$_{-0.31}^{+0.23}$ & 2.14$_{-0.31}^{+0.23}$\\
33 & GMF38.1$-$32.4a & 3.46$_{-0.28}^{+0.22}$ & $-$1.61$_{-0.59}^{+0.43}$ & $-$1.35$_{-0.59}^{+0.43}$ & -0.28$_{-0.21}^{+0.18}$ & 1.12$_{-0.20}^{+0.15}$ & 5.82$_{-0.46}^{+0.31}$ & 5.39$_{-0.46}^{+0.31}$ & 1.34$_{-0.45}^{+0.28}$ & 1.08$_{-0.45}^{+0.28}$ & 3.47$_{-0.32}^{+0.23}$ & 3.05$_{-0.32}^{+0.23}$\\
35 & GMF41.0$-$41.3 & 2.19$_{-0.31}^{+0.25}$ & $-$1.77$_{-0.59}^{+0.44}$ & $-$1.53$_{-0.59}^{+0.44}$ & -0.35$_{-0.25}^{+0.22}$ & 0.48$_{-0.23}^{+0.20}$ & 4.47$_{-0.40}^{+0.32}$ & 2.90$_{-0.40}^{+0.32}$ & 1.43$_{-0.44}^{+0.28}$ & 1.19$_{-0.44}^{+0.28}$ & 2.77$_{-0.31}^{+0.23}$ & 1.20$_{-0.31}^{+0.23}$\\
39 & CFG047.06$+$0.26 & 2.01$_{-0.30}^{+0.25}$ & $-$2.16$_{-0.67}^{+0.44}$ & $-$1.92$_{-0.67}^{+0.44}$ & -0.40$_{-0.25}^{+0.22}$ & 0.37$_{-0.23}^{+0.20}$ & 4.55$_{-0.45}^{+0.31}$ & 4.25$_{-0.45}^{+0.31}$ & 1.76$_{-0.43}^{+0.29}$ & 1.52$_{-0.43}^{+0.29}$ & 2.92$_{-0.32}^{+0.23}$ & 2.62$_{-0.32}^{+0.23}$\\
41 & CFG049.21$-$0.34 & 2.75$_{-0.11}^{+0.10}$ & $-$1.85$_{-0.54}^{+0.40}$ & $-$1.51$_{-0.54}^{+0.40}$ & -0.17$_{-0.11}^{+0.10}$ & 0.66$_{-0.11}^{+0.10}$ & 5.09$_{-0.27}^{+0.19}$ & 4.85$_{-0.27}^{+0.19}$ & 1.69$_{-0.41}^{+0.28}$ & 1.35$_{-0.41}^{+0.28}$ & 3.01$_{-0.27}^{+0.18}$ & 2.77$_{-0.27}^{+0.18}$\\
42 & F42 & 2.83$_{-0.24}^{+0.19}$ & $-$1.98$_{-0.54}^{+0.41}$ & $-$1.87$_{-0.54}^{+0.41}$ & -0.42$_{-0.16}^{+0.16}$ & 1.03$_{-0.14}^{+0.13}$ & 5.38$_{-0.45}^{+0.32}$ & 5.08$_{-0.45}^{+0.32}$ & 1.57$_{-0.43}^{+0.29}$ & 1.46$_{-0.43}^{+0.29}$ & 3.60$_{-0.32}^{+0.23}$ & 3.30$_{-0.32}^{+0.23}$\\
43 & GMF54.0$-$52.0 & 2.61$_{-0.28}^{+0.24}$ & $-$1.42$_{-0.64}^{+0.43}$ & $-$1.17$_{-0.64}^{+0.43}$ & 0.28$_{-0.25}^{+0.20}$ & 1.00$_{-0.23}^{+0.19}$ & 4.41$_{-0.40}^{+0.31}$ & 4.04$_{-0.40}^{+0.31}$ & 1.69$_{-0.42}^{+0.30}$ & 1.44$_{-0.42}^{+0.30}$ & 2.80$_{-0.31}^{+0.23}$ & 2.43$_{-0.31}^{+0.23}$\\
44 & GMF307.2$-$305.4 & 3.44$_{-0.29}^{+0.24}$ & $-$1.79$_{-0.53}^{+0.42}$ & $-$1.56$_{-0.53}^{+0.42}$ & -0.45$_{-0.24}^{+0.20}$ & 1.10$_{-0.22}^{+0.18}$ & 5.99$_{-0.47}^{+0.31}$ & 5.63$_{-0.47}^{+0.31}$ & 1.35$_{-0.43}^{+0.28}$ & 1.12$_{-0.43}^{+0.28}$ & 3.65$_{-0.32}^{+0.23}$ & 3.29$_{-0.32}^{+0.23}$\\
45 & GMF309.5$-$308.7 & 3.04$_{-0.30}^{+0.25}$ & $-$2.18$_{-0.59}^{+0.45}$ & $-$2.07$_{-0.59}^{+0.45}$ & -0.75$_{-0.24}^{+0.21}$ & 0.98$_{-0.23}^{+0.18}$ & 5.94$_{-0.43}^{+0.31}$ & 5.66$_{-0.43}^{+0.31}$ & 1.44$_{-0.45}^{+0.28}$ & 1.33$_{-0.45}^{+0.28}$ & 3.87$_{-0.33}^{+0.23}$ & 3.60$_{-0.33}^{+0.23}$\\
46 & GMF319.0$-$318.7 & 3.09$_{-0.30}^{+0.25}$ & $-$2.02$_{-1.35}^{+0.42}$ & $-$1.89$_{-1.35}^{+0.42}$ & -0.52$_{-0.26}^{+0.21}$ & 1.09$_{-0.24}^{+0.19}$ & 5.77$_{-0.42}^{+0.32}$ & 5.55$_{-0.42}^{+0.32}$ & 1.49$_{-0.42}^{+0.29}$ & 1.37$_{-0.42}^{+0.29}$ & 3.78$_{-0.32}^{+0.23}$ & 3.55$_{-0.32}^{+0.23}$\\
47 & GMF324.5$-$321.4b & 2.27$_{-0.29}^{+0.25}$ & $-$1.76$_{-0.57}^{+0.42}$ & $-$1.57$_{-0.57}^{+0.42}$ & -0.08$_{-0.24}^{+0.20}$ & 0.77$_{-0.22}^{+0.19}$ & 4.42$_{-0.42}^{+0.31}$ & 4.02$_{-0.42}^{+0.31}$ & 1.69$_{-0.41}^{+0.29}$ & 1.50$_{-0.41}^{+0.29}$ & 2.92$_{-0.30}^{+0.23}$ & 2.52$_{-0.30}^{+0.23}$\\
50 & G327.157$-$0.256 & 3.30$_{-0.25}^{+0.21}$ & $-$1.62$_{-0.60}^{+0.42}$ & $-$1.47$_{-0.60}^{+0.42}$ & -0.19$_{-0.18}^{+0.16}$ & 1.30$_{-0.17}^{+0.13}$ & 5.58$_{-0.46}^{+0.32}$ & 5.25$_{-0.46}^{+0.32}$ & 1.44$_{-0.45}^{+0.29}$ & 1.28$_{-0.45}^{+0.29}$ & 3.58$_{-0.32}^{+0.24}$ & 3.24$_{-0.32}^{+0.24}$\\
51 & GMF335.6$-$333.6b & 3.21$_{-0.31}^{+0.24}$ & $-$1.49$_{-0.52}^{+0.41}$ & $-$1.31$_{-0.52}^{+0.41}$ & 0.06$_{-0.26}^{+0.20}$ & 1.34$_{-0.23}^{+0.17}$ & 5.26$_{-0.44}^{+0.31}$ & 4.96$_{-0.44}^{+0.31}$ & 1.55$_{-0.43}^{+0.28}$ & 1.38$_{-0.43}^{+0.28}$ & 3.39$_{-0.32}^{+0.23}$ & 3.08$_{-0.32}^{+0.23}$\\
52 & GMF335.6$-$333.6a & 3.97$_{-0.28}^{+0.23}$ & $-$1.63$_{-0.57}^{+0.42}$ & $-$1.43$_{-0.57}^{+0.42}$ & -0.15$_{-0.22}^{+0.17}$ & 1.56$_{-0.20}^{+0.15}$ & 6.35$_{-0.45}^{+0.32}$ & 6.18$_{-0.45}^{+0.32}$ & 1.48$_{-0.43}^{+0.29}$ & 1.27$_{-0.43}^{+0.29}$ & 3.94$_{-0.32}^{+0.23}$ & 3.77$_{-0.32}^{+0.23}$\\
53 & Nessie & 2.64$_{-0.33}^{+0.27}$ & $-$1.38$_{-1.09}^{+0.31}$ & $-$1.06$_{-1.09}^{+0.31}$ & 0.24$_{-0.28}^{+0.22}$ & 0.87$_{-0.27}^{+0.21}$ & 4.46$_{-0.39}^{+0.28}$ & 4.27$_{-0.38}^{+0.28}$ & 1.62$_{-0.31}^{+0.21}$ & 1.31$_{-0.31}^{+0.21}$ & 2.67$_{-0.25}^{+0.19}$ & 2.49$_{-0.26}^{+0.19}$\\
54 & GMF341.9$-$337.1 & 3.68$_{-0.27}^{+0.22}$ & $-$1.66$_{-0.56}^{+0.45}$ & $-$1.53$_{-0.56}^{+0.45}$ & -0.22$_{-0.22}^{+0.18}$ & 1.51$_{-0.19}^{+0.15}$ & 6.07$_{-0.49}^{+0.32}$ & 5.84$_{-0.49}^{+0.32}$ & 1.45$_{-0.43}^{+0.30}$ & 1.31$_{-0.43}^{+0.30}$ & 3.90$_{-0.33}^{+0.24}$ & 3.67$_{-0.33}^{+0.24}$\\
57 & GMF358.9$-$357.4 & 2.47$_{-0.24}^{+0.20}$ & $-$1.92$_{-1.32}^{+0.44}$ & $-$1.78$_{-1.32}^{+0.44}$ & -0.27$_{-0.23}^{+0.19}$ & 0.87$_{-0.23}^{+0.19}$ & 4.86$_{-0.29}^{+0.21}$ & 4.55$_{-0.29}^{+0.21}$ & 1.66$_{-0.43}^{+0.28}$ & 1.52$_{-0.43}^{+0.28}$ & 3.27$_{-0.29}^{+0.19}$ & 2.95$_{-0.29}^{+0.19}$\\
\end{longtable}
\tablefoot{Columns are (1) GMF ID; (2) GMF name; (3) star formation rate; (4) star formation rate per free-fall time obtained based on the spherical morphology assumption; (5) star formation rate per free-fall time obtained based on the filamentary morphology assumption; (6) surface density of star formation rate; (7) star formation rate per length; (8) gas mass obtained through integrating down to A$_V$=3\,mag in the CO-based extinction map; (9) dense gas mass obtained through integrating down to A$_V$=7\,mag in the CO-based extinction map; (10) gas surface density per free-fall time obtained based on the spherical morphology assumption; (11) gas surface density per free-fall time obtained based on the filamentary morphology assumption; (12) gas mass per length or line-mass; (13) dense gas mass per length. Note that the associated uncertainties of the parameters listed in this table do include the distance errors (also see Appendix~\ref{ap4} for more information).}
\end{landscape}
}

\clearpage
\begin{appendix}
\section{Obtaining column density maps of GMFs}\label{ap1}
\citet{sample-gmf}, \citet{sample-wang15}, and \citet{sample-agmf} used the GRS and ThrUMMS data to investigate the velocity structure of their samples and present the radial velocity range for each giant filament. Thus for the GMFs from \citet{sample-gmf}, \citet{sample-wang15}, and \citet{sample-agmf}, we adopt the velocity ranges offered by the corresponding papers. \citet{sample-bones} also use the GRS and ThrUMMS data to identify the velocity coherent filaments. However, they only offered the radial central velocities for their samples. \citet{sample-li16} and \citet{sample-wang16} used several different gas tracers such as NH$_3$ and HCO$^{+}$ to investigate the velocity structures of their samples. Therefore, for the GMFs from \citet{sample-bones}, \citet{sample-li16}, and \citet{sample-wang16}, we re-checked them in the $^{13}$CO data cubes and obtained the radial velocity ranges. 

Figure~\ref{g26-avg} shows this process of an example, GMF\,20 (i.e., BC\_026.94$-$0.30) identified originally by \citet{sample-bones}: firstly, we obtain its $^{13}$CO average spectrum (black line); secondly, we use the multiple gaussian functions to fit the average spectrum (red line) and then obtain the positions and FWHMs for most of strong velocity components; thirdly, the closest velocity component to the central velocity offered by the corresponding paper is adopted as the velocity range for this object. During this process, we found that three GMFs identified by \citet{sample-li16} or \citet{sample-wang16} show the indication of self-absorption in $^{13}$CO lines. Thus we exclude these three filaments from our samples.

Based on the velocity range of each GMF, we collapsed the data cube to obtain the 
$^{13}$CO intensity map and velocity dispersion map. 
Due to the lack of $^{12}$CO data in GRS survey and different coverage of $^{12}$CO and $^{13}$CO data in ThrUMMS survey, we do not try to obtain the excitation temperature maps based on $^{12}$CO data. Alternatively, we simply adopt a value of 10\,K as the excitation temperature ($T_{\textrm{ex}}$) for all GMFs.

The $^{13}$CO column density can be obtained via \citep{ho82,garden91,bourke97}:
\begin{equation}\label{ap:eqcolmass}
N(^{13} \textrm{CO})=2.42 \times 10^{14}\frac{1+0.88/T_{\textrm{ex}}}{1-e^{-5.29/T_{\textrm{ex}}}}\int T_{\textrm{mb}}^{*}(^{13}\textrm{CO}) ~\textrm{d}v
\end{equation}
where $T_{\textrm{mb}}^{*}(^{13}\textrm{CO})$ is the main-beam temperature in $^{13}$CO emission and $v$ is the line velocity. 
We adopt the $^{12}$C/$^{13}$C ratio of 77 \citep{wilson94} and $N$(H$_{\textrm{2}}$)/$N(^{12}\textrm{CO}) = 1.1\times 10^4$ \citep{frerking82} to convert $N(^{13}\textrm{CO})$ to the molecular hydrogen column density. 
And then we convert $N$(H$_{\textrm{2}}$) to the visual extinction units with the relation of $N$(H$_{\textrm{2}}$) (cm$^{-2}$) $=$ 0.94$\times$10$^{21}A_V$ (mag) \citep{bohlin78}.

Note that some GMFs were identified several times by different papers. Excluding the redundant GMFs and four GMFs that are not fully covered by $^{13}$CO data, we can obtain the column density maps for {56} GMFs.

The special case is the Nessie. The sensitivity of the ThrUMMS data in the region with Nessie is not enough to trace the main filamentary part of Nessie. Thus we decide to use the extinction map that is constructed with the technique improved by \citet{kt13} for Nessie. This extinction mapping technique is based on combining extinction maps made at two wavelength regimes: in near-infrared using NICER \citep[Near-Infrared Color Excess Revisited,][]{nicer} and in mid-infrared using the absorption against the Galactic background \citep{irdc-spitzercatalog,but12}. Using this method, we obtain a high resolution ($\sim$2\arcsec) extinction map of the Nessie. {The details about the high-resolution extinction map of Nessie can be found in \citet{mattern18}}.

\begin{figure}
\centering
\includegraphics[width=1.0\linewidth]{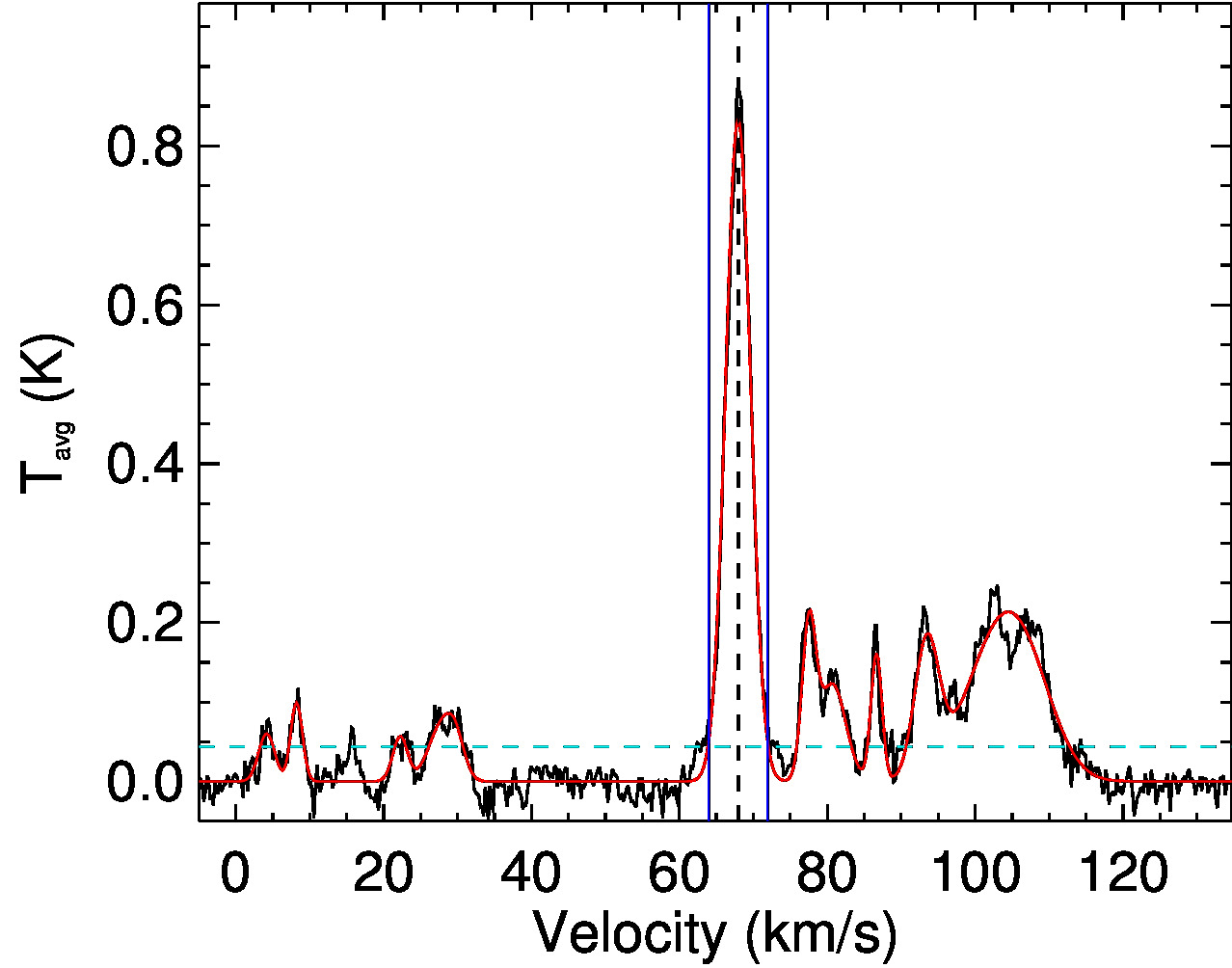}
\caption{The $^{13}$CO average spectrum of GMF 20 (i.e., BC\_026.94$-$0.30) that is identified originally by \citet{sample-bones}. The red curve shows the best fitting using the multiple gaussian components. The dashed line marks the central velocity offered by \citet{sample-bones} and the blue solid lines show the velocity range of this component.}
\label{g26-avg}
\end{figure}

\section{Measuring physical parameters of GMFs}\label{ap2}
All the pixels of the column density map inside the GMF boundary are fitted with an ellipse as shown in Figs.~\ref{fig:CFG47}-\ref{fig:GMF341} of the main text. The coordinates and length of each GMF are defined as the center and major axis of the ellipse, respectively. The area of each GMF is the area of pixels inside the boundary ($A_V = $~3\,mag) of the column density map. The distance of each GMF is adopted from the corresponding reference paper that is given in the last column of Table~\ref{table1}. The Galactocentric radius ($R_{\textrm{gal}}$) and the height above the Galactic midplane ($z$) are calculated based on the coordinates and distances of the GMFs.

We also obtain the $^{13}$CO line width distribution map (the second moment map) for each GMF. Figure~\ref{cfg47-m2map} shows an example of GMF {39} (CFG047.06$+$0.26). The mean line width ($\Delta v$) of a GMF is defined as the average value of the line widths inside the GMF boundary.

We integrate the column density maps of the GMFs down to $A_V =$ 3\,mag and $A_V =$ 7\,mag to obtain the total mass of the cloud ($M_{\textrm{cloud}}$) and the mass of dense gas ($M_{\textrm{dense}}$). 
The average gas surface densities ($\Sigma_{\textrm{gas}}$) and the {cloud mass per unit length} ($\frac{M_{\textrm{cloud}}}{L}$) can be also obtained with $M_{\textrm{cloud}}$, areas and lengths of the GMFs.

Assuming the spherical morphology, the free-fall time of a GMF can be estimated with the formula \citep{binney87,krumholz12}:
\begin{equation*}
t_{\textrm{ff,s}} = \sqrt{\frac{3\pi}{32G\rho}}
\end{equation*}
where $\rho$ is the average volume density. Note that we use an additional subscript `s' in $t_{\textrm{ff,s}}$ to emphasize that this free-fall time is estimated based on the spherical morphology assumption.

Of course, the assumption of spherical morphology is not reasonable for the filamentary structures, especially the filaments with large aspect ratio of length to radius. \citet{tff-filament} computed the free-fall time for finite, uniform, and self-gravitating filamentary clouds and suggested that the free-fall time of a filament whose volume density remains constant during the collapse can be obtained with the formula:
\begin{equation*}
t_{\textrm{ff,f}} = \frac{2}{\pi}\sqrt{\frac{8A}{3}}t_{\textrm{ff,s}}^{\prime}
\end{equation*}
where $A$ is the aspect ratio of the filament and $t_{\textrm{ff,s}}^{\prime}$ is the free-fall time of an spherical cloud with the same volume density as the filamentary cloud. Note that here we also added an subscript `f' in $t_{\textrm{ff,f}}$ to emphasize that this free-fall time is obtained based on the {filamentary} morphology assumption. {Here we also note as a caveat that we could still underestimate the free-fall time because we could significantly underestimate the aspect ratio of GMFs (see Sect. \ref{sect:gasper}).}

\begin{figure}
\includegraphics[width=1.0\linewidth]{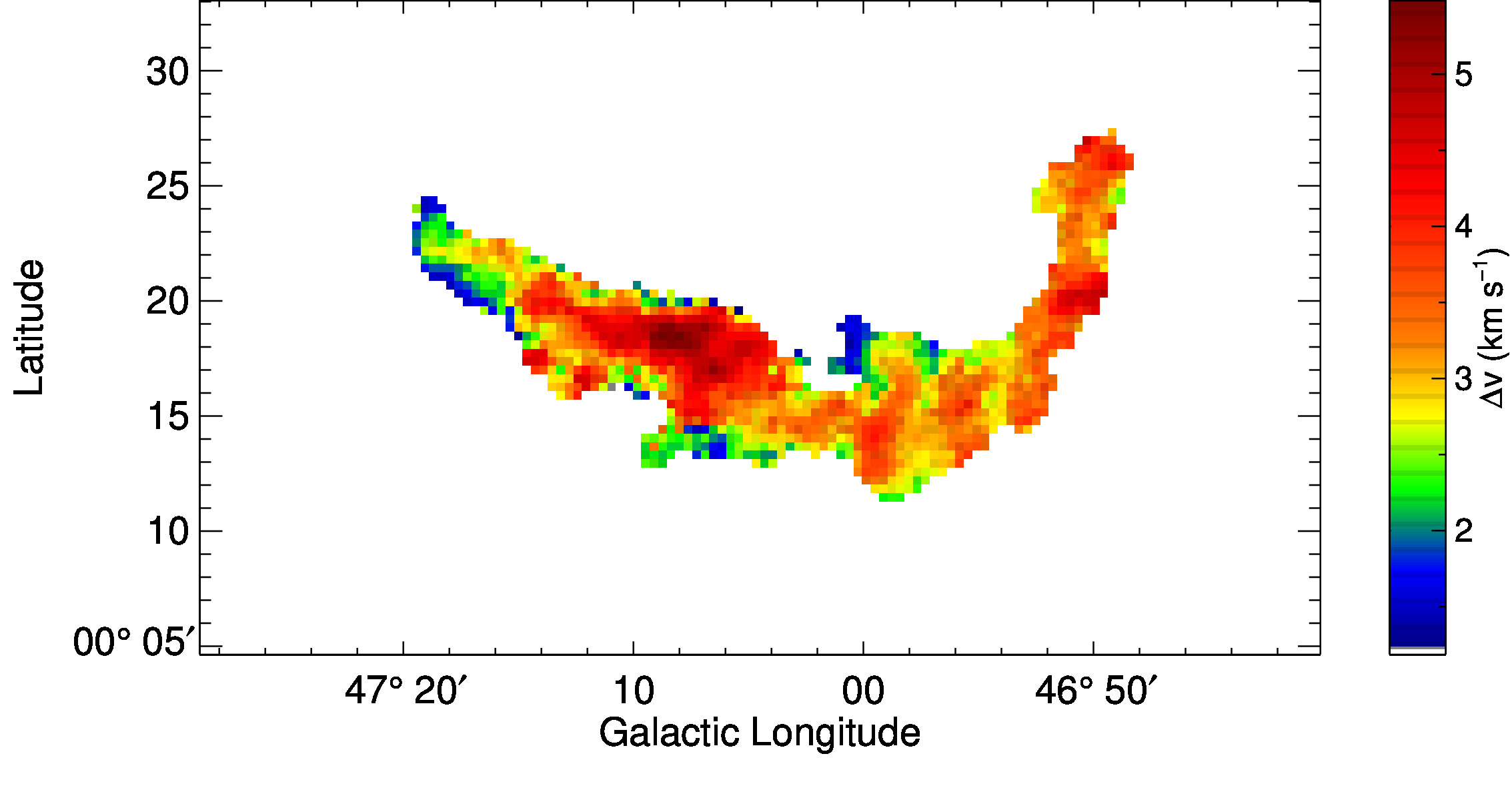}
\caption{$^{13}$CO line width distribution map (obtained from the moment 2 map) of GMF {39} (CFG047.06$+$0.26).}
\label{cfg47-m2map}
\end{figure}

\section{Method used for YSO identification}\label{ap3}

There are six steps in the process of YSO identification. The method is mainly based on the YSO classification schemes presented by \citet{gutermuth09} and \citet{koenig14}. We also combine several color criteria from \citet{robitaille08}, \citet{ven13}, and \citet{saral15} to isolate the AGB stars.

\textit{Step 1}: we use the method suggested by \citet{gutermuth08,gutermuth09} to identify YSOs. The details about this multiphase source classification scheme can be found in the appendix of \citet{gutermuth09}. Here we just summarize our process (also see the description in \citealt{rapson14} and \citealt{mypub15}).

There are three phases in the YSO selection scheme suggested by \citet{gutermuth09}. Phase 1 is applied to the sources that are detected in all four IRAC bands with photometric uncertainties of $<$ 0.2 mag. After removing the contaminants of star-forming galaxies, broad-line active galactic nuclei (AGNs), unresolved knots of shock emission, and sources that have polycyclic aromatic hydrocarbon-contaminated apertures, the sources with 
infrared excess are identified as Class I and Class II sources (see Fig.~\ref{phase1}). The remaining sources are classified as Class III/field sources.

Phase 2 is applied to the sources that lack detections at either 5.8 or 8.0 $\mu$m, but have high-quality ($\sigma < $ 0.05 mag) near-infrared detections in $J$, $H$, and $K_s$ bands. We also require the detections at 3.6 and 4.5 $\mu$m with the uncertainties of $\sigma <$ 0.13 mag in order to minimize the contamination of field stars in the crowded regions. To distinguish the sources with IR excess from those that are simply reddened by dust along the line of sight, we deredden the photometry of sources based on the extinction law presented in \citet{xue16}. The dereddened Ks-[3.6] and [3.6]-[4.5] colors are used to identify the sources with infrared excess at 3.6 and 4.5 $\mu$m, accounting for photometric uncertainty (see panel a of Fig.~\ref{phase23}). Sources with no IR excess at 3.6 and 4.5 $\mu$m are presumed as Class III/field sources.

Phase 3 is applied to the sources that have detections in the MIPS 24 $\mu$m band with the photometric uncertainties of $<$ 0.2 mag. The Class III/field sources that were classified in the previous two phases are re-examined, and 
the transition disks are isolated. The sources that lack detections in some IRAC bands, but are very bright at 24 $\mu$m 
are classified as deeply embedded Class I protostars (see panels b and c of Fig.~\ref{phase23}). 
The AGNs and shock-emission-dominated sources that were classified as contaminants in phase 1 are also re-examined and identified as Class I sources if they have bright MIPS 24 $\mu$m photometry. 
Finally, the Class I sources that were classified in all three phases are re-analyzed to ensure their nature. 
Figure~\ref{phase1} and \ref{phase23} show the multi-color criteria mentioned above that are used to identify YSOs in GMF {39} (CFG047.06$+$0.26).

\begin{figure}
\includegraphics[width=0.5\linewidth]{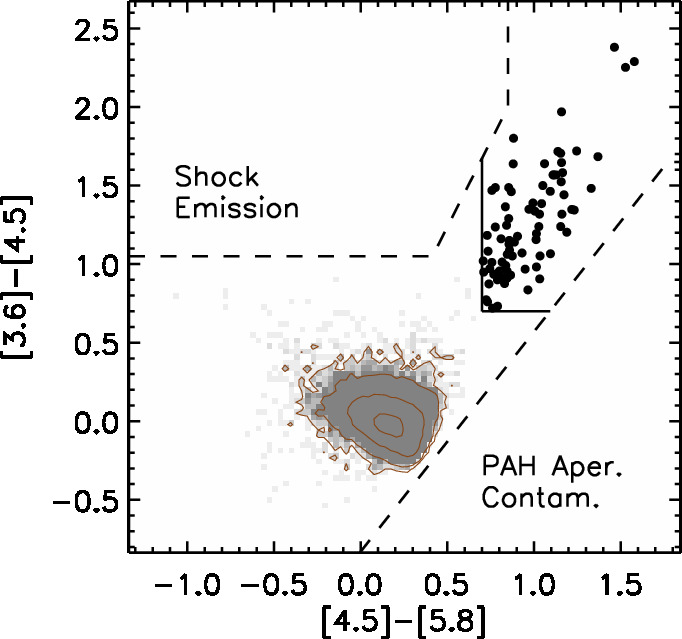}%
\hspace{0.1cm}
\includegraphics[width=0.48\linewidth]{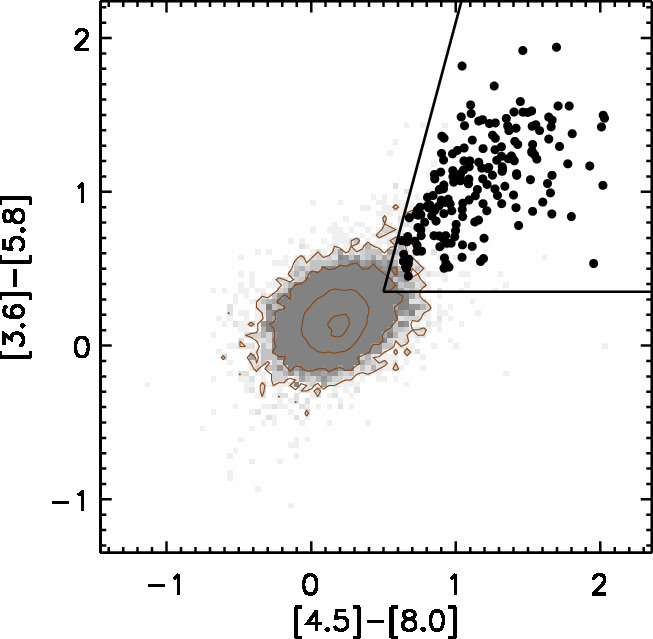}
\caption{The color-color diagrams used for identification of Class I (\textit{left}), and Class II (\textit{right}) sources in \textit{phase 1} of the classification scheme presented by \citet{gutermuth09} for an example of GMF {39} (CFG047.06$+$0.26). The black filled circles represent the selected Class I and Class II sources. The remaining Class III/field sources are marked with the gray dots. The black lines show the color criteria of the classification scheme.}
\label{phase1}
\end{figure}

\begin{figure}
\includegraphics[width=0.43\linewidth]{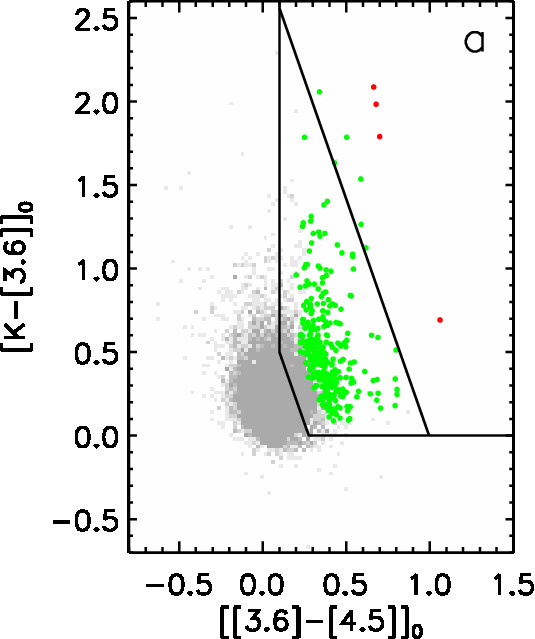}%
\hspace{0.1cm}
\includegraphics[width=0.545\linewidth]{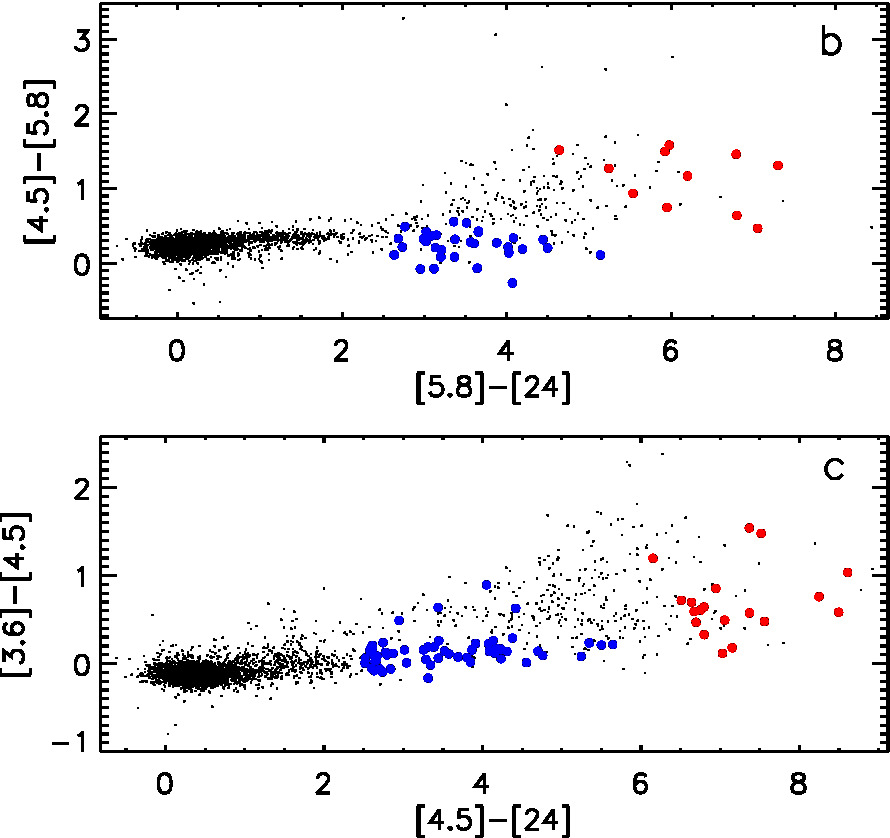}
\caption{The color-color diagrams used for YSO identification in \textit{phase 2} (a) and \textit{phase 3} (b,c) classification schemes suggested by \citet{gutermuth09} for an example of GMF {39} (CFG047.06$+$0.26). The Class I sources identified in \textit{phase 2} and the protostars identified in \textit{phase 3} are marked with red dots. The Class II sources isolated in \textit{phase 2} are labeled with green dots while the transition disks classified in \textit{phase 3} are shown with blue dots.}
\label{phase23}
\end{figure}

\textit{Step 2}: The 70\,$\mu$m photometry of the \textit{Herschel} Hi-Gal catalog \citep{higal-catalog} is used to identify the protostellar objects. Due to the variation of background in the Galactic plane, \citet{higal-catalog} found that there is no good parameter that can be uniquely taken as a measure of the reliability of a source detection and they suggested to use cross-matching sources in different bands to eliminate the spurious detections. On the other hand, \citet{vankempen09} and \citet{heiderman15} suggested that the detection of dense gas around a protostar candidate was a good indicator of a bona fide protostar. Therefore, in the present work only the compact sources with the detections at [70, 160, 250, 350]\,$\mu$m or [70, 250, 350, 500]\,$\mu$m, simultaneously, are identified as the protostars, which means that these protostars have the counterparts of dense cores and thus have high probabilities to be the bona fide protostars. We also crossmatch the protostars identified in this step with the YSOs identified in \textit{Step 1} and only keep the additional protostars in our final YSO catalog.

\textit{Step 3}: All the YSOs identified in the previous steps are re-examined to isolate the possible AGB stars. We combined several classification criteria from \citet{robitaille08}, \citet{ven13}, and \citet{saral15}. The YSOs are classified as AGBs if they follow any of the following criteria as shown in Fig.~\ref{agb_all}:
\begin{itemize}
\item[]{$3 < [3.6] < 9.5$ and $0.2 < [3.6]-[4.5] < 1.25$,}
\item[or]{}
\item[]{$3.5 < [3.6] < 9.5$ and $0.4 < [3.6]-[8.0] < 2.6$,}
\item[or]{}
\item[]{$[4.5] > 7.8$ and $[8.0]-[24] < 2.5$,}
\item[or]{}
\item[]{$0.2 < \frac{\textrm{log}(\frac{S_{[70]}}{S_{[160]}})}{\textrm{log}(\frac{S_{[70]}}{S_{[250]}})} < 0.9$ and $1.2 < \frac{\textrm{log}(\frac{S_{[70]}}{S_{[350]}})}{\textrm{log}(\frac{S_{[160]}}{S_{[350]}})} < 3.3$}
\end{itemize}
where $S_{[x]}$ is the flux at [X] band.
\begin{figure}
\includegraphics[width=1.0\linewidth]{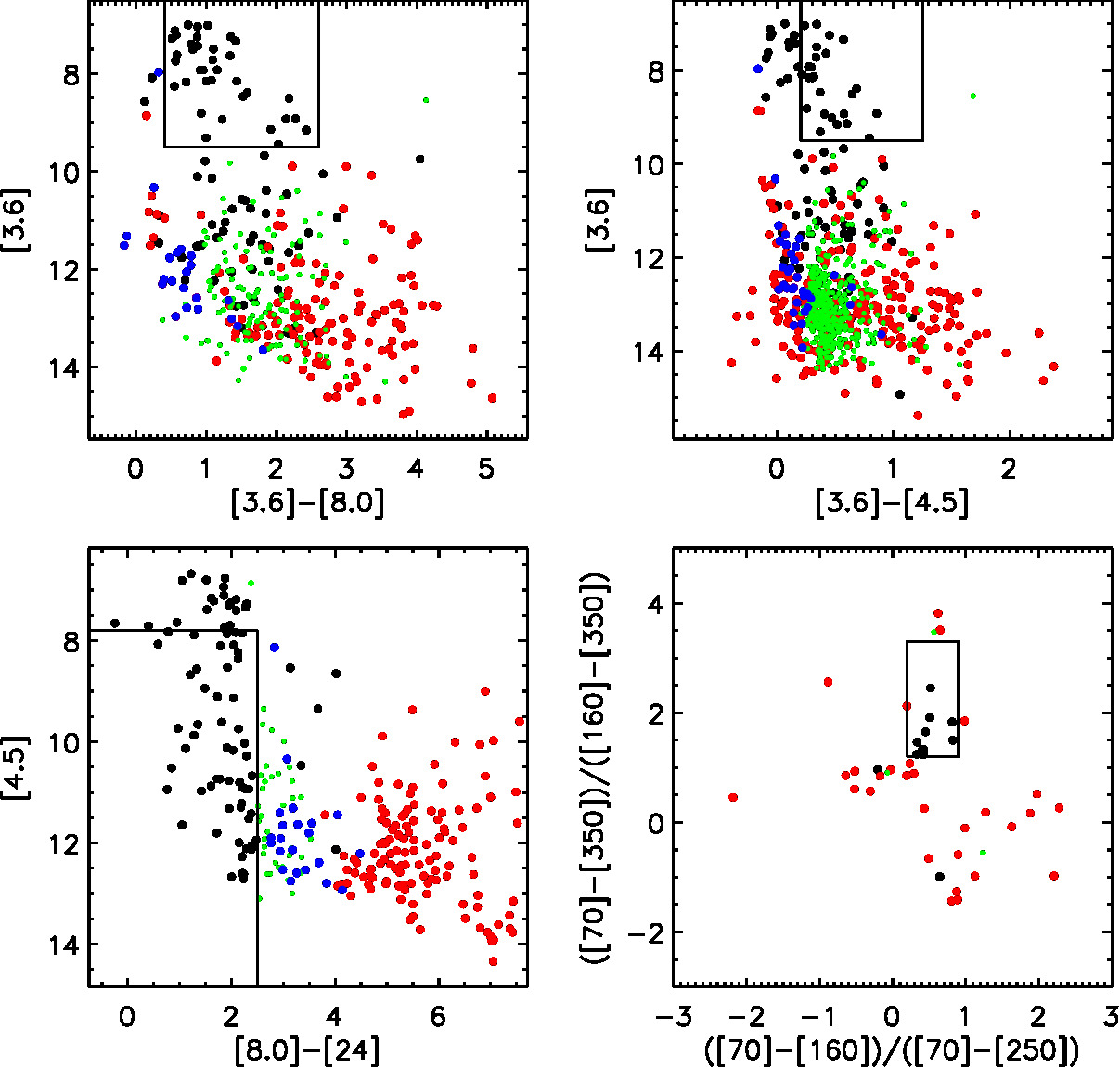}
\caption{The color criteria used to isolate the possible AGB stars from the YSOs in \textit{Step 3} for an example of GMF {39} (CFG047.06$+$0.26). The red dots mark the Class I sources or protostars while the green and blue dots label the Class II and transition disks, respectively. The AGBs are marked with black dots. The black solid lines show the boundaries of the classification criteria.}
\label{agb_all}
\end{figure}

\textit{Step 4}: We use the selection scheme suggested by \citet{koenig14} to identify bright YSOs that are saturated on \textit{Spitzer} images based on the AllWISE$+$NIR catalog. \citet{koenig14} found that the fake source contamination in the Galactic plane can be up to 30\% in the W1 band and even 97\% in the W4 band. Thus they firstly used the signal-to-noise and reduced chi-squared parameters given in the AllWISE photometric catalog to eliminate the spurious detections, which can suppress the contamination rate in any band down to $<$ 7\%.

YSO identification and classification scheme resembles the one described in \citet{koenig12} and the details can be found in \citet{koenig14}. Here we also give a short description about this process.

Firstly, the contaminations of star-forming galaxies and AGNs are filtered out based on their photometry in W1, W2, and W3 bands. Then the Class I and Class II sources are identified using their colors in W1, W2, and W3 bands (see the \textit{top left} panel of Fig.~\ref{ysowise}). Secondly, the remaining sources with the $H$ and $K_s$ detections are identified as YSOs using the color criteria shown in the \textit{top right} panel of Fig.~\ref{ysowise}. Thirdly, the W4 (22\,$\mu$m) photometry is used to identify transition disks and to retrieve possible protostars from the AGN candidates with the color criteria shown in the \textit{bottom} panels of Fig.~\ref{ysowise}. Finally, all YSOs identified in this step are re-examined to exclude the possible AGB stars and CBe stars (see Fig.~\ref{agbwise}). We crossmatch the YSOs identified in this step with the YSOs identified in the previous steps and only the additional YSOs identified in this step are kept in our final YSO catalog.

\begin{figure}
\includegraphics[width=0.55\linewidth]{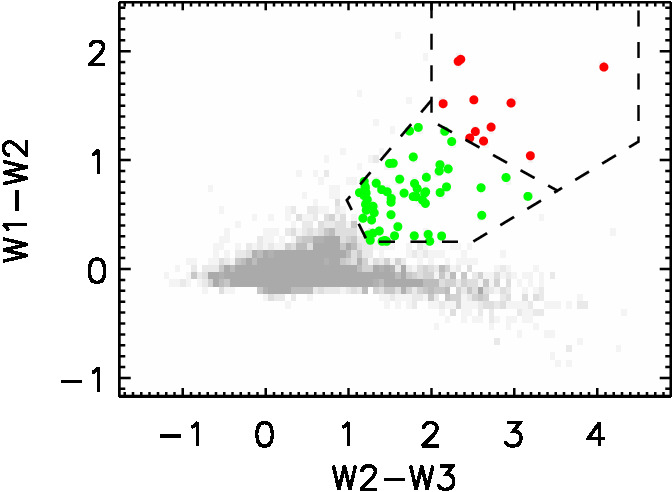}%
\includegraphics[width=0.4\linewidth]{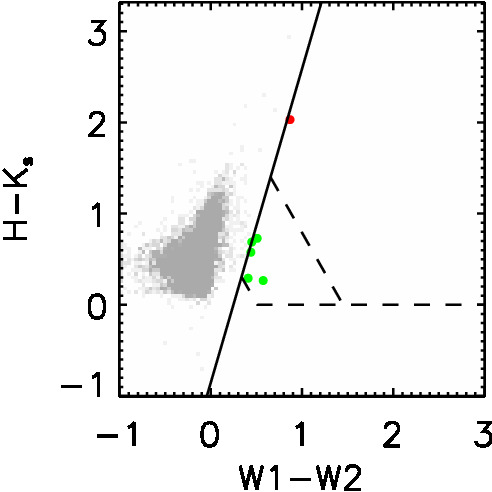}
\vspace{0.1cm}
\includegraphics[width=0.5\linewidth]{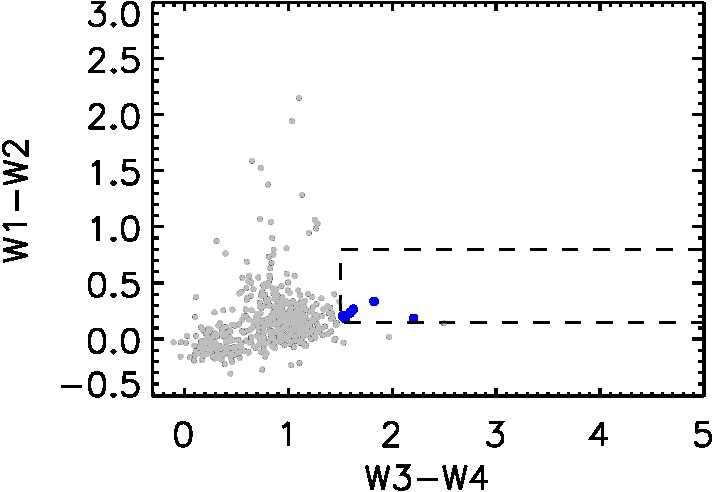}%
\includegraphics[width=0.45\linewidth]{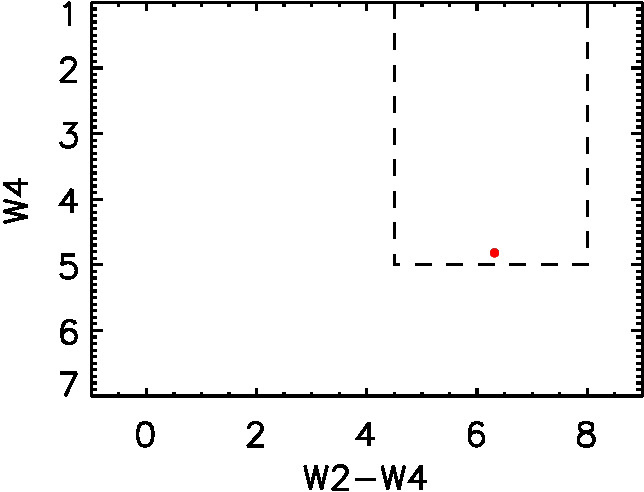}
\caption{The color-color diagrams and the color-magnitude diagram used in \textit{Step 4} for a example of GMF {39} (CFG047.06$+$0.26). The red dots mark the Class I sources and protostars identified in the YSO classification scheme suggested by \citet{koenig14} while the green and blue dots label the Class II sources and transition disks, respectively.}
\label{ysowise}
\end{figure}

\begin{figure}
\includegraphics[width=1.0\linewidth]{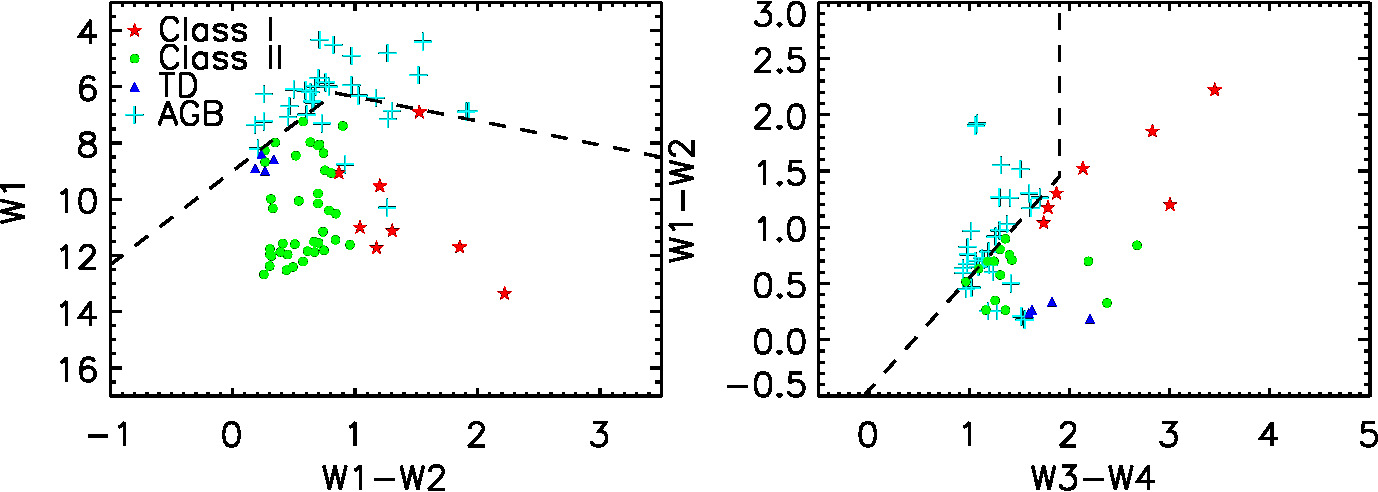}
\caption{The color-magnitude and color-color diagrams used to isolate the possible AGBs from YSOs in \textit{Step 4} for an example of GMF {39} (CFG047.06$+$0.26).}
\label{agbwise}
\end{figure}

\textit{Step 5}: All the YSOs identified in the above steps are re-examined in the $JHK_s$ color-color diagram as shown in Fig.~\ref{ccd} of the main text. The sources located in region 1 are classified as possible giants (e.g., AGB stars). Note that we only plot the YSOs after excluding the giants in Fig.~\ref{ccd}. Because we account the photometric uncertainties of sources, there are still several YSOs locating in region 1 of Fig~\ref{ccd}.

\textit{Step 6}: The sources classified as `YSO', `HII regions' or `HII/YSO' in the RMS catalog \citep{rms-survey} are selected as MYSOs and matched with the YSOs identified in the above steps. We keep the additional MYSOs in our final YSO catalog.

\section{{Estimating uncertainties of the GMF parameters with the Monte Carlo method}}\label{ap4}

\subsection{{Uncertainty of SFR}}\label{ap:d1}
Assuming that the mass completeness ($M_{\textrm{comp}}^{\textrm{c2}}$) of Class II population in each GMF follows the Gaussian distribution while the number ($N_{\textrm{ClassII}}$) of Class II sources brighter than the completeness limit follows the Poisson distribution, we generate one random number of $M_{\textrm{comp}}^{\textrm{c2}}$ and one random number of $N_{\textrm{ClassII}}$. Here we must subtract the contaminants estimated with five nearby control fields from $N_{\textrm{ClassII}}$ (see Section~\ref{sect:contamination}). Each control field can give a contamination fraction for Class II sources. We calculate the mean and standard deviation of five contamination fractions based on five control fields. Assuming the contamination fraction follows a Gaussian distribution, we can also generate one random number of contamination fraction for Class II sources. Then the total number ($N_{\textrm{tot}}^{\textrm{c2}}$) and mass ($M_{\textrm{tot}}^{\textrm{c2}}$) of Class II population are obtained with the Kroupa IMF as described in Section~\ref{sect:massc2}. 

Assuming that the number ($N_{\textrm{ClassI}}$) of Class I sources brighter than the luminosity completeness in each GMF follows the Poisson distribution, we can generate a random number of $N_{\textrm{ClassI}}$. We also generate a random number of contamination fraction for Class I sources as mentioned above and then obtain the total number ($N_{\textrm{tot}}^{\textrm{c1}}$) and mass ($M_{\textrm{tot}}^{\textrm{c1}}$) of Class I population as described in Section~\ref{sect:massc1}.

Finally, the SFR and SFE can be obtained using the method described in Section~\ref{sect:sfrsfe}. Repeating the above process 10, 000 times, we can  obtain the probability distributions for the parameters such as $M_{\textrm{tot}}^{\textrm{c2}}$, $M_{\textrm{tot}}^{\textrm{c1}}$, and SFR as shown in Fig.~\ref{fig_ap4}. The blue solid lines represent the {median values which we adopt as the final values of the corresponding parameters.} The red dashed lines show the 1$\sigma$ uncertainties. 68\%~scores lie within the two red dashed lines.

\begin{figure*}
\includegraphics[width=1.0\linewidth]{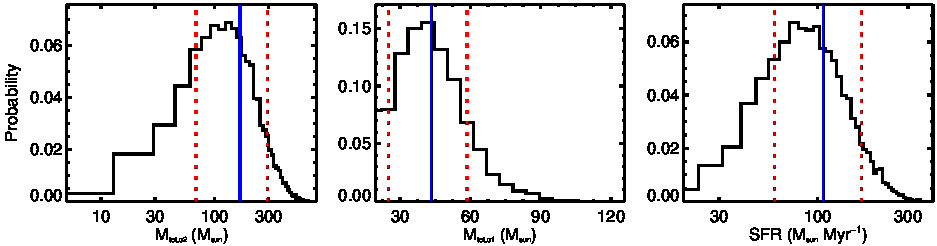}
\caption{The probability distributions of $M_{\textrm{tot}}^{\textrm{c2}}$, $M_{\textrm{tot}}^{\textrm{c1}}$, and SFR in the GMF {39} (CFG047.06$+$0.26). The vertical blue lines show the {median} values and the red dashed lines mark the 1$\sigma$ uncertainties.}
\label{fig_ap4}
\end{figure*}

\subsection{{Uncertainty of cloud mass}}\label{ap:d2}
{The uncertainty of GMF H$_2$ column density is mainly from the observational random errors and the variations of $T_{\textrm{ex}}$ and $^{13}$CO abundance. Compared with the last two components, the observational random errors are small and thus ignored.}

{13 GMFs are also covered or partially covered by ThrUMMS $^{12}$CO survey data. For these GMFs we collapsed the corresponding data cube to obtain the $^{12}$CO intensity maps. Assuming the optically thick for $^{12}$CO (J = 1$-$0) emission lines, the excitation temperature can be obtained using the following formula \citep{radiotools}:}
\begin{equation*}
T_{\textrm{ex}}=\frac{5.53}{\textrm{ln}(1+\frac{5.53}{T_{\textrm{mb}}^{*} (^{12}\textrm{CO})+0.819})}
\end{equation*}
{where $T_{\textrm{mb}}^{*} (^{12}\textrm{CO})$ is obtained from the peak of the $^{12}$CO emission.}
{We obtain the excitation temperature map for the region covered by $^{12}$CO data in each of these GMFs. Combining 13 excitation temperature maps together, Figure~\ref{ap:figtex} shows the excitation temperature distribution. The mean value and standard deviation are $\sim$10\,K and 2.5\,K, respectively, with the minimum and maximum of 2.4\,K and 33.2\,K, individually. We found that the very low excitation temperatures of $<$4\,K are mainly due to the high noise level. Thus finally we adopt a range of [4,35]\,K for the GMF excitation temperature. For the $^{13}$CO abundance of [$^{13}$CO/H$_2$], previous measurements in different interstellar clouds with different conditions vary from $\sim$0.9 to 3.5$\times$10$^{-6}$ with a scatter of $\sim$30$-$50\% \citep{dickman78,frerking82,nice,harjunp04,pineda08,pineda10,ripple13}. We ultimately assume a 50\%~uncertainty for the convention factor of [H$_2$/$^{13}$CO] that we adopted in Appendix~\ref{ap1}.}

{Assuming that $T_{\textrm{ex}}$ and [H$_2$/$^{13}$CO] follow the truncated gaussian distributions, }
\begin{equation*}
T_{\textrm{ex}} \sim N(10, 2.5^2)~~~~~~~~~~~~~~~~~~~~~~~4\,K<T_{\textrm{ex}}<35\,K
\end{equation*}
\begin{equation*}
\frac{[\textrm{H}_{\textrm{2}}/^{\textrm{13}}\textrm{CO}]}{\textrm{10}^{\textrm{5}}} \sim N(8.5,4.25^2)~~~~~~~1<\frac{[\textrm{H}_{\textrm{2}}/^{\textrm{13}}\textrm{CO}]}{\textrm{10}^{\textrm{5}}}<100
\end{equation*}
{we generate one random number of $T_{\textrm{ex}}$ and one random number of [H$_2$/$^{13}$CO]. The column density map for each GMF can be obtained based on Eq.~\ref{ap:eqcolmass} and then the cloud mass, dense gas mass and other physical parameters of each GMF can be also obtained using the methods described in Appendix~\ref{ap1} and \ref{ap2}. Repeating the above process 10, 000 times, we can obtain the probability distributions for the parameters such as $M_{\textrm{cloud}}$ and $\Sigma_{\textrm{gas}}$/t$_{\textrm{ff,s}}$ as shown in Fig.~\ref{ap:figmasserr}. We adopt the median values of the probability distributions as the final values of the corresponding parameters and 68\%~confidential intervals as 1$\sigma$ uncertainties.} 

{The special case is Nessie. For Nessie, we use the extinction map constructed with the technique by \citet{kt13} which combines a near-infrared NICER dust extinction map and a mid-infrared dust extinction map obtained based on mid-infrared absorption against the Galactic background. \citet{kt13} suggested that the uncertainty of the combined extinction map could be mainly from the uncertainty of NIR-MIR dust opacity ratio and they estimated that the uncertainty of this ratio is about 36\%~based on different dust models. Therefore, here we simply adopt an uncertainty of 36\%~for Nessie's extinction map, based on which the uncertainties of other parameters such as gas mass can be obtained.}

\begin{figure}
\includegraphics[width=1.0\linewidth]{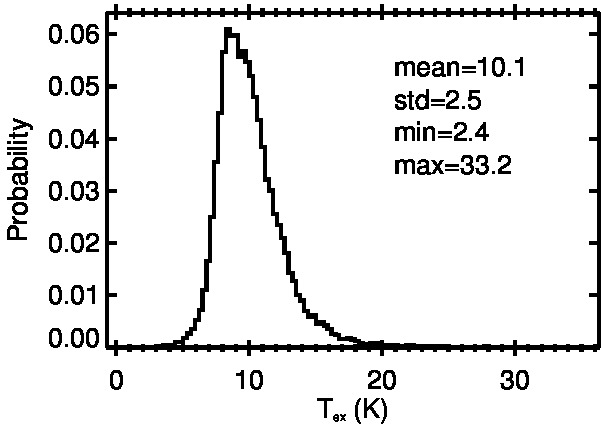}
\caption{{Histogram of $T_{\textrm{ex}}$ obtained in several GMFs that are covered or partially covered by $^{12}$CO data.}}
\label{ap:figtex}
\end{figure}

\begin{figure*}
\includegraphics[width=1.0\linewidth]{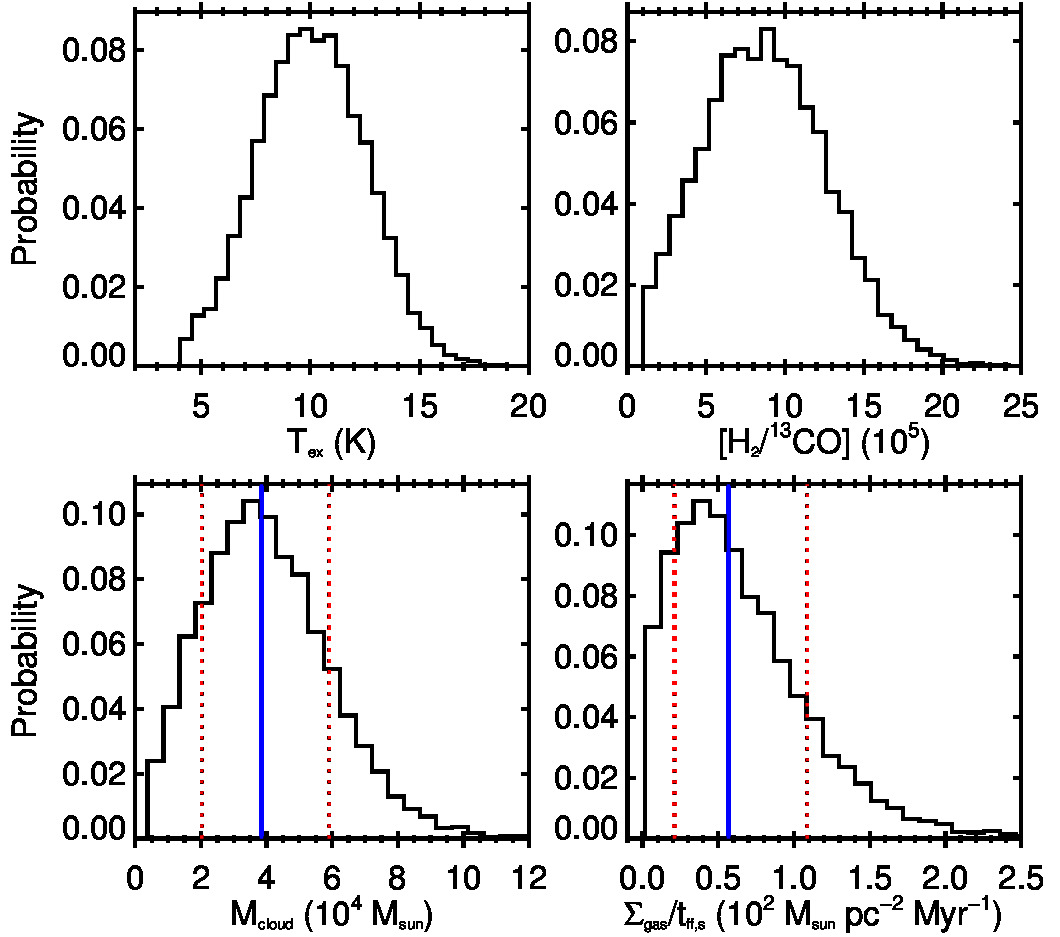}
\caption{{The probability distributions of $T_{\textrm{ex}}$, [H$_2$/$^{13}$CO], $M_{\textrm{cloud}}$ and $\Sigma_{\textrm{gas}}$/t$_{\textrm{ff,s}}$ in the GMF {39} (CFG047.06$+$0.26). The vertical blue lines show the median values and the red dotted lines mark the 1$\sigma$ uncertainties.}}
\label{ap:figmasserr}
\end{figure*}

\subsection{{Including distance errors}}

{The uncertainties estimated in above two sections do not include the distance errors. The uncertainties of parameters listed in Table~\ref{table1} and \ref{table2} do not include the distance errors, either, because distance cancels out in some quantities such as gas surface density ($\Sigma_{\textrm{gas}}$).}

{The distances of GMFs listed in Table~\ref{table1} are directly adopted from the corresponding papers. Most of them are kinematic distances except for the GMFs from \citet{sample-wang16}.}

{\citet{sample-wang16} used three methods to calculate the distances of their giant filaments, including trigonometric parallax measurements, maximum likelihood kinematic distance, and kinematic distance. The different methods give different uncertainties of $\sim$10$-$30\%~and they finally used a conservative 30\%~distance uncertainty. Therefore, we also decide to adopt a 30\%~distance uncertainty for the GMFs from \citet{sample-wang16}.}

{The kinematic distances of GMFs from other reference papers are estimated using different Galactic models. \citet{anderson12} analyzed the effect of different sources of the kinematic distance uncertainty. They found that the uncertainties caused by different rotation curves, non-circular motions, and the change of solar rotation parameters are about $<$5$-$20\%, 5$-$10\%, and $\sim$10\%, respectively, and the combined uncertainty is about 20\%. \citet{roman09} estimated the kinematic distance uncertainty through adding perturbations to the radial velocities of molecular clouds at the Galactic longitudes of 20\degr~and 40\degr~and obtained an uncertainty of $<$30\%. Recently, \citet{ram18} investigated the gas kinematics in a spiral arm using high-resolution smoothed particle hydrodynamics simulations. They found that the kinematic distance error can be as large as $\pm$2\,kpc due to the gas streaming motions. \citet{ram18} also compared their results with \citet{roman09} and found that the relative distance errors are not larger than 30\% for the clouds at galactic longitudes of $\pm$20\degr. Considering these, we finally use a 30\%~distance uncertainty for the GMFs that are not from \citet{sample-wang16}. Therefore, we actually adopt a 30\%~distance uncertainty for all the GMFs.}

{Assuming the distance of each GMF, $D$, follows the truncated gaussian distribution:}
\begin{equation*}
D \sim N(D_0, (0.3D_0)^2)~~~~~~~~~~D > 1 ~\textrm{kpc}
\end{equation*}
{where $D_0$ represents the distance value that is listed in Table~\ref{table1}, we generate one random number of $D$. The physical parameters of each GMF can be obtained using the dependence of parameters on distance. For example, we have obtained the probability distribution of $M_{\textrm{cloud}}$ without considering the distance error in Appendix~\ref{ap:d2} and thus, we can generate one random number, $m_{\textrm{cloud}}$ that follows this distribution. The final total gas mass can be obtained with the relation of}
\begin{equation*}
M_{\textrm{cloud}} = \frac{m_{\textrm{cloud}}}{D_0^2}\times D^2.
\end{equation*}
{Repeating this process 10, 000 times, we can obtain the probability distribution of $M_{\textrm{cloud}}$ after including the distance error as shown in Fig.~\ref{fig:disterr} (top-right panel). The median value of this distribution is adopted as the final value of $M_{\textrm{cloud}}$ while the 68\%~confidential interval is adopted as the 1$\sigma$ uncertainty.}

{For the SFR of each GMF, it is difficult to obtain the analytical dependence on distance. We re-calculate the SFRs using different distances for each GMF and the results are shown in Fig.~\ref{fig:sfrdist}. There is roughly a linear relation between Log(SFR) and Log(Distance) (blue curves). The linear fitting results (red dashed lines) give the slope in range of [1.2, 1.4] with a mean value of 1.3. Therefore, we assume a simple relation between SFR and distance:}
\begin{equation*}
\textrm{SFR} \propto D^{1.3}.
\end{equation*}
{With this relation, we can also obtain the probability distribution of SFR after including distance errors for each GMF as shown in Fig.~\ref{fig:disterr} (bottom-left panel).}

{For other more complex parameters, the dependence on distance is as follows:}
\begin{equation*}
\frac{\Sigma_{\textrm{gas}}}{{t}_{\textrm{ff,s}}} \propto D^{-0.5},
\end{equation*}
\begin{equation*}
\frac{\Sigma_{\textrm{gas}}}{{t}_{\textrm{cross}}} \propto D^{-1},
\end{equation*}
\begin{equation*}
\textrm{SFR}_{\textrm{ff,s}} \propto D^{-0.2},
\end{equation*}
\begin{equation*}
\textrm{SFE}=\frac{1}{1+\frac{m_{\textrm{cloud}}}{2\textrm{sfr}} D^{0.7}},
\end{equation*}
{where SFR=sfr$D^{1.3}$ and $M_{\textrm{cloud}}$=$m_{\textrm{cloud}}D^2$.}
{Table~\ref{table4} lists the parameters and associated uncertainties after including distance error for 34 GMFs that are located within 5.5\,kpc.}

\begin{figure*}
\includegraphics[width=1.0\linewidth]{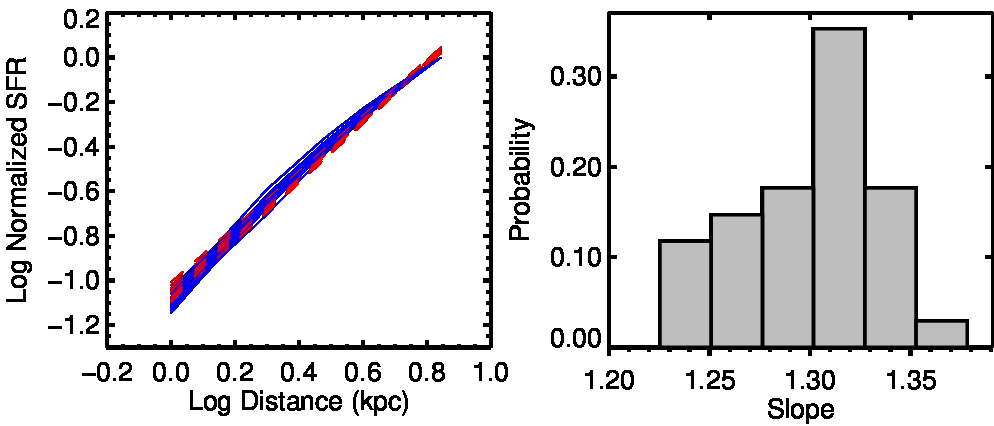}
\caption{{\textit{Left:} The relations between SFR and distance for 34 GMFs. Each blue curve represents one GMF. The red dashed lines show the linear fits to the blue curves; \textit{Right:} the distribution of fitting slopes.}}
\label{fig:sfrdist}
\end{figure*}

\begin{figure*}
\includegraphics[width=1.0\linewidth]{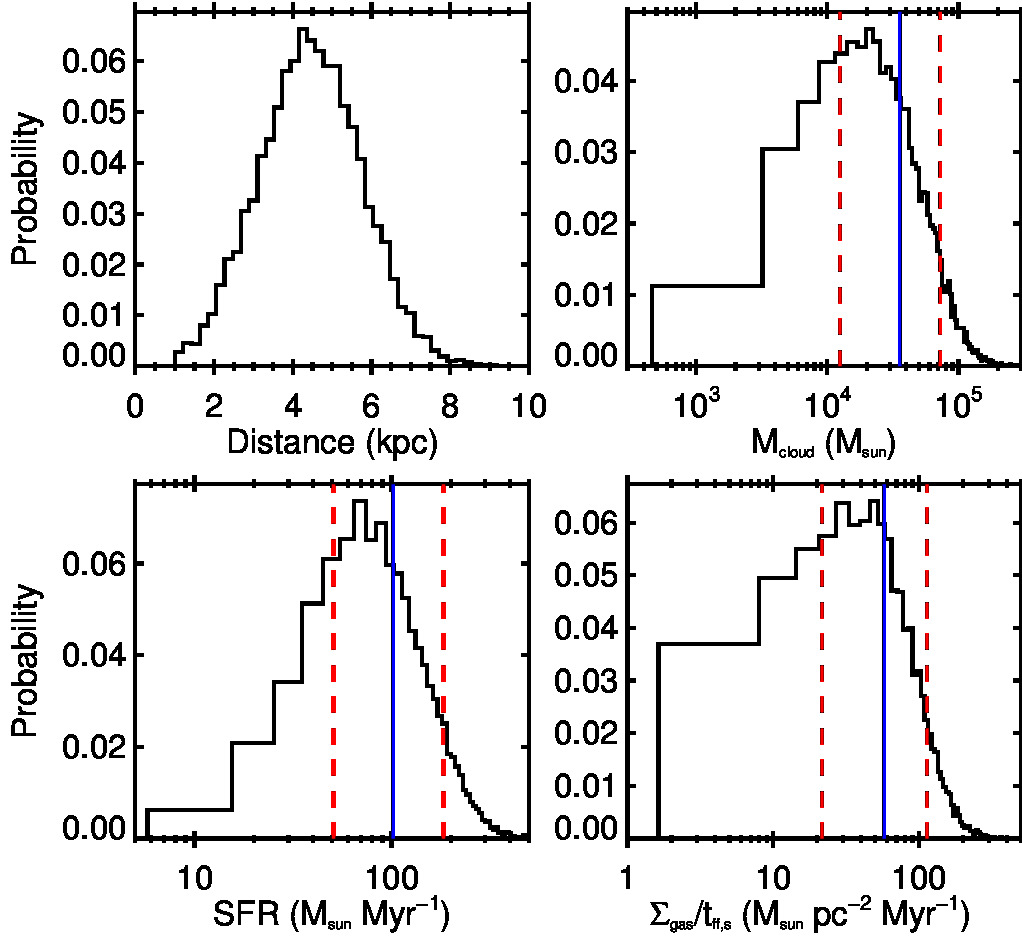}
\caption{{The probability distributions of distance, $M_{\textrm{cloud}}$, SFR, and $\Sigma_{\textrm{cloud}}$/$t_{\textrm{ff,s}}$ in the GMF {39} (CFG047.06$+$0.26) after including 30\%~distance uncertainty. The vertical blue lines show the median values and the red dotted liens mark the 1$\sigma$ uncertainties.}}
\label{fig:disterr}
\end{figure*}


\section{{Investigating SFR$-M_{\textrm{dense}}$ relation with different sub-samples}}\label{apsubsamples}

{In Section~\ref{sect:sfrsfe}, we mentioned that we could underestimate the SFRs for the distant GMFs. Other parameters such as $M_{\textrm{dense}}$ and length of GMFs could be also distance-biased due to the selection effects. These biases could affect the slope of SFR$-$$M_{\textrm{dense}}$ relation. To investigate this, we firstly test the correlations between the parameters such as SFR, $M_{\textrm{dense}}$, and length and distances of GMFs. Figure~\ref{fig:sflaw2corrcheck} shows the result. We do not find any significant correlations between SFR, $M_{\textrm{dense}}$, and length versus distance for GMFs. However, due to the large uncertainties of the parameters, we can not exclude the possibility that there are strong correlations between these parameter sets, which means that we can not exclude the possibility that SFRs and $M_{\textrm{dense}}$ of GMFs are distance-biased.}
{We} {also} {note that there are two points away from other points in Fig.~\ref{sflaw2}. These two GMFs (GMF 6 and 35) have $R$ value of $<$~1, which means that there are other relatively strong velocity components in the line of sight towards them and we could overestimate their SFRs.} 
{Considering above issues, how reliable is the detection of a approximate sub-linear SFR$-M_{\textrm{dense}}$ relation in Fig.~\ref{sflaw2}?}

\begin{figure*}
\includegraphics[width=1.0\linewidth]{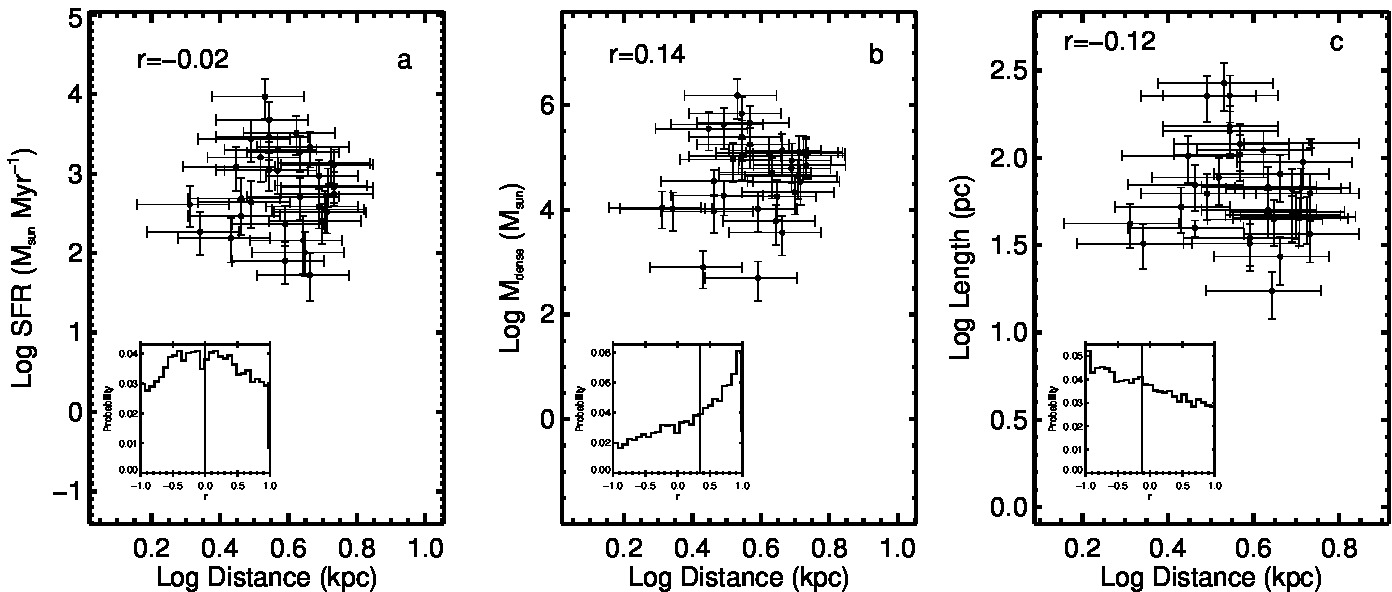}
\caption{{SFRs (a), $M_{\textrm{dense}}$ (b), and length (c) versus distance of GMFs that are located within 5.5 kpc in the log-log plane. The correlation coefficients between different parameter sets obtained without considering uncertainties are marked on top region of each panel. The insert plots in each panel show the probability distributions of correlation coefficients ($r$) obtained with Bayesian linear regression method by \citet{kelly07} and the vertical lines in the insert plots mark the median values.}}
\label{fig:sflaw2corrcheck}
\end{figure*}

{To answer this question, we investigate the slope of SFR$-M_{\textrm{dense}}$ relation in the subsequent {12} samples:}
\begin{itemize}
\item[] {\makebox[2cm]{{Sample 1:}}{GMFs (i.e., sample used in the main text)}.} 
\item[] {\makebox[2cm]{{Sample 2:}}{GMFs$+$nearby clouds}.}
\item[] {\makebox[2cm]{{Sample 3:}}{GMFs excluding GMF 6 and 35.}}
\item[] {\makebox[2cm]{{Sample 4:}}{GMFs excluding GMF 6 and 35 plus nearby clouds}}
\item[] {\makebox[2cm]{{Sample 5:}}{GMFs with $R>$~1.}}
\item[] {\makebox[2cm]{{Sample 6:}}{GMFs with $R>$~1 plus nearby clouds.}}
\item[] {\makebox[2cm]{{Sample 7:}}{GMFs with distances of $<$3.5 kpc.}}
\item[] {\makebox[2cm]{{Sample 8:}}{GMFs with distances of $<$3.5 kpc plus nearby clouds.}}
\item[] {\makebox[2cm]{{Sample~9:}}{GMFs with distances of $<$4 kpc.}}
\item[] {\makebox[2cm]{{Sample 10:}}{GMFs with distances of $<$4 kpc plus nearby clouds.}}
\item[] {\makebox[2cm]{{Sample 11:}}{GMFs with distances of $<$5 kpc.}}
\item[] {\makebox[2cm]{{Sample 12:}}{GMFs with distances of $<$5 kpc plus nearby clouds.}}
\end{itemize}

{Figure~\ref{fig:sflaw2ap} and {\ref{fig:sflaw2dist}} shows the results. 
Assuming the relation of} 
\begin{equation*}
\textrm{SFR} \propto M_{\textrm{dense}}^{\alpha},
\end{equation*}
{the slopes of $\alpha$ 
obtained with Bayesian linear regression \citep{kelly07} in different samples are shown in Fig.~\ref{fig:slope}. 
Although we can not exclude the possibility of $\alpha =$~1, SFR$-M_{\textrm{dense}}$ relation is sub-linear within $\sim$68\%~confidence (1$\sigma$) in all {12} samples.} 

\begin{figure*}
\includegraphics[width=1.0\linewidth]{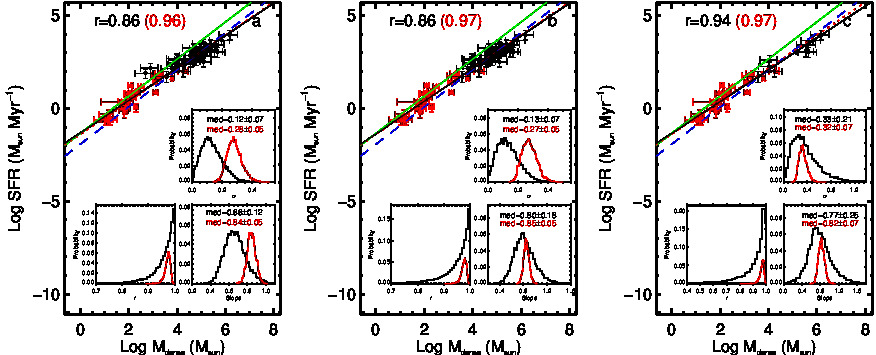}
\caption{{SFR as a function of $M_{\textrm{dense}}$ for (a) \comobj; (b) GMFs excluding two objects (GMF 6 and 35) plus nearby clouds; (c) GMFs with $R>$~1 plus nearby clouds. The GMFs are marked with black filled circles while the nearby clouds are labeled with red triangles. The straight lines represent Lada's equation \citep[green solid line,][]{lada10}, the relation from \citet{wu05} (blue dashed line), fitting result from \citet{evans14} (red dotted line), and our linear fitting results without considering uncertainties (black solid line). The correlation coefficients between SFR and $M_{\textrm{dense}}$ obtained without considering uncertainties are marked on top region of each panel and the black fonts are for GMFs while the red fonts are for \comobj. The insert plots in each panel show the probability distributions of correlation coefficients ($r$), fitting slopes, and intrinsic scatters ($\sigma$) obtained with Bayesian linear regression method by \citet{kelly07} for GMFs (black line) and \comobj~(red line).}}
\label{fig:sflaw2ap}
\end{figure*}

\begin{figure*}
\includegraphics[width=1.0\linewidth]{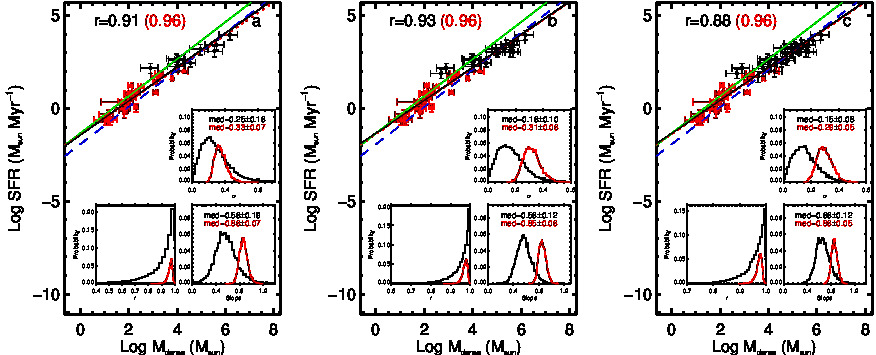}
\caption{{SFR as a function of $M_{\textrm{dense}}$ for (a) GMFs that are located within 3.5 kpc plus nearby clouds; (b) GMFs that are located within 4 kpc plus nearby clouds; (c) GMFs that are located within 5 kpc plus nearby clouds. The GMFs are marked with black filled circles while the nearby clouds are labeled with red triangles. The straight lines represent Lada's equation \citep[green solid line,][]{lada10}, the relation from \citet{wu05} (blue dashed line), fitting result from \citet{evans14} (red dotted line), and our linear fitting results without considering uncertainties (black solid line). The correlation coefficients between SFR and $M_{\textrm{dense}}$ obtained without considering uncertainties are marked on top region of each panel and the black fonts are for GMFs while the red fonts are for \comobj. The insert plots in each panel show the probability distributions of correlation coefficients ($r$), fitting slopes, and intrinsic scatters ($\sigma$) obtained with Bayesian linear regression method by \citet{kelly07} for GMFs (black line) and \comobj~(red line).}}
\label{fig:sflaw2dist}
\end{figure*}

\begin{figure}
\includegraphics[width=1.0\linewidth]{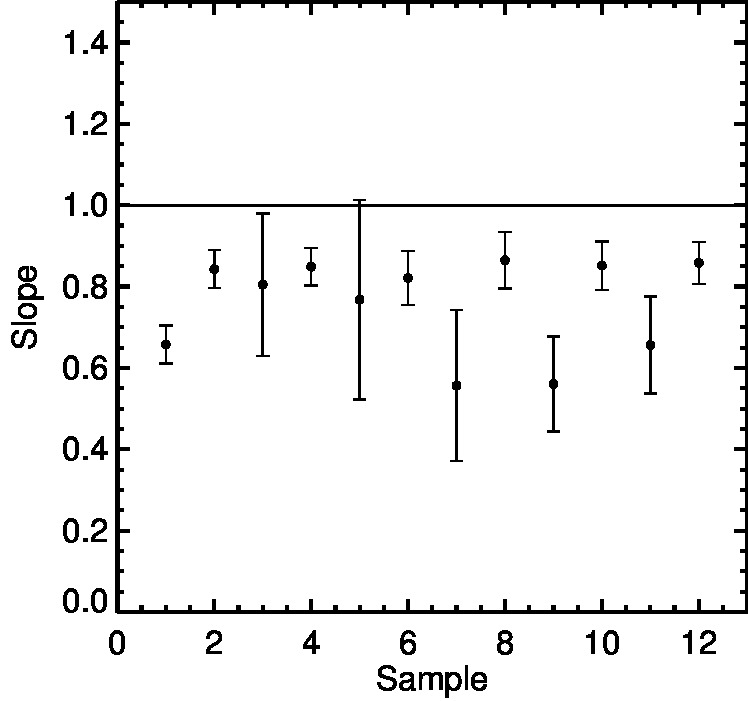}
\caption{{Slopes of $\alpha$ (black filled circles) 
in different samples. The black line marks the slope$=$1.}}
\label{fig:slope}
\end{figure}

\section{Plots of SFR$-M$ relations without error bars}\label{ap4p5}
\begin{figure*}
\includegraphics[width=1.0\linewidth]{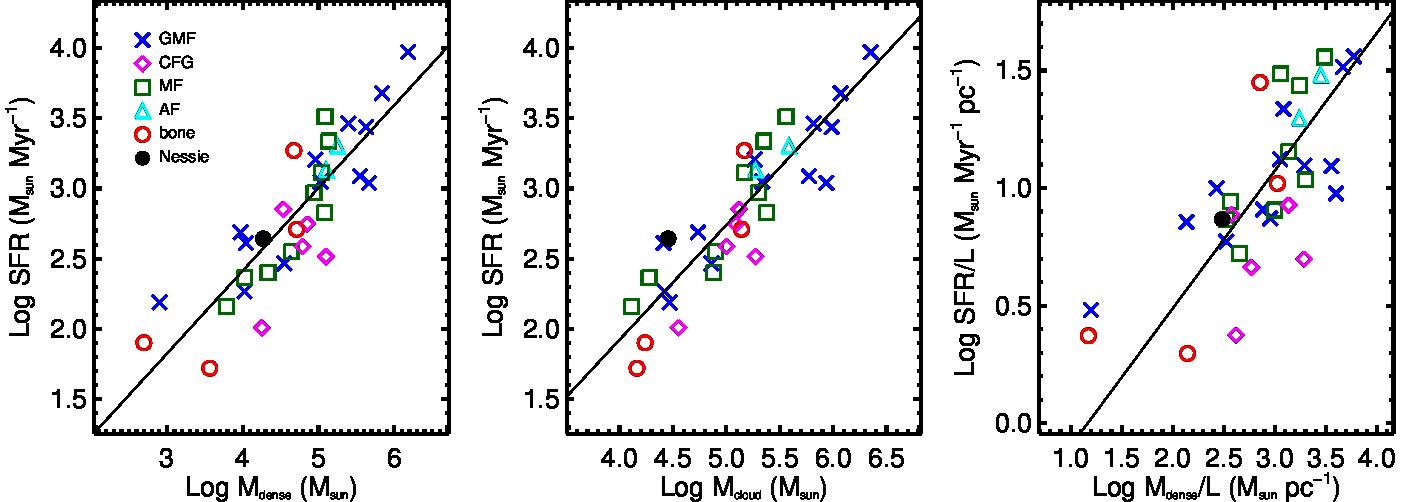}
\caption{\jkfour{SFR and SFR/$L$ as a function of dense gas mass (\textit{left panel}), cloud mass (\textit{middle panel}), and dense gas mass per unit length (\textit{right panel}) for GMFs from different original publications. The black solid lines represent the linear fitting results for all data points. We did not plot error bars with the aim to show different symbols more clearly.}}
\label{fig:sflaw_sample}
\end{figure*}

\section{{Image gallery}}\label{ap5}

\begin{figure*}
\includegraphics[width=1.0\linewidth]{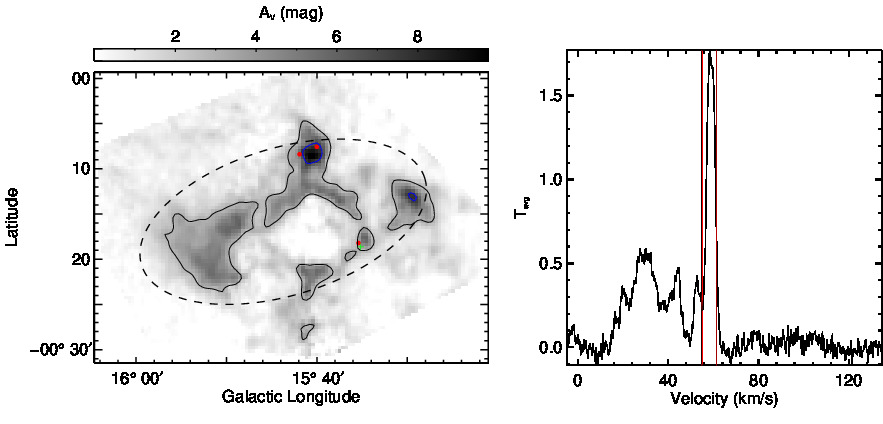}
\caption{Overview of GMF 1 (G015.653-0.224). \textit{Left:} the background is CO-based extinction map. The black and blue contours represent the visual extinction of $A_V=$3 and 7\,mag, individually. The dashed ellipses are obtained through fitting the pixels inside the regions with $A_V>$3\,mag. The identified YSOs are labeled with red filled circles (Class I) and green pluses (Class II); \textit{right:} the $^{13}$CO average spectrum for the region with $A_V>$3\,mag. The red vertical lines mark the GMF velocity range.}
\label{fig:LG015p65}
\end{figure*}

\begin{figure*}
\includegraphics[width=1.0\linewidth]{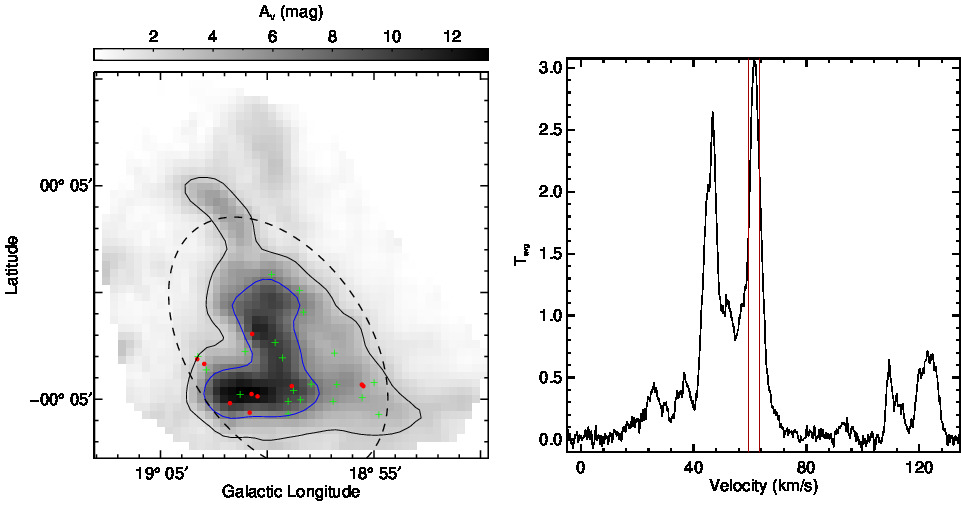}
\caption{Overview of GMF 2 (F19). Others are same as Fig.~\ref{fig:LG015p65}.}
\label{fig:F19}
\end{figure*}
\clearpage

\begin{figure*}
\includegraphics[width=1.0\linewidth]{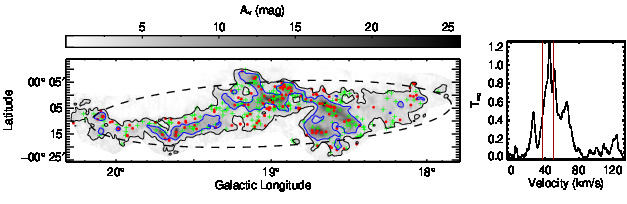}
\caption{Overview of GMF 3 (GMF20.0-17.9). Others are same as Fig.~\ref{fig:LG015p65}.}
\label{fig:GMF20}
\end{figure*}

\begin{figure*}
\includegraphics[width=1.0\linewidth]{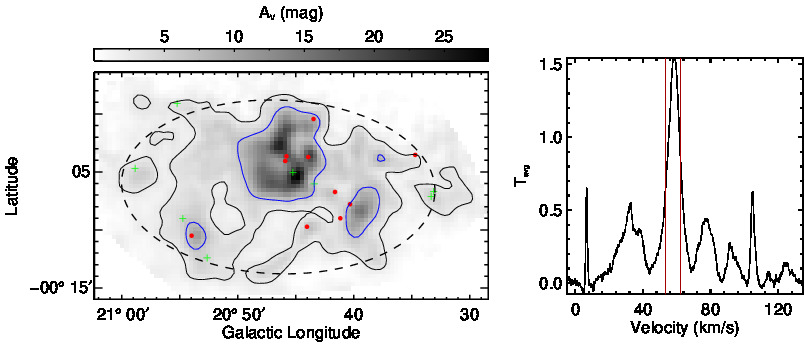}
\caption{Overview of GMF 4 (F20). Others are same as Fig.~\ref{fig:LG015p65}.}
\label{fig:F20}
\end{figure*}
\clearpage

\begin{figure*}
\includegraphics[width=1.0\linewidth]{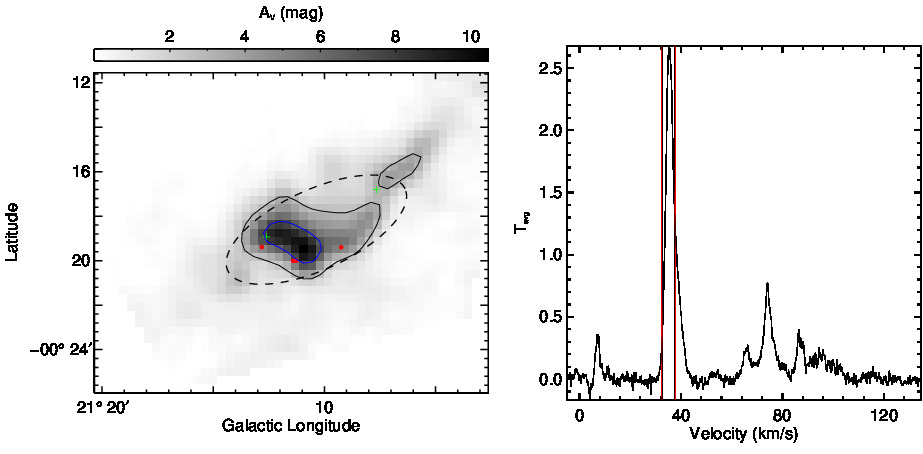}
\caption{Overview of GMF 5 (G021.173-0.312). Others are same as Fig.~\ref{fig:LG015p65}.}
\label{fig:LG021p17}
\end{figure*}

\begin{figure*}
\includegraphics[width=1.0\linewidth]{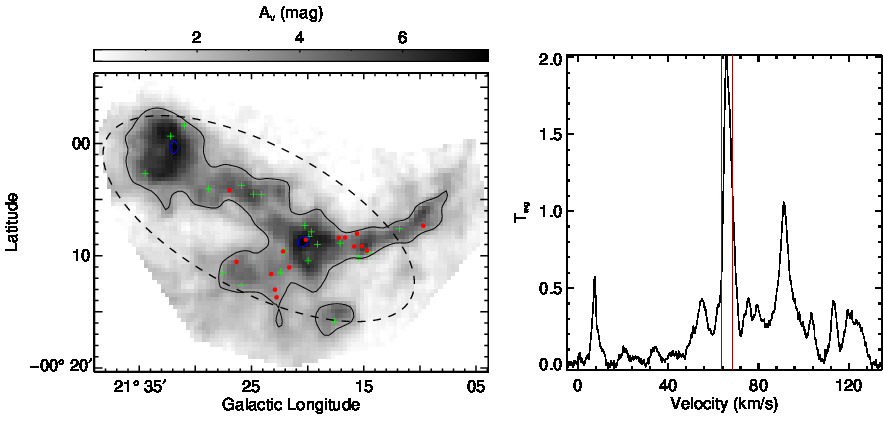}
\caption{Overview of GMF 6 (BC\_021.25-0.15). Others are same as Fig.~\ref{fig:LG015p65}.}
\label{fig:g21}
\end{figure*}
\clearpage

\begin{figure*}
\includegraphics[width=1.0\linewidth]{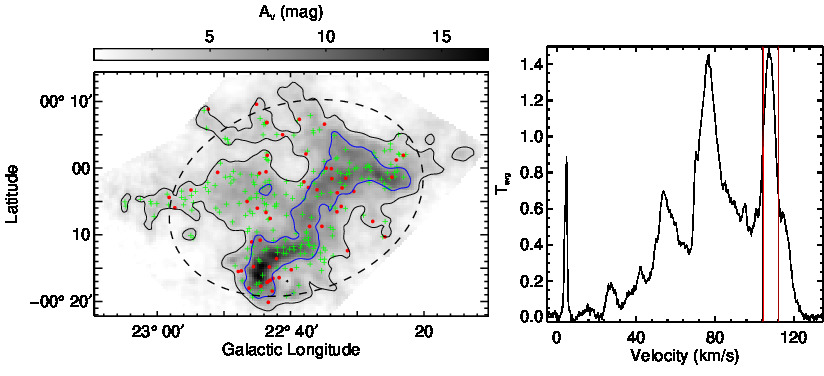}
\caption{Overview of GMF 7 (G022.519-0.025). Others are same as Fig.~\ref{fig:LG015p65}.}
\label{fig:LG022p51}
\end{figure*}

\begin{figure*}
\includegraphics[width=1.0\linewidth]{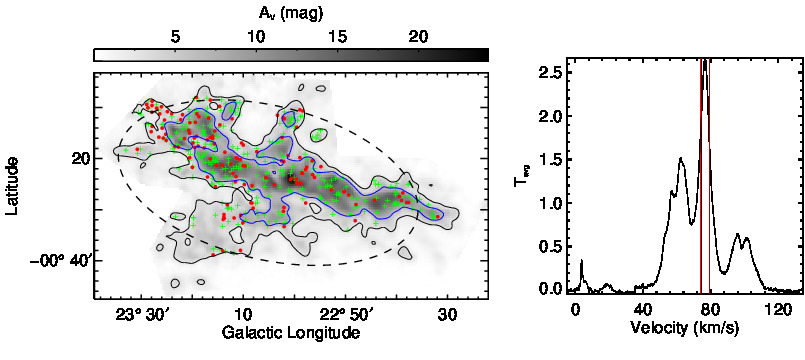}
\caption{Overview of GMF 8 (F21). Others are same as Fig.~\ref{fig:LG015p65}.}
\label{fig:F21}
\end{figure*}
\clearpage

\begin{figure*}
\includegraphics[width=1.0\linewidth]{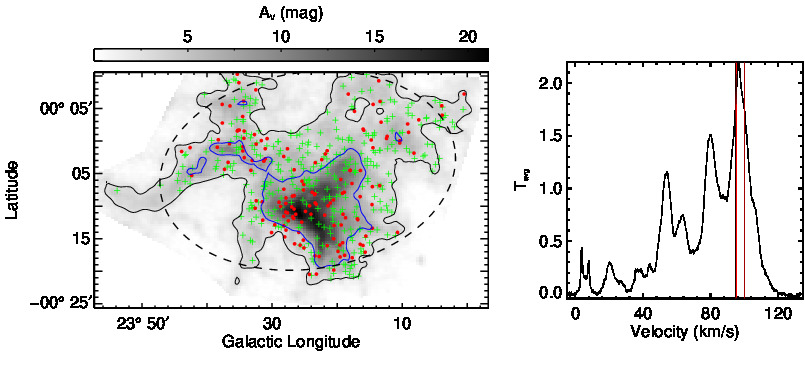}
\caption{Overview of GMF 9 (F23). Others are same as Fig.~\ref{fig:LG015p65}.}
\label{fig:F23}
\end{figure*}

\begin{figure*}
\includegraphics[width=1.0\linewidth]{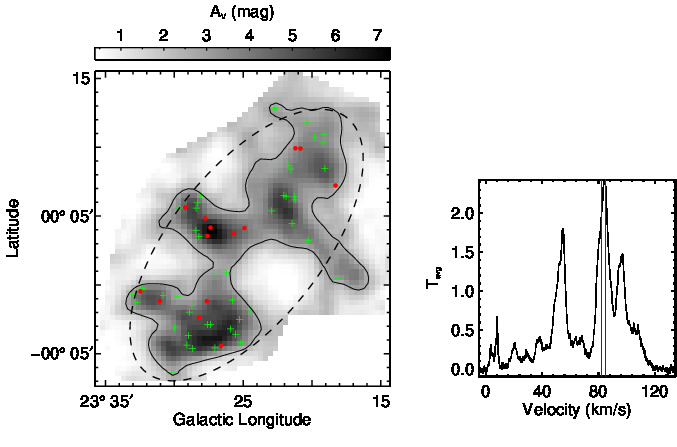}
\caption{Overview of GMF 10 (F24). Others are same as Fig.~\ref{fig:LG015p65}.}
\label{fig:F24}
\end{figure*}
\clearpage

\begin{figure*}
\includegraphics[width=1.0\linewidth]{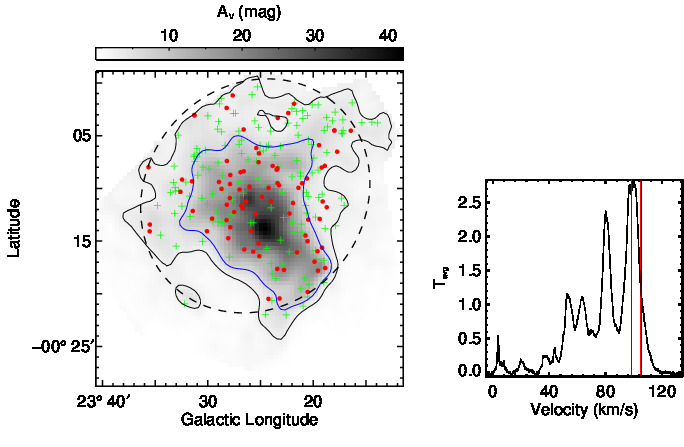}
\caption{Overview of GMF 11 (F22). Others are same as Fig.~\ref{fig:LG015p65}.}
\label{fig:F22}
\end{figure*}

\begin{figure*}
\includegraphics[width=1.0\linewidth]{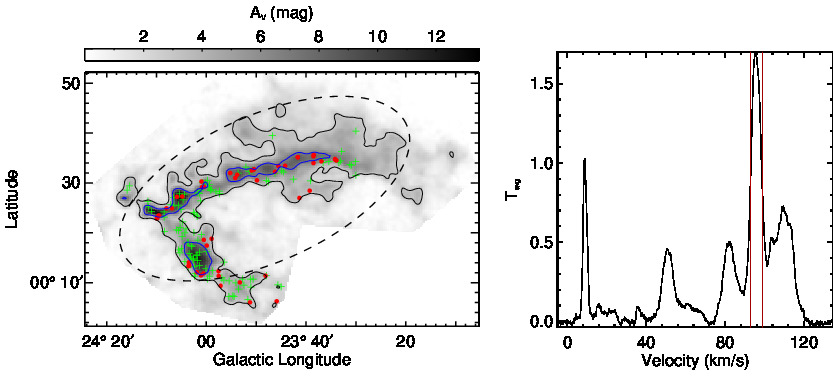}
\caption{Overview of GMF 12 (CFG024.00+0.48). Others are same as Fig.~\ref{fig:LG015p65}.}
\label{fig:CFG24}
\end{figure*}
\clearpage

\begin{figure*}
\includegraphics[width=1.0\linewidth]{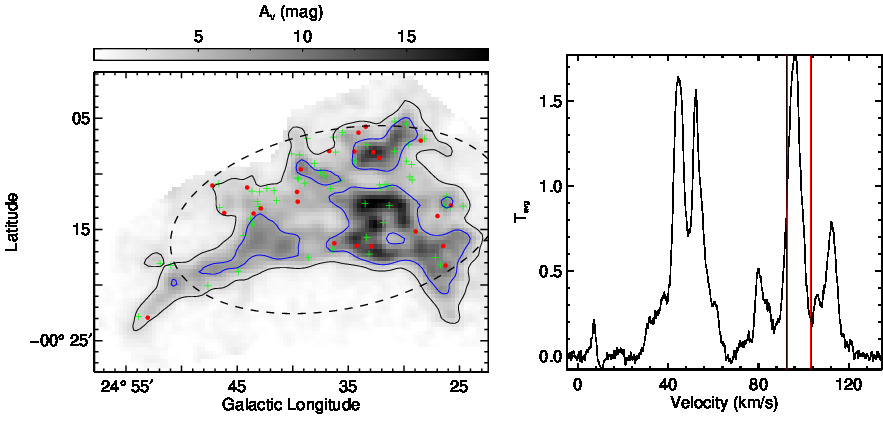}
\caption{Overview of GMF 13 (F26). Others are same as Fig.~\ref{fig:LG015p65}.}
\label{fig:F26}
\end{figure*}

\begin{figure*}
\includegraphics[width=1.0\linewidth]{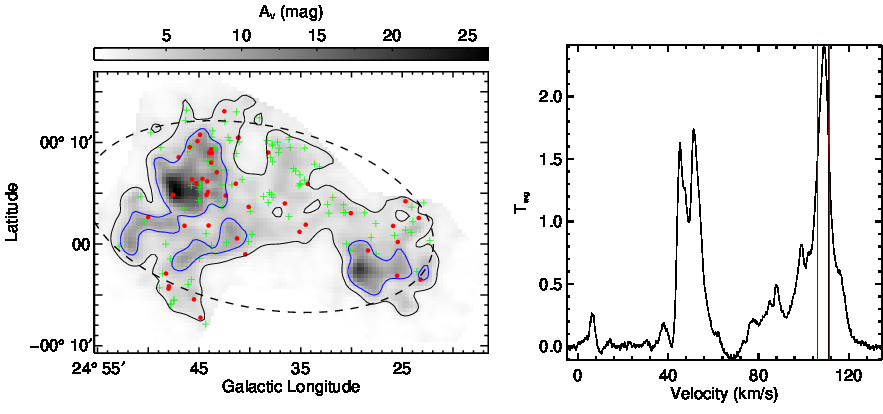}
\caption{Overview of GMF 14 (F27). Others are same as Fig.~\ref{fig:LG015p65}.}
\label{fig:F27}
\end{figure*}
\clearpage

\begin{figure*}
\includegraphics[width=1.0\linewidth]{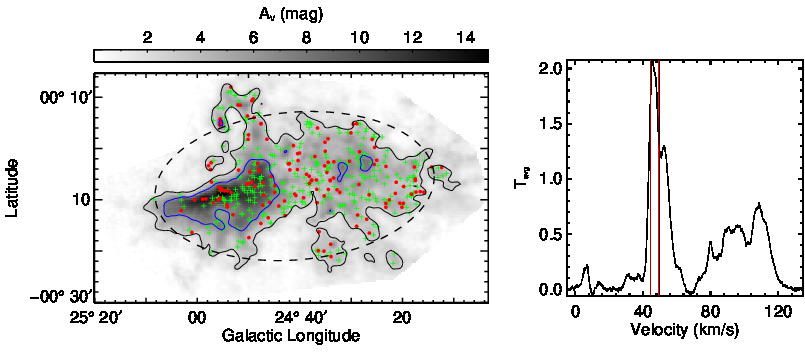}
\caption{Overview of GMF 15 (BC\_24.95-0.17). Others are same as Fig.~\ref{fig:LG015p65}.}
\label{fig:g24}
\end{figure*}

\begin{figure*}
\includegraphics[width=1.0\linewidth]{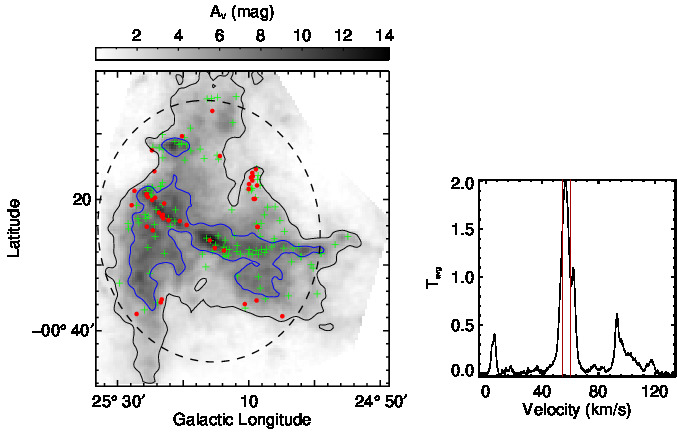}
\caption{Overview of GMF 16 (BC\_025.24-0.45). Others are same as Fig.~\ref{fig:LG015p65}.}
\label{fig:g25}
\end{figure*}
\clearpage

\begin{figure*}
\includegraphics[width=1.0\linewidth]{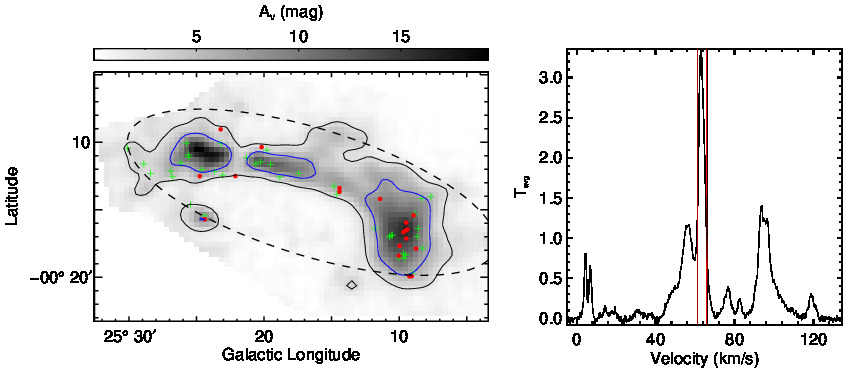}
\caption{Overview of GMF 17 (F28). Others are same as Fig.~\ref{fig:LG015p65}.}
\label{fig:F28}
\end{figure*}

\begin{figure*}
\includegraphics[width=1.0\linewidth]{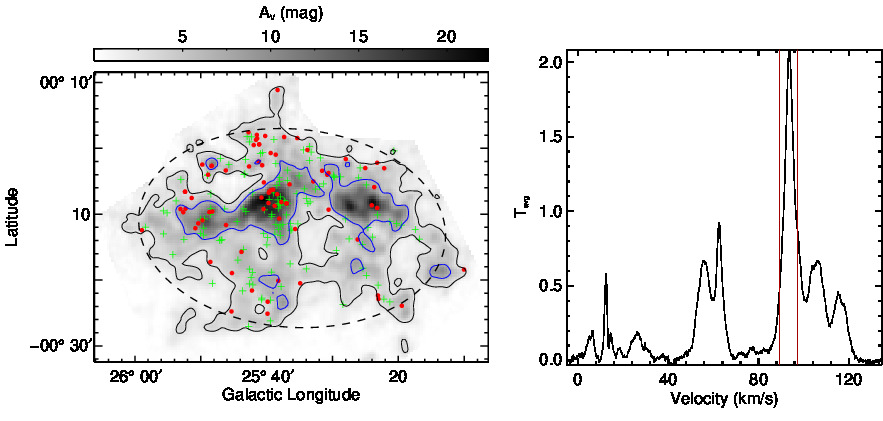}
\caption{Overview of GMF 18 (F29). Others are same as Fig.~\ref{fig:LG015p65}.}
\label{fig:F29}
\end{figure*}
\clearpage

\begin{figure*}
\includegraphics[width=1.0\linewidth]{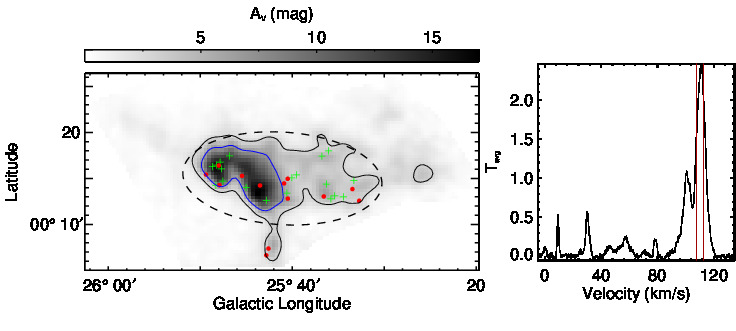}
\caption{Overview of GMF 19 (G025.762+0.241). Others are same as Fig.~\ref{fig:LG015p65}.}
\label{fig:LG025p76}
\end{figure*}

\begin{figure*}
\includegraphics[width=1.0\linewidth]{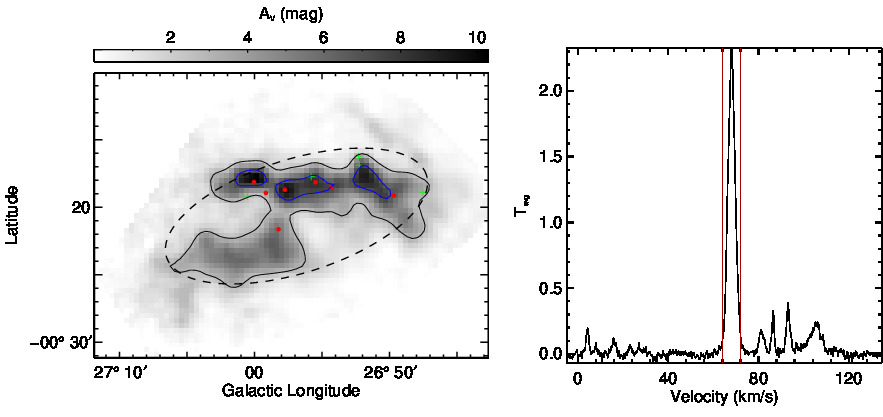}
\caption{Overview of GMF 20 (BC\_26.94-0.30). Others are same as Fig.~\ref{fig:LG015p65}.}
\label{fig:g26}
\end{figure*}
\clearpage

\begin{figure*}
\includegraphics[width=1.0\linewidth]{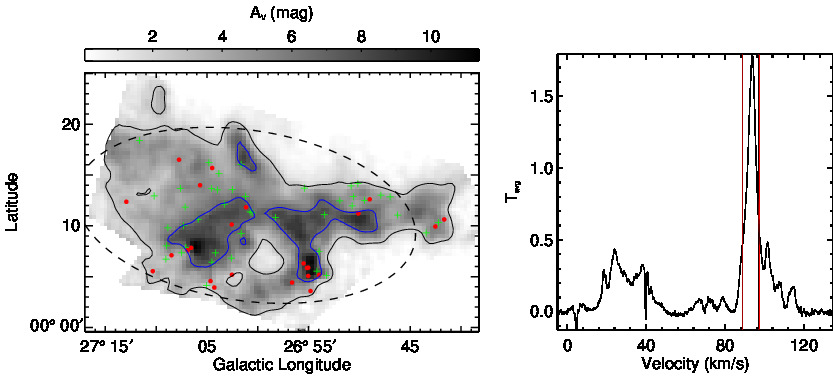}
\caption{Overview of GMF 21 (F30). Others are same as Fig.~\ref{fig:LG015p65}.}
\label{fig:F30}
\end{figure*}

\begin{figure*}
\includegraphics[width=1.0\linewidth]{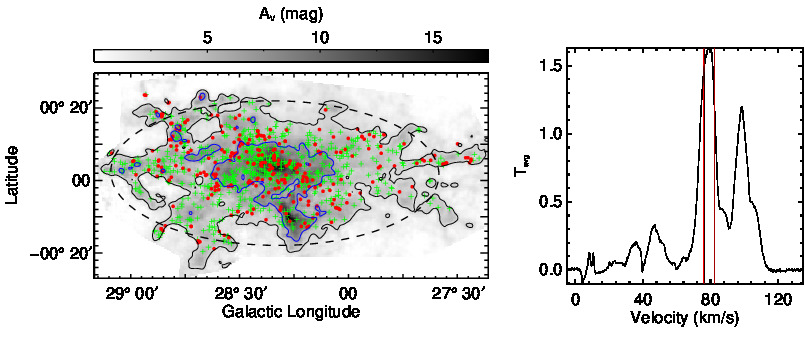}
\caption{Overview of GMF 22 (F31). Others are same as Fig.~\ref{fig:LG015p65}.}
\label{fig:F31}
\end{figure*}
\clearpage

\begin{figure*}
\includegraphics[width=1.0\linewidth]{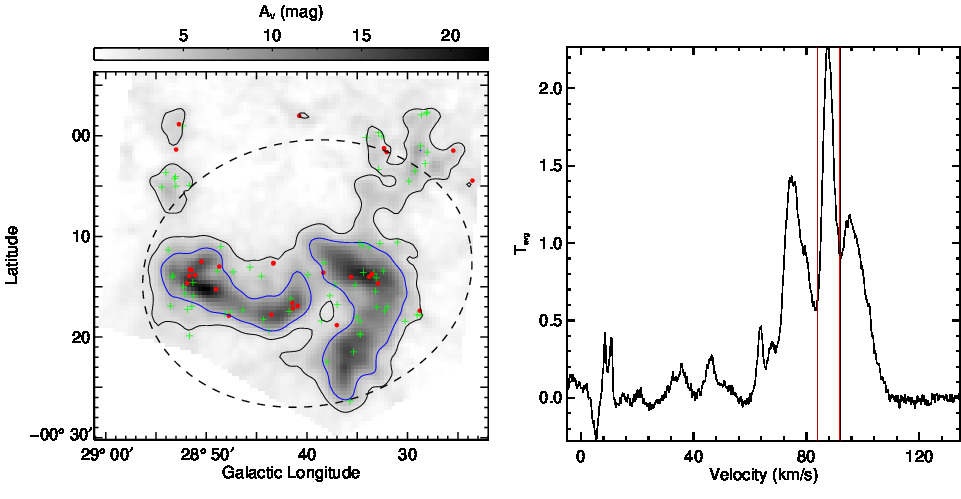}
\caption{Overview of GMF 23 (CFG028.68-0.28). Others are same as Fig.~\ref{fig:LG015p65}.}
\label{fig:CFG28}
\end{figure*}

\begin{figure*}
\includegraphics[width=1.0\linewidth]{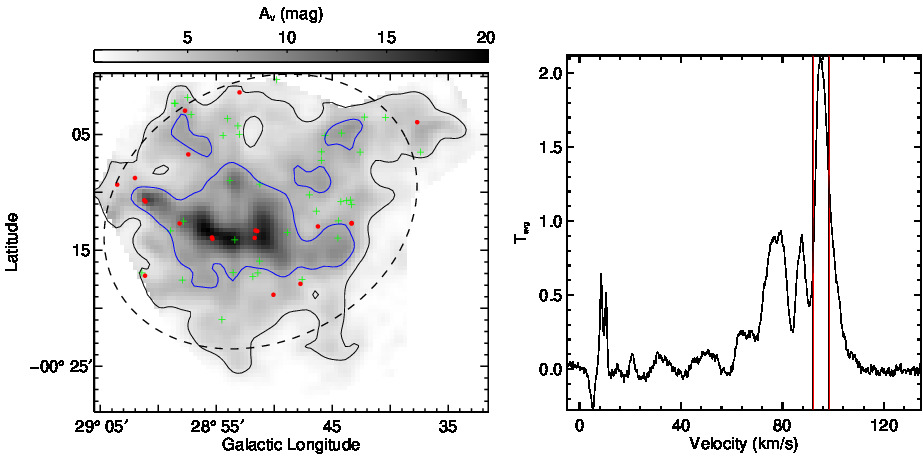}
\caption{Overview of GMF 24 (G028.854-0.238). Others are same as Fig.~\ref{fig:LG015p65}.}
\label{fig:LG028p85}
\end{figure*}
\clearpage

\begin{figure*}
\includegraphics[width=1.0\linewidth]{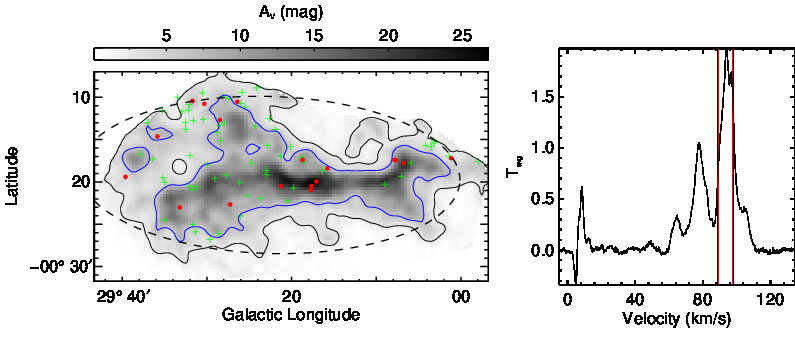}
\caption{Overview of GMF 25 (CFG029.18-0.34). Others are same as Fig.~\ref{fig:LG015p65}.}
\label{fig:CFG29}
\end{figure*}

\begin{figure*}
\includegraphics[width=1.0\linewidth]{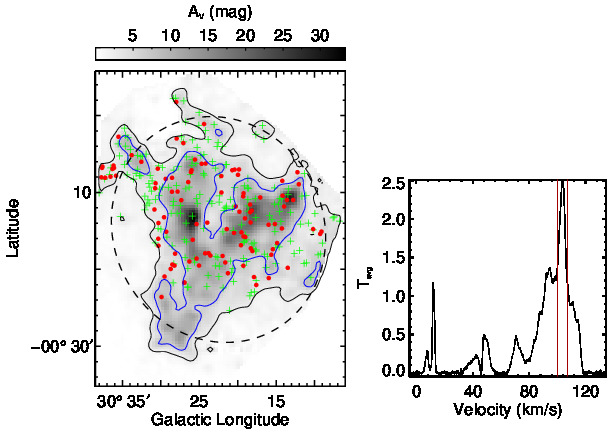}
\caption{Overview of GMF 26 (G030.315-0.154). Others are same as Fig.~\ref{fig:LG015p65}.}
\label{fig:LG030p31}
\end{figure*}
\clearpage

\begin{figure*}
\includegraphics[width=1.0\linewidth]{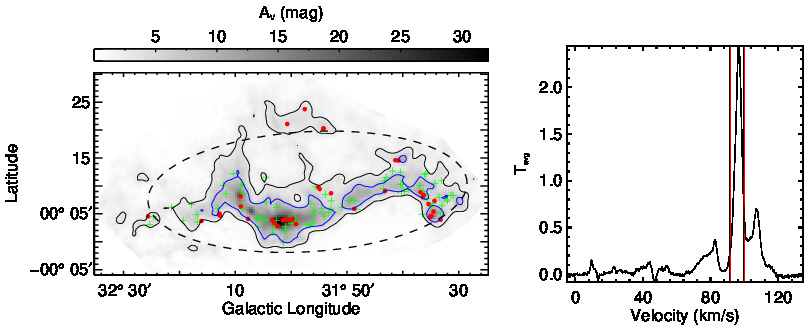}
\caption{Overview of GMF 27 (G32.02+0.06). Others are same as Fig.~\ref{fig:LG015p65}.}
\label{fig:G32.02}
\end{figure*}

\begin{figure*}
\includegraphics[width=1.0\linewidth]{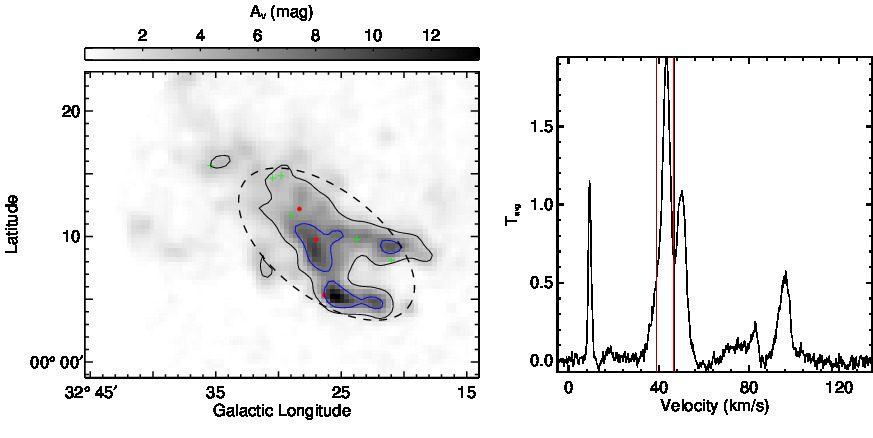}
\caption{Overview of GMF 28 (G032.401+0.082). Others are same as Fig.~\ref{fig:LG015p65}.}
\label{fig:LG032p40}
\end{figure*}
\clearpage

\begin{figure*}
\includegraphics[width=1.0\linewidth]{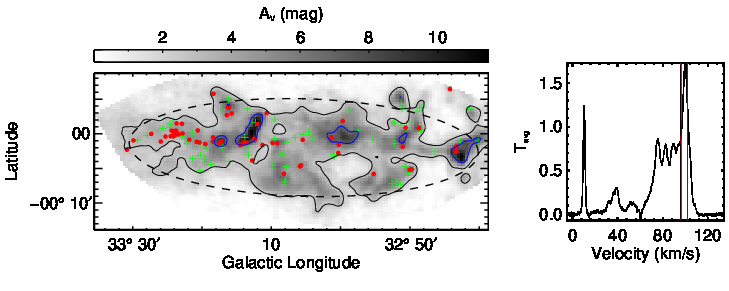}
\caption{Overview of GMF 29 (F34). Others are same as Fig.~\ref{fig:LG015p65}.}
\label{fig:F34}
\end{figure*}

\begin{figure*}
\includegraphics[width=1.0\linewidth]{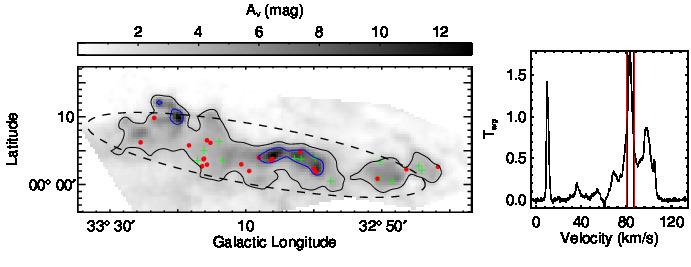}
\caption{Overview of GMF 30 (G033.104+0.068). Others are same as Fig.~\ref{fig:LG015p65}.}
\label{fig:LG033p10}
\end{figure*}
\clearpage

\begin{figure*}
\includegraphics[width=1.0\linewidth]{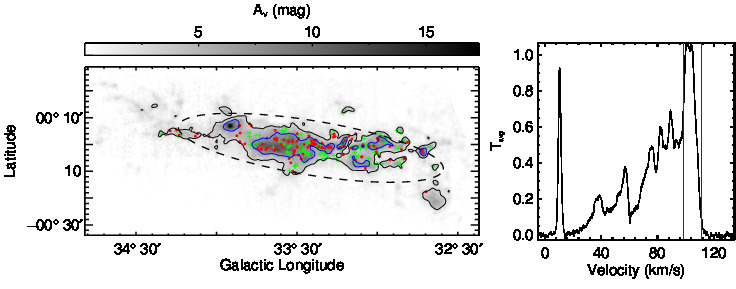}
\caption{Overview of GMF 31 (G033.685-0.020). Others are same as Fig.~\ref{fig:LG015p65}.}
\label{fig:LG033p68}
\end{figure*}

\begin{figure*}
\includegraphics[width=1.0\linewidth]{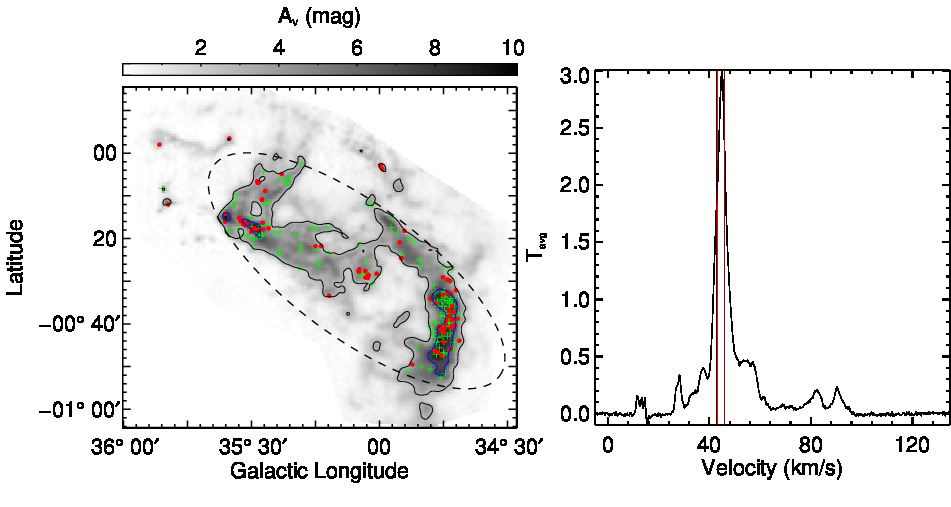}
\caption{Overview of GMF 32 (GMF38.1-32.4b). Others are same as Fig.~\ref{fig:LG015p65}.}
\label{fig:GMF38b}
\end{figure*}
\clearpage

\begin{figure*}
\includegraphics[width=1.0\linewidth]{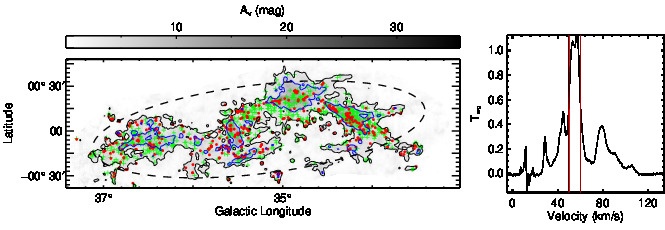}
\caption{Overview of GMF 33 (GMF38.1-32.4a). Others are same as Fig.~\ref{fig:LG015p65}.}
\label{fig:GMF38a}
\end{figure*}

\begin{figure*}
\includegraphics[width=1.0\linewidth]{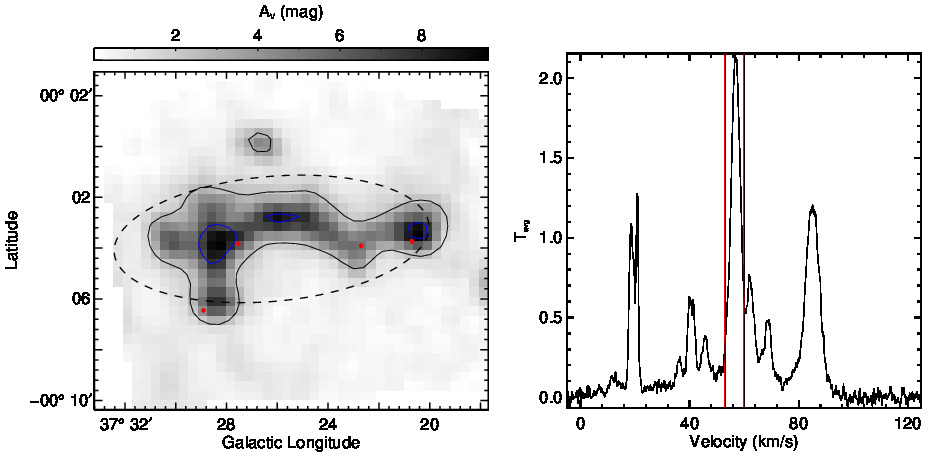}
\caption{Overview of GMF 34 (G037.410-0.070). Others are same as Fig.~\ref{fig:LG015p65}.}
\label{fig:LG037p41}
\end{figure*}
\clearpage

\begin{figure*}
\includegraphics[width=1.0\linewidth]{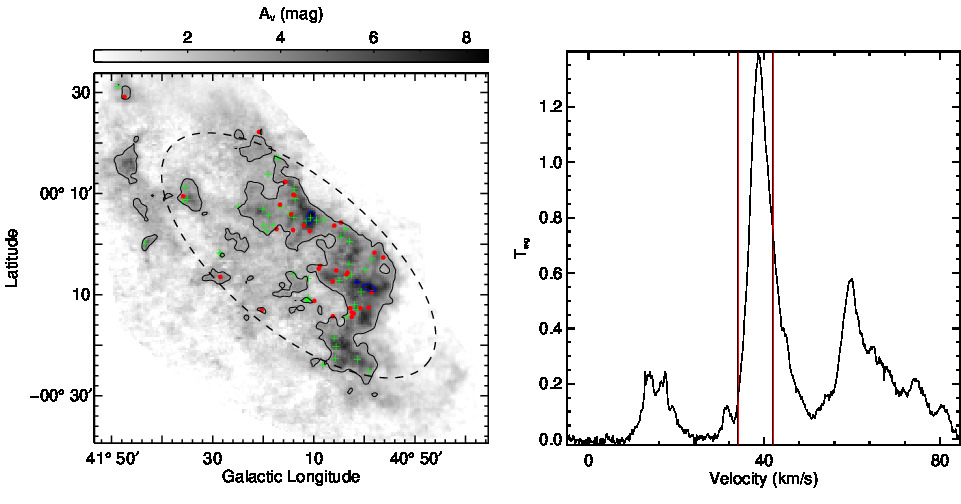}
\caption{Overview of GMF 35 (GMF41.0-41.3). Others are same as Fig.~\ref{fig:LG015p65}.}
\label{fig:GMF41}
\end{figure*}

\begin{figure*}
\includegraphics[width=1.0\linewidth]{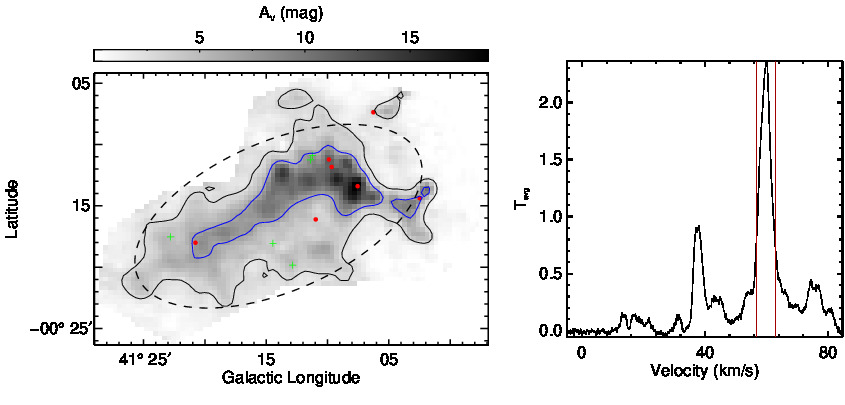}
\caption{Overview of GMF 36 (F38). Others are same as Fig.~\ref{fig:LG015p65}.}
\label{fig:F38}
\end{figure*}
\clearpage

\begin{figure*}
\includegraphics[width=1.0\linewidth]{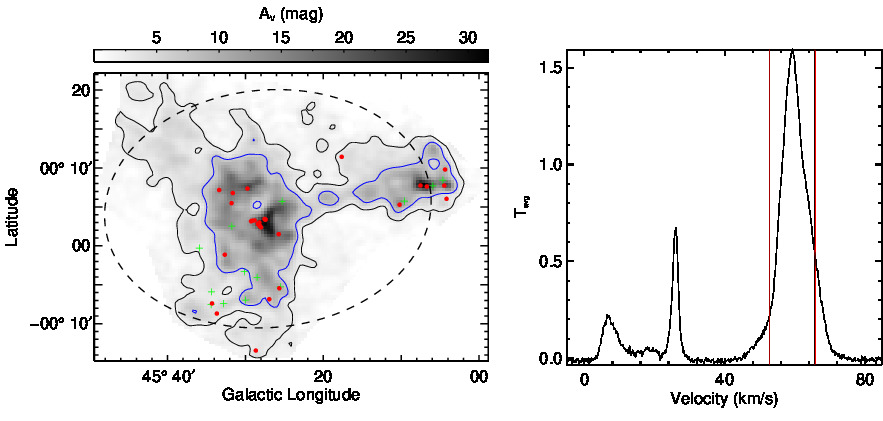}
\caption{Overview of GMF 37 (F39). Others are same as Fig.~\ref{fig:LG015p65}.}
\label{fig:F39}
\end{figure*}

\begin{figure*}
\includegraphics[width=1.0\linewidth]{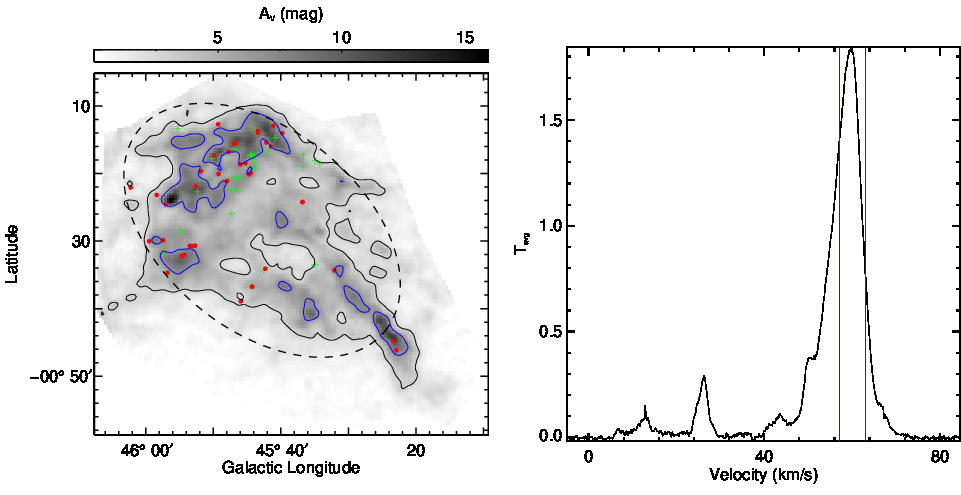}
\caption{Overview of GMF 38 (F40). Others are same as Fig.~\ref{fig:LG015p65}.}
\label{fig:F40}
\end{figure*}
\clearpage

\begin{figure*}
\includegraphics[width=1.0\linewidth]{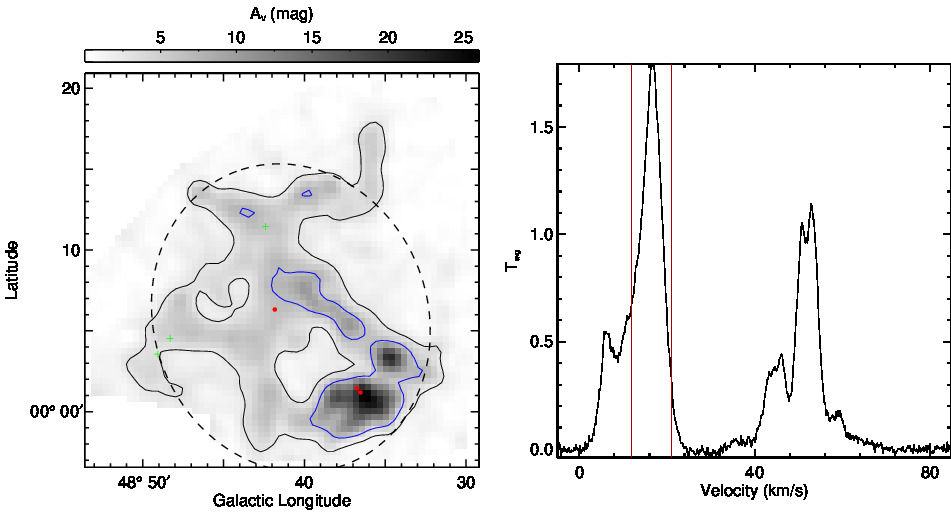}
\caption{Overview of GMF 40 (G048.629+0.096). Others are same as Fig.~\ref{fig:LG015p65}.}
\label{fig:LG048p62}
\end{figure*}

\begin{figure*}
\includegraphics[width=1.0\linewidth]{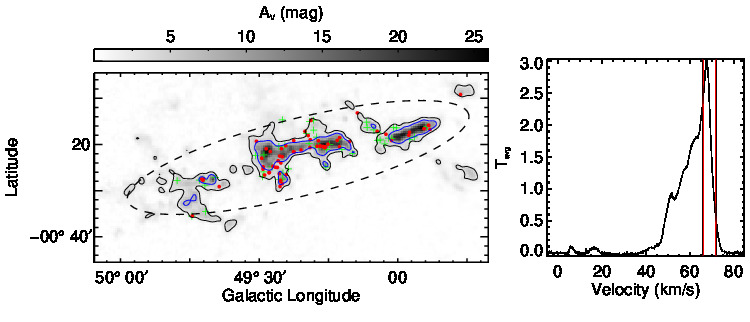}
\caption{Overview of GMF 41 (CFG049.21-0.34). Others are same as Fig.~\ref{fig:LG015p65}.}
\label{fig:CFG49}
\end{figure*}
\clearpage

\begin{figure*}
\includegraphics[width=1.0\linewidth]{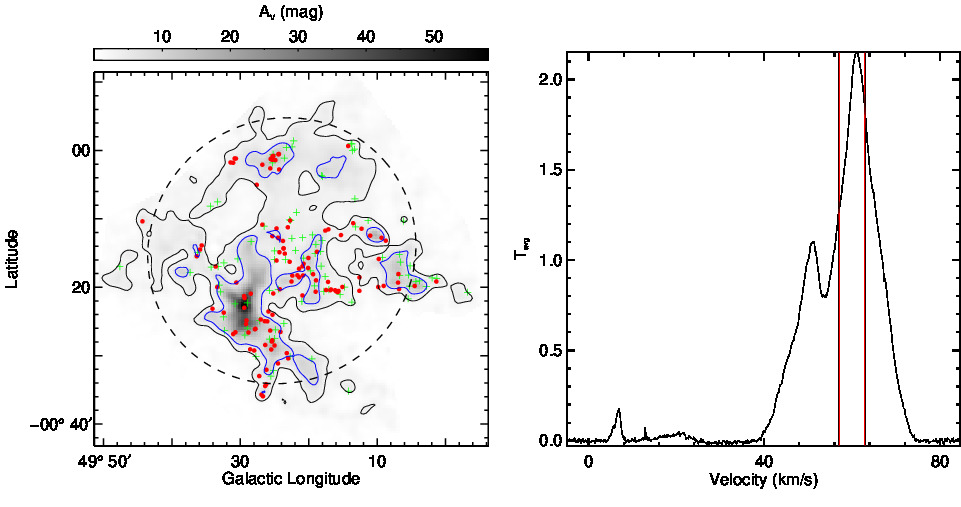}
\caption{Overview of GMF 42 (F42). Others are same as Fig.~\ref{fig:LG015p65}.}
\label{fig:F42}
\end{figure*}

\begin{figure*}
\includegraphics[width=1.0\linewidth]{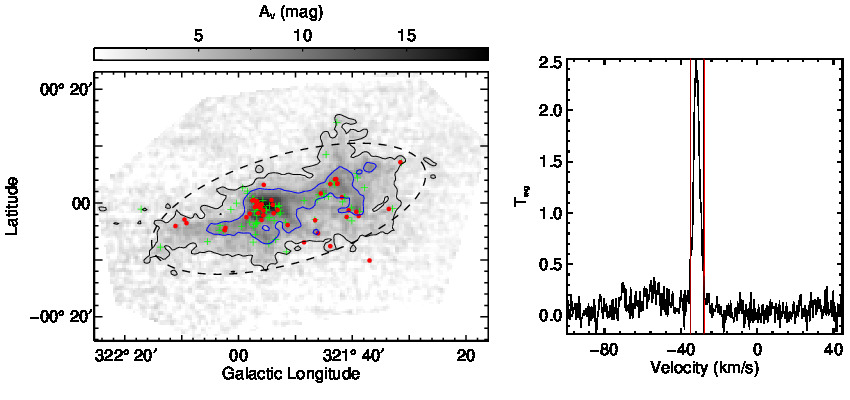}
\caption{Overview of GMF 47 (GMF324.5-321.4b). Others are same as Fig.~\ref{fig:LG015p65}.}
\label{fig:GMF324b}
\end{figure*}
\clearpage

\begin{figure*}
\includegraphics[width=1.0\linewidth]{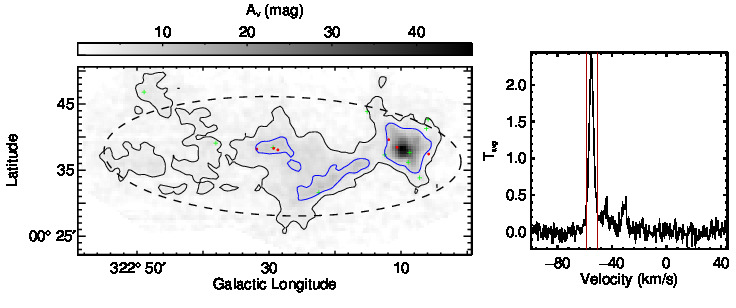}
\caption{Overview of GMF 48 (G322.363+0.542). Others are same as Fig.~\ref{fig:LG015p65}.}
\label{fig:LG322p36}
\end{figure*}

\begin{figure*}
\includegraphics[width=1.0\linewidth]{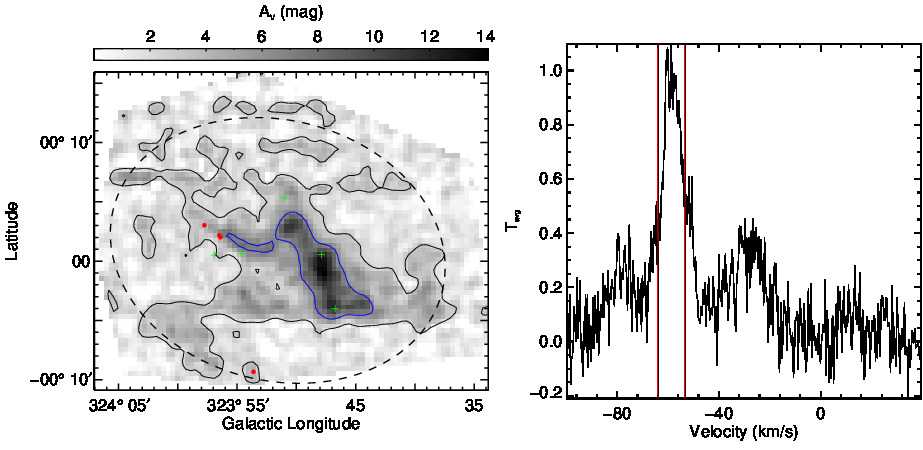}
\caption{Overview of GMF 49 (G323.929+0.036). Others are same as Fig.~\ref{fig:LG015p65}.}
\label{fig:LG323p92}
\end{figure*}
\clearpage

\begin{figure*}
\includegraphics[width=1.0\linewidth]{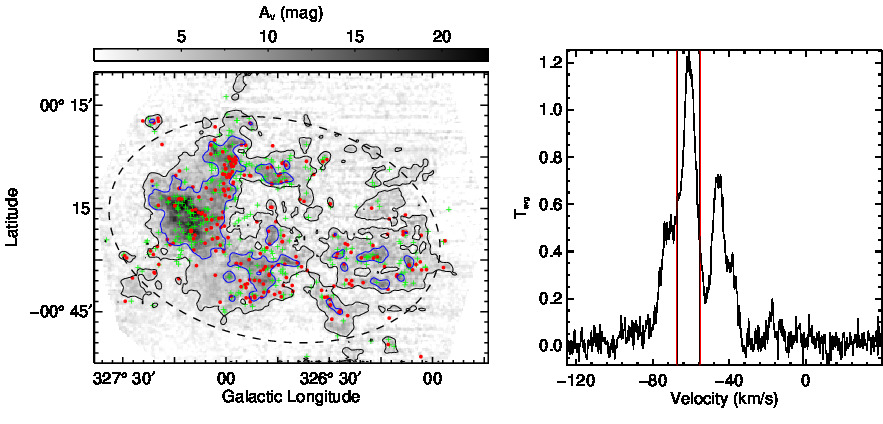}
\caption{Overview of GMF 50 (G327.157-0.256). Others are same as Fig.~\ref{fig:LG015p65}.}
\label{fig:LG327p15}
\end{figure*}

\begin{figure*}
\includegraphics[width=1.0\linewidth]{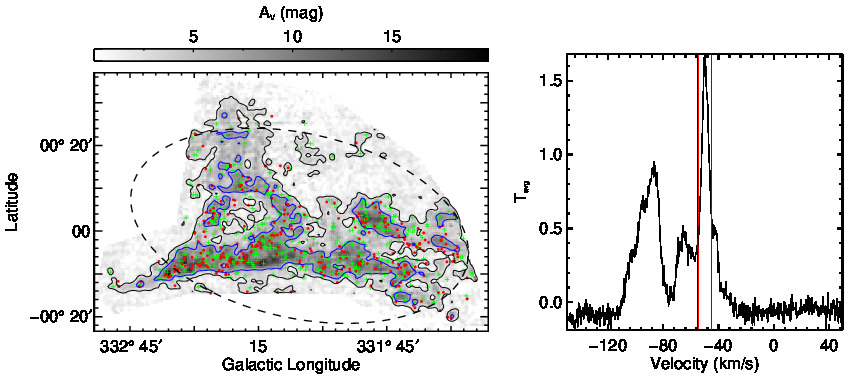}
\caption{Overview of GMF 51 (GMF335.6-333.6b). Others are same as Fig.~\ref{fig:LG015p65}.}
\label{fig:GMF335b}
\end{figure*}
\clearpage

\begin{figure*}
\includegraphics[width=1.0\linewidth]{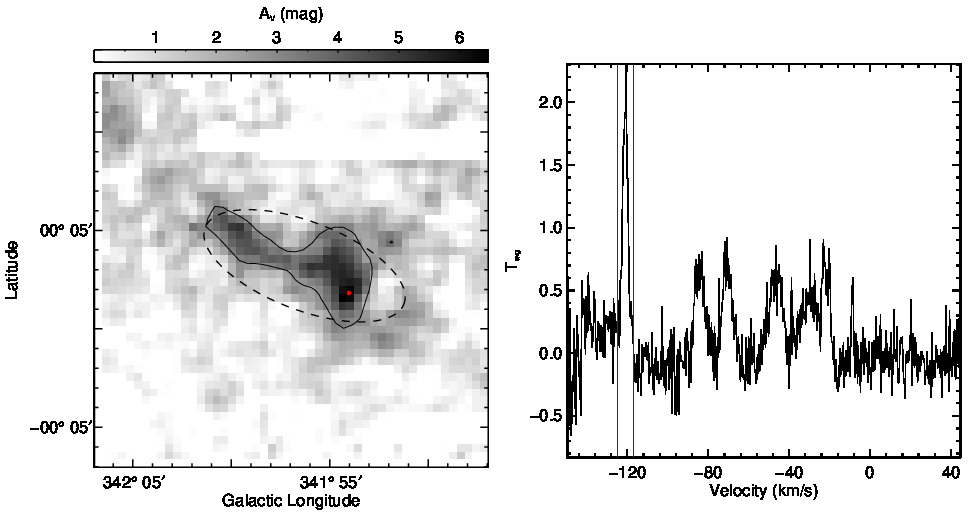}
\caption{Overview of GMF 55 (G341.938+0.054). Others are same as Fig.~\ref{fig:LG015p65}.}
\label{fig:LG341p93}
\end{figure*}
\clearpage

\begin{figure*}
\includegraphics[width=1.0\linewidth]{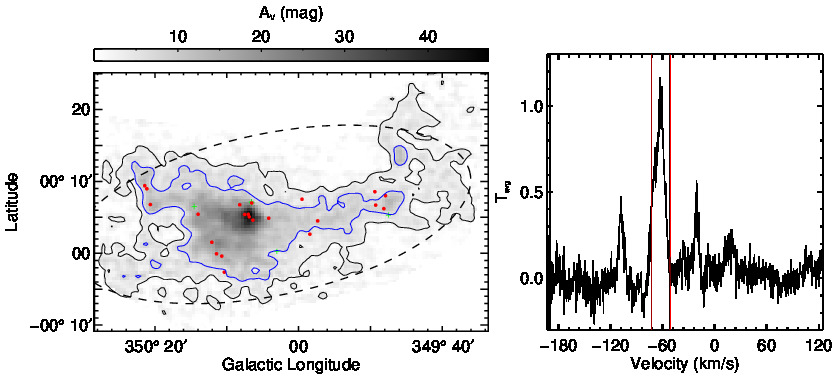}
\caption{Overview of GMF 56 (G349.876+0.099). Others are same as Fig.~\ref{fig:LG015p65}.}
\label{fig:LG349p87}
\end{figure*}

\begin{figure*}
\includegraphics[width=1.0\linewidth]{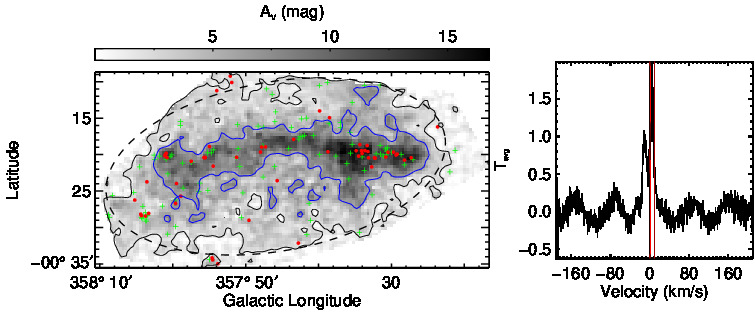}
\caption{Overview of GMF 57 (GMF358.9-357.4). Others are same as Fig.~\ref{fig:LG015p65}.}
\label{fig:GMF358}
\end{figure*}
\clearpage

\end{appendix}
\end{document}